# Kind Inference for Datatypes: Technical Supplement

NINGNING XIE, The University of Hong Kong, China
RICHARD A. EISENBERG, Bryn Mawr College, USA and Tweag I/O
BRUNO C. D. S. OLIVEIRA, The University of Hong Kong, China

This technical supplement to *Kind Inference for Datatypes* serves to expand upon the text in the main paper. It contains detailed typing rules, proofs, and connections to the Glasgow Haskell Compiler (GHC). Sections in this document are meant to connect to sections in the main paper. There are many hyperlinks throughout, especially those highlighting the connections to GHC; you may wish to read on a computer instead of on paper.

## A  OTHER LANGUAGE EXTENSIONS

This section accompanies Section 8 of the main paper, including discussion about more related language extensions. These extensions affect kind inference, but not in a fundamental way.

### A.1  Visible Dependent Quantification

Besides specified type variables for which users can optionally provide type arguments, Haskell also incorporates *visible dependent quantification (VDQ)*[1], e.g., **type** $T :: \forall(k :: \star) \to k \to \star$, with which users are forced to provide type arguments to $T$. That is, one would use $T$ with, e.g., $T \star Int$ and $T (\star \to \star)$ *Maybe*, never just $T$ *Int*. Visible dependent quantification is Haskell's equivalent to routine dependent quantification in dependently typed languages.

To support VDQ, rule DT-TT needs to be extended, as VDQ brings variables into scope for later reference. For example, given

**type** $T :: \forall(k :: \star) \to k \to \star$
**data** $T\ k\ a = MkT$

We should get a context $k :: \star, a :: k$ when checking $MkT$.

VDQ opens an interesting design choice: should unannotated type variables be able to introduce VDQ? For example, in the definition of $P$ below, we use $f$ and $a$ as the arguments to $T$. To make it type-check, we need to infer $P :: \forall(f :: \star) \to f \to \star$.

---

[1]In GHC 8.8, GHC infers kinds using VDQ, but users are not allowed to write VDQ explicitly. This has been rectified for the GHC 8.10 release, as described in this proposal.

Authors' addresses: Ningning Xie, The University of Hong Kong, Department of Computer Science, Hong Kong, China, nnxie@cs.hku.hk; Richard A. Eisenberg, Bryn Mawr College, Department of Computer Science, Bryn Mawr, PA, USA, Tweag I/O, rae@richarde.dev; Bruno C. d. S. Oliveira, The University of Hong Kong, Department of Computer Science, Hong Kong, China, bruno@cs.hku.hk.

**data** $P\ f\ a = MkP\ (T\ f\ a)$

However, the tricky part with inferring the kind of $P$ is that we cannot have a fixed initial form of the kind of $P$, i.e., $\widehat{\alpha} \to \widehat{\beta} \to \star$ or $\forall (f : \widehat{\alpha}) \to \widehat{\beta} \to \star$, when type-checking the **rec** group of $P$, until we type-check $P$'s body. In order to avoid this challenge, we support GHC's current ruling on the matter: *dependent variables must be manifestly so*. That is, the initial kind of a datatype includes VDQ only for those variables that appear, lexically, in the kind of a variable; other type parameters are reflected in a datatype's initial kind with a regular (non-dependent) arrow. This guideline rejects $P$ as an example of non-manifest dependency.

### A.2 Datatype Promotion

Haskellers can use datatypes as kinds and can write data constructors in types [Yorgey et al. 2012]. In the PolyKinds system, types and kinds are mixed (allowing datatypes to be used as kinds), but there is no facility to use a data constructor in a type.

To support such usage, the kinding judgment must now use the term context to fetch the type of data constructors. Moreover, dependency analysis needs to take dependencies on data constructors into account.

**Definition A.1** (Dependency Analysis with Type-Level Data). *We extend Definition 6.1 with*

(iii) *The definition of $T_1$ depends on the definition of $T_2$ if $T_1$ uses data constructors of $T_2$.*

While the appearance of data constructors in types enriches the type language considerably, they do not pose a particular challenge for inference; the rest of our presentation would remain unaffected.

### A.3 Partial Type Signatures

For quite some time, GHC has supported kind signatures on a subset of a datatype's parameters, much like the partial type signatures described by **?**. For example, *App*, below, does not have a signature but still has a kind annotation for $f$.

**data** $App\ (f :: \star \to \star)\ a = A\ (f\ a)$

To deal with such a construct we first need to amend the syntax of a datatype declaration to support kind annotations for variables.

$$\text{datatype decl.} \quad \mathcal{T} \quad ::= \quad \textbf{data}\ T\ \phi = \overline{\mathcal{D}_j}^j$$

Kind annotations can also contain free variables, which need to be generalized in a similar way as signatures. For example, *T2* has kind $\forall \{k :: \star\}.\ \forall (f :: k).\ \star$.

**data** $T2\ (f :: k) = MkT2$

Supporting these partial signatures adds complication to rule PGM-DT-TT (and its algorithmic counterpart) to bring the kind variables into scope. However, and critically, a partial signature will still go via rule PGM-DT-TT, never rule PGM-DT-TTS, used for full signatures only. This means that a partial type signature does *not* unlock polymorphic recursion: the datatype will considered monomorphic and ungeneralized within its own recursive group.

## B  TODAY'S GHC

This paper describes, in depth, how kind inference can work for datatype declarations. Here, we review how our work relates to GHC. To make the claims concrete, this section contains references to specific stretches of code within GHC.

## B.1 Constraint-Based Type Inference

Type inference in GHC is based on generating and solving constraints [Pottier and Rémy 2005; Vytiniotis et al. 2011], distinct from our approach here, where we unify on the fly. Despite this different architecture, our results carry over to the constraint-based style. Instead of using eager unification, we can imagine accumulating constraints in output contexts Θ, and then invoking a solver to extend the context with solutions. This approach is taken by Eisenberg [2016].

In thinking about the change from eager unification to delayed constraints, one might worry about information loss around any place where we apply a context as a substitution, as these substitutions would be empty in a constraint-solving approach without eager unification. At top-level (Figure 6), a constraint-solving approach would run the constraint solver, and the substitutions would contain the same mappings as our approach provides. Conversely, the relations in Figure 8 would become part of the constraint solver, so substituting here is safe, too. A potential problem arises in rule rule A-KTT-APP (Figure 7), where we substitute in the function's kind before running the kind-directed ⊩$^{\text{kapp}}$ judgment. However, our system is predicative: it never unifies a type variable with a polytype. Thus, the substitution in rule A-KTT-APP can never trigger a new usage of rule A-KAPP-TT-FORALL. It *can* distinguish between rule A-KAPP-TT-ARROW and rule A-KAPP-TT-KUVAR, but we conjecture that the choice between these rules is irrelevant: both will lead to equivalent substitutions in the end.

## B.2 Contexts

A typing context is *not* maintained in much of GHC's inference algorithm. Instead, a variable's kind is stored in the data structure representing the variable. This is very convenient, as it means that looking up a variable's type or kind is a pure, fast operation. One downside is that the compiler must maintain an extra invariant that all occurrences of a variable store the same kind; this is straightforward to maintain in practice.

Beyond just storing variables' kinds, the typing context in this paper also critically stores variables' ordering. Lacking contexts, GHC uses a different mechanism: *level numbers*, originally invented to implement untouchability [Vytiniotis et al. 2011, Section 5.1]. Every type variable in GHC is assigned a level number during inference. Type variables contain a structure that includes level numbers. Roughly, the level number of a type variable $a$ corresponds to the number of type variables in scope before $a$. Accordingly, we can tell the relative order (in a hypothetical context, according to the systems in this paper) of two variables simply by comparing their level numbers. One of GHC's invariants is that a unification variable at level $n$ is never unified with a type that mentions a variable with a level number $m > n$; this is much like the extra checks in the unification judgments in our paper.

The *local scopes* of this paper are also tracked by GHC. All the variables in the same local scope are assigned the same level number, and they are flagged as reorderable. After inference is complete, GHC does a topological sort to get the final order.

A final role that contexts play in our formalism is that they store solutions for unification variables; we apply contexts as a substitution. In GHC, unification variables store mutable cells that get filled in. It has a process called *zonking*,[2] which is exactly analogous to our use of contexts as substitutions. Zonking a unification variable replaces the variable with its solution, if any.

---

[2]There are actually two variants of zonking in GHC: we can zonk during type-checking or at the end. The difference between the variants is chiefly what to do for an unfilled unification variable. The former leaves them alone, while the latter has to default them somehow; details are beyond our scope here.

## B.3 Unification

The solver in GHC still has to carry out unification, much along the lines of the unification judgment we present here. This algorithm has to deal with the heterogeneous unification problems we consider, as well. Indeed, GHC's unification algorithm recurs into the kinds of a unification variable and the type it is unifying with, just as ours does. As implied by our focus on decidability of unification, there have been a number of bugs in GHC's implementation that led to loops in the type checker; the most recent is #16902.

GHC actually uses several unification algorithms internally. It has an eager unifier, much like the one we describe. When that unifier fails, it generates the constraint that is sent to the solver. (The eager unifier is meant solely to be an optimization.) There is also a unifier meant to work after type inference is complete; it checks for instance overlap, for example. All the unifiers recur into kinds:

- The eager unifier recurs into kinds.
- The unifier in the solver recurs into kinds.
- The pure unifier uses an invariant that the kinds are related before looking at the types. It must recur when decomposing applications.

In addition, GHC also has an overlap problem within unification, as exhibited in our paper by the overlap between rules a-u-kvarL and a-u-kvarR in Figure 3. Both the eager unifier and the constraint-solver unifier deal with this ambiguity by using heuristics to choose which variable might be more suitable for unification. This particular issue—which variable to unify when there is a choice—has been the subject of some amount of churn over the years.

## B.4 Promotion

The promotion operation, too, is present in GHC, though its form is quite different than what we have presented. Instead of promoting during unification, GHC simply refuses to solve a unification variable if any of the free variables of its supposed solution lives to the right of the variable in the context. Because GHC is working with constraints, it just leaves the unification problem as an unsolved constraint. If there remain unsolved constraints, GHC then promotes the variables it can: some cannot be promoted because they depend on locally bound quantified (not unification) type variables.

## B.5 Complete User-Supplied Kinds

Along with stand-alone kind signatures, as described in this paper, GHC supports *complete user-supplied kinds*, or CUSKs. A datatype has a CUSK when certain syntactic conditions are satisfied; GHC detects these conditions *before* doing any kind inference. These CUSKs are a poor substitute for proper kind signatures, as the syntactic cues are fragile and unexpected: users sometimes write a CUSK without meaning to, and also sometimes leave out a necessary part of a CUSK when they intend to specify the kind. Stand-alone kind signatures are a new feature; they begin with the keyword **type** instead of **data**, as we have used in our paper.

Interestingly, it would be wrong to support CUSKs in a system without polymorphic kinds. Consider this example:

**data** *S1 a* = *MkT1 S2*
**data** *S2* = *MkS2* (*S1 Maybe*)

The types *S1* and *S2* form a group. We put *S2* (which has a CUSK) into the context with kind $\star$. When we check *S1*, we find no constraints on *a* (in the constraint-generation pass; see the general approach below). The kind of *S1* is then defaulted to $\star \to \star$. Checking *S2* fails. Instead, we wish to pretend that *S2* does not have a CUSK. This would mean that constraint-generation happens for all the constructors in both *S1* and *S2*, and *S1* would get its correct kind ($\star \to \star) \to \star$.

With kind-polymorphism, we have no problem because the kind of *T1* will be generalized to $\forall (k :: \star).\ k \to \star$.

This was reported as bug #16609.

### B.6 Dependency Analysis

The algorithm implemented in GHC for processing datatype declarations starts with dependency analysis, as ours does. The dependency analysis is less fine-grained than what we have proposed in this paper: signatures are ignored in the dependency analysis, and so datatypes with signatures are processed alongside all the others. This means that the kinds in the example below have more restrictive kinds in GHC than they do in our system:

**data** *S1* :: $\forall k.\ k \to \star$
**data** *S1 a* = *MkS1* (*S2 Int*)
**data** *S2 a* = *MkS2* (*S3 Int*)
**data** *S3 a* = *MkS3* (*S1 Int*)

A naïve dependency analysis would put all three definitions in the same group. The kind for *S1* is given; it would indeed have that kind. The parameters of *S2* and *S3* would initially have an unknown kind, but when occurrences of *S2* and *S3* are processed (in the definitions of *S1* and *S2*, respectively), this unknown kind would become $\star$. Neither *S2* nor *S3* would be generalized.

There is a ticket to improve the dependency analysis: #9427.

### B.7 Approach to Kind-Checking Datatypes

GHC's approach is summarized in this comment. Overall kind-checking is orchestrated by this function.

After dependency analysis, so-called *initial kinds* are produced for all the datatypes in the group. These either come from a datatype's CUSK or from a simple analysis of the header of the datatype (without looking at constructors). This step corresponds to our algorithm's placing a binding for the datatype in the context, either with the kind signature or with a unification variable (rules A-PGM-DT-TTS and A-PGM-DT-TT).

If there is no CUSK, GHC then passes over all the datatype's constructors, collecting constraints on unification variables. After solving these constraints, GHC generalizes the datatype kind.

For all datatypes, now with generalized kinds, all data constructors are checked (again, for non-CUSK types). Because the kinds of the types are now generalized, this pass infers any invisible parameters to polykinded types. For non-CUSK types, this second pass using generalized kinds replaces the $T_i \mapsto T_i @\phi_i^c$ substitution in the context in the last premise to rule A-PGM-DT-TT. Performing a substitution—instead of re-generating and solving constraints—may be an opportunity for improvement in GHC.

### B.8 Syntax for GADTs

Haskell's *syntax* for GADT declarations is very troublesome. Consider these examples:

**data** *R a* **where**
   *MkR* :: $b \to R\ b$
**data** *S a* **where**
   *MkS* :: *S b*
**data** *T a* **where**
   *MkT* :: $\forall (k :: \star)\ (b :: k).\ T\ b$

In GHC's implementation of GADTs, any variables declared in the header (between **data** and **where**) *do not scope*. In all the examples above, the type variable *a* does not scope over the constructor declarations. This is why we have written the variable *b* in those types, to make it clear that *b* is distinct from *a*. We could have written *a*—it would still be a distinct *a* from that in the header—but it would be more confusing.

The question is: how do we determine the kind of the parameter to the datatype? One possibility is to look only in the header. In all cases above, we would infer no constraints and would give each type a kind of $\forall (k :: \star). k \to \star$. This is unfortunate, as it would make *R* a kind-indexed GADT: the *MkR* constructor would carry a proof that the kind of its type parameter is $\star$. This, in turn, wreaks havoc with type inference, as it is hard to infer the result type of a pattern-match against a GADT [Vytiniotis et al. 2011].

Furthermore, this approach might accept *more* programs than the user wants. Consider this definition:

**data** *P a* **where**
  *MkP1* :: $b \to P\ b$
  *MkP2* :: $f\ a \to P\ f$

Does the user want a kind-indexed GADT, noting that *b* and *f* have different kinds? Or would the user want this rejected? If we make the fully general kind $\forall k. k \to \star$ for *P*, this would be accepted, perhaps surprising users.

It thus seems we wish to look at the data constructors when inferring the kind of the datatype. The challenge in looking at data constructors is that their variables are *locally* bound. In *MkR* and *MkS*, we implicitly quantify over *b*. In *MkR*, we discover that $b :: \star$, and thus that *R* must have kind $\star \to \star$. In *MkS*, we find no constraints on *b*'s kind, and thus no constraints on *S*'s argument's kind, and so we can generalize to get $S :: \forall (k :: \star). k \to \star$. Let us now examine *MkT*: it explicitly brings *k* and *b* into scope. Thus, the argument to *T* has *local* kind *k*. It would be impossible to unify the kind of *T*'s argument—call it $\widehat{\alpha}$—with *k*, because *k* would be bound to the *right* of $\widehat{\alpha}$ in an inference context. Thus it seems we would reject *T*.

This result is also dissatisfying. In practice, GHC implements an ad-hoc algorithm, described in Section B.9.

Our conclusion here is that the design of GADTs in GHC/Haskell is flawed: the type variables mentioned in the header should indeed scope over the constructors. This would mean we could reject *T*: if the user wanted to explicitly make *T* polykinded, they could do so right in the header. We recognize that it would be hard to make this change today, but one result of this work is the interplay between scoping (order in the context) and unification; the current state of affairs will always require ad-hoc support.

### B.9 Polymorphic Recursion

One challenge in kind inference is in the handling of polymorphic recursion. Although non-CUSK types are indeed monomorphic during the constraint-generation pass, some limited form of polymorphic recursion can get through. This is because all type variables are represented by a special form of unification variable called a TyVarTv. TyVarTvs can unify only with other type variables. This design is motivated by the following examples:

**data** *T1* ($a :: k$) $b = MkT1\ (T2\ a\ b)$
**data** *T2* ($c :: j$) $d = MkT2\ (T1\ c\ d)$
**data** *T3 a* **where**
  *MkT3* :: $\forall (k :: \star)\ (b :: k). T3\ b$

We want to accept all of these definitions. The first two, *T1* and *T2*, form a mutually recursive group. Neither has a CUSK. However, the recursive occurrences are not polymorphically recursive: both recursive occurrences are at the *same* kind as the definition. Yet the first parameter to *T1* is declared to have kind *k* while the first parameter to *T2* is declared to have kind *j*. The solution: allow *k* to unify with *j* during the constraint-generation pass. We would *not* want to allow either *k* or *j* to unify with a non-variable, as that would seem to go against the user's wishes. But they must be allowed to unify with each other to accept this example.

With *T3* (identical to *T* from Section B.8), we have a different motivation. During inference, we will guess the kind of *a*; call it $\hat{\alpha}$. When checking the *MkT3* constructor, we will need to unify $\hat{\alpha}$ with the locally bound *k*. We cannot set $\hat{\alpha} := k$, as that will fill $\hat{\alpha}$ with a *k*, bound to $\hat{\alpha}$'s *right* in the context. Instead, we must set $k := \hat{\alpha}$. This is possible only if *k* is represented by a unification variable.

There are two known problems with this approach:

(1) It sometimes accepts polymorphic recursion, even without a CUSK. Here is an example:

   **data** *T4 a* = ∀(*k* :: ★) (*b* :: *k*). *MkT4* (*T4 b*)

   The definition of *T4* is polymorphically recursive: the occurrence *T4 b* is specialized to a kind other than the kind of *a*. Yet this definition is accepted. The two kinds unify (as *k* becomes a unification variable, set to the guessed kind of *a*) during the constraint-generation pass. Then, *T4* is generalized to get the kind ∀*k*. *k* → ★, at which point the last pass goes through without a hitch.

   The reason this acceptance is troublesome is not that *T4* is somehow dangerous or unsafe. It is that we know that polymorphic recursion cannot be inferred [Henglein 1993], and yet GHC does it. Invariably, this must mean that GHC's algorithm will be hard to specify beyond its implementation.

   This wrinkle is described on the GHC wiki.

(2) In rare cases, the constraint-generation pass will succeed, while the final pass—meant to be redundant—will fail. Here is an example:

   **data** *SameKind* :: *k* → *k* → *Type*
   **data** *Bad a* **where**
     *MkBad* :: ∀*k*$_1$ *k*$_2$ (*a* :: *k*$_1$) (*b* :: *k*$_2$). *Bad* (*SameKind a b*)

   During the constraint-generation pass, the kinds *k*$_1$ and *k*$_2$ are allowed to unify, accepting the definition of *Bad*. During the final pass, however, *k*$_1$ and *k*$_2$ are proper quantified type variables, always distinct. Thus the *SameKind a b* type is ill-kinded and rejected.

   The fact that this final pass can fail means that we cannot implement it via a simple substitution, as we do in rule A-PGM-DT-TT. One possible solution is our suggestion to change the scoping of type parameters to GADT-syntax datatype declarations. With that change, our second motivation above for TyVarTvs would disappear. GHC could then use TyVarTvs only for kind variables in the head of a datatype declaration, using proper quantified type variables in constructors. Of course, this change would break much code in the wild, and we do not truly expect it to ever be adopted.

   This problem is documented in this comment.

## B.10 The Quantification Check

Our quantification check (Section 7.2) also has a parallel in GHC, but GHC's solution to the problem differs from ours. Instead of rejecting programs that fail the quantification check, GHC accepts them, replacing the variables that would be (but cannot be) quantified with its constant *Any* :: ∀*k*. *k*. The *Any* type is uninhabited, but exists at all kinds. As such, it is an appropriate replacement for unquantifiable, unconstrained unification variables. Yet this decision in GHC has unfortunate consequences: the *Any* type can appear in error messages, and its introduction induces hard-to-understand type errors.

The GHC developers are questioning their approach to this problem. See this comment and this ticket.

Another design alternative is to generalize the variable to the leftmost position where it is still well-formed. Recall the example in Section 7.2:

**data** *Proxy* :: ∀$k$. $k$ → ★
**data** *Relate* :: ∀$a$ ($b$ :: $a$). $a$ → *Proxy* $b$ → ★
**data** *T* :: ∀($a$ :: ★) ($b$ :: $a$) ($c$ :: $a$) $d$. *Relate* $b$ $d$ → ★

We have $d$ :: $\widehat{\alpha}$, and $\widehat{\alpha}$ = *Proxy* $\widehat{\beta}$, with $\widehat{\beta}$ :: $a$. As there are no further constraints on $\widehat{\beta}$, the definition of *T* is rejected by the quantification check.

Instead of rejecting the program, or solving $\widehat{\beta}$ using *Any*, we can generalize over $\widehat{\beta}$ as a fresh variable $f$, which is put after $a$ to make it well-kinded. Namely, we get

**data** *T* :: ∀($a$ :: ★) {$f$ :: $a$} ($b$ :: $a$) ($c$ :: $a$) ($d$ :: *Proxy* $f$). *Relate* @$a$ @$f$ $b$ $d$ → ★

However, this ordering of the variables violates our declarative specification. Moreover, this type requires an inferred variable to be between specified variables. With higher-rank polymorphism, due to the fact that GHC does not support first-class type-level abstraction (i.e., Λ in types), this type cannot be instantiated to

∀($a$ :: ★) ($b$ :: $a$) ($c$ :: $a$) ($d$ :: *Proxy* $f$). *Relate* @$a$ @$b$ $b$ $d$ → ★

or

∀($a$ :: ★) ($b$ :: $a$) ($c$ :: $a$) ($d$ :: *Proxy* $f$). *Relate* @$a$ @$c$ $b$ $d$ → ★

which makes the generalization less useful.

### B.11 ScopedSort

When GHC deals with a local scope—a set of variables that may be reordered—it does a topological sort on the variables at the end. However, not any topological sort will do: it must use one that preserves the left-to-right ordering of the variables as much as possible. This is because GHC considers these implicitly bound variables to be *specified*: they are available for visible type application. For example, recall the example from Section 2.2, modified slightly:

**data** *Q* ($a$ :: ($f$ $b$)) ($c$ :: $k$) ($x$ :: $f$ $c$)

Inference will tell us that $k$ must come before $f$ and $b$, but the order of $f$ and $b$ is immaterial. Our approach here is to make $f$, $b$, and $k$ *inferred* variables: users of *Q* will not be able to instantiate these parameters with visible type application. However, GHC takes a different view: because the user has written the names of $f$, $b$, and $k$, they will be *specified*. This choice means that the precise sorting algorithm GHC uses to fix the order of local scopes becomes part of the *specification* of the language. Indeed, GHC documents the precise algorithm in its manual. If we followed suit, the algorithm would have to appear in our declarative specification, which goes against the philosophy of a declarative system.

Some recent debate led to a conclusion that we would change the interpretation of the *Q* example from the main paper, meaning that its kind variables would indeed become *inferred*. However, the problem with ScopedSort still exists in type signatures, where type variables may be implicitly bound.

## B.12 The "Forall-or-Nothing" Rule

GHC implements the so-called *forall-or-nothing* rule, which states that either *all* variables are quantified by a user-written forall, or none are. These examples illustrate the effect:

*ex1* :: $a \to b \to a$
*ex2* :: $\forall a\, b.\, a \to b \to a$
*ex3* :: $\forall a.\, a \to b \to a$
*ex4* :: $(\forall a.\, a \to b \to a)$

The signatures for both *ex1* and *ex2* are accepted: *ex1* quantifies none, while *ex2* quantifies all. The signature for *ex3* is rejected, as GHC rejects a mixed economy. However, and perhaps surprisingly, *ex4* is accepted. The only difference between *ex3* and *ex4* is the seemingly-redundant parentheses. However, because the forall-or-nothing rule applies only at the top level of a signature, the rule is not in effect for the $\forall$ in *ex4*.

This rule interacts with the main paper only in that our formalism (and some of our examples) does not respect it. This may be the cause of differing behavior between GHC and the examples we present.

## C COMPLETE SET OF RULES

In this section we include the complete set of rules. Some of the rules are repeated from those in the paper.

### C.1 Declarative Haskell98

$\boxed{\Sigma \vdash^k \sigma : \kappa}$ *(Kinding for Polymorphic Types)*

$$\frac{\Sigma, a : \kappa \vdash^k \sigma : \star}{\Sigma \vdash^k \forall a : \kappa.\sigma : \star} \text{{\sc k-forall}}$$

$\boxed{\Sigma \vdash \Psi}$ *(Well-formed Term Contexts)*

$$\frac{}{\Sigma \vdash \bullet} \text{{\sc ectx-empty}} \qquad \frac{\Sigma \vdash \Psi \quad \Sigma \vdash^k \sigma : \star}{\Sigma \vdash \Psi, D : \sigma} \text{{\sc ectx-dcon}}$$

### C.2 Algorithmic Haskell98

$\boxed{\Delta \Vdash^k \sigma : \kappa \dashv \Theta}$ *(Kinding for Polymorphic Types)*

$$\frac{\Delta \Vdash^{kv} \kappa \quad \Delta, a : \kappa \Vdash^k \sigma : \kappa_2 \dashv \Theta, a : \kappa \quad [\Theta]\kappa_2 = \star}{\Delta \Vdash^k \forall a : \kappa.\sigma : \star \dashv \Theta} \text{{\sc a-k-forall}}$$

$\boxed{\Delta \Vdash^{kc} \sigma \Leftarrow \kappa}$ *(Checking)*

$$\frac{\Delta \Vdash^k \sigma : \kappa_1 \dashv \Delta \quad [\Delta]\kappa_1 = [\Delta]\kappa_2}{\Delta \Vdash^{kc} \sigma \Leftarrow \kappa_2} \text{{\sc a-kc-eq}}$$

$\boxed{\Delta \Vdash^{kv} \kappa}$ *(Well-formed Kinds)*

$$\frac{}{\Delta \Vdash^{kv} \star} \text{{\sc a-kv-star}} \qquad \frac{\Delta \Vdash^{kv} \kappa_1 \quad \Delta \Vdash^{kv} \kappa_2}{\Delta \Vdash^{kv} \kappa_1 \to \kappa_2} \text{{\sc a-kv-arrow}} \qquad \frac{\widehat{\alpha} \in \Delta}{\Delta \Vdash^{kv} \widehat{\alpha}} \text{{\sc a-kv-kuvar}}$$

$\boxed{\Delta \text{ ok}}$ *(Well-formed Type Contexts)*

$$\frac{}{\bullet \text{ ok}} \text{ a-tctx-empty} \qquad \frac{\Delta \text{ ok} \quad \Delta \Vdash^{\mathsf{kv}} \kappa}{\Delta, a : \kappa \text{ ok}} \text{ a-tctx-tvar} \qquad \frac{\Delta \text{ ok} \quad \Delta \Vdash^{\mathsf{kv}} \kappa}{\Delta, T : \kappa \text{ ok}} \text{ a-tctx-tcon} \qquad \frac{\Delta \text{ ok}}{\Delta, \widehat{\alpha} \text{ ok}} \text{ a-tctx-kuvar}$$

$$\frac{\Delta \text{ ok} \quad \Delta \Vdash^{\mathsf{kv}} \kappa}{\Delta, \widehat{\alpha} = \kappa \text{ ok}} \text{ a-tctx-kuvarSolved}$$

$\boxed{\Delta \Vdash^{\mathsf{ectx}} \Gamma}$ *(Well-formed Term Contexts)*

$$\frac{}{\Delta \Vdash^{\mathsf{ectx}} \bullet} \text{ a-ectx-empty} \qquad \frac{\Delta \Vdash^{\mathsf{ectx}} \Gamma \quad \Delta \Vdash^{\mathsf{kc}} \sigma \Leftarrow \star}{\Delta \Vdash^{\mathsf{ectx}} \Gamma, D : \sigma} \text{ a-ectx-dcon}$$

$\boxed{\Delta \longrightarrow\!\!\!\!\rightarrow \Omega}$ *(Defaulting)*

$$\frac{}{\bullet \longrightarrow\!\!\!\!\rightarrow \bullet} \text{ a-ctxde-empty} \qquad \frac{\Delta \longrightarrow\!\!\!\!\rightarrow \Omega}{\Delta, a : \kappa \longrightarrow\!\!\!\!\rightarrow \Omega, a : \kappa} \text{ a-ctxde-tvar} \qquad \frac{\Delta \longrightarrow\!\!\!\!\rightarrow \Omega}{\Delta, T : \kappa \longrightarrow\!\!\!\!\rightarrow \Omega, T : \kappa} \text{ a-ctxde-tcon} \qquad \frac{\Delta \longrightarrow\!\!\!\!\rightarrow \Omega}{\Delta, \widehat{\alpha} = \kappa \longrightarrow\!\!\!\!\rightarrow \Omega, \widehat{\alpha} = \kappa} \text{ a-ctxde-kuvarSolved}$$

$$\frac{\Delta \longrightarrow\!\!\!\!\rightarrow \Omega}{\Delta, \widehat{\alpha} \longrightarrow\!\!\!\!\rightarrow \Omega, \widehat{\alpha} = \star} \text{ a-ctxde-solve}$$

## C.3 Context Application in Haskell98

$[\Delta]\kappa$ applies $\Delta$ as a substitution to $\kappa$.
$$\begin{aligned}
[\Delta]\star &= \star \\
[\Delta]\kappa_1 \to \kappa_2 &= [\Delta]\kappa_1 \to [\Delta]\kappa_2 \\
[\Delta[\widehat{\alpha}]]\widehat{\alpha} &= \widehat{\alpha} \\
[\Delta[\widehat{\alpha} = \kappa]]\widehat{\alpha} &= [\Delta[\widehat{\alpha} = \kappa]]\kappa
\end{aligned}$$

$[\Delta]\Gamma$ applies $\Delta$ as a substitution to $\Gamma$.
$$\begin{aligned}
[\Delta]\bullet &= \bullet \\
[\Delta](\Gamma, D : \sigma) &= [\Delta]\Gamma, D : [\Delta]\sigma
\end{aligned}$$

$[\Omega]\Delta$ applies $\Omega$ as a substitution to $\Delta$.
$$\begin{aligned}
[\bullet]\bullet &= \bullet \\
[\Omega, a : \kappa](\Delta, a : \kappa) &= [\Omega]\Delta, a : [\Omega]\kappa \\
[\Omega, T : \kappa](\Delta, T : \kappa) &= [\Omega]\Delta, T : [\Omega]\kappa \\
[\Omega, \widehat{\alpha} = \kappa](\Delta, \widehat{\alpha}) &= [\Omega]\Delta \\
[\Omega, \widehat{\alpha} = \kappa](\Delta, \widehat{\alpha} = \kappa') &= [\Omega]\Delta \quad \text{if } [\Omega]\kappa = [\Omega]\kappa' \\
[\Omega, \widehat{\alpha} = \kappa]\Delta &= [\Omega]\Delta \quad \text{if } \widehat{\alpha} \notin \Delta
\end{aligned}$$

## C.4 Context Extension in Haskell98

$\boxed{\Delta \longrightarrow \Theta}$ *(Context Extension)*

$$\frac{}{\bullet \longrightarrow \bullet} \text{ a-ctxe-empty} \qquad \frac{\Delta \longrightarrow \Theta}{\Delta, a : \kappa \longrightarrow \Theta, a : \kappa} \text{ a-ctxe-tvar} \qquad \frac{\Delta \longrightarrow \Theta}{\Delta, T : \kappa \longrightarrow \Theta, T : \kappa} \text{ a-ctxe-tcon} \qquad \frac{\Delta \longrightarrow \Theta}{\Delta, \widehat{\alpha} \longrightarrow \Theta, \widehat{\alpha}} \text{ a-ctxe-kuvar}$$

$$\frac{\Delta \longrightarrow \Theta \quad [\Theta]\kappa_1 = [\Theta]\kappa_2}{\Delta, \widehat{\alpha} = \kappa_1 \longrightarrow \Theta, \widehat{\alpha} = \kappa_2} \text{ a-ctxe-kuvarSolved} \qquad \frac{\Delta \longrightarrow \Theta \quad \Theta \Vdash^{\mathsf{kv}} \kappa}{\Delta, \widehat{\alpha} \longrightarrow \Theta, \widehat{\alpha} = \kappa} \text{ a-ctxe-solve} \qquad \frac{\Delta \longrightarrow \Theta}{\Delta \longrightarrow \Theta, \widehat{\alpha}} \text{ a-ctxe-add} \qquad \frac{\Delta \longrightarrow \Theta \quad \Theta \Vdash^{\mathsf{kv}} \kappa}{\Delta \longrightarrow \Theta, \widehat{\alpha} = \kappa} \text{ a-ctxe-addSolved}$$

## C.5 Declarative PolyKinds

$\boxed{\rceil \sigma \lceil}$ *(Kind results in $\star$)*

$$\frac{}{\rceil \star \lceil} \text{SR-STAR} \qquad \frac{\rceil \kappa_2 \lceil}{\rceil \kappa_1 \to \kappa_2 \lceil} \text{SR-ARROW} \qquad \frac{\rceil \sigma \lceil}{\rceil \forall \phi. \sigma \lceil} \text{SR-FORALL}$$

$\boxed{\Sigma \vdash^{\text{inst}} \mu_1 : \eta \sqsubseteq \omega \leadsto \mu_2}$ *(Instantiation)*

$$\frac{}{\Sigma \vdash^{\text{inst}} \mu : \omega \sqsubseteq \omega \leadsto \mu} \text{INST-REFL}$$

$$\frac{\Sigma \vdash^{\text{ela}} \rho : \omega_1 \qquad \Sigma \vdash^{\text{inst}} \mu_1 @ \rho : \eta[a \mapsto \rho] \sqsubseteq \omega_2 \leadsto \mu_2}{\Sigma \vdash^{\text{inst}} \mu_1 : \forall a : \omega_1.\eta \sqsubseteq \omega_2 \leadsto \mu_2} \text{INST-FORALL}$$

$$\frac{\Sigma \vdash^{\text{ela}} \rho : \omega_1 \qquad \Sigma \vdash^{\text{inst}} \mu_1 @ \rho : \eta[a \mapsto \rho] \sqsubseteq \omega_2 \leadsto \mu_2}{\Sigma \vdash^{\text{inst}} \mu_1 : \forall \{a : \omega_1\}.\eta \sqsubseteq \omega_2 \leadsto \mu_2} \text{INST-FORALL-INFER}$$

$\boxed{\Sigma \vdash^{\text{kc}} \sigma \Leftarrow \omega \leadsto \mu}$ *(Kind Checking)*

$$\frac{\Sigma \vdash^{\text{k}} \sigma : \eta \leadsto \mu_1 \qquad \Sigma \vdash^{\text{inst}} \mu_1 : \eta \sqsubseteq \omega \leadsto \mu_2}{\Sigma \vdash^{\text{kc}} \sigma \Leftarrow \omega \leadsto \mu_2} \text{KC-SUB}$$

$\boxed{\Sigma \vdash^{\text{k}} \sigma : \eta \leadsto \mu}$ *(Kinding)*

$$\frac{}{\Sigma \vdash^{\text{k}} \star : \star \leadsto \star} \text{KTT-STAR} \qquad \frac{}{\Sigma \vdash^{\text{k}} \text{Int} : \star \leadsto \text{Int}} \text{KTT-NAT} \qquad \frac{(a : \omega) \in \Sigma}{\Sigma \vdash^{\text{k}} a : \omega \leadsto a} \text{KTT-VAR} \qquad \frac{}{\Sigma \vdash^{\text{k}} \to : \star \to \star \to \star \leadsto \to} \text{KTT-ARROW}$$

$$\frac{(T : \eta) \in \Sigma}{\Sigma \vdash^{\text{k}} T : \eta \leadsto T} \text{KTT-TCON} \qquad \frac{\Sigma \vdash^{\text{k}} \tau_1 : \eta_1 \leadsto \rho_1 \qquad \Sigma \vdash^{\text{inst}} \rho_1 : \eta_1 \sqsubseteq (\omega_1 \to \omega_2) \leadsto \rho_2 \qquad \Sigma \vdash^{\text{kc}} \tau_2 \Leftarrow \omega_1 \leadsto \rho_3}{\Sigma \vdash^{\text{k}} \tau_1\,\tau_2 : \omega_2 \leadsto \rho_2\,\rho_3} \text{KTT-APP}$$

$$\frac{\Sigma \vdash^{\text{k}} \kappa_1 : \forall a : \omega.\eta \leadsto \rho_1 \qquad \Sigma \vdash^{\text{kc}} \kappa_2 \Leftarrow \omega \leadsto \rho_2}{\Sigma \vdash^{\text{k}} \kappa_1 @ \kappa_2 : \eta[a \mapsto \rho_2] \leadsto \rho_1 @ \rho_2} \text{KTT-KAPP}$$

$$\frac{\Sigma \vdash^{\text{k}} \kappa_1 : \forall \{\overline{a_i : \omega_i}^i\}.\forall a : \omega.\eta \leadsto \rho_1' \qquad \overline{\Sigma \vdash^{\text{ela}} \rho_i : \omega_i[\overline{a_i \mapsto \rho_i}^i]}^i \qquad \Sigma \vdash^{\text{kc}} \kappa_2 \Leftarrow \omega[\overline{a_i \mapsto \rho_i}^i] \leadsto \rho_2'}{\Sigma \vdash^{\text{k}} \kappa_1 @ \kappa_2 : \eta[\overline{a_i \mapsto \rho_i}^i][a \mapsto \rho_2] \leadsto \rho_1' @ \overline{\rho_i}^i @ \rho_2'} \text{KTT-KAPP-INFER}$$

$$\frac{\Sigma \vdash^{\text{kc}} \kappa \Leftarrow \star \leadsto \omega \qquad \Sigma, a : \omega \vdash^{\text{kc}} \sigma \Leftarrow \star \leadsto \mu}{\Sigma \vdash^{\text{k}} \forall a : \kappa.\sigma : \star \leadsto \forall a : \omega.\mu} \text{KTT-FORALL} \qquad \frac{\Sigma \vdash^{\text{ela}} \omega : \star \qquad \Sigma, a : \omega \vdash^{\text{kc}} \sigma \Leftarrow \star \leadsto \mu}{\Sigma \vdash^{\text{k}} \forall a.\sigma : \star \leadsto \forall a : \omega.\mu} \text{KTT-FORALLI}$$

$\boxed{\Sigma \vdash^{\text{ela}} \mu : \eta}$ *(Elaborated Kinding)*

$$\frac{}{\Sigma \vdash^{\text{ela}} \star : \star} \text{ELA-STAR} \qquad \frac{}{\Sigma \vdash^{\text{ela}} \text{Int} : \star} \text{ELA-NAT} \qquad \frac{(a : \omega) \in \Sigma}{\Sigma \vdash^{\text{ela}} a : \omega} \text{ELA-VAR} \qquad \frac{(T : \eta) \in \Sigma}{\Sigma \vdash^{\text{ela}} T : \eta} \text{ELA-TCON} \qquad \frac{}{\Sigma \vdash^{\text{ela}} \to : \star \to \star \to \star} \text{ELA-ARROW}$$

$$\frac{\Sigma \vdash^{\text{ela}} \rho_1 : \omega_1 \to \omega_2 \qquad \Sigma \vdash^{\text{ela}} \rho_2 : \omega_1}{\Sigma \vdash^{\text{ela}} \rho_1\,\rho_2 : \omega_2} \text{ELA-APP} \qquad \frac{\Sigma \vdash^{\text{ela}} \rho_1 : \forall a : \omega.\eta \qquad \Sigma \vdash^{\text{ela}} \rho_2 : \omega}{\Sigma \vdash^{\text{ela}} \rho_1 @ \rho_2 : \eta[a \mapsto \rho_2]} \text{ELA-KAPP}$$

ELA-KAPP-INFER
$$\frac{\Sigma \vdash^{\text{ela}} \rho_1 : \forall\{a : \omega\}.\eta \quad \Sigma \vdash^{\text{ela}} \rho_2 : \omega}{\Sigma \vdash^{\text{ela}} \rho_1 @\rho_2 : \eta[a \mapsto \rho_2]}$$

ELA-FORALL
$$\frac{\Sigma \vdash^{\text{ela}} \omega : \star \quad \Sigma, a : \omega \vdash^{\text{ela}} \mu : \star}{\Sigma \vdash^{\text{ela}} \forall a : \omega.\mu : \star}$$

ELA-FORALL-INFER
$$\frac{\Sigma \vdash^{\text{ela}} \omega : \star \quad \Sigma, a : \omega \vdash^{\text{ela}} \mu : \star}{\Sigma \vdash^{\text{ela}} \forall\{a : \omega\}.\mu : \star}$$

$\boxed{\Sigma \text{ ok}}$ *(Well-formed Type Contexts)*

TCTX-EMPTY
$$\frac{}{\bullet \text{ ok}}$$

TCTX-TVAR-TT
$$\frac{\Sigma \text{ ok} \quad \Sigma \vdash^{\text{ela}} \rho : \star}{\Sigma, a : \rho \text{ ok}}$$

TCTX-TCON-TT
$$\frac{\Sigma \text{ ok} \quad \Sigma \vdash^{\text{ela}} \eta : \star}{\Sigma, T : \eta \text{ ok}}$$

$\boxed{\Sigma \vdash \Psi}$ *(Well-formed Term Contexts)*

ECTX-EMPTY
$$\frac{}{\Sigma \vdash \bullet}$$

ECTX-DCON-TT
$$\frac{\Sigma \vdash \Psi \quad \Sigma \vdash^{\text{ela}} \mu : \star}{\Sigma \vdash \Psi, D : \mu}$$

## C.6 Algorithmic PolyKinds

$\boxed{\Delta \Vdash^{\text{inst}} \mu_1 : \eta \sqsubseteq \omega \leadsto \mu_2 \dashv \Theta}$ *(Instantiation)*

A-INST-REFL
$$\frac{\Delta \Vdash^{\text{u}} \omega_1 \approx \omega_2 \dashv \Theta}{\Delta \Vdash^{\text{inst}} \mu : \omega_1 \sqsubseteq \omega_2 \leadsto \mu \dashv \Theta}$$

A-INST-FORALL
$$\frac{\Delta, \widehat{\alpha} : \omega_1 \Vdash^{\text{inst}} \mu_1 @\widehat{\alpha} : \eta[a \mapsto \widehat{\alpha}] \sqsubseteq \omega_2 \leadsto \mu_2 \dashv \Theta}{\Delta \Vdash^{\text{inst}} \mu_1 : \forall a : \omega_1.\eta \sqsubseteq \omega_2 \leadsto \mu_2 \dashv \Theta}$$

A-INST-FORALL-INFER
$$\frac{\Delta, \widehat{\alpha} : \omega_1 \Vdash^{\text{inst}} \mu_1 @\widehat{\alpha} : \eta[a \mapsto \widehat{\alpha}] \sqsubseteq \omega_2 \leadsto \mu_2 \dashv \Theta}{\Delta \Vdash^{\text{inst}} \mu_1 : \forall\{a : \omega_1\}.\eta \sqsubseteq \omega_2 \leadsto \mu_2 \dashv \Theta}$$

$\boxed{\Delta \Vdash^{\text{kc}} \sigma \Leftarrow \eta \leadsto \mu \dashv \Theta}$ *(Kind Checking)*

A-KC-SUB
$$\frac{\Delta \Vdash^{\text{k}} \sigma : \eta \leadsto \mu_1 \dashv \Delta_1 \quad \Delta_1 \Vdash^{\text{inst}} \mu_1 : [\Delta_1]\eta \sqsubseteq [\Delta_1]\omega \leadsto \mu_2 \dashv \Delta_2}{\Delta \Vdash^{\text{kc}} \sigma \Leftarrow \omega \leadsto \mu_2 \dashv \Delta_2}$$

$\boxed{\Delta \Vdash^{\text{k}} \sigma : \eta \leadsto \mu \dashv \Theta}$ *(Kinding)*

A-KTT-STAR
$$\frac{}{\Delta \Vdash^{\text{k}} \star : \star \leadsto \star \dashv \Delta}$$

A-KTT-NAT
$$\frac{}{\Delta \Vdash^{\text{k}} \text{Int} : \star \leadsto \text{Int} \dashv \Delta}$$

A-KTT-VAR
$$\frac{(a : \omega) \in \Delta}{\Delta \Vdash^{\text{k}} a : \omega \leadsto a \dashv \Delta}$$

A-KTT-TCON
$$\frac{(T : \eta) \in \Delta}{\Delta \Vdash^{\text{k}} T : \eta \leadsto T \dashv \Delta}$$

A-KTT-ARROW
$$\frac{}{\Delta \Vdash^{\text{k}} \to : \star \to \star \to \star \leadsto \to \dashv \Delta}$$

A-KTT-FORALL
$$\frac{\Delta \Vdash^{\text{kc}} \kappa \Leftarrow \star \leadsto \omega \dashv \Delta_1 \quad \Delta_1, a : \omega \Vdash^{\text{kc}} \sigma \Leftarrow \star \leadsto \mu \dashv \Delta_2, a : \omega, \Delta_3 \quad \Delta_3 \hookrightarrow a}{\Delta \Vdash^{\text{k}} \forall a : \kappa.\sigma : \star \leadsto \forall a : \omega.[\Delta_3]\mu \dashv \Delta_2, \text{unsolved}(\Delta_3)}$$

A-KTT-APP
$$\frac{\Delta \Vdash^{\text{k}} \tau_1 : \eta_1 \leadsto \rho_1 \dashv \Delta_1 \quad \Delta_1 \Vdash^{\text{kapp}} (\rho_1 : [\Delta_1]\eta_1) \bullet \tau_2 : \omega \leadsto \rho \dashv \Theta}{\Delta \Vdash^{\text{k}} \tau_1 \tau_2 : \omega \leadsto \rho \dashv \Theta}$$

A-KTT-FORALLI
$$\frac{\Delta, \widehat{\alpha} : \star, a : \widehat{\alpha} \Vdash^{\text{kc}} \sigma \Leftarrow \star \leadsto \mu \dashv \Delta_2, a : \widehat{\alpha}, \Delta_3 \quad \Delta_3 \hookrightarrow a}{\Delta \Vdash^{\text{k}} \forall a.\sigma : \star \leadsto \forall a : \widehat{\alpha}.[\Delta_3]\mu \dashv \Delta_2, \text{unsolved}(\Delta_3)}$$

$$\frac{\text{A-KTT-KAPP}}{\Delta \Vdash^{\text{k}} \tau_1 : \eta \leadsto \rho_1 \dashv \Delta_1 \quad [\Delta_1]\eta = \forall a : \omega.\eta_2 \quad \Delta_1 \Vdash^{\text{kc}} \tau_2 \Leftarrow \omega \leadsto \rho_2 \dashv \Delta_2}{\Delta \Vdash^{\text{k}} \tau_1 @ \tau_2 : \eta_2[a \mapsto \rho_2] \leadsto \rho_1 @ \rho_2 \dashv \Delta_2}$$

$$\frac{\text{A-KTT-KAPP-INFER}}{[\Delta_1]\eta = \forall \overline{\{a_i : \omega_i\}}^i.\forall a : \omega.\eta_2 \quad \Delta_1, \overline{\widehat{\alpha}_i : \omega_i[\overline{a_i \mapsto \widehat{\alpha}_i}^i]}^i \Vdash^{\text{kc}} \tau_2 \Leftarrow \omega[\overline{a_i \mapsto \widehat{\alpha}_i}^i] \leadsto \rho_2 \dashv \Delta_2}{\Delta \Vdash^{\text{k}} \tau_1 @ \tau_2 : \eta_2[\overline{a_i \mapsto \widehat{\alpha}_i}^i][a \mapsto \rho_2] \leadsto \rho_1 @\overline{\widehat{\alpha}_i}^i @\rho_2 \dashv \Delta_2}$$

$$\boxed{\Delta \Vdash^{\text{kapp}} (\rho_1 : \eta) \bullet \tau : \omega \leadsto \rho_2 \dashv \Theta} \qquad \text{(Application Kinding)}$$

$$\frac{\text{A-KAPP-TT-ARROW}}{\Delta \Vdash^{\text{kc}} \tau \Leftarrow \omega_1 \leadsto \rho_2 \dashv \Theta}{\Delta \Vdash^{\text{kapp}} (\rho_1 : \omega_1 \to \omega_2) \bullet \tau : \omega_2 \leadsto \rho_1 \rho_2 \dashv \Theta}$$

$$\frac{\text{A-KAPP-TT-FORALL}}{\Delta, \widehat{\alpha} : \omega_1 \Vdash^{\text{kapp}} (\rho_1 @\widehat{\alpha} : \eta[a \mapsto \widehat{\alpha}]) \bullet \tau : \omega \leadsto \rho \dashv \Theta}{\Delta \Vdash^{\text{kapp}} (\rho_1 : \forall a : \omega_1.\eta) \bullet \tau : \omega \leadsto \rho \dashv \Theta}$$

$$\frac{\text{A-KAPP-TT-FORALL-INFER}}{\Delta, \widehat{\alpha} : \omega_1 \Vdash^{\text{kapp}} (\rho_1 @\widehat{\alpha} : \eta[a \mapsto \widehat{\alpha}]) \bullet \tau : \omega \leadsto \rho \dashv \Theta}{\Delta \Vdash^{\text{kapp}} (\rho_1 : \forall \{a : \omega_1\}.\eta) \bullet \tau : \omega \leadsto \rho \dashv \Theta}$$

$$\frac{\text{A-KAPP-TT-KUVAR}}{\Delta_1, \widehat{\alpha}_1 : \star, \widehat{\alpha}_2 : \star, \widehat{\alpha} : \omega = (\widehat{\alpha}_1 \to \widehat{\alpha}_2), \Delta_2 \Vdash^{\text{kc}} \tau \Leftarrow \widehat{\alpha}_1 \leadsto \rho_2 \dashv \Theta}{\Delta_1, \widehat{\alpha} : \omega, \Delta_2 \Vdash^{\text{kapp}} (\rho_1 : \widehat{\alpha}) \bullet \tau : \widehat{\alpha}_2 \leadsto \rho_1 \rho_2 \dashv \Theta}$$

$$\boxed{\Delta \Vdash^{\text{ela}} \mu : \eta} \qquad \text{(Elaborated Kinding)}$$

$$\frac{\text{A-ELA-STAR}}{\Delta \Vdash^{\text{ela}} \star : \star} \quad \frac{\text{A-ELA-KUVAR}}{(\widehat{\alpha} : \omega) \in \Delta}{\Delta \Vdash^{\text{ela}} \widehat{\alpha} : [\Delta]\omega} \quad \frac{\text{A-ELA-NAT}}{\Delta \Vdash^{\text{ela}} \text{Int} : \star} \quad \frac{\text{A-ELA-VAR}}{(a : \omega) \in \Delta}{\Delta \Vdash^{\text{ela}} a : [\Delta]\omega} \quad \frac{\text{A-ELA-TCON}}{(T : \eta) \in \Delta}{\Delta \Vdash^{\text{ela}} T : [\Delta]\eta}$$

$$\frac{\text{A-ELA-ARROW}}{\Delta \Vdash^{\text{ela}} \to : \star \to \star \to \star} \quad \frac{\text{A-ELA-FORALL}}{\Delta \Vdash^{\text{ela}} \omega : \star \quad \Delta, a : \omega \Vdash^{\text{ela}} \mu : \star}{\Delta \Vdash^{\text{ela}} \forall a : \omega.\mu : \star}$$

$$\frac{\text{A-ELA-FORALL-INFER}}{\Delta \Vdash^{\text{ela}} \omega : \star \quad \Delta, a : \omega \Vdash^{\text{ela}} \mu : \star}{\Delta \Vdash^{\text{ela}} \forall \{a : \omega\}.\mu : \star} \quad \frac{\text{A-ELA-APP}}{\Delta \Vdash^{\text{ela}} \rho_1 : \omega_1 \to \omega_2 \quad \Delta \Vdash^{\text{ela}} \rho_2 : \omega_1}{\Delta \Vdash^{\text{ela}} \rho_1 \rho_2 : \omega_2}$$

$$\frac{\text{A-ELA-KAPP}}{\Delta \Vdash^{\text{ela}} \rho_1 : \forall a : \omega.\eta \quad \Delta \Vdash^{\text{ela}} \rho_2 : \omega}{\Delta \Vdash^{\text{ela}} \rho_1 @\rho_2 : \eta[a \mapsto [\Delta]\rho_2]} \quad \frac{\text{A-ELA-KAPP-INFER}}{\Delta \Vdash^{\text{ela}} \rho_1 : \forall \{a : \omega\}.\eta \quad \Delta \Vdash^{\text{ela}} \rho_2 : \omega}{\Delta \Vdash^{\text{ela}} \rho_1 @\rho_2 : \eta[a \mapsto [\Delta]\rho_2]}$$

$$\boxed{\Delta \Vdash^{\text{gen}}_{\phi^{\text{c}}} \Gamma_1 \leadsto \Gamma_2} \qquad \text{(Generalization)}$$

$$\frac{\text{A-GEN}}{\overline{\widehat{\phi^{\text{c}}_i} = \text{unsolved}(\mu_i)}^i}{\Delta \Vdash^{\text{gen}}_{\phi^{\text{c}}} \overline{D_i : \mu_i}^i \leadsto \overline{D : \forall\{\phi^{\text{c}}\}.\forall\{\phi^{\text{c}}_i\}.(\mu[\widehat{\phi^{\text{c}}_i} \mapsto \phi^{\text{c}}_i])}^i}$$

$\boxed{\Delta \text{ ok}}$　　　　　　　　　　　　　　　　　　　　　　　　　　　　　　*(Well-formed Type Contexts)*

$$\text{A-TCTX-EMPTY} \quad \frac{\Delta \text{ ok} \quad \Delta \Vdash^{\text{ela}} \omega : \star}{\bullet \text{ ok}} \quad \text{A-TCTX-TVAR-TT} \quad \frac{\Delta \text{ ok} \quad \Delta \Vdash^{\text{ela}} \eta : \star}{\Delta, a : \omega \text{ ok}} \quad \text{A-TCTX-TCON-TT} \quad \frac{\Delta \text{ ok} \quad \Delta \Vdash^{\text{ela}} \omega : \star}{\Delta, T : \eta \text{ ok}} \quad \text{A-TCTX-KUVAR-TT} \quad \frac{}{\Delta, \widehat{\alpha} : \omega \text{ ok}}$$

$$\text{A-TCTX-KUVARSOLVED-TT} \quad \frac{\Delta \text{ ok} \quad \Delta \Vdash^{\text{ela}} \omega_2 : [\Delta]\omega_1}{\Delta, \widehat{\alpha} : \omega_1 = \omega_2 \text{ ok}} \qquad \text{A-TCTX-LO} \quad \frac{\Delta, \Delta^{\text{lo}} \text{ ok}}{\Delta, \{\Delta^{\text{lo}}\} \text{ ok}} \qquad \text{A-TCTX-MARKER} \quad \frac{\Delta \text{ ok}}{\Delta, \blacktriangleright_D \text{ ok}}$$

$\boxed{\Delta \Vdash^{\text{ectx}} \Gamma}$　　　　　　　　　　　　　　　　　　　　　　　　　　　*(Well-formed Term Contexts)*

$$\text{A-ECTX-EMPTY} \quad \frac{}{\Delta \Vdash^{\text{ectx}} \bullet} \qquad \text{A-ECTX-DCON-TT} \quad \frac{\Delta \Vdash^{\text{ectx}} \Gamma \quad \Delta \Vdash^{\text{ela}} \mu : \star}{\Delta \Vdash^{\text{ectx}} \Gamma, D : \mu}$$

$\boxed{\Delta \Vdash^{\mathsf{u}} \omega_1 \approx \omega_2 \dashv \Theta}$　　　　　　　　　　　　　　　　　　　　　　　　　　　　*(Unification)*

$$\text{A-U-REFL-TT} \quad \frac{}{\Delta \Vdash^{\mathsf{u}} \omega \approx \omega \dashv \Delta}$$

$$\text{A-U-APP} \quad \frac{\Delta \Vdash^{\mathsf{u}} \rho_1 \approx \rho_3 \dashv \Delta_1 \quad \Delta_1 \Vdash^{\mathsf{u}} [\Delta_1]\rho_2 \approx [\Delta_1]\rho_4 \dashv \Theta}{\Delta \Vdash^{\mathsf{u}} \rho_1 \rho_2 \approx \rho_3 \rho_4 \dashv \Theta}$$

$$\text{A-U-KAPP} \quad \frac{\Delta \Vdash^{\mathsf{u}} \rho_1 \approx \rho_3 \dashv \Delta_1 \quad \Delta_1 \Vdash^{\mathsf{u}} [\Delta_1]\rho_2 \approx [\Delta_1]\rho_4 \dashv \Theta}{\Delta \Vdash^{\mathsf{u}} \rho_1 @ \rho_2 \approx \rho_3 @ \rho_4 \dashv \Theta}$$

$$\text{A-U-KVARL-TT} \quad \frac{\Delta \Vdash^{\text{pr}}_{\widehat{\alpha}} \rho_1 \rightsquigarrow \rho_2 \dashv \Theta_1, \widehat{\alpha} : \omega_1, \Theta_2 \quad \Theta_1 \Vdash^{\text{ela}} \rho_2 : \omega_2 \quad \Theta_1 \Vdash^{\mathsf{u}} [\Theta_1]\omega_1 \approx \omega_2 \dashv \Theta_3}{\Delta \Vdash^{\mathsf{u}} \widehat{\alpha} \approx \rho_1 \dashv \Theta_3, \widehat{\alpha} : \omega_1 = \rho_2, \Theta_2}$$

$$\text{A-U-KVARR-TT} \quad \frac{\Delta \Vdash^{\text{pr}}_{\widehat{\alpha}} \rho_1 \rightsquigarrow \rho_2 \dashv \Theta_1, \widehat{\alpha} : \omega_1, \Theta_2 \quad \Theta_1 \Vdash^{\text{ela}} \rho_2 : \omega_2 \quad \Theta_1 \Vdash^{\mathsf{u}} [\Theta_1]\omega_1 \approx \omega_2 \dashv \Theta_3}{\Delta \Vdash^{\mathsf{u}} \rho_1 \approx \widehat{\alpha} \dashv \Theta_3, \widehat{\alpha} : \omega_1 = \rho_2, \Theta_2}$$

$$\text{A-U-KVARL-LO-TT} \quad \frac{\Delta_1, \Delta_2 \Vdash^{\text{mv}} \widehat{\alpha} : \omega_1 \rightsquigarrow \Theta \quad \Delta[\{\Theta\}] \Vdash^{\text{pr}}_{\widehat{\alpha}} \rho_1 \rightsquigarrow \rho_2 \dashv \Theta_1, \{\Theta_2, \widehat{\alpha} : \omega_1, \Theta_3\}, \Theta_4 \quad \Theta_1, \{\Theta_2\} \Vdash^{\text{ela}} \rho_2 : \omega_2 \quad \Theta_1, \{\Theta_2\} \Vdash^{\mathsf{u}} [\Theta_1, \Theta_2]\omega_1 \approx \omega_2 \dashv \Theta_5, \{\Theta_6\}}{\Delta[\{\Delta_1, \widehat{\alpha} : \omega_1, \Delta_2\}] \Vdash^{\mathsf{u}} \widehat{\alpha} \approx \rho_1 \dashv \Theta_5, \{\Theta_6, \widehat{\alpha} : \omega_1 = \rho_2, \Theta_3\}, \Theta_4}$$

$$\text{A-U-KVARR-LO-TT} \quad \frac{\Delta_1, \Delta_2 \Vdash^{\text{mv}} \widehat{\alpha} : \omega_1 \rightsquigarrow \Theta \quad \Delta[\{\Theta\}] \Vdash^{\text{pr}}_{\widehat{\alpha}} \rho_1 \rightsquigarrow \rho_2 \dashv \Theta_1, \{\Theta_2, \widehat{\alpha} : \omega_1, \Theta_3\}, \Theta_4 \quad \Theta_1, \{\Theta_2\} \Vdash^{\text{ela}} \rho_2 : \omega_2 \quad \Theta_1, \{\Theta_2\} \Vdash^{\mathsf{u}} [\Theta_1, \Theta_2]\omega_1 \approx \omega_2 \dashv \Theta_5, \{\Theta_6\}}{\Delta[\{\Delta_1, \widehat{\alpha} : \omega_1, \Delta_2\}] \Vdash^{\mathsf{u}} \rho_1 \approx \widehat{\alpha} \dashv \Theta_5, \{\Theta_6, \widehat{\alpha} : \omega_1 = \rho_2, \Theta_3\}, \Theta_4}$$

$\boxed{\Delta \Vdash^{\text{pr}}_{\widehat{\alpha}} \omega_1 \rightsquigarrow \omega_2 \dashv \Theta}$　　　　　　　　　　　　　　　　　　　　　　　　　　　　*(Promotion)*

$$\text{A-PR-STAR} \quad \frac{}{\Delta \Vdash^{\text{pr}}_{\widehat{\alpha}} \star \rightsquigarrow \star \dashv \Delta} \qquad \text{A-PR-ARR} \quad \frac{}{\Delta \Vdash^{\text{pr}}_{\widehat{\alpha}} \rightarrow \rightsquigarrow \rightarrow \dashv \Delta} \qquad \text{A-PR-TCON} \quad \frac{}{\Delta[T][\widehat{\alpha}] \Vdash^{\text{pr}}_{\widehat{\alpha}} T \rightsquigarrow T \dashv \Delta[T][\widehat{\alpha}]}$$

$$\text{A-PR-NAT} \quad \frac{}{\Delta \Vdash^{\text{pr}}_{\widehat{\alpha}} \text{Int} \rightsquigarrow \text{Int} \dashv \Delta} \qquad \text{A-PR-APP} \quad \frac{\Delta \Vdash^{\text{pr}}_{\widehat{\alpha}} \omega_1 \rightsquigarrow \rho_1 \dashv \Delta_1 \quad \Delta_1 \Vdash^{\text{pr}}_{\widehat{\alpha}} [\Delta_1]\omega_2 \rightsquigarrow \rho_2 \dashv \Theta}{\Delta \Vdash^{\text{pr}}_{\widehat{\alpha}} \omega_1 \omega_2 \rightsquigarrow \rho_1 \rho_2 \dashv \Theta}$$

A-PR-KAPP
$$\frac{\Delta \Vdash^{pr}_{\widehat{\alpha}} \omega_1 \rightsquigarrow \rho_1 \dashv \Delta_1 \quad \Delta_1 \Vdash^{pr}_{\widehat{\alpha}} [\Delta_1]\omega_2 \rightsquigarrow \rho_2 \dashv \Theta}{\Delta \Vdash^{pr}_{\widehat{\alpha}} \omega_1 @ \omega_2 \rightsquigarrow \rho_1 @ \rho_2 \dashv \Theta}$$

A-PR-TVAR
$$\overline{\Delta[a][\widehat{\alpha}] \Vdash^{pr}_{\widehat{\alpha}} a \rightsquigarrow a \dashv \Delta[a][\widehat{\alpha}]}$$

A-PR-KUVARL
$$\overline{\Delta[\widehat{\beta}][\widehat{\alpha}] \Vdash^{pr}_{\widehat{\alpha}} \widehat{\beta} \rightsquigarrow \widehat{\beta} \dashv \Delta[\widehat{\beta}][\widehat{\alpha}]}$$

A-PR-KUVARR-TT
$$\frac{\Delta \Vdash^{pr}_{\widehat{\alpha}} [\Delta]\rho \rightsquigarrow \rho_1 \dashv \Theta[\widehat{\alpha}][\widehat{\beta} : \rho]}{\Delta[\widehat{\alpha}][\widehat{\beta} : \rho] \Vdash^{pr}_{\widehat{\alpha}} \widehat{\beta} \rightsquigarrow \widehat{\beta}_1 \dashv \Theta[\widehat{\beta}_1 : \rho_1, \widehat{\alpha}][\widehat{\beta} : \rho = \widehat{\beta}_1]}$$

$\boxed{\Delta_1 \mathbin{+\!\!+}^{mv} \Delta_2 \rightsquigarrow \Theta}$ *(Moving)*

A-MV-EMPTY
$$\overline{\bullet \mathbin{+\!\!+}^{mv} \Delta \rightsquigarrow \Delta}$$

A-MV-KUVAR
$$\frac{\text{vars}(\omega) \mathbin{\sharp} \text{dom}(\Delta_2) \quad \Delta_1 \mathbin{+\!\!+}^{mv} \Delta_2 \rightsquigarrow \Theta}{\widehat{\alpha} : \omega, \Delta_1 \mathbin{+\!\!+}^{mv} \Delta_2 \rightsquigarrow \widehat{\alpha} : \omega, \Theta}$$

A-MV-KUVARM
$$\frac{\neg(\text{vars}(\omega) \mathbin{\sharp} \text{dom}(\Delta_2)) \quad \Delta_1 \mathbin{+\!\!+}^{mv} \Delta_2, \widehat{\alpha} : \omega \rightsquigarrow \Theta}{\widehat{\alpha} : \omega, \Delta_1 \mathbin{+\!\!+}^{mv} \Delta \rightsquigarrow \Theta}$$

A-MV-TVAR
$$\frac{\text{vars}(\omega) \mathbin{\sharp} \text{dom}(\Delta_2) \quad \Delta_1 \mathbin{+\!\!+}^{mv} \Delta_2 \rightsquigarrow \Theta}{a : \omega, \Delta_1 \mathbin{+\!\!+}^{mv} \Delta_2 \rightsquigarrow a : \omega, \Theta}$$

A-MV-TVARM
$$\frac{\neg(\text{vars}(\omega) \mathbin{\sharp} \text{dom}(\Delta_2)) \quad \Delta_1 \mathbin{+\!\!+}^{mv} \Delta_2, a : \omega \rightsquigarrow \Theta}{a : \omega, \Delta_1 \mathbin{+\!\!+}^{mv} \Delta_2 \rightsquigarrow \Theta}$$

## C.7 Context Application in PolyKinds

$[\Delta]\eta$ applies $\Delta$ as a substitution to $\eta$.

$$
\begin{aligned}
[\Delta]\star &= \star \\
[\Delta]\text{Int} &= \text{Int} \\
[\Delta]a &= a \\
[\Delta]T &= T \\
[\Delta]\to &= \to \\
[\Delta]\forall a : \omega.\eta &= \forall a : [\Delta]\omega.[\Delta]\eta \\
[\Delta]\forall\{a : \omega\}.\eta &= \forall\{a : [\Delta]\omega\}.[\Delta]\eta \\
[\Delta](\rho_1 \rho_2) &= ([\Delta]\rho_1)([\Delta]\rho_2) \\
[\Delta](\rho_1 @ \rho_2) &= ([\Delta]\rho_1) @ ([\Delta]\rho_2) \\
[\Delta[\widehat{\alpha}]]\widehat{\alpha} &= \widehat{\alpha} \\
[\Delta[\widehat{\alpha} : \omega = \rho]]\widehat{\alpha} &= [\Delta[\widehat{\alpha} : \omega = \rho]]\rho
\end{aligned}
$$

$[\Delta]\Gamma$ applies $\Delta$ as a substitution to $\Gamma$.

$$
\begin{aligned}
[\Omega]\bullet &= \bullet \\
[\Omega](\Gamma, D : \mu) &= [\Omega]\Gamma, D : [\Omega]\mu
\end{aligned}
$$

$[\Omega]\Delta$ applies $\Omega$ as a substitution to $\Delta$.

$$
\begin{aligned}
[\Omega]\bullet &= \bullet \\
[\Omega, a : \omega](\Delta, a : \omega) &= [\Omega]\Delta, a : [\Omega]\omega \\
[\Omega, T : \omega](\Delta, T : \omega) &= [\Omega]\Delta, T : [\Omega]\omega \\
[\Omega, \widehat{\alpha} : \omega = \rho](\Delta, \widehat{\alpha} : \omega) &= [\Omega]\Delta \\
[\Omega, \widehat{\alpha} : \omega = \rho_1](\Delta, \widehat{\alpha} : \omega = \rho_2) &= [\Omega]\Delta \quad \text{if } [\Omega]\rho_1 = [\Omega]\rho_2 \\
[\Omega, \widehat{\alpha} : \omega = \rho]\Delta &= [\Omega]\Delta \quad \text{if } \widehat{\alpha} \notin \Delta \\
[\Omega, \blacktriangleright_D](\Delta, \blacktriangleright_D) &= [\Omega]\Delta \\
[\Omega, \{\Omega_1\}](\Delta, \{\Delta_1\}) &= [\Omega, \Omega_1](\Delta, \Delta') \\
&\quad \text{where } \Delta' = \text{topo}(\Delta_1)
\end{aligned}
$$

## C.8 Context Extension in PolyKinds

$\boxed{\Delta \longrightarrow \Theta}$ *(Context Extension)*

$$\text{A-CTXE-EMPTY} \quad \frac{}{\bullet \longrightarrow \bullet}$$

$$\text{A-CTXE-TVAR-TT} \quad \frac{\Delta \longrightarrow \Theta}{\Delta, a : \omega \longrightarrow \Theta, a : \omega}$$

$$\text{A-CTXE-TCON-TT} \quad \frac{\Delta \longrightarrow \Theta}{\Delta, T : \eta \longrightarrow \Theta, T : \eta}$$

$$\text{A-CTXE-KUVAR-TT} \quad \frac{\Delta \longrightarrow \Theta}{\Delta, \widehat{\alpha} : \omega \longrightarrow \Theta, \widehat{\alpha} : \omega}$$

$$\text{A-CTXE-KUVARSOLVED-TT} \quad \frac{\Delta \longrightarrow \Theta \quad [\Theta]\rho_1 = [\Theta]\rho_2}{\Delta, \widehat{\alpha} : \omega = \rho_1 \longrightarrow \Theta, \widehat{\alpha} : \omega = \rho_2}$$

$$\text{A-CTXE-SOLVE-TT} \quad \frac{\Delta \longrightarrow \Theta \quad \Theta \Vdash^{\text{ela}} \rho : [\Theta]\omega}{\Delta, \widehat{\alpha} : \omega \longrightarrow \Theta, \widehat{\alpha} : \omega = \rho}$$

$$\text{A-CTXE-ADD-TT} \quad \frac{\Delta \longrightarrow \Theta \quad \Theta \Vdash^{\text{ela}} \omega : \star}{\Delta \longrightarrow \Theta, \widehat{\alpha} : \omega}$$

$$\text{A-CTXE-ADDSOLVED-TT} \quad \frac{\Delta \longrightarrow \Theta \quad \Theta \Vdash^{\text{ela}} \rho : [\Theta]\omega}{\Delta \longrightarrow \Theta, \widehat{\alpha} : \omega = \rho}$$

$$\text{A-CTXE-MARKER} \quad \frac{\Delta \longrightarrow \Theta}{\Delta, \blacktriangleright_D \longrightarrow \Theta, \blacktriangleright_D}$$

$$\text{A-CTXE-LO} \quad \frac{\Delta \longrightarrow \Theta \quad \Delta, \text{topo}(\Delta_1) \longrightarrow \Theta, \Theta_1}{\Delta, \{\Delta_1\} \longrightarrow \Theta, \{\Theta_1\}}$$

## D PROOF FOR HASKELL98

### D.1 List of Lemmas

*D.1.1 Well-formedness of Declarative Type System.*

**Lemma D.1** (Well-formedness of Declarative Typing Data Constructor Declaration). *If* $\Sigma \vdash^{\text{dc}}_{\tau_1} \mathcal{D} \rightsquigarrow \tau_2$, *then* $\Sigma \vdash^{\text{k}} \tau_2 : \star$.

**Lemma D.2** (Well-formedness of Declarative Typing Datatype Declaration). *If* $\Sigma \vdash^{\text{dt}} \mathcal{T} \rightsquigarrow \Psi$, *then* $\Sigma \vdash \Psi$.

*D.1.2 Well-formedness of Algorithmic Type System.*

**Lemma D.3** (Well-formedness of Promotion). *If* $\Delta_1, \widehat{\alpha}, \Delta_2$ *ok, and* $\Delta_1, \widehat{\alpha}, \Delta_2 \Vdash^{\text{pr}}_{\widehat{\alpha}} \kappa_1 \rightsquigarrow \kappa_2 \dashv \Theta$, *then* $\Theta = \Theta_1, \widehat{\alpha}, \Theta_2$, *and* $\Delta_1, \widehat{\alpha}, \Delta_2 \longrightarrow \Theta$, *and* $\Theta_1 \Vdash^{\text{kv}} \kappa_2$, *and* $\Theta$ *ok. By weakening, there is also* $\Theta \Vdash^{\text{kv}} \kappa_2$.

**Lemma D.4** (Well-formedness of Unification). *If* $\Delta$ *ok, and* $\Delta \Vdash^{\text{u}} \kappa_1 \approx \kappa_2 \dashv \Theta$, *then* $\Delta \longrightarrow \Theta$, *and* $\Theta$ *ok.*

**Lemma D.5** (Well-formedness of Application Kinding). *If* $\Delta$ *ok, and* $\Delta \Vdash^{\text{kapp}} \kappa_1 \bullet \kappa_2 : \kappa \dashv \Theta$, *then* $\Delta \longrightarrow \Theta$, *and* $\Theta$ *ok. Moreover, if* $\Delta \Vdash^{\text{kv}} \kappa_1$, *then we have* $\Theta \Vdash^{\text{kv}} \kappa$.

**Lemma D.6** (Well-formedness of Kinding). *If* $\Delta$ *ok, and* $\Delta \Vdash^{\text{k}} \sigma : \kappa \dashv \Theta$, *then* $\Delta \longrightarrow \Theta$, *and* $\Theta$ *ok, and* $\Theta \Vdash^{\text{kv}} \kappa$ *and* $\Theta \Vdash^{\text{kc}} \sigma \Leftarrow \kappa$.

**Lemma D.7** (Well-formedness of Typing Data Constructor Declarations). *If* $\Delta$ *ok, and* $\Delta \Vdash^{\text{dc}}_{\tau'} \mathcal{D} \rightsquigarrow \tau \dashv \Theta$, *then* $\Delta \longrightarrow \Theta$, *and* $\Theta$ *ok. and* $\Theta \Vdash^{\text{kc}} \tau \Leftarrow \star$.

**Lemma D.8** (Well-formedness of Typing Datatype Declaration). *If* $\Delta$ *ok, and* $\Delta \Vdash^{\text{dt}} \mathcal{T} \rightsquigarrow \Gamma \dashv \Theta$, *then* $\Delta \longrightarrow \Theta$, *and* $\Theta$ *ok, and* $\Theta \Vdash^{\text{ectx}} \Gamma$.

*D.1.3 Properties of Context Extension.*

**Lemma D.9** (Declaration Preservation). *If* $\Delta \longrightarrow \Theta$, *if a type constructor or a type variable or a kind unification variable is declared in* $\Delta$, *then it is declared in* $\Theta$.

**Lemma D.10** (Extension Weakening). *Given* $\Delta \longrightarrow \Theta$,
- *if* $\Delta \Vdash^{\text{kv}} \kappa$, *then* $\Theta \Vdash^{\text{kv}} \kappa$;

- *if* $\Delta \Vdash^{kc} \sigma \Leftarrow \kappa$, *then* $\Theta \Vdash^{kc} \sigma \Leftarrow \kappa$.

**Definition D.11** (Contextual Size).

| $\Delta \vdash \star$ | | = | 1 |
| $\Delta \vdash \kappa_1 \to \kappa_2$ | | = | $1 + \mid \Delta \vdash \kappa_1 \mid + \mid \Delta \vdash \kappa_2 \mid$ |
| $\Delta[\widehat{\alpha}] \vdash \widehat{\alpha}$ | | = | 1 |
| $\Delta[\widehat{\alpha} = \kappa] \vdash \widehat{\alpha}$ | | = | $1 + \mid \Delta[\widehat{\alpha} = \kappa] \vdash \kappa \mid$ |

**Lemma D.12** (Substitution Kinding). *If $\Delta$ ok, and $\Delta \Vdash^{kv} \kappa$, then $\Delta \Vdash^{kv} [\Delta]\kappa$.*

**Lemma D.13** (Context Extension with Defaulting is Context Extension). *If $\Delta \longrightarrow\!\!\!\!\!\rightarrow \Theta$, then $\Delta \longrightarrow \Theta$.*

**Lemma D.14** (Reflexivity of Context Extension). *If $\Delta$ ok, then $\Delta \longrightarrow \Delta$.*

**Lemma D.15** (Well-formedness of Context Extension). *If $\Delta$ ok, and $\Delta \longrightarrow \Theta$, then $\Theta$ ok.*

**Definition D.16** (Softness). *A context $\Delta$ is soft iff it contains only of $\widehat{\alpha}$ and $\widehat{\alpha} = \kappa$ declarations.*

**Lemma D.17** (Extension Order).
  (1) *If $\Delta_1, a : \kappa, \Delta_2 \longrightarrow \Theta$, then $\Theta = \Theta_1, a : \kappa, \Theta_2$, where $\Delta_1 \longrightarrow \Theta_1$. Moreover, if $\Delta_2$ soft, then $\Theta_2$ soft.*
  (2) *If $\Delta_1, T : \kappa, \Delta_2 \longrightarrow \Theta$, then $\Theta = \Theta_1, T : \kappa, \Theta_2$, where $\Delta_1 \longrightarrow \Theta_1$. Moreover, if $\Delta_2$ soft, then $\Theta_2$ soft.*
  (3) *If $\Delta_1, \widehat{\alpha}, \Delta_2 \longrightarrow \Theta$, then $\Theta = \Theta_1, \Theta', \Theta_2$, where $\Delta_1 \longrightarrow \Theta_1$, and $\Theta'$ is either $\widehat{\alpha}$ or $\widehat{\alpha} = \kappa$ for some $\kappa$. Moreover, if $\Delta_2$ soft, then $\Theta_2$ soft.*
  (4) *If $\Delta_1, \widehat{\alpha} = \kappa_1, \Delta_2 \longrightarrow \Theta$, then $\Theta = \Theta_1, \widehat{\alpha} = \kappa_2, \Theta_2$, where $\Delta_1 \longrightarrow \Theta_1$, and $[\Theta_1]\kappa_1 = [\Theta_1]\kappa_2$. Moreover, if $\Delta_2$ soft, then $\Theta_2$ soft.*

**Lemma D.18** (Substitution Extension Invariance). *If $\Delta$ ok, and $\Delta \Vdash^{kv} \kappa$, and $\Delta \longrightarrow \Theta$, then $[\Theta]\kappa = [\Theta]([\Delta]\kappa)$ and $[\Theta]\kappa = [\Delta]([\Theta]\kappa)$. As a corollary, if $\Delta \Vdash^{kv} \kappa_1$, $\Delta \Vdash^{kv} \kappa_2$, and $[\Delta]\kappa_1 = [\Delta]\kappa_2$, then $[\Theta]\kappa_1 = [\Theta]\kappa_2$.*

**Lemma D.19** (Substitution Stability). *If $\Delta_1, \Delta_2$ ok, and $\Delta_1 \Vdash^{kv} \kappa$, then $[\Delta_1, \Delta_2]\kappa = [\Delta_1]\kappa$.*

**Lemma D.20** (Transitivity of Context Extension). *If $\Delta'$ ok, and $\Delta' \longrightarrow \Delta$, and $\Delta \longrightarrow \Theta$, then $\Delta' \longrightarrow \Theta$.*

**Lemma D.21** (Solution Admissibility for Extension). *If $\Delta_1, \widehat{\alpha}, \Delta_2$ ok and $\Delta_1 \Vdash^{kv} \kappa$, then $\Delta_1, \widehat{\alpha}, \Delta_2 \longrightarrow \Delta_1, \widehat{\alpha} = \kappa, \Delta_2$.*

**Lemma D.22** (Solved Variable Addition for Extension). *If $\Delta_1, \Delta_2$ ok and $\Delta_1 \Vdash^{kv} \kappa$, then $\Delta_1, \Delta_2 \longrightarrow \Delta_1, \widehat{\alpha} = \kappa, \Delta_2$.*

**Lemma D.23** (Unsolved Variable Addition). *If $\Delta_1, \Delta_2$ ok then $\Delta_1, \Delta_2 \longrightarrow \Delta_1, \widehat{\alpha}, \Delta_2$.*

**Lemma D.24** (Parallel Admissibility). *If $\Delta_1 \longrightarrow \Theta_1$, and $\Delta_1, \Delta_2$ ok, and $\Delta_1, \Delta_2 \longrightarrow \Theta_1, \Theta_2$, and $\Delta_2$ is fresh w.r.t. $\Theta_1$, then:*
  - $\Delta_1, \widehat{\alpha}, \Delta_2 \longrightarrow \Theta_1, \widehat{\alpha}, \Theta_2$

- If $\Theta_1 \Vdash^{kv} \kappa$, then $\Delta_1, \widehat{\alpha}, \Delta_2 \longrightarrow \Theta_1, \widehat{\alpha} = \kappa, \Theta_2$
- If $[\Theta_1]\kappa_1 = [\Theta_1]\kappa_2$, then $\Delta_1, \widehat{\alpha} = \kappa_1, \Delta_2 \longrightarrow \Theta_1, \widehat{\alpha} = \kappa_2, \Theta_2$

**Lemma D.25** (Parallel Extension Solution). *If $\Delta_1, \widehat{\alpha}, \Delta_2 \longrightarrow \Theta_1, \widehat{\alpha} = \kappa_2, \Theta_2$, and $[\Theta_1]\kappa_1 = [\Theta_1]\kappa_2$, then $\Delta_1, \widehat{\alpha} = \kappa_1, \Delta_2 \longrightarrow \Theta_1, \widehat{\alpha} = \kappa_2, \Theta_2$.*

**Lemma D.26** (Parallel Variable Update). *If $\Delta_1, \widehat{\alpha}, \Delta_2 \longrightarrow \Theta_1, \widehat{\alpha} = \kappa, \Theta_2$, and $\Delta_1 \Vdash^{kv} \kappa_1$, and $\Theta_1 \Vdash^{kv} \kappa_2$, and $[\Theta_1]\kappa = [\Theta_1]\kappa_1 = [\Theta_1]\kappa_2$, then $\Delta_1, \widehat{\alpha} = \kappa_1, \Delta_2 \longrightarrow \Theta_1, \widehat{\alpha} = \kappa_2, \Theta_2$*

### D.1.4 Properties of Complete Context.

**Lemma D.27** (Type Constructor Preservation). *If $\Delta$ ok, and $(T : \kappa) \in \Delta$, and $\Delta \longrightarrow \Omega$, then $(T : [\Omega]\kappa) \in [\Omega]\Delta$.*

**Lemma D.28** (Type Variable Preservation). *If $\Delta$ ok, and $(a : \kappa) \in \Delta$, and $\Delta \longrightarrow \Omega$, then $(a : [\Omega]\kappa) \in [\Omega]\Delta$.*

**Lemma D.29** (Finishing Kinding). *If $\Omega$ ok, and $\Omega \Vdash^{kv} \kappa$, and $\Omega \longrightarrow \Omega'$, then $[\Omega]\kappa = [\Omega']\kappa$.*

**Lemma D.30** (Finishing Term Contexts). *If $\Omega$ ok, and $\Omega \Vdash^{ectx} \Gamma$, and $\Omega \longrightarrow \Omega'$, then $[\Omega']\Gamma = [\Omega]\Gamma$.*

**Lemma D.31** (Stability of Complete Contexts). *If $\Delta \longrightarrow \Omega$, then $[\Omega]\Delta = [\Omega]\Omega$.*

**Lemma D.32** (Softness Goes Away). *If $\Delta_1, \Delta_2 \longrightarrow \Omega_1, \Omega_2$ where $\Delta_1 \longrightarrow \Omega_1$, and $\Delta_2$ soft, then $[\Omega_1, \Omega_2](\Delta_1, \Delta_2) = [\Omega_1]\Delta_1$.*

**Lemma D.33** (Confluence of Completeness). *If $\Delta_1 \longrightarrow \Omega$, and $\Delta_2 \longrightarrow \Omega$, then $[\Omega]\Delta_1 = [\Omega]\Delta_2$.*

**Lemma D.34** (Finishing Completions). *If $\Omega$ ok, and $\Omega \longrightarrow \Omega'$, then $[\Omega]\Omega = [\Omega']\Omega'$.*

### D.1.5 Soundness of Algorithm.

**Lemma D.35** (Soundness of Kind Validating). *If $\Omega$ ok, and $\Omega \Vdash^{kv} \kappa$, then $[\Omega]\kappa$ is a validate kind in the declarative system.*

**Lemma D.36** (Soundness of Well-formed Type Context). *If $\Delta$ ok, and $\Delta \longrightarrow \Omega$, then $[\Omega]\Delta$ is a valid type context in the declarative system.*

**Lemma D.37** (Soundness of Well-formed Term Context). *If $\Delta$ ok, and $\Delta \Vdash^{ectx} \Gamma$, and $\Delta \longrightarrow \Omega$, then $[\Omega]\Delta \vdash [\Omega]\Gamma$.*

**Lemma D.38** (Soundness of Promotion). *If $\Delta$ ok, and $\Delta \Vdash^{pr}_{\widehat{\alpha}} \kappa_1 \rightsquigarrow \kappa_2 \dashv \Theta$, then $[\Theta]\kappa_1 = [\Theta]\kappa_2$. Moreover, if $\Delta \Vdash^{kv} \kappa_1$, and $\Theta \longrightarrow \Omega$, then $[\Omega]\kappa_1 = [\Omega]\kappa_2$.*

**Lemma D.39** (Soundness of Unification). *If $\Delta$ ok, and $\Delta \Vdash^{kv} \kappa_1$, and $\Delta \Vdash^{kv} \kappa_2$, and $\Delta \Vdash^{\mu} \kappa_1 \approx \kappa_2 \dashv \Theta$, then $[\Theta]\kappa_1 = [\Theta]\kappa_2$. If $\Theta \longrightarrow \Omega$, then $[\Omega]\kappa_1 = [\Omega]\kappa_2$.*

**Lemma D.40** (Soundness of Application Kinding). *If $\Delta$ ok, and $\Delta \Vdash^{kv} \kappa_1$, and $\Delta \Vdash^{kv} \kappa_2$, and $\Delta \Vdash^{kapp} \kappa_1 \bullet \kappa_2 : \kappa_3 \dashv \Theta$, then $[\Theta]\kappa_1 = [\Theta]\kappa_2 \rightarrow [\Theta]\kappa_3$. If $\Theta \longrightarrow \Omega$, then $[\Omega]\kappa_1 = [\Omega]\kappa_2 \rightarrow [\Omega]\kappa_3$.*

**Lemma D.41** (Soundness of Kinding). *If $\Delta$ ok, and $\Delta \Vdash^{k} \sigma : \kappa \dashv \Theta$, and $\Theta \longrightarrow \Omega$, then $[\Omega]\Delta \vdash^{k} [\Omega]\sigma : [\Omega]\kappa$.*

**Lemma D.42** (Soundness of Typing Data Constructor Declaration). *If $\Delta$ ok, and $\Delta \Vdash^{dc}_{\tau'} \mathcal{D} \rightsquigarrow \tau \dashv \Theta$, and $\Theta \longrightarrow \Omega$, then $[\Omega]\Delta \vdash^{dc}_{\tau'} \mathcal{D} \rightsquigarrow \tau$.*

**Lemma D.43** (Soundness of Typing Datatype Declaration). *If $\Delta$ ok, and $\Delta \Vdash^{dt} \mathcal{T} \rightsquigarrow \Gamma \dashv \Theta$, and $\Theta \longrightarrow \Omega$, then $[\Omega]\Delta \vdash^{dt} \mathcal{T} \rightsquigarrow [\Omega]\Gamma$.*

**Lemma D.44** (Soundness of Typing Program). *If $\Omega$ ok, and $\Omega \Vdash^{ectx} \Gamma$, and $\Omega; \Gamma \Vdash^{pgm} pgm : \sigma$, then $[\Omega]\Omega; [\Omega]\Gamma \vdash^{pgm} pgm : \sigma$.*

### D.1.6 Completeness of Algorithm.

**Lemma D.45** (Completeness of Promotion). *Given $\Delta$ ok, and $\Delta \longrightarrow \Omega$, and $\Delta \Vdash^{kv} \widehat{\alpha}$, and $\Delta \Vdash^{kv} \kappa$, and $[\Delta]\widehat{\alpha} = \widehat{\alpha}$, and $[\Delta]\kappa = \kappa$, if $\kappa$ is free of $\widehat{\alpha}$, then there exists $\kappa_2, \Theta$ and $\Omega'$ such that $\Theta \longrightarrow \Omega'$, and $\Omega \longrightarrow \Omega'$, and $\Delta \Vdash^{pr}_{\widehat{\alpha}} \kappa \rightsquigarrow \kappa_2 \dashv \Theta$.*

**Lemma D.46** (Completeness of Unification). *Given $\Delta$ ok, and $\Delta \longrightarrow \Omega$, and $\Delta \Vdash^{kv} \kappa_1$ and $\Delta \Vdash^{kv} \kappa_2$, and $[\Delta]\kappa_1 = \kappa_1$ and $[\Delta]\kappa_2 = \kappa_2$, if $[\Omega]\kappa_1 = [\Omega]\kappa_2$, then there exists $\Theta$ and $\Omega'$ such that $\Theta \longrightarrow \Omega'$, and $\Omega \longrightarrow \Omega'$ and $\Delta \Vdash^{u} \kappa_1 \approx \kappa_2 \dashv \Theta$.*

**Lemma D.47** (Completeness of Application Kinding). *Given $\Delta$ ok, and $\Delta \longrightarrow \Omega$, and $\Delta \Vdash^{kv} \kappa$ and $\Delta \Vdash^{kv} \kappa'$, and $[\Delta]\kappa = \kappa$ and $[\Delta]\kappa' = \kappa'$, if $[\Omega]\kappa = [\Omega]\kappa' \to \kappa_1$, then there exists $\kappa_2, \Theta$ and $\Omega'$ such that $\Theta \longrightarrow \Omega'$, and $\Omega \longrightarrow \Omega'$ and $\Delta \Vdash^{kapp} \kappa \bullet \kappa' : \kappa_2 \dashv \Theta$, and $[\Omega']\kappa_2 = \kappa_1$.*

**Lemma D.48** (Completeness of Kinding). *Given $\Delta$ ok and $\Delta \longrightarrow \Omega$, if $[\Omega]\Delta \vdash^{k} [\Omega]\sigma : \kappa$, then there exists $\Theta$ and $\Omega'$ such that $\Theta \longrightarrow \Omega'$, and $\Omega \longrightarrow \Omega'$ and $\Delta \Vdash^{k} \sigma : \kappa' \dashv \Theta$, and $[\Omega']\kappa' = \kappa$.*

**Lemma D.49** (Completeness of Typing Data Constructor Declaration). *Given $\Delta$ ok and $\Delta \longrightarrow \Omega$, if $[\Omega]\Delta \vdash^{dc}_{\tau'} \mathcal{D} \rightsquigarrow \tau$, then there exists $\Theta$ and $\Omega'$ such that $\Theta \longrightarrow \Omega'$, and $\Omega \longrightarrow \Omega'$ and $\Delta \Vdash^{dc}_{\tau'} \mathcal{D} \rightsquigarrow \tau \dashv \Theta$.*

**Lemma D.50** (Completeness of Typing Datatype Declaration). *Given $\Delta$ ok, and $\Delta \longrightarrow \Omega$, if $[\Omega]\Delta \vdash^{dt} \mathcal{T} \rightsquigarrow \Psi$, then there exists $\Theta$ and $\Omega'$ such that $\Theta \longrightarrow \Omega'$, and $\Omega \longrightarrow \Omega'$ and $\Delta \Vdash^{dt} \mathcal{T} \rightsquigarrow \Gamma \dashv \Theta$ and $\Psi = [\Omega']\Gamma$.*

**Theorem D.51** (Completeness of Typing a Group). *Given $\Omega$ ok, if $[\Omega]\Omega \vdash^{grp} \mathbf{rec}\ \overline{\mathcal{T}_i}^{i} \rightsquigarrow \overline{\kappa_i}^{i}; \overline{\Psi_i}^{i}$, then there exists $\overline{\kappa'_i}^{i}, \overline{\Gamma_i}^{i}, \Theta$, and $\Omega'$, such that $\Omega \Vdash^{grp} \mathbf{rec}\ \overline{\mathcal{T}_i}^{i} \rightsquigarrow \overline{\kappa'_i}^{i}; \overline{\Gamma_i}^{i} \dashv \Theta$, where $\Theta \longrightarrow \Omega'$, and $\overline{[\Omega']\kappa'_i = \kappa_i}^{i}$, and $\overline{\Psi_i = [\Omega']\Gamma_i}^{i}$.*

## D.2 Proofs

### D.2.1 Well-formedness of Declarative Type System.

**Lemma D.1** (Well-formedness of Declarative Typing Data Constructor Declaration). *If $\Sigma \vdash^{dc}_{\tau_1} \mathcal{D} \rightsquigarrow \tau_2$, then $\Sigma \vdash^{k} \tau_2 : \star$.*

Proof. We have

$$\frac{\Sigma \vdash^{\mathsf{k}} \overline{\tau_i}^i \to \tau : \star}{\Sigma \vdash^{\mathsf{dc}}_{\tau} D \overline{\tau_i}^i \rightsquigarrow \overline{\tau_i}^i \to \tau} \text{\scriptsize DC-DECL}$$

The goal follows trivially.

□

**Lemma D.2** (Well-formedness of Declarative Typing Datatype Declaration). *If $\Sigma \vdash^{\mathsf{dt}} \mathcal{T} \rightsquigarrow \Psi$, then $\Sigma \vdash \Psi$.*

Proof. We have

$$\frac{(T : \overline{\kappa_i}^i \to \star) \in \Sigma \qquad \overline{\Sigma, \overline{a_i : \kappa_i}^i \vdash^{\mathsf{dc}}_{T \overline{a_i}^i} \mathcal{D}_j \rightsquigarrow \tau_j}^j}{\Sigma \vdash^{\mathsf{dt}} \mathbf{data}\ T\ \overline{a_i}^i = \overline{\mathcal{D}_j}^j \rightsquigarrow \overline{D_j : \forall \overline{a_i : \kappa_i}^i.\tau_j}^j} \text{\scriptsize DT-DECL}$$

| | |
|---|---|
| $\overline{\Sigma, \overline{a_i : \kappa_i}^i \vdash^{\mathsf{k}} \tau_j : \star}^j$ | By Lemma D.1 |
| $\overline{\Sigma \vdash^{\mathsf{k}} \forall \overline{a_i : \kappa_i}^i.\tau_j : \star}^j$ | By rule K-FORALL |
| $\Sigma \vdash \overline{D_j : \forall \overline{a_i : \kappa_i}^i.\tau_j}^j$ | By rule ECTX-DCON |

□

*D.2.2 Well-formedness of Algorithmic Type System.* By Lemma D.15 we know that if $\Delta$ ok, and $\Delta \longrightarrow \Theta$, it follows that $\Theta$ ok. Therefore, in the following lemma when we have $\Delta$ ok and $\Delta \longrightarrow \Theta$, we always implicitly derive that $\Theta$ ok.

**Lemma D.3** (Well-formedness of Promotion). *If $\Delta_1, \widehat{\alpha}, \Delta_2$ ok, and $\Delta_1, \widehat{\alpha}, \Delta_2 \Vdash^{\mathsf{pr}}_{\widehat{\alpha}} \kappa_1 \rightsquigarrow \kappa_2 \dashv \Theta$, then $\Theta = \Theta_1, \widehat{\alpha}, \Theta_2$, and $\Delta_1, \widehat{\alpha}, \Delta_2 \longrightarrow \Theta$, and $\Theta_1 \Vdash^{\mathsf{kv}} \kappa_2$, and $\Theta$ ok. By weakening, there is also $\Theta \Vdash^{\mathsf{kv}} \kappa_2$.*

Proof. By induction on promotion.

- Case

$$\overline{\Delta \Vdash^{\mathsf{pr}}_{\widehat{\alpha}} \star \rightsquigarrow \star \dashv \Delta} \text{\scriptsize A-PR-STAR}$$

The goals hold trivially.

- Case

$$\frac{\Delta \Vdash^{\mathsf{pr}}_{\widehat{\alpha}} \kappa_1 \rightsquigarrow \kappa_3 \dashv \Delta_1 \qquad \Delta_1 \Vdash^{\mathsf{pr}}_{\widehat{\alpha}} [\Delta_1]\kappa_2 \rightsquigarrow \kappa_4 \dashv \Theta}{\Delta \Vdash^{\mathsf{pr}}_{\widehat{\alpha}} \kappa_1 \to \kappa_2 \rightsquigarrow \kappa_3 \to \kappa_4 \dashv \Theta} \text{\scriptsize A-PR-ARROW}$$

| | |
|---|---|
| $\Delta \longrightarrow \Delta_1 \land \Delta_1 = \Delta_{11}, \widehat{\alpha}, \Delta_{12} \land \Delta_{11} \Vdash^{\mathsf{kv}} \kappa_3$ | I.H. |
| $\Delta_1 \longrightarrow \Theta \land \Theta = \Theta_1, \widehat{\alpha}, \Theta_2 \land \Theta_1 \Vdash^{\mathsf{kv}} \kappa_4$ | I.H. |
| $\Delta \longrightarrow \Theta$ | By Lemma D.20 |
| $\Delta_{11} \longrightarrow \Theta_1$ | By Lemma D.17 |
| $\Theta_1 \Vdash^{\mathsf{kv}} \kappa_3$ | By Lemma D.10 |
| $\Theta_1 \Vdash^{\mathsf{kv}} \kappa_3 \to \kappa_4$ | By rule A-KV-ARROW |

- Case

$$\text{A-PR-KUVARL}$$
$$\overline{\Delta[\widehat{\beta}][\widehat{\alpha}] \Vdash^{\text{pr}}_{\widehat{\alpha}} \widehat{\beta} \rightsquigarrow \widehat{\beta} \dashv \Delta[\widehat{\beta}][\widehat{\alpha}]}$$

The goals hold trivially.

- Case

$$\text{A-PR-KUVARR}$$
$$\overline{\Delta[\widehat{\alpha}][\widehat{\beta}] \Vdash^{\text{pr}}_{\widehat{\alpha}} \widehat{\beta} \rightsquigarrow \widehat{\beta}_1 \dashv \Delta[\widehat{\beta}_1, \widehat{\alpha}][\widehat{\beta} = \widehat{\beta}_1]}$$

Most goals hold trivially. By Lemmas D.21 and D.23 and transitivity (Lemma D.20) we can prove $\Delta[\widehat{\alpha}][\widehat{\beta}] \longrightarrow \Delta[\widehat{\beta}_1, \widehat{\alpha}][\widehat{\beta} = \widehat{\beta}_1]$.

□

**Lemma D.4** (Well-formedness of Unification). *If $\Delta$ ok, and $\Delta \Vdash^{\text{u}} \kappa_1 \approx \kappa_2 \dashv \Theta$, then $\Delta \longrightarrow \Theta$, and $\Theta$ ok.*

Proof. By induction on the derivation of kind unification.

- Case

$$\text{A-U-REFL}$$
$$\overline{\Delta \Vdash^{\text{u}} \kappa \approx \kappa \dashv \Delta}$$

$\Delta \longrightarrow \Delta$ | By Lemma D.14

- Case

$$\text{A-U-ARROW}$$
$$\frac{\Delta \Vdash^{\text{u}} \kappa_1 \approx \kappa_3 \dashv \Theta_1 \qquad \Theta_1 \Vdash^{\text{u}} [\Theta_1]\kappa_2 \approx [\Theta_1]\kappa_4 \dashv \Theta}{\Delta \Vdash^{\text{u}} \kappa_1 \rightarrow \kappa_2 \approx \kappa_3 \rightarrow \kappa_4 \dashv \Theta}$$

$\Delta \longrightarrow \Theta_1$ | By I.H.
$\Theta_1 \longrightarrow \Theta$ | By I.H.
$\Delta \longrightarrow \Theta$ | By Lemma D.20

- Case

$$\text{A-U-KVARL}$$
$$\frac{\Delta \Vdash^{\text{pr}}_{\widehat{\alpha}} \kappa \rightsquigarrow \kappa_2 \dashv \Theta[\widehat{\alpha}]}{\Delta[\widehat{\alpha}] \Vdash^{\text{u}} \widehat{\alpha} \approx \kappa \dashv \Theta[\widehat{\alpha} = \kappa_2]}$$

$\Theta = \Theta_1, \widehat{\alpha}, \Theta_2 \wedge \Delta \longrightarrow \Theta[\widehat{\alpha}] \wedge \Theta_1 \Vdash^{\text{kv}} \kappa_2$ | By Lemma D.3
$\Theta \longrightarrow \Theta[\widehat{\alpha} = \kappa_2]$ | By Lemma D.21
$\Delta \longrightarrow \Theta[\widehat{\alpha} = \kappa_2]$ | By Lemma D.20

- Case

$$\text{A-U-KVARR}$$
$$\frac{\Delta \Vdash^{\text{pr}}_{\widehat{\alpha}} \kappa \rightsquigarrow \kappa_2 \dashv \Theta[\widehat{\alpha}]}{\Delta[\widehat{\alpha}] \Vdash^{\text{u}} \kappa \approx \widehat{\alpha} \dashv \Theta[\widehat{\alpha} = \kappa_2]}$$

Similar to the previous case.

□

**Lemma D.5** (Well-formedness of Application Kinding). *If $\Delta$ ok, and $\Delta \Vdash^{\text{kapp}} \kappa_1 \bullet \kappa_2 : \kappa \dashv \Theta$, then $\Delta \longrightarrow \Theta$, and $\Theta$ ok. Moreover, if $\Delta \Vdash^{\text{kv}} \kappa_1$, then we have $\Theta \Vdash^{\text{kv}} \kappa$.*

Proof. By induction on the derivation of application kinding.

- Case

$$\frac{\Delta[\widehat{\alpha}_1, \widehat{\alpha}_2, \widehat{\alpha} = \widehat{\alpha}_1 \rightarrow \widehat{\alpha}_2] \Vdash^{\mathsf{u}} \widehat{\alpha}_1 \approx \kappa \dashv \Theta}{\Delta[\widehat{\alpha}] \Vdash^{\mathsf{kapp}} \widehat{\alpha} \bullet \kappa : \widehat{\alpha}_2 \dashv \Theta} \text{ a-kapp-kuvar}$$

| | |
|---|---|
| $\Delta[\widehat{\alpha}] \longrightarrow \Delta[\widehat{\alpha}_1, \widehat{\alpha}_2, \widehat{\alpha}]$ | By Lemma D.23 |
| $\Delta[\widehat{\alpha}_1, \widehat{\alpha}_2, \widehat{\alpha}] \longrightarrow \Delta[\widehat{\alpha}_1, \widehat{\alpha}_2, \widehat{\alpha} = \widehat{\alpha}_1 \rightarrow \widehat{\alpha}_2]$ | By Lemma D.21 |
| $\Delta[\widehat{\alpha}_1, \widehat{\alpha}_2, \widehat{\alpha} = \widehat{\alpha}_1 \rightarrow \widehat{\alpha}_2] \longrightarrow \Theta$ | By Lemma D.4 |
| $\Delta \longrightarrow \Theta$ | By Lemma D.20 |
| $\Delta[\widehat{\alpha}_1, \widehat{\alpha}_2, \widehat{\alpha} = \widehat{\alpha}_1 \rightarrow \widehat{\alpha}_2] \Vdash^{\mathsf{kv}} \widehat{\alpha}_2$ | By rule a-kv-kuvar |
| $\Theta \Vdash^{\mathsf{kv}} \widehat{\alpha}_2$ | By Lemma D.10 |

- Case

$$\frac{\Delta \Vdash^{\mathsf{u}} \kappa_1 \approx \kappa \dashv \Theta}{\Delta \Vdash^{\mathsf{kapp}} \kappa_1 \rightarrow \kappa_2 \bullet \kappa : \kappa_2 \dashv \Theta} \text{ a-kapp-arrow}$$

| | |
|---|---|
| $\Delta \longrightarrow \Theta$ | By Lemma D.4 |
| $\Delta \Vdash^{\mathsf{kv}} \kappa_1 \rightarrow \kappa_2$ | Given |
| $\Delta \Vdash^{\mathsf{kv}} \kappa_2$ | By inversion |
| $\Theta \Vdash^{\mathsf{kv}} \kappa_2$ | By Lemma D.10 |

□

**Lemma D.6** (Well-formedness of Kinding). *If $\Delta$ ok, and $\Delta \Vdash^{\mathsf{k}} \sigma : \kappa \dashv \Theta$, then $\Delta \longrightarrow \Theta$, and $\Theta$ ok, and $\Theta \Vdash^{\mathsf{kv}} \kappa$ and $\Theta \Vdash^{\mathsf{kc}} \sigma \Leftarrow \kappa$.*

Proof. By induction on the derivation of kinding.

- Case for rules a-k-nat, a-k-var, a-k-tcon, and a-k-arrow follows trivially.
- Case

$$\frac{\Delta \Vdash^{\mathsf{kv}} \kappa \quad \Delta, a : \kappa \Vdash^{\mathsf{k}} \sigma : \kappa_2 \dashv \Theta, a : \kappa \quad [\Theta]\kappa_2 = \star}{\Delta \Vdash^{\mathsf{k}} \forall a : \kappa.\sigma : \star \dashv \Theta} \text{ a-k-forall}$$

| | |
|---|---|
| $\Delta, a : \kappa \longrightarrow \Theta, a : \kappa \wedge \Theta, a : \kappa \Vdash^{\mathsf{kc}} \sigma \Leftarrow \star$ | By I.H. |
| $\Delta \longrightarrow \Theta$ | By inversion |
| $\Theta \Vdash^{\mathsf{kv}} \star$ | By rule a-kv-star |
| $\Delta \Vdash^{\mathsf{kv}} \kappa$ | Given |
| $\Theta \Vdash^{\mathsf{kv}} \kappa$ | By Lemma D.10 |
| $\Theta \Vdash^{\mathsf{kc}} \forall a : \kappa.\sigma \Leftarrow \star$ | By rules a-kc-eq and a-k-forall |

- Case

$$\frac{\Delta \Vdash^{\mathsf{k}} \tau_1 : \kappa_1 \dashv \Theta_1 \quad \Theta_1 \Vdash^{\mathsf{k}} \tau_2 : \kappa_2 \dashv \Theta_2 \quad \Theta_2 \Vdash^{\mathsf{kapp}} [\Theta_2]\kappa_1 \bullet [\Theta_2]\kappa_2 : \kappa_3 \dashv \Theta}{\Delta \Vdash^{\mathsf{k}} \tau_1 \tau_2 : \kappa_3 \dashv \Theta} \text{ a-k-app}$$

| | |
|---|---|
| $\Delta \longrightarrow \Theta_1 \wedge \Theta_1 \Vdash^{\mathsf{kv}} \kappa_1 \wedge \Theta_1 \Vdash^{\mathsf{kc}} \tau_1 \Leftarrow \kappa_1$ | I.H. |
| $\Theta_1 \longrightarrow \Theta_2 \wedge \Theta_2 \Vdash^{\mathsf{kv}} \kappa_2 \wedge \Theta_2 \Vdash^{\mathsf{kc}} \tau_2 \Leftarrow \kappa_2$ | I.H. |
| $\Theta_2 \longrightarrow \Theta$ | By Lemma D.5 |
| $\Theta_1 \longrightarrow \Theta$ | By Lemma D.20 |
| $\Delta \longrightarrow \Theta$ | By Lemma D.20 |
| $\Theta_2 \Vdash^{\mathsf{kv}} \kappa_1$ | By Lemma D.10 |
| $\Theta_2 \Vdash^{\mathsf{kv}} [\Theta_2]\kappa_1$ | By Lemma D.12 |
| $\Theta \Vdash^{\mathsf{kv}} \kappa_3$ | By Lemma D.5 |
| $\Theta \Vdash^{\mathsf{kc}} \tau_1 \Leftarrow \kappa_1$ | By Lemma D.10 |
| $\Theta \Vdash^{\mathsf{kc}} \tau_2 \Leftarrow \kappa_2$ | By Lemma D.10 |
| $[\Theta]([\Theta_2]\kappa_1) = [\Theta]([\Theta_2]\kappa_2) \to [\Theta]\kappa_3$ | By Lemma D.40 |
| $[\Theta]\kappa_1 = [\Theta]\kappa_2 \to [\Theta]\kappa_3$ | By Lemma D.18 |
| $\Theta \Vdash^{\mathsf{kapp}} [\Theta]\kappa_1 \bullet [\Theta]\kappa_2 : [\Theta]\kappa_3 \dashv \Theta$ | By rule A-KAPP-ARROW and rule A-U-REFL |
| $\Theta \Vdash^{\mathsf{kc}} \tau_1\, \tau_2 \Leftarrow \kappa_3$ | By rules A-KC-EQ and A-K-APP |

□

**Lemma D.7** (Well-formedness of Typing Data Constructor Declarations). *If $\Delta$ ok, and $\Delta \Vdash^{\mathsf{dc}}_{\tau'} \mathcal{D} \rightsquigarrow \tau \dashv \Theta$, then $\Delta \longrightarrow \Theta$, and $\Theta$ ok. and $\Theta \Vdash^{\mathsf{kc}} \tau \Leftarrow \star$.*

Proof.

$$\frac{\Delta \Vdash^{\mathsf{k}} \overline{\tau_i}^{\,i} \to \tau : \star \dashv \Theta}{\Delta \Vdash^{\mathsf{dc}}_{\tau} D\, \overline{\tau_i}^{\,i} \rightsquigarrow \overline{\tau_i}^{\,i} \to \tau \dashv \Theta} \text{A-DC-DECL}$$

Follows directly from Lemma D.6.

□

**Lemma D.8** (Well-formedness of Typing Datatype Declaration). *If $\Delta$ ok, and $\Delta \Vdash^{\mathsf{dt}} \mathcal{T} \rightsquigarrow \Gamma \dashv \Theta$, then $\Delta \longrightarrow \Theta$, and $\Theta$ ok, and $\Theta \Vdash^{\mathsf{ectx}} \Gamma$.*

Proof.
A-DT-DECL
$$\frac{(T:\kappa) \in \Delta \quad \Delta, \overline{\widehat{\alpha_i}}^{\,i} \Vdash^{\mu} [\Delta]\kappa \approx (\overline{\widehat{\alpha_i}}^{\,i} \to \star) \dashv \Theta_1, \overline{\widehat{\alpha_i} = \kappa_i}^{\,i} \quad \overline{\Theta_j, \overline{a_i:\kappa_i}^{\,i} \Vdash^{\mathsf{dc}}_{T\,\overline{a_i}^{\,i}} \mathcal{D}_j \rightsquigarrow \tau_j \dashv \Theta_{j+1}, \overline{a_i:\kappa_i}^{\,i}}^{\,j}}{\Delta \Vdash^{\mathsf{dt}} \text{data } T\, \overline{a_i}^{\,i} = \overline{\mathcal{D}_j}^{\,j \in 1..n} \rightsquigarrow \overline{D_j : \forall \overline{a_i:\kappa_i}^{\,i}.\tau_j}^{\,j} \dashv \Theta_{n+1}}$$

| | |
|---|---|
| $\Delta \longrightarrow \Delta, \overline{\widehat{\alpha_i}}^{\,i}$ | By rule A-CTXE-ADD |
| $\Delta, \overline{\widehat{\alpha_i}}^{\,i} \longrightarrow \Theta_1, \overline{\widehat{\alpha_i} = \kappa_i}^{\,i}$ | By Lemma D.4 |
| $\Delta \longrightarrow \Theta_1$ | By Lemma D.17 |
| $\Theta_1, \overline{a_i:\kappa_i}^{\,i} \longrightarrow \Theta_{n+1}, \overline{a_i:\kappa_i}^{\,i}$ | By Lemma D.7 and Lemma D.20 |
| $\Theta_1 \longrightarrow \Theta_{n+1}$ | By inversion |
| $\Delta \longrightarrow \Theta$ | By Lemma D.20 |
| $\overline{\Theta_{j+1}, \overline{a_i:\kappa_i}^{\,i} \Vdash^{\mathsf{kc}} \tau_j \Leftarrow \star}^{\,j \in 1..n}$ | By Lemma D.7 |
| $\overline{\Theta_n, \overline{a_i:\kappa_i}^{\,i} \Vdash^{\mathsf{kc}} \tau_j \Leftarrow \star}^{\,j \in 1..n}$ | By Lemma D.10 |
| $\overline{\Theta_n \Vdash^{\mathsf{kc}} \forall \overline{a_i:\kappa_i}^{\,i}.\tau_j \Leftarrow \star}^{\,j \in 1..n}$ | By rules A-KC-EQ and A-K-FORALL |
| $\overline{\Theta_n \Vdash^{\mathsf{ectx}} D_j : \forall \overline{a_i:\widehat{\alpha_i}}^{\,i}.\tau_j}^{\,j \in 1..n}$ | By rule A-ECTX-DCON |

□

### D.2.3 Properties of Context Extension.

**Lemma D.9** (Declaration Preservation). *If $\Delta \longrightarrow \Theta$, if a type constructor or a type variable or a kind unification variable is declared in $\Delta$, then it is declared in $\Theta$.*

Proof. By a straightforward induction on $\Delta \longrightarrow \Theta$. □

**Lemma D.10** (Extension Weakening). *Given $\Delta \longrightarrow \Theta$,*
- *if $\Delta \Vdash^{kv} \kappa$, then $\Theta \Vdash^{kv} \kappa$;*
- *if $\Delta \Vdash^{kc} \sigma \Leftarrow \kappa$, then $\Theta \Vdash^{kc} \sigma \Leftarrow \kappa$.*

Proof. **Part 1** By induction on $\Delta \Vdash^{kv} \kappa$.
- Case

$$\frac{}{\Delta \Vdash^{kv} \star} \text{a-kv-star}$$

The goal holds trivially.
- Case

$$\frac{\Delta \Vdash^{kv} \kappa_1 \qquad \Delta \Vdash^{kv} \kappa_2}{\Delta \Vdash^{kv} \kappa_1 \to \kappa_2} \text{a-kv-arrow}$$

The goal holds directly from I.H..
- Case

$$\frac{\widehat{\alpha} \in \Delta}{\Delta \Vdash^{kv} \widehat{\alpha}} \text{a-kv-kuvar}$$

The goal holds directly from Lemma D.9.

**Part 2** By induction on $\Delta \Vdash^{k} \sigma : \kappa \dashv \Delta$.
- The case for rules a-k-nat and a-k-arrow holds trivially.
- The case for rules a-k-var and a-k-tcon holds from Lemma D.7 and Lemma D.18.
- The case for rule a-k-forall holds from I.H. and Lemma D.18.
- The case for rule a-k-app depends on the extension weakening of application kinding. Given the hypothesis, it's impossible for the derivation to ever use rule a-kapp-kuvar. The extension weakening on rule a-kapp-arrow then depends on the extension weakening of kind unification. Given the hypothesis, it's impossible for the derivation to ever use rules a-u-kvarL and a-u-kvarR. The case for rule a-u-rrefl holds trivially, and the case for rule a-u-arrow holds directly from I.H..

□

**Lemma D.12** (Substitution Kinding). *If $\Delta$ ok, and $\Delta \Vdash^{kv} \kappa$, then $\Delta \Vdash^{kv} [\Delta]\kappa$.*

Proof. By induction on $|\Delta \vdash \kappa|$. We then case analyze $\kappa$.
- $\kappa = \star$. The goal holds trivially.
- $\kappa = \kappa_1 \to \kappa_2$. The goal directly from I.H..
- $\kappa = \widehat{\alpha}$. If $\widehat{\alpha}$ is unsolved in $\Delta$, then the goal holds directly. Or otherwise we have $\Delta = \Delta_1, \widehat{\alpha} = \kappa, \Delta_2$. Because $\Delta$ ok, we have $\Delta_1 \Vdash^{kv} \kappa$ and $|\Delta_1 \vdash \kappa| = |\Delta \vdash \kappa|$, which is less then $|\Delta \vdash \widehat{\alpha}|$. So we apply I.H. to get the goal.

□

**Lemma D.13** (Context Extension with Defaulting is Context Extension). *If $\Delta \longrightarrow\!\!\!\!\!\rightarrow \Theta$, then $\Delta \longrightarrow \Theta$.*

Proof. By straightforward induction on $\Delta \longrightarrow\!\!\!\!\!\rightarrow \Theta$. □

**Lemma D.14** (Reflexivity of Context Extension). *If $\Delta$ ok, then $\Delta \longrightarrow \Delta$.*

Proof. By straightforward induction on $\Delta$ ok. The conclusion follows directly from the definition. □

**Lemma D.15** (Well-formedness of Context Extension). *If $\Delta$ ok, and $\Delta \longrightarrow \Theta$, then $\Theta$ ok.*

Proof. By induction on $\Delta \longrightarrow \Theta$.

- Case
$$\frac{}{\bullet \longrightarrow \bullet} \text{a-ctxe-empty}$$
Follows directly by rule a-tctx-empty.

- Case
$$\frac{\Delta \longrightarrow \Theta}{\Delta, a : \kappa \longrightarrow \Theta, a : \kappa} \text{a-ctxe-tvar}$$

| | |
|---|---|
| $\Delta, a : \kappa$ ok | Given |
| $\Delta \Vdash^{kv} \kappa$ | By inversion |
| $\Delta \longrightarrow \Theta$ | Given |
| $\Theta \Vdash^{kv} \kappa$ | By lemma D.10 |
| $\Theta$ ok | I.H. |
| $\Theta, a : \kappa$ ok | By rule a-tctx-tvar |

- Case
$$\frac{\Delta \longrightarrow \Theta}{\Delta, T : \kappa \longrightarrow \Theta, T : \kappa} \text{a-ctxe-tcon}$$
This case is similar to the case for rule a-tctx-tvar.

- Case
$$\frac{\Delta \longrightarrow \Theta}{\Delta, \widehat{\alpha} \longrightarrow \Theta, \widehat{\alpha}} \text{a-ctxe-kuvar}$$
The goal holds directly from I.H. and rule a-tctx-kuvar.

- Case
$$\frac{\Delta \longrightarrow \Theta \quad [\Theta]\kappa_1 = [\Theta]\kappa_2}{\Delta, \widehat{\alpha} = \kappa_1 \longrightarrow \Theta, \widehat{\alpha} = \kappa_2} \text{a-ctxe-kuvarSolved}$$

| | |
|---|---|
| $\Theta$ ok | I.H. |
| $\Delta, \widehat{\alpha} = \kappa_1$ ok | Given |
| $\Delta \Vdash^{kv} \kappa_1$ | By inversion |
| $\Delta \longrightarrow \Theta$ | Given |
| $\Theta \Vdash^{kv} \kappa_1$ | By lemma D.10 |

Suppose $\kappa_2$ is not well-formed under $\Theta$, then it must contain kind unification variables that are not in $\Theta$. Then it is impossible to have $[\Theta]\kappa_1 = [\Theta]\kappa_2$ given $\Theta \Vdash^{kv} \kappa_1$. Thus by contradiction we have $\Theta \Vdash^{kv} \kappa_2$. Then $\Theta, \widehat{\alpha} = \kappa_2$ ok by rule a-tctx-tcon.

- Case

  $$\frac{\Delta \longrightarrow \Theta \qquad \Theta \Vdash^{kv} \kappa}{\Delta, \widehat{\alpha} \longrightarrow \Theta, \widehat{\alpha} = \kappa} \text{ a-ctxe-solve}$$

  The goal holds directly from I.H. and rule a-tctx-kuvarSolved.

- Case

  $$\frac{\Delta \longrightarrow \Theta}{\Delta \longrightarrow \Theta, \widehat{\alpha}} \text{ a-ctxe-add}$$

  The goal holds directly from I.H. and rule a-tctx-kuvar.

- Case

  $$\frac{\Delta \longrightarrow \Theta \qquad \Theta \Vdash^{kv} \kappa}{\Delta \longrightarrow \Theta, \widehat{\alpha} = \kappa} \text{ a-ctxe-addSolved}$$

  The goal holds directly from I.H. and rule a-tctx-kuvarSolved.

□

**Lemma D.17** (Extension Order).

(1) If $\Delta_1, a : \kappa, \Delta_2 \longrightarrow \Theta$, then $\Theta = \Theta_1, a : \kappa, \Theta_2$, where $\Delta_1 \longrightarrow \Theta_1$. Moreover, if $\Delta_2$ **soft**, then $\Theta_2$ **soft**.
(2) If $\Delta_1, T : \kappa, \Delta_2 \longrightarrow \Theta$, then $\Theta = \Theta_1, T : \kappa, \Theta_2$, where $\Delta_1 \longrightarrow \Theta_1$. Moreover, if $\Delta_2$ **soft**, then $\Theta_2$ **soft**.
(3) If $\Delta_1, \widehat{\alpha}, \Delta_2 \longrightarrow \Theta$, then $\Theta = \Theta_1, \Theta', \Theta_2$, where $\Delta_1 \longrightarrow \Theta_1$, and $\Theta'$ is either $\widehat{\alpha}$ or $\widehat{\alpha} = \kappa$ for some $\kappa$. Moreover, if $\Delta_2$ **soft**, then $\Theta_2$ **soft**.
(4) If $\Delta_1, \widehat{\alpha} = \kappa_1, \Delta_2 \longrightarrow \Theta$, then $\Theta = \Theta_1, \widehat{\alpha} = \kappa_2, \Theta_2$, where $\Delta_1 \longrightarrow \Theta_1$, and $[\Theta_1]\kappa_1 = [\Theta_1]\kappa_2$. Moreover, if $\Delta_2$ **soft**, then $\Theta_2$ **soft**.

Proof. We give the detailed proof for the first part. The proof for the rest parts is similar.
By induction on $\Delta_1, a : \kappa, \Delta_2 \longrightarrow \Theta$.

- Case $\Delta = \bullet$ by rule a-ctxe-empty. This case is impossible.
- Case $\Delta_1, a : \kappa \longrightarrow \Theta', a : \kappa$ by rule a-ctxe-tvar when $\Delta_2$ is empty. In this case, let $\Theta_1 = \Theta'$ and $\Theta_2$ be empty. All goals follow directly.
- Case $\Delta_1, a : \kappa, \Delta', b : \kappa_2 \longrightarrow \Theta', b : \kappa_2$ by rule a-ctxe-tvar where $\Delta_2 = \Delta', b : \kappa_2$ and $\Delta_1, a : \kappa, \Delta' \longrightarrow \Theta'$. By I.H. we have $\Theta' = \Theta_1, a : \kappa, \Theta'_2$ and $\Delta_1 \longrightarrow \Theta_1$. Let $\Theta_2 = \Theta'_2, b : \kappa_2$ and all goals follow directly.
- Case $\Delta_1, a : \kappa, \Delta', T : \kappa_2 \longrightarrow \Theta', T : \kappa_2$ by rule a-ctxe-tcon where $\Delta_2 = \Delta', T : \kappa_2$ and $\Delta_1, a : \kappa, \Delta' \longrightarrow \Theta'$. This case is similar to the above case.
- Case $\Delta_1, a : \kappa, \Delta', \widehat{\alpha} \longrightarrow \Theta', \widehat{\alpha}$ by rule a-ctxe-kuvar where $\Delta_2 = \Delta', \widehat{\alpha}$ and $\Delta_1, a : \kappa, \Delta' \longrightarrow \Theta'$. By I.H. we have $\Theta' = \Theta_1, a : \kappa, \Theta'_2$ and $\Delta_1 \longrightarrow \Theta_1$. Let $\Theta_2 = \Theta'_2, \widehat{\alpha}$ and all goals follow directly. And if $\Delta'$ **soft**, by I.H. we have $\Theta'_2$ **soft**. By definition we have $\Theta_2$ **soft**.
- Case for rules a-ctxe-kuvarSolved, a-ctxe-solve, a-ctxe-add, and a-ctxe-addSolved are similar to the above case.

□

**Lemma D.18** (Substitution Extension Invariance). *If $\Delta$ ok, and $\Delta \Vdash^{kv} \kappa$, and $\Delta \longrightarrow \Theta$, then $[\Theta]\kappa = [\Theta]([\Delta]\kappa)$ and $[\Theta]\kappa = [\Delta]([\Theta]\kappa)$. As a corollary, if $\Delta \Vdash^{kv} \kappa_1$, $\Delta \Vdash^{kv} \kappa_2$, and $[\Delta]\kappa_1 = [\Delta]\kappa_2$, then $[\Theta]\kappa_1 = [\Theta]\kappa_2$.*

Proof. Because $\Delta \Vdash^{\mathsf{kv}} \kappa$, so every solved kind unification variable in $\Delta$ is solved in $\Theta$. Therefore $[\Theta]\kappa = [\Delta]([\Theta]\kappa)$.

To show that $[\Theta]\kappa = [\Theta]([\Delta]\kappa)$, we do induction on $\mid \Delta \vdash \kappa \mid$.

- $$\frac{}{\Delta \Vdash^{\mathsf{kv}} \star} \text{{\scriptsize A-KV-STAR}}$$

  The goal follows trivially.

- $$\frac{\Delta \Vdash^{\mathsf{kv}} \kappa_1 \quad \Delta \Vdash^{\mathsf{kv}} \kappa_2}{\Delta \Vdash^{\mathsf{kv}} \kappa_1 \to \kappa_2} \text{{\scriptsize A-KV-ARROW}}$$

  The goal follows directly from I.H..

- $$\frac{\widehat{\alpha} \in \Delta}{\Delta \Vdash^{\mathsf{kv}} \widehat{\alpha}} \text{{\scriptsize A-KV-KUVAR}}$$

  There are two subcases. Firstly, $\widehat{\alpha}$ is unsolved in $\Delta$. Then $[\Theta]([\Delta]\widehat{\alpha}) = [\Theta]\widehat{\alpha}$ follows directly. Or we have $\Delta = \Delta_1, \widehat{\alpha} = \kappa, \Delta_2$. Then by Lemma D.17 we have $\Theta = \Theta_1, \widehat{\alpha} = \kappa', \Theta_2$ and $[\Theta_1]\kappa = [\Theta_1]\kappa'$. Because $\mid \Delta \vdash \kappa \mid < \mid \Delta \vdash \widehat{\alpha} \mid$, by I.H., we know that $[\Theta]\kappa = [\Theta]([\Delta]\kappa)$. Therefore, $[\Theta]\widehat{\alpha} = [\Theta]\kappa' = [\Theta]\kappa = [\Theta]([\Delta]\kappa) = [\Theta]([\Delta]\widehat{\alpha})$.

For the corollary, we have $[\Theta]\kappa_1 = [\Theta]([\Delta]\kappa_1) = [\Theta]([\Delta]\kappa_2) = [\Theta]\kappa_2$. □

**Lemma D.19** (Substitution Stability). *If $\Delta_1, \Delta_2$ ok, and $\Delta_1 \Vdash^{\mathsf{kv}} \kappa$, then $[\Delta_1, \Delta_2]\kappa = [\Delta_1]\kappa$.*

Proof. Follows directly as $\kappa$ and $\Delta_1$ do not contain kind variables in $\Delta_2$. □

**Lemma D.20** (Transitivity of Context Extension). *If $\Delta'$ ok, and $\Delta' \longrightarrow \Delta$, and $\Delta \longrightarrow \Theta$, then $\Delta' \longrightarrow \Theta$.*

Proof. By induction on $\Delta \longrightarrow \Theta$.

- Case
  $$\frac{}{\bullet \longrightarrow \bullet} \text{{\scriptsize A-CTXE-EMPTY}}$$

  We have $\Delta' \longrightarrow \bullet$ as given.

- Case
  $$\frac{\Delta \longrightarrow \Theta}{\Delta, a : \kappa \longrightarrow \Theta, a : \kappa} \text{{\scriptsize A-CTXE-TVAR}}$$

| | |
|---|---|
| $\Delta' \longrightarrow \Delta, a : \kappa$ | Given |
| $\Delta' = \Delta_1, a : \kappa \wedge \Delta_1 \longrightarrow \Delta$ | By inversion |
| $\Delta_1 \longrightarrow \Theta$ | I.H. |
| $\Delta_1, a : \kappa \longrightarrow \Theta, a : \kappa$ | By rule A-CTXE-TVAR |

- Case
$$\frac{\Delta \longrightarrow \Theta}{\Delta, T : \kappa \longrightarrow \Theta, T : \kappa} \text{ a-ctxe-tcon}$$

This case is similar to the case for rule a-ctxe-tvar.

- Case
$$\frac{\Delta \longrightarrow \Theta}{\Delta, \widehat{\alpha} \longrightarrow \Theta, \widehat{\alpha}} \text{ a-ctxe-kuvar}$$

Since $\Delta' \longrightarrow \Delta, \widehat{\alpha}$, the derivation must conclude with either rule a-ctxe-kuvar or rule a-ctxe-add.
  – By rule a-ctxe-kuvar.

| | |
|---|---|
| $\Delta' = \Delta_1, \widehat{\alpha} \wedge \Delta_1 \longrightarrow \Delta$ | Given |
| $\Delta_1 \longrightarrow \Theta$ | I.H. |
| $\Delta_1, \widehat{\alpha} \longrightarrow \Theta, \widehat{\alpha}$ | By rule a-ctxe-kuvar |

  – By rule a-ctxe-add.

| | |
|---|---|
| $\Delta' \longrightarrow \Delta$ | Given |
| $\Delta' \longrightarrow \Theta$ | I.H. |
| $\Delta' \longrightarrow \Theta, \widehat{\alpha}$ | By rule a-ctxe-add |

- Case
$$\frac{\Delta \longrightarrow \Theta \quad [\Theta]\kappa_1 = [\Theta]\kappa_2}{\Delta, \widehat{\alpha} = \kappa_1 \longrightarrow \Theta, \widehat{\alpha} = \kappa_2} \text{ a-ctxe-kuvarSolved}$$

Since $\Delta' \longrightarrow \Delta, \widehat{\alpha} = \kappa_1$, the derivation must conclude with either rule a-ctxe-kuvarSolved or rule a-ctxe-addSolved.
  – By rule a-ctxe-kuvarSolved.

| | |
|---|---|
| $\Delta' = \Delta_1, \widehat{\alpha} = \kappa_0 \wedge \Delta_1 \longrightarrow \Delta \wedge [\Delta]\kappa_0 = [\Delta]\kappa_1$ | Given |
| $\Delta_1 \longrightarrow \Theta$ | I.H. |
| $[\Theta]\kappa_0$ | |
| $= [\Theta]\kappa_1$ | By Lemma D.18 |
| $= [\Theta]\kappa_2$ | Given |
| $\Delta_1, \widehat{\alpha} = \kappa_0 \longrightarrow \Theta, \widehat{\alpha} = \kappa_2$ | By rule a-ctxe-kuvar |

  – By rule a-ctxe-addSolved.

| | |
|---|---|
| $\Delta' \longrightarrow \Delta$ | Given |
| $\Delta' \longrightarrow \Theta$ | I.H. |
| $\Delta' \longrightarrow \Theta, \widehat{\alpha} = \kappa_2$ | By rule a-ctxe-addSolved |

- Case
$$\frac{\Delta \longrightarrow \Theta \quad \Theta \Vdash^{\mathsf{kv}} \kappa}{\Delta, \widehat{\alpha} \longrightarrow \Theta, \widehat{\alpha} = \kappa} \text{ a-ctxe-solve}$$

Since $\Delta' \longrightarrow \Delta, \widehat{\alpha}$, the derivation must conclude with either rule a-ctxe-kuvar or rule a-ctxe-add.
  – By rule a-ctxe-kuvar.

$\Delta' = \Delta_1, \widehat{\alpha} \wedge \Delta_1 \longrightarrow \Delta$ | Given
$\Delta_1 \longrightarrow \Theta$ | I.H.
$\Delta_1, \widehat{\alpha} \longrightarrow \Theta, \widehat{\alpha} = \kappa$ | By rule A-CTXE-SOLVE

– By rule A-CTXE-ADD.

$\Delta' \longrightarrow \Delta$ | Given
$\Delta' \longrightarrow \Theta$ | I.H.
$\Delta' \longrightarrow \Theta, \widehat{\alpha} = \kappa$ | By rule A-CTXE-ADDSOLVED

- Case

$$\frac{\Delta \longrightarrow \Theta}{\Delta \longrightarrow \Theta, \widehat{\alpha}} \text{ A-CTXE-ADD}$$

$\Delta' \longrightarrow \Theta$ | I.H.
$\Delta' \longrightarrow \Theta, \widehat{\alpha}$ | By rule A-CTXE-ADD

- Case

$$\frac{\Delta \longrightarrow \Theta \quad \Theta \Vdash^{\mathsf{kv}} \kappa}{\Delta \longrightarrow \Theta, \widehat{\alpha} = \kappa} \text{ A-CTXE-ADDSOLVED}$$

$\Delta' \longrightarrow \Theta$ | I.H.
$\Delta' \longrightarrow \Theta, \widehat{\alpha} = \kappa$ | By rule A-CTXE-ADDSOLVED

□

**Lemma D.21** (Solution Admissibility for Extension). *If $\Delta_1, \widehat{\alpha}, \Delta_2$ ok and $\Delta_1 \Vdash^{\mathsf{kv}} \kappa$, then $\Delta_1, \widehat{\alpha}, \Delta_2 \longrightarrow \Delta_1, \widehat{\alpha} = \kappa, \Delta_2$.*

PROOF. By induction on $\Delta_2$.
- Case $\Delta_2$ is empty. Then $\Delta_1 \longrightarrow \Delta_1$ by Lemma D.14, and $\Delta_1, \widehat{\alpha} \longrightarrow \Delta_1, \widehat{\alpha} = \kappa$ holds by rule A-CTXE-SOLVE.
- Case $\Delta_2 = \Delta_2', a : \kappa$. By I.H., we $\Delta_1, \widehat{\alpha}, \Delta_2' \longrightarrow \Delta_1, \widehat{\alpha} = \kappa, \Delta_2'$. Then by rule A-CTXE-TVAR we are done.
- Case $\Delta_2 = \Delta_2', T : \kappa$. By I.H. and rule A-CTXE-TCON.
- Case $\Delta_2 = \Delta_2', \widehat{\alpha}$. By I.H. and rule A-CTXE-KUVAR.
- Case $\Delta_2 = \Delta_2', \widehat{\alpha} = \kappa$. By I.H. and rule A-CTXE-KUVARSOLVED.

□

**Lemma D.22** (Solved Variable Addition for Extension). *If $\Delta_1, \Delta_2$ ok and $\Delta_1 \Vdash^{\mathsf{kv}} \kappa$, then $\Delta_1, \Delta_2 \longrightarrow \Delta_1, \widehat{\alpha} = \kappa, \Delta_2$.*

PROOF. The proof is exactly the same as the one for Lemma D.21. Except for the case when $\Delta_2$ is empty, we use rule A-CTXE-ADDSOLVED.

□

**Lemma D.23** (Unsolved Variable Addition). *If $\Delta_1, \Delta_2$ ok then $\Delta_1, \Delta_2 \longrightarrow \Delta_1, \widehat{\alpha}, \Delta_2$.*

PROOF. The proof is exactly the same as the one for Lemma D.21. Except for the case when $\Delta_2$ is empty, we use rule A-CTXE-ADD.

□

**Lemma D.24** (Parallel Admissibility). *If $\Delta_1 \longrightarrow \Theta_1$, and $\Delta_1, \Delta_2$ ok, and $\Delta_1, \Delta_2 \longrightarrow \Theta_1, \Theta_2$, and $\Delta_2$ is fresh w.r.t. $\Theta_1$, then:*

- $\Delta_1, \widehat{\alpha}, \Delta_2 \longrightarrow \Theta_1, \widehat{\alpha}, \Theta_2$
- *If $\Theta_1 \Vdash^{\mathsf{kv}} \kappa$, then $\Delta_1, \widehat{\alpha}, \Delta_2 \longrightarrow \Theta_1, \widehat{\alpha} = \kappa, \Theta_2$*
- *If $[\Theta_1]\kappa_1 = [\Theta_1]\kappa_2$, then $\Delta_1, \widehat{\alpha} = \kappa_1, \Delta_2 \longrightarrow \Theta_1, \widehat{\alpha} = \kappa_2, \Theta_2$*

Proof. **Part 1** By induction on $\Theta_2$.
- $\Theta_2 = \bullet$. Because $\Delta_2$ is fresh w.r.t. $\Theta_1$, we must have $\Delta_2 = \bullet$. We have $\Delta_1, \widehat{\alpha} \longrightarrow \Theta_1, \widehat{\alpha}$ by rule a-ctxe-kuvar.
- $\Theta_2 = \Theta_2', a : \kappa$. Then the derivation of $\Delta_1, \Delta_2 \longrightarrow \Theta_1, \Theta_2$ must conclude with rule a-ctxe-tvar. It must be $\Delta_2 = \Delta_2', a : \kappa$. (Or otherwise if $(a : \kappa) \in \Delta_1$, then we must have $(a : \kappa) \in \Theta_1$ by Lemma D.9, and $\Theta_1, \Theta_2$ is no longer well-formed.)

| | |
|---|---|
| $\Delta_1, \Delta_2', a : \kappa \longrightarrow \Theta_1, \Theta_2', a : \kappa$ | Given |
| $\Delta_1, \Delta_2' \longrightarrow \Theta_1, \Theta_2'$ | By inversion |
| $\Delta_1, \widehat{\alpha}, \Delta_2' \longrightarrow \Theta_1, \widehat{\alpha}, \Theta_2'$ | I.H. |
| $\Delta_1, \widehat{\alpha}, \Delta_2', a : \kappa \longrightarrow \Theta_1, \widehat{\alpha}, \Theta_2', a : \kappa$ | By rule a-ctxe-tvar |

- $\Theta_2 = \Theta_2', T : \kappa$ This case is similar to the case when $\Theta_2 = \Theta_2', a : \kappa$, except that we reason using rule a-ctxe-tcon.
- $\Theta_2 = \Theta_2', \widehat{\alpha}_1$ Then the derivation of $\Delta_1, \Delta_2 \longrightarrow \Theta_1, \Theta_2$ must conclude with either rule a-ctxe-kuvar or rule a-ctxe-add.
  - Subcase: the derivation concludes with rule a-ctxe-kuvar. It must be $\Delta_2 = \Delta_2', \widehat{\alpha}_1$.

| | |
|---|---|
| $\Delta_1, \Delta_2' \longrightarrow \Theta_1, \Theta_2'$ | Given |
| $\Delta_1, \widehat{\alpha}, \Delta_2' \longrightarrow \Theta_1, \widehat{\alpha}, \Theta_2'$ | I.H. |
| $\Delta_1, \widehat{\alpha}, \Delta_2', \widehat{\alpha}_1 \longrightarrow \Theta_1, \widehat{\alpha}, \Theta_2', \widehat{\alpha}_1$ | By rule a-ctxe-kuvar |

  - Subcase: the derivation concludes with rule a-ctxe-add.

| | |
|---|---|
| $\Delta_1, \Delta_2 \longrightarrow \Theta_1, \Theta_2'$ | Given |
| $\Delta_1, \widehat{\alpha}, \Delta_2 \longrightarrow \Theta_1, \widehat{\alpha}, \Theta_2'$ | I.H. |
| $\Delta_1, \widehat{\alpha}, \Delta_2 \longrightarrow \Theta_1, \widehat{\alpha}, \Theta_2', \widehat{\alpha}_1$ | By rule a-ctxe-add |

- $\Theta_2 = \Theta_2', \widehat{\alpha}_1 = \kappa$. Then the derivation of $\Delta_1, \Delta_2 \longrightarrow \Theta_1, \Theta_2$ must conclude with either rule a-ctxe-kuvarSolved or rule a-ctxe-addSolved or rule a-ctxe-solve. In either case, the reasoning is similar to the above case.

**Part 2** Similar to Part 1, except that when $\Theta_2 = \bullet$, we apply rule a-ctxe-solve.

**Part 3** Similar to Part 1, except that when $\Theta_2 = \bullet$, we apply rule a-ctxe-kuvarSolved.

□

**Lemma D.25** (Parallel Extension Solution). *If $\Delta_1, \widehat{\alpha}, \Delta_2 \longrightarrow \Theta_1, \widehat{\alpha} = \kappa_2, \Theta_2$, and $[\Theta_1]\kappa_1 = [\Theta_1]\kappa_2$, then $\Delta_1, \widehat{\alpha} = \kappa_1, \Delta_2 \longrightarrow \Theta_1, \widehat{\alpha} = \kappa_2, \Theta_2$.*

Proof. By induction on $\Theta_2$.
- Case $\Theta_2$ is empty. Then $\Delta_2$ must be empty. Then $\Delta_1, \widehat{\alpha} \longrightarrow \Theta_1, \widehat{\alpha} = \kappa_2$. By inversion we have $\Delta_1 \longrightarrow \Theta_1$. And $\Delta_1, \widehat{\alpha} = \kappa_1 \longrightarrow \Theta_1, \widehat{\alpha} = \kappa_2$ holds by rule a-ctxe-kuvarSolved.
- Case $\Theta_2 = \Theta_2', a : \kappa$. Then $\Delta_2 = \Delta_2', a : \kappa$. By I.H., we $\Delta_1, \widehat{\alpha} = \kappa_1, \Delta_2' \longrightarrow \Theta_1, \widehat{\alpha} = \kappa_2, \Theta_2'$. Then by rule a-tctxe-tvar we are done.
- Case $\Theta_2 = \Theta_2', T : \kappa$. By I.H. and rule a-tctxe-tcon.

- Case $\Theta_2 = \Theta_2', \widehat{\alpha}_2$. Then the derivation of $\Delta_1, \widehat{\alpha}, \Delta_2 \longrightarrow \Theta_1, \widehat{\alpha} = \kappa_2, \Theta_2', \widehat{\alpha}_2$ must conclude with either rule a-ctxe-kuvar or rule a-ctxe-add.
  - Subcase: the derivation concludes with rule a-ctxe-kuvar. It must be $\Delta_2 = \Delta_2', \widehat{\alpha}_1$.

  $\Delta_1, \widehat{\alpha}, \Delta_2' \longrightarrow \Theta_1, \widehat{\alpha} = \kappa_2, \Theta_2'$ | Given
  $\Delta_1, \widehat{\alpha} = \kappa_1, \Delta_2' \longrightarrow \Theta_1, \widehat{\alpha} = \kappa_2, \Theta_2'$ | I.H.
  $\Delta_1, \widehat{\alpha} = \kappa_1, \Delta_2', \widehat{\alpha}_2 \longrightarrow \Theta_1, \widehat{\alpha} = \kappa_2, \Theta_2', \widehat{\alpha}_2$ | By rule a-ctxe-kuvar

  - Subcase: the derivation concludes with rule a-ctxe-add.

  $\Delta_1, \widehat{\alpha}, \Delta_2 \longrightarrow \Theta_1, \widehat{\alpha} = \kappa_2, \Theta_2'$ | Given
  $\Delta_1, \widehat{\alpha} = \kappa_1, \Delta_2 \longrightarrow \Theta_1, \widehat{\alpha} = \kappa_2, \Theta_2'$ | I.H.
  $\Delta_1, \widehat{\alpha} = \kappa_1, \Delta_2 \longrightarrow \Theta_1, \widehat{\alpha} = \kappa_2, \Theta_2', \widehat{\alpha}_1$ | By rule a-ctxe-add

- Case $\Theta_2 = \Theta_2, \widehat{\alpha}_2 = \kappa$. This case is similar to the last one.

□

**Lemma D.26** (Parallel Variable Update). *If $\Delta_1, \widehat{\alpha}, \Delta_2 \longrightarrow \Theta_1, \widehat{\alpha} = \kappa, \Theta_2$, and $\Delta_1 \Vdash^{kv} \kappa_1$, and $\Theta_1 \Vdash^{kv} \kappa_2$, and $[\Theta_1]\kappa = [\Theta_1]\kappa_1 = [\Theta_1]\kappa_2$, then $\Delta_1, \widehat{\alpha} = \kappa_1, \Delta_2 \longrightarrow \Theta_1, \widehat{\alpha} = \kappa_2, \Theta_2$*

Proof. The proof is exactly the same as the one for Lemma D.25. Except for the case when $\Theta_2$ is empty, we use rule a-ctxe-solve.

□

### D.2.4 Properties of Complete Context.

**Lemma D.27** (Type Constructor Preservation). *If $\Delta$ ok, and $(T : \kappa) \in \Delta$, and $\Delta \longrightarrow \Omega$, then $(T : [\Omega]\kappa) \in [\Omega]\Delta$.*

Proof. Suppose $\Delta = \Delta_1, T : \kappa, \Delta_2$. Then by Lemma D.17 we know $\Omega = \Omega_1, T : \kappa, \Omega_2, \Delta_1 \longrightarrow \Omega_1$. So $(T : [\Omega_1]\kappa) \in [\Omega]\Delta$ according to the definition of context application. Because $\Delta$ ok, and $\Delta \longrightarrow \Omega$, by Lemma D.15 we have $\Omega$ ok. So by inversion we have $\Omega_1 \Vdash^{kv} \kappa$. By Lemma D.19 we have $[\Omega]\kappa = [\Omega_1]\kappa$. Therefore $(T : [\Omega]\kappa) \in [\Omega]\Delta$.

□

**Lemma D.28** (Type Variable Preservation). *If $\Delta$ ok, and $(a : \kappa) \in \Delta$, and $\Delta \longrightarrow \Omega$, then $(a : [\Omega]\kappa) \in [\Omega]\Delta$.*

Proof. This lemma is similar to Lemma D.27.

□

**Lemma D.29** (Finishing Kinding). *If $\Omega$ ok, and $\Omega \Vdash^{kv} \kappa$, and $\Omega \longrightarrow \Omega'$, then $[\Omega]\kappa = [\Omega']\kappa$.*

Proof. By Lemma D.18 we know $[\Omega']\kappa = [\Omega']([\Omega]\kappa)$. Because $[\Omega]\kappa$ contains no unsolved kind unification variable, we have $[\Omega']([\Omega]\kappa) = [\Omega]\kappa$. Therefore $[\Omega']\kappa = [\Omega]\kappa$.

□

**Lemma D.30** (Finishing Term Contexts). *If $\Omega$ ok, and $\Omega \Vdash^{ectx} \Gamma$, and $\Omega \longrightarrow \Omega'$, then $[\Omega']\Gamma = [\Omega]\Gamma$.*

Proof. By $\Omega \Vdash^{ectx} \Gamma$, we have that any kind $\kappa$ that appears in $\Gamma$ has $\Omega \Vdash^{kv} \kappa$. So our goal follows directly from Lemma D.29.

□

**Lemma D.31** (Stability of Complete Contexts). *If $\Delta \longrightarrow \Omega$, then $[\Omega]\Delta = [\Omega]\Omega$.*

Proof. By induction on $\Delta \longrightarrow \Omega$.

- Case

$$\frac{}{\bullet \longrightarrow \bullet} \text{\textsc{a-ctxe-empty}}$$

The goal follows trivially.
- Case

$$\frac{\Delta \longrightarrow \Theta}{\Delta, a : \kappa \longrightarrow \Theta, a : \kappa} \text{\textsc{a-ctxe-tvar}}$$

We have $\Delta = \Delta', a : \kappa$, $\Omega = \Omega', a : \kappa$ and $\Delta' \longrightarrow \Omega'$.

| $[\Omega]\Delta$ | |
|---|---|
| $= [\Omega', a : \kappa](\Delta', a : \kappa)$ | By definition |
| $= [\Omega']\Delta', a : [\Omega']\kappa$ | By definition |
| $= [\Omega']\Omega', a : [\Omega']\kappa$ | By I.H. |
| $= [\Omega', a : \kappa](\Omega', a : \kappa)$ | By definition |

- Case

$$\frac{\Delta \longrightarrow \Theta}{\Delta, T : \kappa \longrightarrow \Theta, T : \kappa} \text{\textsc{a-ctxe-tcon}}$$

This case is similar to the case for rule \textsc{a-ctxe-tvar}.
- Case

$$\frac{\Delta \longrightarrow \Theta}{\Delta, \widehat{\alpha} \longrightarrow \Theta, \widehat{\alpha}} \text{\textsc{a-ctxe-kuvar}}$$

This case is impossible as $\Omega$ is a complete context.
- Case

$$\frac{\Delta \longrightarrow \Theta \quad [\Theta]\kappa_1 = [\Theta]\kappa_2}{\Delta, \widehat{\alpha} = \kappa_1 \longrightarrow \Theta, \widehat{\alpha} = \kappa_2} \text{\textsc{a-ctxe-kuvarSolved}}$$

We have $\Delta = \Delta', \widehat{\alpha} = \kappa_1$, $\Omega = \Omega', \widehat{\alpha} = \kappa_2$ and $[\Omega']\kappa_1 = [\Omega']\kappa_2$.

| $[\Omega]\Delta$ | |
|---|---|
| $= [\Omega', \widehat{\alpha} = \kappa_2](\Delta', \widehat{\alpha} = \kappa_1)$ | By definition |
| $= [\Omega']\Delta'$ | By definition |
| $= [\Omega']\Omega'$ | By I.H. |
| $= [\Omega', \widehat{\alpha} = \kappa_2](\Omega', \widehat{\alpha} = \kappa_2)$ | By definition |

- Case

$$\frac{\Delta \longrightarrow \Theta \quad \Theta \Vdash^{\text{kv}} \kappa}{\Delta, \widehat{\alpha} \longrightarrow \Theta, \widehat{\alpha} = \kappa} \text{\textsc{a-ctxe-solve}}$$

This case is similar to the case for rule \textsc{a-ctxe-kuvarSolved}.
- Case

$$\frac{\Delta \longrightarrow \Theta}{\Delta \longrightarrow \Theta, \widehat{\alpha}} \text{\textsc{a-ctxe-add}}$$

This case is impossible as $\Omega$ is a complete context.

- Case

$$\frac{\Delta \longrightarrow \Theta \qquad \Theta \Vdash^{\mathrm{kv}} \kappa}{\Delta \longrightarrow \Theta, \widehat{\alpha} = \kappa} \text{\small A-CTXE-ADDSOLVED}$$

This case is similar to the case for rule A-CTXE-KUVARSOLVED.

□

**Lemma D.32** (Softness Goes Away). *If $\Delta_1, \Delta_2 \longrightarrow \Omega_1, \Omega_2$ where $\Delta_1 \longrightarrow \Omega_1$, and $\Delta_2$ **soft**, then $[\Omega_1, \Omega_2](\Delta_1, \Delta_2) = [\Omega_1]\Delta_1$.*

PROOF. By induction on $\Delta_2$ and the goal follows directly from the definition of context application.

□

**Lemma D.33** (Confluence of Completeness). *If $\Delta_1 \longrightarrow \Omega$, and $\Delta_2 \longrightarrow \Omega$, then $[\Omega]\Delta_1 = [\Omega]\Delta_2$.*

PROOF. By Lemma D.31 we have $[\Omega]\Delta_1 = [\Omega]\Omega$ and $[\Omega]\Delta_2 = [\Omega]\Omega$. Therefore $[\Omega]\Delta_1 = [\Omega]\Delta_2$.

□

**Lemma D.34** (Finishing Completions). *If $\Omega$ ok, and $\Omega \longrightarrow \Omega'$, then $[\Omega]\Omega = [\Omega']\Omega'$.*

PROOF. By induction on $\Omega \longrightarrow \Omega'$.

- Case

$$\frac{}{\bullet \longrightarrow \bullet} \text{\small A-CTXE-EMPTY}$$

The goal follows trivially.

- Cases for rules A-CTXE-KUVAR and A-CTXE-ADD are impossible as $\Omega$ and $\Omega'$ are complete contexts.
- Case

$$\frac{\Delta \longrightarrow \Theta}{\Delta, a : \kappa \longrightarrow \Theta, a : \kappa} \text{\small A-CTXE-TVAR}$$

So we have $\Omega = \Omega_1, a : \kappa$, and $\Omega' = \Omega'_1, a : \kappa$.

| | |
|---|---|
| $[\Omega_1]\Omega_1 = [\Omega'_1]\Omega'_1$ | By I.H. |
| $[\Omega_1, a : \kappa](\Omega_1, a : \kappa) = [\Omega_1]\Omega_1, a : [\Omega_1]\kappa$ | By definition |
| $[\Omega'_1, a : \kappa](\Omega'_1, a : \kappa) = [\Omega'_1]\Omega'_1, a : [\Omega'_1]\kappa$ | By definition |
| $\Omega$ ok | Given |
| $\Omega_1$ ok $\wedge$ $\Omega_1 \Vdash^{\mathrm{kv}} \kappa$ | By inversion |
| $[\Omega_1]\kappa = [\Omega'_1]\kappa$ | By Lemma D.29 |
| $[\Omega_1, a : \kappa](\Omega_1, a : \kappa) = [\Omega'_1, a : \kappa](\Omega'_1, a : \kappa)$ | Follows from the equations |

- The rest cases are similar to the above case.

□

### D.2.5 Soundness of Algorithm.

**Lemma D.35** (Soundness of Kind Validating). *If $\Omega$ ok, and $\Omega \Vdash^{\mathrm{kv}} \kappa$, then $[\Omega]\kappa$ is a validate kind in the declarative system.*

PROOF. By induction on the size of $|\Omega \vdash \kappa|$. Then case analyze on $\kappa$.

- Case $\kappa = \star$. Follows trivially by $[\Omega]\star = \star$.

- Case $\kappa = \kappa_1 \to \kappa_2$. Follows directly from I.H..
- Case $\kappa = \widehat{\alpha}$. $\Omega$ must be $\Omega_1, \widehat{\alpha} = \kappa, \Omega_2$, and $[\Omega]\widehat{\alpha} = [\Omega]\kappa$. By I.H., we know $[\Omega]\kappa$ is a well-formed kind.

□

**Lemma D.36** (Soundness of Well-formed Type Context). *If $\Delta$ ok, and $\Delta \longrightarrow \Omega$, then $[\Omega]\Delta$ is a valid type context in the declarative system.*

Proof. By induction on the well-formedness of type context.

- Case

  $$\frac{}{\bullet \text{ ok}} \text{ a-tctx-empty}$$

  Holds trivially.

- Case

  $$\frac{\Delta \text{ ok} \quad \Delta \Vdash^{\text{kv}} \kappa}{\Delta, a : \kappa \text{ ok}} \text{ a-tctx-tvar}$$

| | |
|---|---|
| $\Delta, a : \kappa \longrightarrow \Omega$ | Given |
| $\Omega = \Omega_1, a : \kappa, \Omega_2 \wedge \Omega_2 \text{ soft} \wedge \Delta \longrightarrow \Omega_1$ | By Lemma D.17 |
| $[\Omega](\Delta, a : \kappa)$ | |
| $= [\Omega_1, a : \kappa, \Omega_2](\Delta, a : \kappa)$ | |
| $= [\Omega_1, a : \kappa](\Delta, a : \kappa)$ | By Lemma D.32 |
| $= [\Omega_1]\Delta, a : [\Omega_1]\kappa$ | By definition |
| $[\Omega_1]\Delta$ is a valid declarative type context | I.H. |
| $[\Omega_1]\kappa$ is a declarative validate kind | By Lemma D.35 |
| $[\Omega_1]\Delta, a : [\Omega_1]\kappa$ is a valid type context | |

- Case

  $$\frac{\Delta \text{ ok} \quad \Delta \Vdash^{\text{kv}} \kappa}{\Delta, T : \kappa \text{ ok}} \text{ a-tctx-tcon}$$

  This case is similar to the case rule a-tctx-tvar.

- Case

  $$\frac{\Delta \text{ ok}}{\Delta, \widehat{\alpha} \text{ ok}} \text{ a-tctx-kuvar}$$

| | |
|---|---|
| $\Delta, \widehat{\alpha} \longrightarrow \Omega$ | Given |
| $\Omega = \Omega_1, \widehat{\alpha} = \kappa, \Omega_2 \wedge \Omega_2 \text{ soft} \wedge \Delta \longrightarrow \Omega_1$ | By Lemma D.17 |
| $[\Omega](\Delta, \widehat{\alpha})$ | |
| $= [\Omega_1, \widehat{\alpha} = \kappa, \Omega_2](\Delta, \widehat{\alpha})$ | |
| $= [\Omega_1]\Delta$ | By Lemma D.32 |
| $[\Omega_1]\Delta$ is a valid declarative type context | I.H. |

- Case

  $$\frac{\Delta \text{ ok} \quad \Delta \Vdash^{\text{kv}} \kappa}{\Delta, \widehat{\alpha} = \kappa \text{ ok}} \text{ a-tctx-kuvarSolved}$$

This case is similar to the case rule A-TCTX-KUVAR.

□

**Lemma D.37** (Soundness of Well-formed Term Context). *If $\Delta$ ok, and $\Delta \Vdash^{\text{ectx}} \Gamma$, and $\Delta \longrightarrow \Omega$, then $[\Omega]\Delta \vdash [\Omega]\Gamma$.*

Proof. By induction on the judgment of well-formed term context.

- Case

$$\frac{}{\Delta \Vdash^{\text{ectx}} \bullet} \text{A-ECTX-EMPTY}$$

Follows trivially by rule ECTX-EMPTY.

- Case

$$\frac{\Delta \Vdash^{\text{ectx}} \Gamma \qquad \Delta \Vdash^{\text{kc}} \sigma \Leftarrow \star}{\Delta \Vdash^{\text{ectx}} \Gamma, x : \sigma} \text{A-ECTX-VAR}$$

| | |
|---|---|
| $[\Omega]\Delta \vdash [\Omega]\Gamma$ | I.H. |
| $[\Omega]\Delta \vdash^{k} [\Omega]\sigma : [\Omega]\star$ | By Lemma D.41 |
| $[\Omega]\Delta \vdash^{k} [\Omega]\sigma : \star$ | By definition |
| $[\Omega](\Gamma, x : \sigma) = [\Omega]\Gamma, x : [\Omega]\sigma$ | By definition |
| $[\Omega]\Delta \vdash [\Omega](\Gamma, x : \sigma)$ | By rule ECTX-VAR |

- Case

$$\frac{\Delta \Vdash^{\text{ectx}} \Gamma \qquad \Delta \Vdash^{\text{kc}} \sigma \Leftarrow \star}{\Delta \Vdash^{\text{ectx}} \Gamma, D : \sigma} \text{A-ECTX-DCON}$$

This case is similar to the case for rule A-ECTX-VAR.

□

**Lemma D.38** (Soundness of Promotion). *If $\Delta$ ok, and $\Delta \Vdash^{\text{pr}}_{\hat{\alpha}} \kappa_1 \leadsto \kappa_2 \dashv \Theta$, then $[\Theta]\kappa_1 = [\Theta]\kappa_2$. Moreover, if $\Delta \Vdash^{\text{kv}} \kappa_1$, and $\Theta \longrightarrow \Omega$, then $[\Omega]\kappa_1 = [\Omega]\kappa_2$.*

Proof. By Lemma D.3 and Lemma D.18, if given $[\Theta]\kappa_1 = [\Theta]\kappa_2$, we can prove $[\Omega]\kappa_1 = [\Omega]\kappa_2$. Thus we only need to prove that $[\Theta]\kappa_1 = [\Theta]\kappa_2$.

By a straightforward induction on the promotion judgment. All cases follow trivially.

□

**Lemma D.39** (Soundness of Unification). *If $\Delta$ ok, and $\Delta \Vdash^{\text{kv}} \kappa_1$, and $\Delta \Vdash^{\text{kv}} \kappa_2$, and $\Delta \Vdash^{u} \kappa_1 \approx \kappa_2 \dashv \Theta$, then $[\Theta]\kappa_1 = [\Theta]\kappa_2$. If $\Theta \longrightarrow \Omega$, then $[\Omega]\kappa_1 = [\Omega]\kappa_2$.*

Proof. By Lemma D.4, Lemma D.10 and Lemma D.18, if given $[\Theta]\kappa_1 = [\Theta]\kappa_2$. we can prove $[\Omega]\kappa_1 = [\Omega]\kappa_2$. Thus we only need to prove that $[\Theta]\kappa_1 = [\Theta]\kappa_2$.

By induction on the unification judgment.

- Case

$$\frac{}{\Delta \Vdash^{u} \kappa \approx \kappa \dashv \Delta} \text{A-U-REFL}$$

$[\Delta]\kappa = [\Delta]\kappa$.

- Case

$$\frac{\Delta \Vdash^\mu \kappa_1 \approx \kappa_3 \dashv \Theta_1 \quad \Theta_1 \Vdash^\mu [\Theta_1]\kappa_2 \approx [\Theta_1]\kappa_4 \dashv \Theta}{\Delta \Vdash^\mu \kappa_1 \to \kappa_2 \approx \kappa_3 \to \kappa_4 \dashv \Theta} \text{a-u-arrow}$$

| | |
|---|---|
| $\Delta \longrightarrow \Theta_1$ | By Lemma D.4 |
| $\Theta_1 \longrightarrow \Theta$ | By Lemma D.4 |
| $\Delta \longrightarrow \Theta$ | By Lemma D.20 |
| $[\Theta_1]\kappa_1 = [\Theta_1]\kappa_3$ | By I.H. |
| $[\Theta]\kappa_1 = [\Theta]\kappa_3$ | By Lemma D.18 |
| $\Delta \Vdash^{\text{kv}} \kappa_2$ | By inversion |
| $\Theta_1 \Vdash^{\text{kv}} [\Theta_1]\kappa_2$ | By Lemma D.10 and Lemma D.12 |
| $\Theta_1 \Vdash^{\text{kv}} [\Theta_1]\kappa_4$ | As above |
| $[\Theta]\kappa_2 = [\Theta]\kappa_4$ | By I.H. and Lemma D.18 |
| $[\Theta](\kappa_1 \to \kappa_2) = [\Theta](\kappa_3 \to \kappa_4)$ | Follows directly |

- Case

$$\frac{\Delta \Vdash^{\text{pr}}_{\widehat{\alpha}} \kappa \rightsquigarrow \kappa_2 \dashv \Theta[\widehat{\alpha}]}{\Delta[\widehat{\alpha}] \Vdash^\mu \widehat{\alpha} \approx \kappa \dashv \Theta[\widehat{\alpha} = \kappa_2]} \text{a-u-kvarL}$$

| | |
|---|---|
| $[\Theta[\widehat{\alpha}]]\kappa = [\Theta[\widehat{\alpha}]]\kappa_2$ | By Lemma D.38 |
| $\Delta \longrightarrow \Theta[\widehat{\alpha}] \land \Theta[\widehat{\alpha}] \Vdash^{\text{kv}} \kappa_2$ | By Lemma D.3 |
| $\Theta[\widehat{\alpha}] \longrightarrow \Theta[\widehat{\alpha} = \kappa_2]$ | By Lemma D.3 and Lemma D.21 |
| $[\Theta[\widehat{\alpha} = \kappa_2]]\kappa = [\Theta[\widehat{\alpha} = \kappa_2]]\kappa_2$ | By Lemma D.18 |

- Case

$$\frac{\Delta \Vdash^{\text{pr}}_{\widehat{\alpha}} \kappa \rightsquigarrow \kappa_2 \dashv \Theta[\widehat{\alpha}]}{\Delta[\widehat{\alpha}] \Vdash^\mu \kappa \approx \widehat{\alpha} \dashv \Theta[\widehat{\alpha} = \kappa_2]} \text{a-u-kvarR}$$

This case is similar to the case for rule a-u-kvarL.

□

**Lemma D.40** (Soundness of Application Kinding). *If $\Delta$ ok, and $\Delta \Vdash^{\text{kv}} \kappa_1$, and $\Delta \Vdash^{\text{kv}} \kappa_2$, and $\Delta \Vdash^{\text{kapp}} \kappa_1 \bullet \kappa_2 : \kappa_3 \dashv \Theta$, then $[\Theta]\kappa_1 = [\Theta]\kappa_2 \to [\Theta]\kappa_3$. If $\Theta \longrightarrow \Omega$, then $[\Omega]\kappa_1 = [\Omega]\kappa_2 \to [\Omega]\kappa_3$.*

Proof. By Lemma D.5, Lemma D.10 and Lemma D.18, we know $[\Omega]\kappa_1 = [\Omega]([\Theta]\kappa_1)$ and $[\Omega]\kappa_2 = [\Omega]([\Theta]\kappa_2)$ and $[\Omega]\kappa_3 = [\Omega]([\Theta]\kappa_3)$. Thus we only need to prove that $[\Theta]\kappa_1 = [\Theta]\kappa_2 \to [\Theta]\kappa_3$.

By induction on the application kinding judgment.

- Case

$$\frac{\Delta[\widehat{\alpha}_1, \widehat{\alpha}_2, \widehat{\alpha} = \widehat{\alpha}_1 \to \widehat{\alpha}_2] \Vdash^\mu \widehat{\alpha}_1 \approx \kappa \dashv \Theta}{\Delta[\widehat{\alpha}] \Vdash^{\text{kapp}} \widehat{\alpha} \bullet \kappa : \widehat{\alpha}_2 \dashv \Theta} \text{a-kapp-kuvar}$$

| | |
|---|---|
| $[\Theta]\widehat{\alpha}_1 = [\Theta]\kappa$ | By Lemma D.39 |
| $\Delta[\widehat{\alpha}_1, \widehat{\alpha}_2, \widehat{\alpha} = \widehat{\alpha}_1 \to \widehat{\alpha}_2] \longrightarrow \Theta$ | By Lemma D.4 |
| $[\Theta]\widehat{\alpha}$ | |
| $= [\Theta]([\Delta[\widehat{\alpha}_1, \widehat{\alpha}_2, \widehat{\alpha} = \widehat{\alpha}_1 \to \widehat{\alpha}_2]]\widehat{\alpha})$ | By Lemma D.18 |
| $= [\Theta]([\Delta[\widehat{\alpha}_1, \widehat{\alpha}_2, \widehat{\alpha} = \widehat{\alpha}_1 \to \widehat{\alpha}_2]]\widehat{\alpha}_1 \to \widehat{\alpha}_2)$ | By definition |

$$
\begin{aligned}
&= [\Theta](\widehat{\alpha}_1 \to \widehat{\alpha}_2) && \text{By Lemma D.18} \\
&= [\Theta]\widehat{\alpha}_1 \to [\Theta]\widehat{\alpha}_2 && \text{By definition} \\
&= [\Theta]\kappa \to [\Theta]\widehat{\alpha}_2 && \text{By substituting the equation}
\end{aligned}
$$

- Case

$$
\frac{\Delta \Vdash^{\mu} \kappa_1 \approx \kappa \dashv \Theta}{\Delta \Vdash^{\mathsf{kapp}} \kappa_1 \to \kappa_2 \bullet \kappa : \kappa_2 \dashv \Theta} \text{ a-kapp-arrow}
$$

$$
\begin{aligned}
&[\Theta]\kappa_1 = [\Theta]\kappa && \text{By Lemma D.39} \\
&[\Theta](\kappa_1 \to \kappa_2) \\
&= [\Theta]\kappa_1 \to [\Theta]\kappa_2 && \text{By definition} \\
&= [\Theta]\kappa \to [\Theta]\kappa_2 && \text{By substituting the equation}
\end{aligned}
$$

□

**Lemma D.41** (Soundness of Kinding). *If $\Delta$ ok, and $\Delta \Vdash^k \sigma : \kappa \dashv \Theta$, and $\Theta \longrightarrow \Omega$, then $[\Omega]\Delta \vdash^k [\Omega]\sigma : [\Omega]\kappa$.*

Proof. By induction on the kinding judgment.

- Case

$$
\frac{}{\Delta \Vdash^k \mathsf{Int} : \star \dashv \Delta} \text{ a-k-nat}
$$

$[\Omega]\Delta \vdash^k [\Omega]\mathsf{Int} : [\Omega]\star$ By rule k-nat.

- Case

$$
\frac{(a : \kappa) \in \Delta}{\Delta \Vdash^k a : \kappa \dashv \Delta} \text{ a-k-var}
$$

$$
\begin{aligned}
&(a : \kappa) \in \Delta && \text{Given} \\
&(a : [\Omega]\kappa) \in [\Omega]\Delta && \text{By Lemma D.28} \\
&[\Omega]\Delta \vdash [\Omega]a : [\Omega]\kappa && \text{By rule k-var}
\end{aligned}
$$

- Case

$$
\frac{(T : \kappa) \in \Delta}{\Delta \Vdash^k T : \kappa \dashv \Delta} \text{ a-k-tcon}
$$

Similar to the rule a-k-var case with Lemma D.27.

- Case

$$
\frac{}{\Delta \Vdash^k \to : \star \to \star \to \star \dashv \Delta} \text{ a-k-arrow}
$$

$[\Omega]\Delta \vdash^k [\Omega] \to : [\Omega](\star \to \star \to \star)$ by rule k-arrow.

- Case

$$
\frac{\Delta \Vdash^{\mathsf{kv}} \kappa \quad \Delta, a : \kappa \Vdash^k \sigma : \kappa_2 \dashv \Theta, a : \kappa \quad [\Theta]\kappa_2 = \star}{\Delta \Vdash^k \forall a : \kappa . \sigma : \star \dashv \Theta} \text{ a-k-forall}
$$

$$
\begin{aligned}
&\Theta \longrightarrow \Omega && \text{Given} \\
&\Theta, a : \kappa \longrightarrow \Omega, a : \kappa && \text{By rule a-ctxe-tvar}
\end{aligned}
$$

| | |
|---|---|
| $[\Omega, a : \kappa](\Delta, a : \kappa) \Vdash^k [\Omega, a : \kappa]\sigma : [\Omega, a : \kappa]\kappa_2$ | I.H. |
| $[\Omega]\Delta, a : [\Omega]\kappa \Vdash^k [\Omega, a : \kappa]\sigma : [\Omega, a : \kappa]([\Theta, a : \kappa]\kappa_2)$ | By Lemma D.18 |
| $[\Omega]\Delta, a : [\Omega]\kappa \Vdash^k [\Omega, a : \kappa]\sigma : [\Omega, a : \kappa]\star$ | By definition |
| $[\Omega]\Delta, a : [\Omega]\kappa \Vdash^k [\Omega]\sigma : [\Omega]\star$ | By property of context application |
| $[\Omega]\Delta \Vdash^k \forall a : [\Omega]\kappa.[\Omega]\sigma : [\Omega]\star$ | By rule K-FORALL |

- Case
$$\frac{\text{A-K-APP}}{\Delta \Vdash^k \tau_1 : \kappa_1 \dashv \Theta_1 \quad \Theta_1 \Vdash^k \tau_2 : \kappa_2 \dashv \Theta_2 \quad \Theta_2 \Vdash^{kapp} [\Theta_2]\kappa_1 \bullet [\Theta_2]\kappa_2 : \kappa_3 \dashv \Theta}{\Delta \Vdash^k \tau_1 \tau_2 : \kappa_3 \dashv \Theta}$$

| | |
|---|---|
| $\Theta_1 \longrightarrow \Theta_2$ | By Lemma D.6 |
| $\Theta_2 \longrightarrow \Theta$ | By Lemma D.4 |
| $\Theta_1 \longrightarrow \Omega$ | By Lemma D.20 |
| $[\Omega]\Delta \Vdash^k [\Omega]\tau_1 : [\Omega]\kappa_1$ | I.H. |
| $[\Omega]\Delta \Vdash^k [\Omega]\tau_2 : [\Omega]\kappa_2$ | Similarly |
| $[\Omega]\kappa_1 = [\Omega](\kappa_2 \rightarrow \kappa_3) = [\Omega]\kappa_2 \rightarrow [\Omega]\kappa_3$ | By Lemma D.40 |
| $[\Omega]\Delta \Vdash^k [\Omega](\tau_1 \tau_2) : [\Omega]\kappa_3$ | By rule K-APP-P |

□

**Lemma D.42** (Soundness of Typing Data Constructor Declaration). *If $\Delta$ ok, and $\Delta \Vdash^{dc}_{\tau'} \mathcal{D} \rightsquigarrow \tau \dashv \Theta$, and $\Theta \longrightarrow \Omega$, then $[\Omega]\Delta \vdash^{dc}_{\tau'} \mathcal{D} \rightsquigarrow \tau$.*

PROOF. We have
$$\frac{\text{A-DC-DECL}}{\Delta \Vdash^{dc}_{\tau} D \overline{\tau_i}^i \rightsquigarrow \overline{\tau_i}^i \rightarrow \tau \dashv \Theta}$$

Follows directly from Lemma D.41 and rule DC-DECL.

□

**Lemma D.43** (Soundness of Typing Datatype Declaration). *If $\Delta$ ok, and $\Delta \Vdash^{dt} \mathcal{T} \rightsquigarrow \Gamma \dashv \Theta$, and $\Theta \longrightarrow \Omega$, then $[\Omega]\Delta \vdash^{dt} \mathcal{T} \rightsquigarrow [\Omega]\Gamma$.*

PROOF. We have

$$\frac{\text{A-DT-DECL}}{(T : \kappa) \in \Delta \quad \Delta, \overline{\widehat{\alpha_i}}^i \Vdash^u [\Delta]\kappa \approx (\overline{\widehat{\alpha_i}}^i \rightarrow \star) \dashv \Theta_1, \overline{\widehat{\alpha_i} = \kappa_i}^i \quad \overline{\Theta_j, \overline{a_i : \kappa_i}^i \Vdash^{dc}_{T \overline{a_i}^i} \mathcal{D}_j \rightsquigarrow \tau_j \dashv \Theta_{j+1}, \overline{a_i : \kappa_i}^i}^j}{\Delta \Vdash^{dt} \textbf{data } T \overline{a_i}^i = \overline{\mathcal{D}_j}^{j \in 1..n} \rightsquigarrow \overline{D_j : \forall \overline{a_i : \kappa_i}^i.\tau_j}^j \dashv \Theta_{n+1}}$$

| | |
|---|---|
| $\Delta, \overline{\widehat{\alpha_i}}^i \longrightarrow \Theta_1, \overline{\widehat{\alpha_i} = \kappa_i}^i$ | By Lemma D.4 |
| $\Delta \longrightarrow \Theta_1$ | By Lemma D.17 |
| $\overline{\Theta_j, \overline{a_i : \kappa_i}^i \longrightarrow \Theta_{j+1}, \overline{a_i : \kappa_i}^i}^j$ | By Lemma D.7 |
| $\overline{\Theta_j \longrightarrow \Theta_{j+1}}^j$ | By Lemma D.17 |
| $\Theta_{n+1} \longrightarrow \Omega$ | Given |
| $\overline{\Theta_j \longrightarrow \Omega}^j$ | By Lemma D.20 |
| $\Delta \longrightarrow \Omega$ | By Lemma D.20 |
| $\Theta_{n+1}, \overline{a_i : \kappa_i}^i \longrightarrow \Omega, \overline{a_i : \kappa_i}^i$ | By rule A-CTXE-TVAR |
| $\overline{\Theta_{j+1}, \overline{a_i : \kappa_i}^i \longrightarrow \Omega, \overline{a_i : \kappa_i}^i}^j$ | By Lemma D.20 |
| $\overline{[\Omega, \overline{a_i : \kappa_i}^i](\Theta_{j+1}, \overline{a_i : \kappa_i}^i) \vdash^{\text{dc}}_{T\,\overline{a_i}^i} \mathcal{D}_j \rightsquigarrow [\Omega]\tau_j}^j$ | By Lemma D.42 |
| $\overline{[\Omega]\Theta_{j+1}, \overline{a_i : [\Omega]\kappa_i}^i \vdash^{\text{dc}}_{T\,\overline{a_i}^i} \mathcal{D}_j \rightsquigarrow [\Omega]\tau_j}^j$ | By definition |
| $[\Omega]\Delta, \overline{a_i : [\Omega]\kappa_i}^i \vdash^{\text{dc}}_{T\,\overline{a_i}^i} \mathcal{D}_j \rightsquigarrow [\Omega]\tau_j$ | (1) By Lemma D.33 |
| $(T : \kappa) \in \Delta$ | Given |
| $(T : [\Omega]\kappa) \in [\Omega]\Delta$ | (2) By Lemma D.27 |
| $\Theta_1, \overline{\widehat{\alpha}_i = \kappa_i}^i \longrightarrow \Omega, \overline{\widehat{\alpha}_i = \kappa_i}^i$ | By rule A-CTXE-KUVARSOLVED |
| $[\Omega, \overline{\widehat{\alpha}_i = \kappa_i}^i]([\Delta]\kappa) = [\Omega, \overline{\widehat{\alpha}_i = \kappa_i}^i](\overline{\widehat{\alpha}_i}^i \to \star)$ | By Lemma D.39 |
| $[\Omega]([\Delta]\kappa) = \overline{[\Omega]\kappa_i}^i \to \star$ | By definition |
| $[\Omega]\kappa = \overline{[\Omega]\kappa_i}^i \to \star$ | (3) By Lemma D.18 |
| $[\Omega]\Delta \vdash^{\text{dt}} \mathcal{T} \rightsquigarrow [\Omega]\Gamma$ | By rule DT-DECL and (1), (2), (3) |

□

**Lemma D.44** (Soundness of Typing Program). *If $\Omega$ ok, and $\Omega \Vdash^{\text{ectx}} \Gamma$, and $\Omega; \Gamma \Vdash^{\text{pgm}} pgm : \sigma$, then $[\Omega]\Omega; [\Omega]\Gamma \vdash^{\text{pgm}} pgm : \sigma$.*

PROOF. By induction on the typing program judgment.

- Case
  $$\frac{[\Omega]\Omega; [\Omega]\Gamma \vdash e : \sigma}{\Omega; \Gamma \Vdash^{\text{pgm}} e : \sigma}\text{A-PGM-EXPR}$$

  The conclusion holds directly from the hypothesis and rule PGM-EXPR.

- Case
  $$\frac{\Theta_1 = \Omega, \overline{\widehat{\alpha}_i}^i, \overline{T_i : \widehat{\alpha}_i}^i \quad \overline{\Theta_i \Vdash^{\text{dt}} \mathcal{T}_i \rightsquigarrow \Gamma_i \dashv \Theta_{i+1}}^i \quad \Theta_{n+1} \longrightarrow \Omega' \quad \Omega'; \Gamma, \overline{\Gamma_i}^i \Vdash^{\text{pgm}} pgm : \sigma}{\Omega; \Gamma \Vdash^{\text{pgm}} \text{rec } \overline{\mathcal{T}_i}^{i \in 1..n}; pgm : \sigma}\text{A-PGM-DT}$$

| | |
|---|---|
| $\overline{\Theta_i \longrightarrow \Theta_{i+1}}^i$ | By Lemma D.8 |
| $\Theta_{n+1} \longrightarrow \Omega'$ | By Lemma D.13 |
| $\overline{\Theta_i \longrightarrow \Omega'}^i$ | By Lemma D.20 |
| $\Omega, \overline{\widehat{\alpha}_i}^{i\in 1..n}, \overline{T_i : \widehat{\alpha}_i}^{i\in 1..n} \longrightarrow \Omega'$ | $\Theta_1 = \Omega, \overline{\widehat{\alpha}_i}^{i\in 1..n}, \overline{T_i : \widehat{\alpha}_i}^{i\in 1..n}$ |
| $\Omega' = \Omega'', \overline{\widehat{\alpha}_i = \kappa_i, \Omega_i}^{i\in 1..n}, \overline{T_i : \widehat{\alpha}_i, \Omega'_i}^{i\in 1..n}$ | By Lemma D.17 |
| $\overline{\Omega_i \text{ soft}}^i \wedge \overline{\Omega'_i \text{ soft}}^i$ | Above |
| $\Omega \longrightarrow \Omega''$ | Above |
| $[\Omega']\Omega'$ | |
| $= [\Omega'', \overline{\widehat{\alpha}_i = \kappa_i, \Omega_i}^{i\in 1..n}, \overline{T_i : \widehat{\alpha}_i, \Omega'_i}^{i\in 1..n}]$ | |
| $\quad (\Omega'', \overline{\widehat{\alpha}_i = \kappa_i, \Omega_i}^{i\in 1..n}, \overline{T_i : \widehat{\alpha}_i, \Omega'_i}^{i\in 1..n})$ | |
| $= [\Omega'']\Omega'', \overline{T_i : \kappa_i}^{i\in 1..n}$ | for some $\kappa_i$, by Lemma D.32 |
| $= [\Omega]\Omega, \overline{T_i : \kappa_i}^{i\in 1..n}$ | By Lemma D.34 |
| $\overline{[\Omega']\Theta_i \vdash^{\text{dt}} \mathcal{T}_i \rightsquigarrow [\Omega']\Gamma_i}^{i\in 1..n}$ | By Lemma D.43 |
| $\overline{[\Omega']\Theta_i = [\Omega']\Omega'}^{i \in 1..n}$ | By Lemma D.31 |
| $\overline{[\Omega']\Omega' \vdash^{dt} \mathcal{T}_i \leadsto [\Omega']\Gamma_i}^{i \in 1..n}$ | By substituting the equation |
| $[\Omega]\Omega, \overline{T_i : \kappa_i}^{i \in 1..n} \vdash^{dt} \mathcal{T}_i \leadsto [\Omega']\Gamma_i^{i \in 1..n}$ | (1) By substituting the equation |
| $[\Omega'](\Gamma, \overline{\Gamma_i}^{i \in 1..n}) = [\Omega']\Gamma, \overline{[\Omega']\Gamma_i}^{i \in 1..n}$ | By definition |
| $[\Omega']\Gamma$ | |
| $= [\Omega'', \overline{\widehat{\alpha_i} = \kappa_i, \Omega_i}^{i \in 1..n}, \overline{T_i : \widehat{\alpha_i}, \Omega'_i}^{i \in 1..n}]\Gamma$ | |
| $= [\Omega'']\Gamma$ | By definition and freshness |
| $= [\Omega]\Gamma$ | By Lemma D.30 |
| $[\Omega']\Omega'; [\Omega'](\Gamma, \overline{\Gamma_i}^{i \in 1..n}) \vdash^{pgm} pgm : \sigma$ | I.H. |
| $[\Omega]\Omega, \overline{T_i : \kappa_i}^{i \in 1..n}; [\Omega]\Gamma, \overline{[\Omega']\Gamma_i}^{i \in 1..n} \vdash^{pgm} pgm : \sigma$ | (2) By substituting equations |
| $[\Omega]\Omega; [\Omega]\Gamma \vdash^{pgm} pgm : \sigma$ | By rule PGM-DT and (1), (2) |

□

### D.2.6 Completeness of Algorithm.

**Lemma D.45** (Completeness of Promotion). *Given $\Delta$ ok, and $\Delta \longrightarrow \Omega$, and $\Delta \Vdash^{kv} \widehat{\alpha}$, and $\Delta \Vdash^{kv} \kappa$, and $[\Delta]\widehat{\alpha} = \widehat{\alpha}$, and $[\Delta]\kappa = \kappa$, if $\kappa$ is free of $\widehat{\alpha}$, then there exists $\kappa_2, \Theta$ and $\Omega'$ such that $\Theta \longrightarrow \Omega'$, and $\Omega \longrightarrow \Omega'$, and $\Delta \Vdash^{pr}_{\widehat{\alpha}} \kappa \leadsto \kappa_2 \dashv \Theta$.*

Proof. By induction on $\kappa$.

- $\kappa = \star$. Then by rule A-PR-STAR, we have $\Theta = \Delta$, and $\Omega' = \Omega$.
- $\kappa = \kappa_1 \to \kappa_2$.

| | |
|---|---|
| $\Delta \Vdash^{pr}_{\widehat{\alpha}} \kappa_1 \leadsto \kappa_3 \dashv \Delta_1 \wedge \Delta_1 \longrightarrow \Omega_1 \wedge \Omega \longrightarrow \Omega_1$ | I.H. |
| $\Delta_1 \Vdash^{kv} \widehat{\alpha}$ | By Lemma D.10 |
| $\Delta_1 \Vdash^{kv} [\Delta_1]\kappa_2$ | By Lemma D.10 and Lemma D.12 |
| $\Delta_1 \Vdash^{pr}_{\widehat{\alpha}} [\Delta_1]\kappa_2 \leadsto \kappa_4 \dashv \Theta \wedge \Theta \longrightarrow \Omega' \wedge \Omega_1 \longrightarrow \Omega'$ | I.H. |
| $\Delta \Vdash^{pr}_{\widehat{\alpha}} \kappa_1 \to \kappa_2 \leadsto \kappa_3 \to \kappa_4 \dashv \Theta$ | By rule A-PR-ARROW |
| $\Omega \longrightarrow \Omega'$ | By Lemma D.20 |

- $\kappa = \widehat{\beta}$.
  - $\widehat{\beta}$ is to the left of $\widehat{\alpha}$. Then by rule A-PR-KUVARL, we have $\Theta = \Delta$, and $\Omega' = \Omega$.
  - $\widehat{\beta}$ is to the right of $\widehat{\alpha}$. Then by rule A-PR-KUVARR, we have $\Theta = \Delta[\widehat{\beta_2}, \widehat{\alpha}][\widehat{\beta} = \widehat{\beta_2}]$.

| | |
|---|---|
| $\Delta[\widehat{\alpha}][\widehat{\beta}] \longrightarrow \Omega$ | Given |
| $\Omega = \Omega[\widehat{\alpha} = \kappa_3][\widehat{\beta} = \kappa_4]$ | By Lemma D.17 |
| $\Delta[\widehat{\beta_2}, \widehat{\alpha}][\widehat{\beta}] \longrightarrow \Omega[\widehat{\beta_2} = [\Omega]\kappa_4, \widehat{\alpha} = \kappa_3][\widehat{\beta} = \kappa_4]$ | By Lemma D.24 |
| $\Delta[\widehat{\beta_2}, \widehat{\alpha}][\widehat{\beta} = \widehat{\beta_2}] \longrightarrow \Omega[\widehat{\beta_2} = [\Omega]\kappa_4, \widehat{\alpha} = \kappa_3][\widehat{\beta} = \kappa_4]$ | By Lemma D.25 |
| $\Omega_1 = \Omega[\widehat{\beta_2} = [\Omega]\kappa_4, \widehat{\alpha} = \kappa_3][\widehat{\beta} = \kappa_4]$ | Let |

□

**Lemma D.46** (Completeness of Unification). *Given $\Delta$ ok, and $\Delta \longrightarrow \Omega$, and $\Delta \Vdash^{kv} \kappa_1$ and $\Delta \Vdash^{kv} \kappa_2$, and $[\Delta]\kappa_1 = \kappa_1$ and $[\Delta]\kappa_2 = \kappa_2$, if $[\Omega]\kappa_1 = [\Omega]\kappa_2$, then there exists $\Theta$ and $\Omega'$ such that $\Theta \longrightarrow \Omega'$, and $\Omega \longrightarrow \Omega'$ and $\Delta \Vdash^{u} \kappa_1 \approx \kappa_2 \dashv \Theta$.*

Proof. By case analysis on $\kappa_1$ on $\kappa_2$.

- $\kappa_1 = \star$ and $\kappa_2 = \star$. Then by rule A-U-REFL, we have $\Theta = \Delta$, and $\Omega' = \Omega$.
- $\kappa_1 = \kappa_{11} \to \kappa_{12}$ and $\kappa_2 = \kappa_{21} \to \kappa_{22}$.

| | |
|---|---|
| $[\Omega](\kappa_{11} \to \kappa_{12})$ | |
| $= [\Omega]\kappa_{11} \to [\Omega]\kappa_{12}$ | By definition |
| $= [\Omega](\kappa_{21} \to \kappa_{22})$ | Given |
| $= [\Omega]\kappa_{21} \to [\Omega]\kappa_{22}$ | By definition |
| $[\Omega]\kappa_{11} = [\Omega]\kappa_{21} \wedge [\Omega]\kappa_{12} = [\Omega]\kappa_{22}$ | Follows directly |
| $\Delta \Vdash^\mu \kappa_{11} \approx \kappa_{12} \dashv \Theta_1 \wedge \Theta_1 \longrightarrow \Omega_1 \wedge \Omega \longrightarrow \Omega_1$ | (1) I.H. |
| $[\Omega_1]([\Theta_1]\kappa_{12}) = [\Omega_1]\kappa_{12} = [\Omega_1]([\Omega]\kappa_{12})$ | By Lemma D.18 |
| $= [\Omega_1]([\Omega]\kappa_{22})$ | Known |
| $= [\Omega_1]\kappa_{22} = [\Omega_1]([\Theta_1]\kappa_{22})$ | By Lemma D.18 |
| $\Theta_1 \Vdash^\mu [\Theta_1]\kappa_{21} \approx [\Theta_1]\kappa_{22} \dashv \Theta \wedge \Theta \longrightarrow \Omega' \wedge \Omega_1 \longrightarrow \Omega'$ | (2) I.H. |
| $\Delta \Vdash^\mu \kappa_{11} \to \kappa_{12} \approx \kappa_{21} \to \kappa_{22} \dashv \Theta$ | By rule A-U-ARROW and (1) (2) |
| $\Omega \longrightarrow \Omega'$ | By Lemma D.20 |

- $\kappa_2 = \widehat{\alpha}$. Then we have $[\Omega]\widehat{\alpha} = [\Omega]\kappa_1$.
  - $\kappa_1 = \widehat{\alpha}$. Then by rule A-U-REFL, we have $\Theta = \Delta$, and $\Omega' = \Omega$.
  - Otherwise $\kappa_1$ must be free of $\widehat{\alpha}$.

| | |
|---|---|
| $\Delta \Vdash^{\text{pr}}_{\widehat{\alpha}} \kappa_1 \rightsquigarrow \kappa_2 \dashv \Theta_1 \wedge \Theta_1 \longrightarrow \Omega' \wedge \Omega \longrightarrow \Omega'$ | By Lemma D.45 |
| $\Delta \Vdash^{\text{kv}} \widehat{\alpha}$ | Given |
| $\Theta_1 = \Theta_{11}, \widehat{\alpha}, \Theta_{12}$ | By Lemma D.3 |
| $\Theta = \Theta_{11}, \widehat{\alpha} = \kappa_2, \Theta_{12}$ | Let |
| $\Delta \Vdash^\mu \kappa_1 \approx \widehat{\alpha} \dashv \Theta$ | By rule A-U-KVARR |
| $[\Omega']\widehat{\alpha}$ | |
| $= [\Omega']\kappa_1$ | By Lemma D.18 |
| $= [\Omega']\kappa_2$ | By Lemma D.38 |
| $\Theta \longrightarrow \Omega'$ | By Lemma D.25 |

- The case when $\kappa_1 = \widehat{\alpha}$ is the same.

□

**Lemma D.47** (Completeness of Application Kinding). *Given $\Delta$ ok, and $\Delta \longrightarrow \Omega$, and $\Delta \Vdash^{\text{kv}} \kappa$ and $\Delta \Vdash^{\text{kv}} \kappa'$, and $[\Delta]\kappa = \kappa$ and $[\Delta]\kappa' = \kappa'$, if $[\Omega]\kappa = [\Omega]\kappa' \to \kappa_1$, then there exists $\kappa_2$, $\Theta$ and $\Omega'$ such that $\Theta \longrightarrow \Omega'$, and $\Omega \longrightarrow \Omega'$ and $\Delta \Vdash^{\text{kapp}} \kappa \bullet \kappa' : \kappa_2 \dashv \Theta$, and $[\Omega']\kappa_2 = \kappa_1$.*

Proof. By induction on $\kappa$.

- $\kappa = \widehat{\alpha}$ for some $\widehat{\alpha}$ and $[\Omega]\widehat{\alpha} = [\Omega]\kappa' \to \kappa_1$.

| | |
|---|---|
| $\Delta = \Delta_1, \widehat{\alpha}, \Delta_2$ | Assume |
| $\Delta_3 = \Delta_1, \widehat{\alpha_1}, \widehat{\alpha_2}, \widehat{\alpha} = \widehat{\alpha_1} \to \widehat{\alpha_2}, \Delta_2$ | Let |
| $\Delta \longrightarrow \Delta_3$ | By Lemma D.23, Lemma D.21, and Lemma D.20 |
| $\Omega = \Omega_1, \widehat{\alpha} = \kappa_3, \Omega_2$ | Assume |
| $\Omega_3 = \Omega_1, \widehat{\alpha_1} = [\Omega]\kappa', \widehat{\alpha_2} = \kappa_1, \widehat{\alpha} = \kappa_3, \Omega_2$ | Let |
| $\Omega \longrightarrow \Omega_3$ | By Lemma D.22, and Lemma D.20 |
| $\Delta \longrightarrow \Omega$ | Given |
| $\Delta_3 \longrightarrow \Omega_3$ | By Lemma D.24 and Lemma D.25 |
| $\Delta_3 \Vdash^\mu \widehat{\alpha_1} \approx \kappa' \dashv \Theta \wedge \Theta \longrightarrow \Omega' \wedge \Omega_3 \longrightarrow \Omega'$ | By Lemma D.46 |
| $\Delta \Vdash^{\text{kapp}} \widehat{\alpha} \bullet \kappa' : \widehat{\alpha_2} \dashv \Theta$ | By rule A-KAPP-KUVAR |

$\Omega \longrightarrow \Omega'$ — By Lemma D.20
$[\Omega']\widehat{\alpha}_2$
$= [\Omega_3]\widehat{\alpha}_2$ — By Lemma D.29
$= \kappa_1$

- Case $\kappa = \kappa_{21} \to \kappa_{22}$.

$[\Omega]\kappa_{21} = [\Omega]\kappa' \land [\Omega]\kappa_{22} = \kappa_1$ — Follows directly
$\Delta \Vdash^\mu \kappa_{21} \approx \kappa' \dashv \Theta \land \Theta \longrightarrow \Omega' \land \Omega \longrightarrow \Omega'$ — By Lemma D.46
$\Delta \Vdash^{\mathsf{kapp}} \kappa \bullet \kappa' : \kappa_{22} \dashv \Theta$ — By rule A-KAPP-ARROW
$[\Omega']\kappa_{22} = [\Omega]\kappa_{22} = \kappa_1$ — By Lemma D.29

□

**Lemma D.48** (Completeness of Kinding). *Given $\Delta$ ok and $\Delta \longrightarrow \Omega$, if $[\Omega]\Delta \vdash^k [\Omega]\sigma : \kappa$, then there exists $\Theta$ and $\Omega'$ such that $\Theta \longrightarrow \Omega'$, and $\Omega \longrightarrow \Omega'$ and $\Delta \Vdash^k \sigma : \kappa' \dashv \Theta$, and $[\Omega']\kappa' = \kappa$.*

Proof. By induction on the kinding judgment.

- Case
$$\frac{}{\Sigma \vdash^k \mathsf{Int} : \star} \text{ K-NAT}$$

$\Delta \Vdash^k \mathsf{Int} : \star \dashv \Delta$ — By rule A-K-NAT
$\Theta = \Delta$ — Let
$\Omega' = \Omega$ — Let

- Case
$$\frac{(a : \kappa) \in \Sigma}{\Sigma \vdash^k a : \kappa} \text{ K-VAR}$$

$(a : \kappa) \in [\Omega]\Delta$ — Given
$(a : \kappa_2) \in \Delta \land [\Omega]\kappa_2 = \kappa$ — By inversion
$\Delta \Vdash^k a : \kappa_2 \dashv \Delta$ — By rule A-K-VAR
$\Theta = \Delta$ — Let
$\Omega' = \Omega$ — Let

- Case
$$\frac{(T : \kappa) \in \Sigma}{\Sigma \vdash^k T : \kappa} \text{ K-TCON}$$

Similar as the case for rule K-VAR.

- Case
$$\frac{}{\Sigma \vdash^k \to : \star \to \star \to \star} \text{ K-ARROW}$$

Similar as the case for rule K-NAT.

- Case

$$\frac{\Sigma, a : \kappa \vdash^k \sigma : \star}{\Sigma \vdash^k \forall a : \kappa.\sigma : \star} \text{\scriptsize K-FORALL}$$

| | |
|---|---|
| $\Delta, a : \kappa \longrightarrow \Omega, a : \kappa$ | By rule A-CTXE-TVAR |
| $[\Omega, a : \kappa](\Delta, a : \kappa) \vdash^k [\Omega]\sigma : \star$ | Given |
| $\Delta, a : \kappa \Vdash^k \sigma : \kappa_1 \dashv \Theta_1 \wedge \Theta_1 \longrightarrow \Omega_1 \wedge \Omega, a : \kappa \longrightarrow \Omega_1 \wedge [\Omega_1]\kappa_1 = \star$ | I.H. |
| $\Theta_1 = \Theta, a : \kappa$ | By inversion |
| $\Delta \Vdash^k \forall a : \kappa.\sigma : \star \dashv \Theta$ | By rule A-K-FORALL |
| $\Omega, a : \kappa \longrightarrow \Omega_1$ | Known |
| $\Omega_1 = \Omega_{11}, a : \kappa, \Omega_{12} \wedge \Omega \longrightarrow \Omega_{11}$ | By Lemma D.17 |
| $\Omega' = \Omega_{11}$ | Let |

- Case

$$\frac{\Sigma \vdash^k \tau_1 : \kappa_1 \to \kappa_2 \quad \Sigma \vdash^k \tau_2 : \kappa_1}{\Sigma \vdash^k \tau_1\,\tau_2 : \kappa_2} \text{\scriptsize K-APP}$$

| | |
|---|---|
| $[\Omega]\Delta \vdash^k [\Omega]\tau_1 : \kappa_1 \to \kappa_2$ | Given |
| $\Delta \Vdash^k \tau_1 : \kappa'_1 \dashv \Theta_1 \wedge \Theta_1 \longrightarrow \Omega_1 \wedge \Omega \longrightarrow \Omega_1 \wedge [\Omega_1]\kappa'_1 = \kappa_1 \to \kappa_2$ | I.H. |
| $\Delta \longrightarrow \Theta_1$ | By Lemma D.4 |
| $\Delta \longrightarrow \Omega_1$ | By Lemma D.20 |
| $[\Omega]\Delta \vdash^k [\Omega]\tau_2 : \kappa_1$ | Given |
| $[\Omega]\tau_2 = [\Omega]([\Omega_1]\tau_2) = [\Omega_1]\tau_2$ | By Lemma D.18 |
| $[\Omega]\Delta \vdash^k [\Omega_1]\tau_2 : \kappa_1$ | Follows directly |
| $[\Omega]\Delta$ | |
| $= [\Omega]\Omega$ | By Lemma D.31 |
| $= [\Omega_1]\Omega_1$ | By Lemma D.34 |
| $= [\Omega_1]\Theta_1$ | By Lemma D.31 |
| $[\Omega_1]\Theta_1 \vdash^k [\Omega_1]\tau_2 : \kappa_1$ | Follows directly |
| $\Theta_1 \Vdash^k \tau_2 : \kappa'_2 \dashv \Theta_2 \wedge \Theta_2 \longrightarrow \Omega_2 \wedge \Omega_1 \longrightarrow \Omega_2 \wedge [\Omega_2]\kappa'_2 = \kappa_1$ | I.H. |
| $[\Omega_2]\kappa'_1 = [\Omega_2]([\Omega_1]\kappa'_1) = \kappa_1 \to \kappa_2$ | By Lemma D.18 |
| $\Theta_1 \Vdash^{kv} \kappa'_1$ | By Lemma D.6 |
| $\Theta_2 \Vdash^{kv} \kappa'_2 \wedge \Theta_1 \longrightarrow \Theta_2$ | By Lemma D.6 |
| $\Theta_2 \Vdash^{kv} \kappa'_1$ | By Lemma D.10 |
| $\Theta_2 \Vdash^{kv} [\Theta_2]\kappa'_1 \wedge \Theta_2 \Vdash^{kv} [\Theta_2]\kappa'_2$ | By Lemma D.12 |
| $\Theta_2 \Vdash^{kapp} [\Theta_2]\kappa'_1 \bullet [\Theta_2]\kappa'_2 : \kappa_3 \dashv \Theta \wedge \Theta \longrightarrow \Omega' \wedge \Omega_2 \longrightarrow \Omega' \wedge [\Omega']\kappa_3 = \kappa_2$ | By Lemma D.47 |
| $\Delta \Vdash^k \tau_1\,\tau_2 : \kappa_3 \dashv \Theta$ | By rule A-K-APP |
| $\Omega \longrightarrow \Omega'$ | By Lemma D.20 |

□

**Lemma D.49** (Completeness of Typing Data Constructor Declaration). *Given $\Delta$ ok and $\Delta \longrightarrow \Omega$, if $[\Omega]\Delta \vdash^{dc}_{\tau'} \mathcal{D} \rightsquigarrow \tau$, then there exists $\Theta$ and $\Omega'$ such that $\Theta \longrightarrow \Omega'$, and $\Omega \longrightarrow \Omega'$ and $\Delta \Vdash^{dc}_{\tau'} \mathcal{D} \rightsquigarrow \tau \dashv \Theta$.*

Proof. Given

$$\frac{\Sigma \vdash^{\mathsf{k}} \overline{\tau_i}^i \to \tau : \star}{\Sigma \vdash^{\mathsf{dc}}_\tau D\,\overline{\tau_i}^i \rightsquigarrow \overline{\tau_i}^i \to \tau} \quad \text{dc-decl}$$

Follows directly from Lemma D.48 and rule a-dc-decl.

□

**Lemma D.50** (Completeness of Typing Datatype Declaration). *Given $\Delta$ ok, and $\Delta \longrightarrow \Omega$, if $[\Omega]\Delta \vdash^{\mathsf{dt}} \mathcal{T} \rightsquigarrow \Psi$, then there exists $\Theta$ and $\Omega'$ such that $\Theta \longrightarrow \Omega'$, and $\Omega \longrightarrow \Omega'$ and $\Delta \Vdash^{\mathsf{dt}} \mathcal{T} \rightsquigarrow \Gamma \dashv \Theta$ and $\Psi = [\Omega']\Gamma$.*

Proof. We have

$$\frac{(T : \overline{\kappa_i}^i \to \star) \in \Sigma \quad \overline{\Sigma, \overline{a_i : \kappa_i}^i \vdash^{\mathsf{dc}}_{T\,\overline{a_i}^i} \mathcal{D}_j \rightsquigarrow \tau_j}^j}{\Sigma \vdash^{\mathsf{dt}} \mathbf{data}\,T\,\overline{a_i}^i = \overline{\mathcal{D}_j}^j \rightsquigarrow \overline{D_j : \forall \overline{a_i : \kappa_i}^i.\tau_j}^j} \quad \text{dt-decl}$$

| | |
|---|---|
| $(T : \overline{\kappa_i}^i \to \star) \in [\Omega]\Delta$ | Given |
| $(T : \kappa) \in \Delta \wedge [\Omega]\kappa = \overline{\kappa_i}^i \to \star$ | By inversion |
| $\Delta \longrightarrow \Omega$ | Given |
| $\Delta, \overline{\widehat{\alpha}_i}^i \longrightarrow \Omega, \overline{\widehat{\alpha}_i = \kappa_i}^i$ | By rule a-ctxe-solve |
| $[\Omega, \overline{\widehat{\alpha}_i = \kappa_i}^i]\kappa = \overline{\kappa_i}^i \to \star = [\Omega, \overline{\widehat{\alpha}_i = \kappa_i}^i](\overline{\widehat{\alpha}_i}^i \to \star)$ | Follows directly |
| $[\Omega, \overline{\widehat{\alpha}_i = \kappa_i}^i]\kappa$ | |
| $= [\Omega, \overline{\widehat{\alpha}_i = \kappa_i}^i]([\Delta, \overline{\widehat{\alpha}_i = \kappa_i}^i]\kappa)$ | By Lemma D.10 |
| $= [\Omega, \overline{\widehat{\alpha}_i = \kappa_i}^i]([\Delta]\kappa)$ | $\widehat{\alpha}_i$ fresh |
| $\Delta, \overline{\widehat{\alpha}_i}^i \Vdash^{\mu} [\Delta]\kappa \approx \overline{\widehat{\alpha}_i}^i \to \star \dashv \Theta_1, \overline{\widehat{\alpha}_i = \kappa'_i}^i$ | By Lemma D.46 and inversion |
| $\wedge \Theta_1, \overline{\widehat{\alpha}_i = \kappa'_i}^i \longrightarrow \Omega_1 \wedge \Omega, \overline{\widehat{\alpha}_i = \kappa_i}^i \longrightarrow \Omega_1$ | Above |
| $[\Omega_1]\kappa_i = [\Omega_1]\kappa'_i = \kappa_i$ | By Lemma D.17 |
| $\Delta, \overline{\widehat{\alpha}_i}^i \longrightarrow \Theta_1, \overline{\widehat{\alpha}_i = \kappa'_i}^i$ | By Lemma D.4 |
| $\Delta \longrightarrow \Theta_1$ | By Lemma D.17 |
| $\Omega_1 = \Omega_{11}, \widehat{\alpha}_1 = \kappa'', \Omega_{12} \wedge \Theta_1 \longrightarrow \Omega_{11}$ | By Lemma D.17 |
| $\Omega \longrightarrow \Omega_{11}$ | By Lemma D.17 |
| $[\Omega_{11}]\kappa'_i = [\Omega_1]\kappa'_i = \kappa_i$ | By Lemma D.19 |
| $\Theta_1, \overline{a_i : \kappa'_i}^i \longrightarrow \Omega_{11}, \overline{a_i : \kappa'_i}^i$ | By rule a-ctxe-tvar |
| $[\Omega_{11}, \overline{a_i : \kappa'_i}^i](\Theta_1, \overline{a_i : \kappa'_i}^i)$ | |
| $= [\Omega_{11}]\Theta_1, \overline{a_i : [\Omega_{11}]\kappa'_i}^i$ | By definition |
| $= [\Omega_{11}]\Theta_1, \overline{a_i : \kappa_i}^i$ | By equations |
| $= [\Omega_{11}]\Omega_{11}, \overline{a_i : \kappa_i}^i$ | By Lemma D.31 |
| $= [\Omega]\Omega, \overline{a_i : \kappa_i}^i$ | By Lemma D.34 |
| $= [\Omega]\Delta, \overline{a_i : \kappa_i}^i$ | By Lemma D.31 |
| $[\Omega]\Delta, \overline{a_i : \kappa_i}^i \vdash^{\mathsf{dc}}_{T\,\overline{a_i}^i} \mathcal{D}_1 \rightsquigarrow \tau_1$ | Given |
| $[\Omega_{11}, \overline{a_i : \kappa'_i}^i](\Theta_1, \overline{a_i : \kappa'_i}^i) \vdash^{\mathsf{dc}}_{T\,\overline{a_i}^i} \mathcal{D}_1 \rightsquigarrow \tau_1$ | By equations |
| $(\Theta_1, \overline{a_i : \kappa'_i}^i) \Vdash^{\mathsf{dc}}_{T\,\overline{a_i}^i} \mathcal{D}_1 \rightsquigarrow \tau_1 \dashv \Theta_2, \overline{a_i : \kappa'_i}^i$ | By Lemma D.49 and inversion |
| $\wedge \Theta_2, \overline{a_i : \kappa'_i}^i \longrightarrow \Omega_2 \wedge \Omega_{11}, \overline{a_i : \kappa'_i}^i \longrightarrow \Omega_2$ | By Lemma D.49 |
| $\Omega_2 = \Omega_{21}, a_1 : \kappa'_1, \Omega_{22} \wedge \Omega_{11} \longrightarrow \Omega_{21}$ | By Lemma D.17 |
| $\Theta_2 \longrightarrow \Omega_{21}$ | By Lemma D.17 |
| $\Theta_2, \overline{a_i : \kappa'_i}^i \longrightarrow \Omega_{21}, \overline{a_i : \kappa'_i}^i$ | By rule A-CTXE-TVAR |
| $\Omega_{11} \longrightarrow \Omega_{21}$ | By Lemma D.17 |
| $\Omega_{11}, \overline{a_i : \kappa'_i}^i \longrightarrow \Omega_{21}, \overline{a_i : \kappa'_i}^i$ | By rule A-CTXE-TVAR |

We repeat the process for each $j$. Let $\Theta_{n+1}, \overline{a_i : \kappa'_i}^i$ and $\Omega_{n+1}, \overline{a_i : \kappa'_i}^i$ be the final output context and the complete context. And $\Theta_{n+1}, \overline{a_i : \kappa'_i}^i \longrightarrow \Omega_{n+1}, \overline{a_i : \kappa'_i}^i$.

By Lemma D.20 we have $\Omega \longrightarrow \Omega_{n+1}$.

By Lemma D.29 we have $[\Omega_{n+1}]\kappa'_i = [\Omega_{11}]\kappa'_i = \kappa_i$.

So collecting all the hypothesis, by rule A-DT-DECL we get $\Delta \Vdash^{\mathsf{dt}} \mathcal{T} \leadsto \Gamma \dashv \Theta$. And $[\Omega_{n+1}]\Gamma = \Psi$.

Let $\Omega' = \Omega_{n+1}$.

□

**Theorem D.51** (Completeness of Typing a Group). *Given $\Omega$ ok, if $[\Omega]\Omega \vdash^{\mathsf{grp}} \mathbf{rec}\ \overline{\mathcal{T}_i}^i \leadsto \overline{\kappa_i}^i; \overline{\Psi_i}^i$, then there exists $\overline{\kappa'_i}^i, \overline{\Gamma_i}^i, \Theta$, and $\Omega'$, such that $\Omega \Vdash^{\mathsf{grp}} \mathbf{rec}\ \overline{\mathcal{T}_i}^i \leadsto \overline{\kappa'_i}^i; \overline{\Gamma_i}^i \dashv \Theta$, where $\Theta \longrightarrow \Omega'$, and $\overline{[\Omega']\kappa'_i = \kappa_i}^i$, and $\overline{\Psi_i = [\Omega']\Gamma_i}^i$.*

PROOF. We have

| | |
|---|---|
| $\Theta_1 = \Omega, \overline{\widehat{\alpha_i}}^{i \in 1..n}, \overline{T_i : \widehat{\alpha_i}}^{i \in 1..n}$ | Let |
| $\Omega_1 = \Omega, \overline{\widehat{\alpha_i} = \kappa_i}^{i \in 1..n}, \overline{T_i : \widehat{\alpha_i}}^{i \in 1..n}$ | Let |
| $\Theta_1 \longrightarrow \Omega_1$ | By Lemma D.21 |
| $[\Omega]\Omega, \overline{T_i : \kappa_i}^i \vdash^{\mathsf{dt}} \mathcal{T}_1 \leadsto \Psi_1$ | Given |
| $[\Omega_1]\Theta_1 \vdash^{\mathsf{dt}} \mathcal{T}_1 \leadsto \Psi_1$ | By definition |
| $\Theta_1 \Vdash^{\mathsf{dt}} \mathcal{T}_1 \leadsto \Gamma_1 \dashv \Theta_2 \wedge \Theta_2 \longrightarrow \Omega_2 \wedge \Omega_1 \longrightarrow \Omega_2 \wedge \Psi_1 = [\Omega_2]\Gamma_1$ | (1) By Lemma D.50 |
| $[\Omega]\Omega, \overline{T_i : \kappa_i}^i \vdash^{\mathsf{dt}} \mathcal{T}_2 \leadsto \Psi_2$ | Given |
| $[\Omega_2]\Theta_2$ | |
| $= [\Omega_2]\Omega_2$ | By Lemma D.31 |
| $= [\Omega_1]\Omega_1$ | By Lemma D.34 |
| $= [\Omega]\Omega, \overline{T_i : \kappa_i}^i$ | By definition |
| $[\Omega_2]\Theta_2 \vdash^{\mathsf{dt}} \mathcal{T}_2 \leadsto \Psi_2$ | Substitute the equation |
| $\Theta_2 \Vdash^{\mathsf{dt}} \mathcal{T}_2 \leadsto \Gamma_2 \dashv \Theta_3 \wedge \Theta_3 \longrightarrow \Omega_3 \wedge \Omega_2 \longrightarrow \Omega_3 \wedge \Psi_2 = [\Omega_3]\Gamma_2$ | (2) By Lemma D.50 |
| By repeating the process from (1) to (2) for each $i$, we can get | |
| $\overline{\Theta_i \Vdash^{\mathsf{dt}} \mathcal{T}_i \leadsto \Gamma_i \dashv \Theta_{i+1}}^{i \in 1..n} \wedge \Theta_{n+1} \longrightarrow \Omega_{n+1} \wedge \overline{\Psi_i = [\Omega_{i+1}]\Gamma_i}^i$ | |
| $\Omega' = \Omega_{n+1}$ | Let |
| $\overline{[\Omega']\widehat{\alpha_i} = [\Omega']([\Omega_1]\widehat{\alpha_i})}^i$ | By Lemma D.18 |
| $\overline{[\Omega']\widehat{\alpha_i} = \kappa_i}^i$ | Namely |
| $\overline{[\Omega']\Gamma_i = [\Omega_{i+1}]\Gamma_i}^i$ | By Lemma D.30 |
| $\overline{[\Omega']\Gamma_i = \Psi_i}^i$ | Namely |

□

# E PROOF FOR HASKELL98 WITH KIND PARAMETERS
## E.1 List of Lemmas

**Theorem E.1** (Principality of Haskell98 with Kind Parameters). *If* $\Sigma \vdash^{\text{grp}} \text{rec } \overline{\mathcal{T}_i}^i \leadsto \overline{\kappa_i}^i; \overline{\Psi_i}^i$, *then there exists some* $\overline{\kappa'_i}^i$ *such that* $\Sigma \vdash \text{rec } \overline{\mathcal{T}_i}^i \leadsto^{\text{p}} \overline{\kappa'_i}^i$.

**Theorem E.2** (Completeness of Typing Programs with Kind Parameters). *Given algorithmic contexts* $\Omega, \Gamma$, *and a program pgm, if* $[\Omega]\Omega; [\Omega]\Gamma \vdash^{\text{pgm}} pgm : \sigma$, *then* $\Omega; \Gamma \Vdash^{\text{pgm}} pgm : \sigma$.

## E.2 Proofs

**Theorem E.1** (Principality of Haskell98 with Kind Parameters). *If* $\Sigma \vdash^{\text{grp}} \text{rec } \overline{\mathcal{T}_i}^i \leadsto \overline{\kappa_i}^i; \overline{\Psi_i}^i$, *then there exists some* $\overline{\kappa'_i}^i$ *such that* $\Sigma \vdash \text{rec } \overline{\mathcal{T}_i}^i \leadsto^{\text{p}} \overline{\kappa'_i}^i$.

Proof. We have

| | |
|---|---|
| $\Sigma \vdash^{\text{grp}} \text{rec } \overline{\mathcal{T}_i}^i \leadsto \overline{\kappa_i}^i; \overline{\Psi_i}^i$ | Given |
| $\Omega = \Sigma$ | Let |
| $\Omega \Vdash^{\text{grp}} \text{rec } \overline{\mathcal{T}_i}^i \leadsto \overline{\kappa'_i}^i; \overline{\Gamma_i}^i \dashv \Theta$ | By Theorem D.51 |
| $\Theta \longrightarrow \Omega' \wedge \overline{[\Omega']\kappa'_i = \kappa_i}^i \wedge \overline{[\Omega']\Gamma_i = \Psi_i}^i$ | Above |

We solve all unsolved kind unification variables in $\Theta$ with fresh kind parameters to get $\Omega_1$. Then we choose $\overline{\kappa''_i = [\Omega_1]\kappa'_i}^i$, and we prove $\Sigma \vdash \text{rec } \overline{\mathcal{T}_i}^i \leadsto^{\text{p}} \overline{\kappa''_i}^i$.

| | |
|---|---|
| $\Theta \longrightarrow \Omega_1$ | By Lemma D.21 |
| $\Omega, \overline{\widehat{\alpha}_i}^i, \overline{T_i : \widehat{\alpha}_i}^i \longrightarrow \Omega_1$ | By Lemma D.20 |
| $\Omega_1 = \Omega_{11}, \widehat{\alpha}_1 = \kappa_{11}, \Omega_{12}, \overline{T_i : \widehat{\alpha}_i, \Omega'_i}^i$ | By Lemma D.17 |
| $\wedge \Omega \longrightarrow \Omega_{11} \wedge \Omega_{12} \text{ soft} \wedge \overline{\Omega'_i \text{ soft}}^i$ | Above |
| $[\Omega_1]\Theta$ | |
| $= [\Omega_1](\Omega, \overline{\widehat{\alpha}_i}^i, \overline{T_i : \widehat{\alpha}_i}^i)$ | By Lemma D.33 |
| $= [\Omega]\Omega, \overline{T_i : [\Omega_1]\widehat{\alpha}_i}^i$ | By definition and Lemma D.32 |
| $= \Sigma, \overline{T_i : \kappa''_i}^i$ | By Lemma D.17 |
| $\Sigma \vdash^{\text{grp}} \text{rec } \overline{\mathcal{T}_i}^i \leadsto \overline{\kappa''_i}^i; \overline{\Psi'_i}^i$ | repeat Lemma D.43 |

For any $\overline{\kappa_i}^i$ such that $\Sigma \vdash^{\text{grp}} \text{rec } \overline{\mathcal{T}_i}^i \leadsto \overline{\kappa_i}^i; \overline{\Psi_i}^i$, by Theorem D.51 we know there exists some $\Omega'$ such that $\Theta \longrightarrow \Omega'$ and $\overline{[\Omega']\kappa'_i = \kappa_i}^i$ and $\overline{[\Omega']\Gamma_i = \Psi_i}^i$. Now we construct a kind parameter substitution $S$. If in $\Theta$, we have an unsolved kind unification variable $\widehat{\alpha}$, which maps to a parameter $P$ in $\Omega_1$, then $S$ maps $P$ to $[\Omega']\widehat{\alpha}$. Because $\Theta \longrightarrow \Omega'$, then $S(\Omega_1) \longrightarrow \Omega'$ by Lemma D.25. So $S(\kappa''_i) = S([\Omega_1]\kappa'_i) = [S(\Omega_1)]\kappa'_i$. By Lemma D.29, we have $[S(\Omega_1)]\kappa'_i = [\Omega']\kappa'_i = \kappa_i$. Similarly we have $S(\Psi'_i) = \Psi_i$.

□

**Theorem E.2** (Completeness of Typing Programs with Kind Parameters). *Given algorithmic contexts* $\Omega, \Gamma$, *and a program pgm, if* $[\Omega]\Omega; [\Omega]\Gamma \vdash^{\text{pgm}} pgm : \sigma$, *then* $\Omega; \Gamma \Vdash^{\text{pgm}} pgm : \sigma$.

Proof. By induction on typing programs.

- Case

$$\frac{\Sigma; \Psi \vdash e : \sigma}{\Sigma; \Psi \vdash^{\text{pgm}} e : \sigma} \text{ pgm-expr}$$

Follows trivially by rule A-PGM-EXPR.

- Case PGM-DTP

$$\frac{\Sigma \vdash^{\text{grp}} \mathbf{rec}\,\overline{\mathcal{T}_i}^i \leadsto \overline{\kappa_i}^i; \overline{\Psi_i}^i \quad \Sigma \vdash \mathbf{rec}\,\overline{\mathcal{T}_i}^i \leadsto^{\text{p}} \overline{\kappa_i}^i \quad \Sigma, \overline{T_i : S^\star(\kappa_i)}^i; \Psi, \overline{S^\star(\Psi_i)}^i \vdash^{\text{pgm}} pgm : \sigma}{\Sigma; \Psi \vdash^{\text{pgm}} \mathbf{rec}\,\overline{\mathcal{T}_i}^i; pgm : \sigma}$$

| | |
|---|---|
| $\Sigma \vdash^{\text{grp}} \mathbf{rec}\,\overline{\mathcal{T}_i}^i \leadsto \overline{\kappa_i}^i; \overline{\Psi_i}^i$ | Given |
| $\Sigma \vdash \mathbf{rec}\,\overline{\mathcal{T}_i}^i \leadsto^{\text{p}} \overline{\kappa_i}^i$ | Given |
| $\Omega \Vdash^{\text{grp}} \mathbf{rec}\,\overline{\mathcal{T}_i}^i \leadsto \overline{\kappa'_i}^i; \overline{\Gamma_i}^i \dashv \Theta$ | By Theorem D.51 |
| $\Theta \longrightarrow \Omega'$ | Above |
| $\overline{[\Omega']\kappa'_i = \kappa_i}^i$ | Above |
| $\overline{[\Omega']\Gamma_i = \Psi_i}^i$ | Above |

Because from Theorem E.1 we know that if we solve all unsolved kind unification variables in $\Theta$ with fresh parameters to get $\Omega_1$, then $\overline{[\Omega_1]\kappa'_i}^i$ are principal kinds. Because $\overline{\kappa_i}^i$ are principal kinds, then $\overline{[\Omega_1]\kappa'_i}^i$ and $\overline{\kappa_i}^i$ are equivalent up to renaming of type parameters. Suppose $\Theta \longrightarrow \Omega_2$, then $\overline{[\Omega_2]\kappa'_i = S^\star(\kappa_i)}^i$. Similarly we can prove $\overline{[\Omega_2]\Gamma_i = S^\star(\Psi_i)}^i$.

| | |
|---|---|
| $\Omega, \overline{\widehat{\alpha_i}}^{i\in 1..n}, \overline{T_i : \widehat{\alpha_i}}^{i\in 1..n} \longrightarrow \Theta$ | By Lemma D.8 |
| $\Theta \longrightarrow \Omega_2$ | By Lemma D.13 |
| $\Omega, \overline{\widehat{\alpha_i}}^{i\in 1..n}, \overline{T_i : \widehat{\alpha_i}}^{i\in 1..n} \longrightarrow \Omega_2$ | By Lemma D.20 |
| $\Omega, \overline{\widehat{\alpha_i} = [\Omega_2]\kappa'_i}^{i\in 1..n}, \overline{T_i : \widehat{\alpha_i}}^{i\in 1..n} \longrightarrow \Omega_2$ | By Lemma D.25 |
| $[\Omega_2]\Omega_2$ | |
| $= [\Omega, \overline{\widehat{\alpha_i} = [\Omega_2]\kappa'_i}^{i\in 1..n}, \overline{T_i : \widehat{\alpha_i}}^{i\in 1..n}]$ | |
| $(\Omega, \overline{\widehat{\alpha_i} = [\Omega_2]\kappa'_i}^{i\in 1..n}, \overline{T_i : \widehat{\alpha_i}}^{i\in 1..n})$ | By Lemma D.34 |
| $= [\Omega]\Omega, \overline{T_i : [\Omega]([\Omega_2]\kappa'_i)}^{i\in 1..n}$ | By definition |
| $= [\Omega]\Omega, \overline{T_i : S^\star(\kappa_i)}^{i\in 1..n}$ | By substituting the equation |
| $[\Omega_2](\Gamma, \overline{\Gamma_i}^i)$ | |
| $= [\Omega_2]\Gamma, \overline{[\Omega_2]\Gamma_i}^i$ | By definition |
| $= [\Omega]\Gamma, \overline{S^\star(\Psi_i)}^i$ | By Lemma D.30 and Lemma D.19 |
| $[\Omega]\Omega, \overline{T_i : S^\star(k_i)}^i; [\Omega]\Gamma, \overline{S^\star(\Psi_i)}^i \vdash_p pgm : \sigma$ | Given |
| $[\Omega_2]\Omega_2; [\Omega_2](\Gamma, \overline{\Gamma_i}^i) \vdash^{\text{pgm}} pgm : \sigma$ | By substituting the equations |
| $\Omega_2; \Gamma, \overline{\Gamma_i}^i \Vdash^{\text{pgm}} pgm : \sigma$ | I.H. |
| $\Omega; \Gamma \Vdash^{\text{pgm}} \mathbf{rec}\,\overline{\mathcal{T}_i}^i; pgm : \sigma$ | By rule A-PGM-DT |

□

# F PROOF FOR POLYKINDS
## F.1 List of Lemmas
### F.1.1 Well-formedness of Declarative Type System.

**Lemma F.1** (Well-formedness of Declarative Instantiation). *If $\Sigma \vdash^{\mathsf{ela}} \mu_1 : \eta_1$, and $\Sigma \vdash^{\mathsf{inst}} \mu_1 : \eta_1 \sqsubseteq \eta_2 \rightsquigarrow \mu_2$, then $\Sigma \vdash^{\mathsf{ela}} \mu_2 : \eta_2$.*

**Lemma F.2** (Well-formedness of Declarative Kinding). *We have:*
- *if $\Sigma \vdash^{\mathsf{k}} \sigma : \eta \rightsquigarrow \mu$, then $\Sigma \vdash^{\mathsf{ela}} \mu : \eta$;*
- *if $\Sigma \vdash^{\mathsf{kc}} \sigma \Leftarrow \eta \rightsquigarrow \mu$, then $\Sigma \vdash^{\mathsf{ela}} \mu : \eta$.*

**Lemma F.3** (Well-formedness of Declarative Elaborated Kinding). *If $\Sigma$ ok, and $\Sigma \vdash^{\mathsf{ela}} \mu : \eta$, then $\Sigma \vdash^{\mathsf{ela}} \eta : \star$.*

**Lemma F.4** (Well-formedness of Declarative Typing Signature). *If $\Sigma$ ok, and $\Sigma \vdash^{\mathsf{sig}} \mathcal{S} \rightsquigarrow T : \eta$, then $\Sigma \vdash^{\mathsf{ela}} \eta : \star$.*

**Lemma F.5** (Well-formedness of Declarative Typing Data Constructor Declaration). *If $\Sigma$ ok, and $\Sigma \vdash^{\mathsf{dc}}_{\rho} \mathcal{D} \rightsquigarrow \mu$, then $\Sigma \vdash^{\mathsf{ela}} \mu : \star$.*

**Lemma F.6** (Well-formedness of Declarative Typing Datatype Declaration). *If $\Sigma$ ok, and $\Sigma \vdash^{\mathsf{dt}} \mathcal{T} \rightsquigarrow \Psi$, then $\Sigma \vdash \Psi$.*

**Lemma F.7** (Well-formedness of Declarative Generalization). *If $\Sigma$ ok, and $\Sigma \vdash \Psi_1$ and $\Sigma \vdash^{\mathsf{gen}}_{\phi^c} \Psi_1 \rightsquigarrow \Psi_2$, then $\Sigma \vdash \Psi_2$.*

### F.1.2 Well-formedness of Algorithmic Type System.

**Lemma F.8** (Well-formedness of Promotion). *If $\Delta_1, \widehat{\alpha} : \omega, \Delta_2$ ok, and $\Delta_1, \widehat{\alpha} : \omega, \Delta_2 \Vdash^{\mathsf{ela}} \rho_1 : \omega_2$, and $\Delta_1, \widehat{\alpha} : \omega, \Delta_2 \Vdash^{\mathsf{pr}}_{\widehat{\alpha}} \rho_1 \rightsquigarrow \rho_2 \dashv \Theta$, then $\Theta = \Theta_1, \widehat{\alpha} : \omega, \Theta_2$, and $\Delta_1, \widehat{\alpha} : \omega, \Delta_2 \longrightarrow \Theta$, and $\Theta_1 \Vdash^{\mathsf{ela}} \rho_2 : [\Theta]\omega_2$, and $\Theta$ ok. By weakening, there is also $\Theta \Vdash^{\mathsf{ela}} \rho_2 : [\Theta]\omega_2$. Similar lemma holds when in the input context, $\widehat{\alpha} : \omega$ is in a local scope.*

**Lemma F.9** (Well-formedness of Moving). *If $\Delta_1 \mathbin{+\!\!+}^{\mathsf{mv}} \Delta_2 \rightsquigarrow \Theta$, then $\mathsf{topo}(\Delta_1, \Delta_2) = \Theta$.*

**Lemma F.10** (Well-formedness of Unification). *If $\Delta$ ok, and $\Delta \Vdash^{\mathsf{u}} \kappa_1 \approx \kappa_2 \dashv \Theta$, then $\Delta \longrightarrow \Theta$, and $\Theta$ ok.*

**Lemma F.11** (Well-formedness of Instantiation). *If $\Delta \Vdash^{\mathsf{inst}} \rho_1 : \eta_1 \sqsubseteq \eta_2 \rightsquigarrow \rho_2 \dashv \Theta$, and $\Delta \Vdash^{\mathsf{ela}} \rho_1 : \eta_1$, then $\Delta \longrightarrow \Theta$, and $\Theta$ ok, and $\Theta \Vdash^{\mathsf{ela}} \rho_2 : [\Theta]\eta_2$.*

**Lemma F.12** (Well-formedness of Quantification Check). *If $\Delta_1, a : \omega, \Delta_2$ ok, and $\Delta_2 \hookleftarrow a$, then $\Delta_1, \Delta_2$ ok.*

**Lemma F.13** (Well-formedness of Unsolved). *If $\Delta_1, \Delta_2$ ok, and $\Delta_2$ **soft**, then $\Delta_1, \mathsf{unsolved}(\Delta_2)$ ok.*

**Lemma F.14** (Well-formedness of topo). *If $\Delta_1, \Delta_2$ ok, then $\Delta_1, \mathsf{topo}(\Delta_2)$ ok.*

**Lemma F.15** (Well-formedness of Kinding). *Given $\Delta$ ok,*

- if $\Delta \Vdash^{\mathsf{k}} \sigma : \eta \rightsquigarrow \mu \dashv \Theta$, then $\Delta \longrightarrow \Theta$, and $\Theta$ ok and $\Theta \Vdash^{\mathsf{ela}} \mu : [\Theta]\eta$;
- if $\Delta \Vdash^{\mathsf{kc}} \sigma \Leftarrow \eta \rightsquigarrow \mu \dashv \Theta$, then $\Delta \longrightarrow \Theta$, and $\Theta$ ok and $\Theta \Vdash^{\mathsf{ela}} \mu : [\Theta]\eta$.
- if $\Delta \Vdash^{\mathsf{kapp}} (\rho_1 : \eta) \bullet \tau : \omega \rightsquigarrow \rho_2 \dashv \Theta$, and $\Delta \Vdash^{\mathsf{ela}} \rho_1 : \eta$, then $\Delta \longrightarrow \Theta$, and $\Theta$ ok, and $\Theta \Vdash^{\mathsf{ela}} \rho_2 : [\Theta]\omega$.

**Lemma F.16** (Well-formedness of Elaborated Kinding). *If $\Delta$ ok, and $\Delta \Vdash^{\mathsf{ela}} \mu : \eta$, then $\Delta \Vdash^{\mathsf{ela}} \eta : \star$, and $[\Delta]\eta = \eta$.*

**Lemma F.17** (Well-formedness of Typing Signature). *If $\Omega$ ok, and $\Omega \Vdash^{\mathsf{sig}} \mathcal{S} \rightsquigarrow T : \mu$, then $\Omega \Vdash^{\mathsf{ela}} \mu : \star$.*

**Lemma F.18** (Well-formedness of Typing Data Constructor Declaration). *If $\Delta$ ok, and $\Delta \Vdash^{\mathsf{dc}}_\rho \mathcal{D} \rightsquigarrow \mu \dashv \Theta$, then $\Delta \longrightarrow \Theta$, and $\Theta \Vdash^{\mathsf{ela}} \mu : \star$.*

**Lemma F.19** (Well-formedness of Typing Datatype Declaration). *If $\Delta$ ok, and $\Delta \Vdash^{\mathsf{dt}} \mathcal{T} \rightsquigarrow \Gamma \dashv \Theta$, then $\Delta \longrightarrow \Theta$, and $\Theta \Vdash^{\mathsf{ectx}} \Gamma$.*

**Lemma F.20** (Well-formedness of Generalization). *If $\Delta$ ok, and $\Delta \Vdash^{\mathsf{ectx}} \Gamma_1$, and $\Delta \Vdash^{\mathsf{gen}}_{\phi^c} \Gamma_1 \rightsquigarrow \Gamma_2$, then $\Delta \Vdash^{\mathsf{ectx}} \Gamma_2$.*

### F.1.3 Properties of Context Extension.

**Lemma F.21** (Declaration Preservation). *If $\Delta \longrightarrow \Theta$, if a type constructor or a type variable or a kind unification variable is declared in $\Delta$, then it is declared in $\Theta$.*

**Lemma F.22** (Extension Weakening). *Given $\Delta \longrightarrow \Theta$, if $\Delta \Vdash^{\mathsf{ela}} \mu : \eta$, then $\Theta \Vdash^{\mathsf{ela}} \mu : [\Theta]\eta$.*

**Definition F.23** (Contextual Size).

| | | |
|---|---|---|
| $\mid \Delta \vdash \star \mid$ | = | $1$ |
| $\mid \Delta \vdash a \mid$ | = | $1$ |
| $\mid \Delta \vdash \mathsf{Int} \mid$ | = | $1$ |
| $\mid \Delta \vdash T \mid$ | = | $1$ |
| $\mid \Delta \vdash \rightarrow \mid$ | = | $1$ |
| $\mid \Delta \vdash \omega_1\, \omega_2 \mid$ | = | $1+ \mid \Delta \vdash \omega_1 \mid + \mid \Delta \vdash \omega_2 \mid$ |
| $\mid \Delta \vdash \omega_1\, @\omega_2 \mid$ | = | $1+ \mid \Delta \vdash \omega_1 \mid + \mid \Delta \vdash \omega_2 \mid$ |
| $\mid \Delta[\widehat{\alpha} : \omega] \vdash \widehat{\alpha} \mid$ | = | $1$ |
| $\mid \Delta[\widehat{\alpha} : \omega = \rho] \vdash \widehat{\alpha} \mid$ | = | $1+ \mid \Delta[\widehat{\alpha} : \omega = \rho] \vdash \omega \mid$ |
| $\mid \Delta \vdash \forall a : \rho.\omega \mid$ | = | $1+ \mid \Delta \vdash \rho \mid + \mid \Delta \vdash \omega \mid$ |
| $\mid \Delta \vdash \forall \{a : \rho\}.\omega \mid$ | = | $1+ \mid \Delta \vdash \rho \mid + \mid \Delta \vdash \omega \mid$ |

**Lemma F.24** (Substitution Kinding). *If $\Delta$ ok, and $\Delta \Vdash^{\mathsf{ela}} \mu : \eta$, then $\Delta \Vdash^{\mathsf{ela}} [\Delta]\mu : \eta$.*

**Lemma F.25** (Soft Substitution Kinding). *If $\Delta_1, \Delta_2$ ok, and $\Delta_2$ **soft**, and $\Delta_1, \Delta_2 \Vdash^{\mathsf{ela}} \mu : \eta$, then $\Delta_1, \mathsf{unsolved}(\Delta_2) \Vdash^{\mathsf{ela}} [\Delta_2]\mu : \eta$.*

**Lemma F.26** (Reflexivity of Context Extension). *If $\Delta$ ok, then $\Delta \longrightarrow \Delta$.*

**Lemma F.27** (Well-formedness of Context Extension). *If $\Delta$ ok, and $\Delta \longrightarrow \Theta$, then $\Theta$ ok.*

**Definition F.28** (Softness). *A context $\Delta$ is soft iff it contains only of $\widehat{\alpha}$ and $\widehat{\alpha} = \kappa$ declarations, including local scopes.*

**Lemma F.29** (Extension Order).

(1) *If $\Delta_1, a : \omega, \Delta_2 \longrightarrow \Theta$, then $\Theta = \Theta_1, a : \omega, \Theta_2$, where $\Delta_1 \longrightarrow \Theta_1$. Moreover, if $\Delta_2$ soft, then $\Theta_2$ soft.*
(2) *If $\Delta_1, T : \eta, \Delta_2 \longrightarrow \Theta$, then $\Theta = \Theta_1, T : \eta, \Theta_2$, where $\Delta_1 \longrightarrow \Theta_1$. Moreover, if $\Delta_2$ soft, then $\Theta_2$ soft.*
(3) *If $\Delta_1, \widehat{\alpha} : \omega, \Delta_2 \longrightarrow \Theta$, then $\Theta = \Theta_1, \Theta', \Theta_2$, where $\Delta_1 \longrightarrow \Theta_1$, and $\Theta'$ is either $\widehat{\alpha} : \omega$ or $\widehat{\alpha} : \omega = \rho$ for some $\rho$. Moreover, if $\Delta_2$ soft, then $\Theta_2$ soft.*
(4) *If $\Delta_1, \widehat{\alpha} : \omega = \rho_1, \Delta_2 \longrightarrow \Theta$, then $\Theta = \Theta_1, \widehat{\alpha} : \omega = \rho_2, \Theta_2$, where $\Delta_1 \longrightarrow \Theta_1$, and $[\Theta_1]\rho_1 = [\Theta_1]\rho_2$. Moreover, if $\Delta_2$ soft, then $\Theta_2$ soft.*
(5) *If $\Delta_1, \{\Delta\}, \Delta_2 \longrightarrow \Theta$, then $\Theta = \Theta_1, \{\Theta\}, \Theta_2$, where $\Delta_1 \longrightarrow \Theta_1$, Moreover, if $\Delta_2$ soft, then $\Theta_2$ soft.*

**Lemma F.30** (Substitution Extension Invariance). *If $\Delta$ ok, and $\Delta \Vdash^{\mathsf{ela}} \mu : \eta$, and $\Delta \longrightarrow \Theta$, then $[\Theta]\kappa = [\Theta]([\Delta]\mu)$ and $[\Theta]\kappa = [\Delta]([\Theta]\mu)$. As a corollary, if $\Delta \Vdash^{\mathsf{ela}} \mu_1 : \eta_1$, $\Delta \Vdash^{\mathsf{ela}} \mu_2 : \eta_2$, and $[\Delta]\mu_1 = [\Delta]\mu_2$, then $[\Theta]\mu_1 = [\Theta]\mu_2$.*

**Lemma F.31** (Substitution Stability). *If $\Delta_1, \Delta_2$ ok, and $\Delta_1 \Vdash^{\mathsf{ela}} \rho : \omega$, then $[\Delta_1]\rho = [\Delta_1, \Delta_2]\rho$.*

**Lemma F.32** (Transitivity of Context Extension). *If $\Delta'$ ok, and $\Delta' \longrightarrow \Delta$, and $\Delta \longrightarrow \Theta$, then $\Delta' \longrightarrow \Theta$.*

**Lemma F.33** (Solution Admissibility for Extension).

- *If $\Delta_1, \widehat{\alpha} : \omega, \Delta_2$ ok and $\Delta_1 \Vdash^{\mathsf{ela}} \rho : [\Delta_1]\omega$, then $\Delta_1, \widehat{\alpha} : \omega, \Delta_2 \longrightarrow \Delta_1, \widehat{\alpha} : \omega = \rho, \Delta_2$.*
- *If $\Delta_1, \{\Delta_3, \widehat{\alpha} : \omega, \Delta_4\}, \Delta_2$ ok and $\Delta_1, \Delta_3 \Vdash^{\mathsf{ela}} \rho : [\Delta_1, \Delta_3]\omega$, then $\Delta_1, \{\Delta_3, \widehat{\alpha} : \omega, \Delta_4\}, \Delta_2 \longrightarrow \Delta_1, \{\Delta_3, \widehat{\alpha} : \omega = \rho, \Delta_4\}, \Delta_2$.*

**Lemma F.34** (Solved Variable Addition for Extension).

- *If $\Delta_1, \Delta_2$ ok and $\Delta_1 \Vdash^{\mathsf{ela}} \rho : [\Delta_1]\omega$, then $\Delta_1, \Delta_2 \longrightarrow \Delta_1, \widehat{\alpha} : \omega = \rho, \Delta_2$.*
- *If $\Delta_1, \{\Delta_2, \Delta_3\}, \Delta_4$ ok and $\Delta_1, \Delta_2 \Vdash^{\mathsf{ela}} \rho : [\Delta_1, \Delta_2]\omega$, then $\Delta_1, \{\Delta_2, \Delta_3\}, \Delta_4 \longrightarrow \Delta_1, \{\Delta_2, \widehat{\alpha} : \omega = \rho, \Delta_3\}, \Delta_4$.*

**Lemma F.35** (Unsolved Variable Addition).

- *If $\Delta_1, \Delta_2$ ok and $\Delta_1 \Vdash^{\mathsf{ela}} \omega : \star$ then $\Delta_1, \Delta_2 \longrightarrow \Delta_1, \widehat{\alpha} : \omega, \Delta_2$.*
- *If $\Delta_1, \{\Delta_2, \Delta_3\}, \Delta_4$ ok and $\Delta_1, \Delta_2 \Vdash^{\mathsf{ela}} \omega : \star$, then $\Delta_1, \{\Delta_2, \Delta_3\}, \Delta_4 \longrightarrow \Delta_1, \{\Delta_2, \widehat{\alpha} : \omega, \Delta_3\}, \Delta_4$.*

**Lemma F.36** (Parallel Admissibility).

- *If $\Delta_1 \longrightarrow \Theta_1$, and $\Delta_1, \Delta_2$ ok, and $\Delta_1, \Delta_2 \longrightarrow \Theta_1, \Theta_2$, and $\Delta_2$ is fresh w.r.t. $\Theta_1$, then:*
  – *if $\Delta_1 \Vdash^{\mathsf{ela}} \omega : \star$, then $\Delta_1, \widehat{\alpha} : \omega, \Delta_2 \longrightarrow \Theta_1, \widehat{\alpha} : \omega, \Theta_2$;*
  – *if $\Theta_1 \Vdash^{\mathsf{ela}} \rho : [\Theta_1]\omega$, then $\Delta_1, \widehat{\alpha} : \omega, \Delta_2 \longrightarrow \Theta_1, \widehat{\alpha} : \omega = \rho, \Theta_2$;*
  – *if $[\Theta_1]\rho_1 = [\Theta_1]\rho_2$, then $\Delta_1, \widehat{\alpha} : \omega = \rho_1, \Delta_2 \longrightarrow \Theta_1, \widehat{\alpha} : \omega = \rho_2, \Theta_2$.*
- *If $\Delta_1, \{\Delta_3\} \longrightarrow \Theta_1, \{\Theta_3\}$, and $\Delta_1, \{\Delta_3, \Delta_4\}, \Delta_2$ ok, and $\Delta_1, \{\Delta_3, \Delta_4\}, \Delta_2 \longrightarrow \Theta_1, \{\Theta_3, \Theta_4\}, \Theta_2$, and $\Delta_2, \Delta_4$ is fresh w.r.t. $\Theta_1, \Theta_3$, then:*
  – *if $\Delta_1, \{\Delta_3\} \Vdash^{\mathsf{ela}} \omega : \star$, then $\Delta_1, \{\Delta_3, \widehat{\alpha} : \omega, \Delta_4\}, \Delta_2 \longrightarrow \Theta_1, \{\Theta_3, \widehat{\alpha} : \omega, \Theta_4\}, \Theta_2$;*
  – *if $\Theta_1, \{\Theta_3\} \Vdash^{\mathsf{ela}} \rho : [\Theta_1, \Theta_3]\omega$, then $\Delta_1, \{\Delta_3, \widehat{\alpha} : \omega, \Delta_4\}, \Delta_2 \longrightarrow \Theta_1, \{\Theta_3, \widehat{\alpha} : \omega = \rho, \Theta_4\}, \Theta_2$;*

- *if* $[\Theta_1, \Theta_3]\rho_1 = [\Theta_1, \Theta_3]\rho_2$, *then* $\Delta_1, \{\Delta_3, \widehat{\alpha} : \omega = \rho_1, \Delta_4\}, \Delta_2 \longrightarrow \Theta_1, \{\Theta_3, \widehat{\alpha} : \omega = \rho_2, \Theta_4\}, \Theta_2$.

**Lemma F.37** (Parallel Extension Solution).
- *If* $\Delta_1, \widehat{\alpha} : \omega, \Delta_2 \longrightarrow \Theta_1, \widehat{\alpha} : \omega = \rho_2, \Theta_2$, *and* $[\Theta_1]\rho_1 = [\Theta_1]\rho_2$, *then* $\Delta_1, \widehat{\alpha} : \omega = \rho_1, \Delta_2 \longrightarrow \Theta_1, \widehat{\alpha} : \omega = \kappa_2, \Theta_2$.
- *If* $\Delta_1, \{\Delta_3, \widehat{\alpha} : \omega, \Delta_4\}, \Delta_2 \longrightarrow \Theta_1, \{\Theta_3, \widehat{\alpha} : \omega = \rho_2, \Theta_4\}, \Theta_2$, *and* $[\Theta_1, \Theta_3]\rho_1 = [\Theta_1, \Theta_3]\rho_2$, *then* $\Delta_1, \{\Delta_3, \widehat{\alpha} : \omega = \rho_1, \Delta_4\}, \Delta_2 \longrightarrow \Theta_1, \{\Theta_3, \widehat{\alpha} : \omega = \rho_2, \Theta_4\}, \Theta_2$.

**Lemma F.38** (Parallel Variable Update).
- *If* $\Delta_1, \widehat{\alpha} : \omega, \Delta_2 \longrightarrow \Theta_1, \widehat{\alpha} : \omega = \rho, \Theta_2$, *and* $\Delta_1 \Vdash^{\text{ela}} \rho_1 : [\Delta_1]\omega$, *and* $\Theta_1 \Vdash^{\text{ela}} \rho_2 : [\Theta_1]\omega$, *and* $[\Theta_1]\rho = [\Theta_1]\rho_1 = [\Theta_1]\rho$, *then* $\Delta_1, \widehat{\alpha} : \omega = \rho_1, \Delta_2 \longrightarrow \Theta_1, \widehat{\alpha} : \omega = \rho_2, \Theta_2$
- *If* $\Delta_1, \{\Delta_3, \widehat{\alpha} : \omega, \Delta_4\}, \Delta_2 \longrightarrow \Theta_1, \{\Theta_3, \widehat{\alpha} : \omega = \rho, \Theta_4\}, \Theta_2$, *and* $\Delta_1, \Delta_3 \Vdash^{\text{ela}} \rho_1 : [\Delta_1, \Delta_3]\omega$, *and* $\Theta_1, \Theta_3 \Vdash^{\text{ela}} \rho_2 : [\Theta_1, \Theta_3]\omega$, *and* $[\Theta_1]\rho = [\Theta_1]\rho_1 = [\Theta_1]\rho$, *then* $\Delta_1, \{\Delta_3, \widehat{\alpha} : \omega = \rho_1, \Delta_4\}, \Delta_2 \longrightarrow \Theta_1, \{\Theta_3, \widehat{\alpha} : \omega = \rho_2, \Theta_4\}, \Theta_2$

### F.1.4 Properties of Complete Context.

**Lemma F.39** (Type Constructor Preservation). *If $\Delta$ ok, then $(T : \eta) \in \Delta$, and $\Delta \longrightarrow \Omega$, then $(T : [\Omega]\eta) \in [\Omega]\Delta$.*

**Lemma F.40** (Type Variable Preservation). *If $(a : \omega) \in \Delta$, and $\Delta \longrightarrow \Omega$, then $(a : [\Omega]\omega) \in [\Omega]\Delta$.*

**Lemma F.41** (Finishing Kinding). *If $\Omega$ ok, and $\Omega \Vdash^{\text{ela}} \rho : \omega$, and $\Omega \longrightarrow \Omega'$, then $[\Omega]\rho = [\Omega']\rho$.*

**Lemma F.42** (Finishing Term Contexts). *If $\Omega$ ok, and $\Omega \Vdash^{\text{ectx}} \Gamma$, and $\Omega \longrightarrow \Omega'$, then $[\Omega']\Gamma = [\Omega]\Gamma$.*

**Lemma F.43** (Stability of Complete Contexts). *If $\Delta \longrightarrow \Omega$, then $[\Omega]\Delta = [\Omega]\Omega$.*

**Lemma F.44** (Softness Goes Away). *If $\Delta_1, \Delta_2 \longrightarrow \Omega_1, \Omega_2$ where $\Delta_1 \longrightarrow \Omega_1$, and $\Delta_2$ **soft**, then $[\Omega_1, \Omega_2](\Delta_1, \Delta_2) = [\Omega_1]\Delta_1$.*

**Lemma F.45** (Confluence of Completeness). *If $\Delta_1 \longrightarrow \Omega$, and $\Delta_2 \longrightarrow \Omega$, then $[\Omega]\Delta_1 = [\Omega]\Delta_2$.*

**Lemma F.46** (Finishing Completions). *If $\Omega$ ok, and $\Omega \longrightarrow \Omega'$, then $[\Omega']\Omega' = topo([\Omega]\Omega)$.*

### F.1.5 Termination.

**Lemma F.47** (Promotion Preserves $\langle \Delta \rangle$). *If $\Delta \Vdash^{\text{pr}}_{\widehat{\alpha}} \omega_1 \rightsquigarrow \omega_2 \dashv \Theta$, then $\langle \Delta \rangle = \langle \Theta \rangle$.*

**Lemma F.48** (Unification Makes Progress). *If $\Delta \Vdash^{\mu} \omega_1 \approx \omega_2 \dashv \Theta$, then either $\Theta = \Delta$, or $\langle \Theta \rangle < \langle \Delta \rangle$.*

**Lemma F.49** (Promotion Preserves $|\rho|$). *Given a context $\Delta[\widehat{\alpha}]$ ok, if $\Delta \Vdash^{\text{pr}}_{\widehat{\alpha}} \omega_1 \rightsquigarrow \omega_2 \dashv \Theta$, then for all $\rho$, we have $|[\Delta]\rho| = |[\Theta]\rho|$.*

**Theorem F.50** (Promotion Terminates). *Given a context $\Delta[\widehat{\alpha}]$ ok, and a kind $\rho_1$ with $[\Delta]\rho_1 = \rho_1$, it is decidable whether there exists $\Theta$ such that $\Delta \Vdash^{\text{pr}}_{\widehat{\alpha}} \omega_1 \rightsquigarrow \omega_2 \dashv \Theta$.*

**Theorem F.51** (Unification Terminates). *Given a context $\Delta$ ok, and kinds $\rho_1$ and $\rho_2$, where $[\Delta]\rho_1 = \rho_1$, and $[\Delta]\rho_2 = \rho_2$, it is decidable whether there exists $\Theta$ such that $\Delta \Vdash^{\mu} \rho_1 \approx \rho_2 \dashv \Theta$.*

#### F.1.6 Source of Unification Variables.

**Lemma F.52** (Source of Unification Variables). *If $\Delta \Vdash^k \sigma : \eta \leadsto \mu \dashv \Theta$, then for any $\widehat{\alpha} \in$ unsolved($\Theta$), either $\widehat{\alpha} \in$ fkv($[\Theta]\mu$), or there exists $\widehat{\beta} \in$ unsolved($\Delta$) such that $\widehat{\alpha} \in$ fkv($[\Theta]\widehat{\beta}$).*

#### F.1.7 Soundness of Algorithm.

**Lemma F.53** (Soundness of Promotion). *If $\Delta$ ok, and $[\Delta]\omega_1 = \omega_1$, and $\Delta \Vdash^{\text{pr}}_{\widehat{\alpha}} \omega_1 \leadsto \omega_2 \dashv \Theta$, then $[\Theta]\omega_1 = [\Theta]\omega_2 = \omega_2$. If $\Theta \longrightarrow \Omega$, then $[\Omega]\omega_1 = [\Omega]\omega_2$.*

**Lemma F.54** (Soundness of Unification). *If $\Delta$ ok, and $\Delta \Vdash^{\mu} \omega_1 \approx \omega_2 \dashv \Theta$, then $[\Theta]\omega_1 = [\Theta]\omega_2$. If $\Theta \longrightarrow \Omega$, then $[\Omega]\omega_1 = [\Omega]\omega_2$.*

**Lemma F.55** (Soundness of Instantiation). *If $\Delta$ ok, and $\Delta \Vdash^{\text{ela}} \mu_1 : \eta$, and $\Delta \Vdash^{\text{ela}} \omega : \star$, and $\Delta \Vdash^{\text{inst}} \mu_1 : \eta \sqsubseteq \omega \leadsto \mu_2 \dashv \Theta$, and $\Theta \longrightarrow \Omega$, then $[\Omega]\Delta \vdash^{\text{inst}} [\Omega]\mu_1 : [\Omega]\eta \sqsubseteq [\Omega]\omega \leadsto [\Omega]\mu_2$.*

**Lemma F.56** (Soundness of Kinding). *If $\Delta$ ok, we have*
- *if $\Delta \Vdash^k \sigma : \eta \leadsto \mu \dashv \Theta$, and $\Theta \longrightarrow \Omega$, then $[\Omega]\Delta \vdash^k \sigma : [\Omega]\eta \leadsto [\Omega]\mu$;*
- *if $\Delta \Vdash^{\text{kc}} \sigma \Leftarrow \eta \leadsto \mu \dashv \Theta$, and $\Theta \longrightarrow \Omega$, then $[\Omega]\Delta \vdash^{\text{kc}} \sigma \Leftarrow [\Omega]\eta \leadsto [\Omega]\mu$.*
- *if $\Delta \Vdash^{\text{kapp}} (\rho_1 : \eta) \bullet \tau : \omega \leadsto \rho_2 \dashv \Theta$, and $\Delta \Vdash^{\text{ela}} \rho_1 : \eta$, and $\Theta \longrightarrow \Omega$, then $[\Omega]\Delta \vdash^{\text{inst}} [\Omega]\rho_1 : [\Omega]\eta \sqsubseteq (\omega_1 \to [\Omega]\omega) \leadsto \rho_3$, and $[\Omega]\Delta \vdash^{\text{kc}} \tau \Leftarrow \omega_1 \leadsto \rho_4$. and $[\Omega]\rho_2 = \rho_3\,\rho_4$.*

**Lemma F.57** (Soundness of Elaborated Kinding). *If $\Delta$ ok, and $\Delta \Vdash^{\text{ela}} \mu : \eta$, and $\Delta \longrightarrow \Omega$, then $[\Omega]\Delta \vdash^{\text{ela}} [\Omega]\mu : [\Omega]\eta$.*

**Lemma F.58** (Soundness of Typing Signature). *If $\Delta$ ok, and $\Omega \Vdash^{\text{sig}} \mathcal{S} \leadsto T : \eta$, then $[\Omega]\Omega \vdash^{\text{sig}} \mathcal{S} \leadsto T : \eta$.*

**Lemma F.59** (Soundness of Typing Data Constructor Decl.). *If $\Delta$ ok, and $\Delta \Vdash^{\text{dc}}_{\rho} \mathcal{D} \leadsto \mu \dashv \Theta$, and $\Theta \longrightarrow \Omega$, then $[\Omega]\Delta \vdash^{\text{dc}}_{([\Omega]\rho)} \mathcal{D} \leadsto [\Omega]\mu$.*

**Lemma F.60** (Soundness of Typing Datatype Decl.). *If $\Delta$ ok, and $\Delta \Vdash^{\text{dt}} \mathcal{T} \leadsto \Gamma \dashv \Theta$, and $\Theta \longrightarrow \Omega$, then $[\Omega]\Delta \vdash^{\text{dt}} \mathcal{T} \leadsto [\Omega]\Gamma$.*

**Lemma F.61** (Soundness of Typing Program). *If $\Omega; \Gamma \Vdash^{\text{pgm}} pgm : \mu$, then $[\Omega]\Omega; [\Omega]\Gamma \vdash^{\text{pgm}} pgm : [\Omega]\mu$.*

#### F.1.8 Principality.

**Lemma F.62** (Completeness of Promotion). *Given $\Delta$ ok, and $\Delta \longrightarrow \Omega$, and $\widehat{\alpha} \in \Delta$, and $\Delta \Vdash^{\text{ela}} \rho : \omega$, and $[\Delta]\widehat{\alpha} = \widehat{\alpha}$, and $[\Delta]\rho = \rho$, if $\rho$ does not depend on $\widehat{\alpha}$ in the dependency graph, then there exists $\rho_2$, $\Theta$ and $\Omega'$ such that $\Theta \longrightarrow \Omega'$, and $\Omega \longrightarrow \Omega'$, and $\Delta \Vdash^{\text{pr}}_{\widehat{\alpha}} \rho \leadsto \rho_2 \dashv \Theta$.*

**Lemma F.63** (Completeness of Unification). *Given $\Delta$ ok, and $\Delta \longrightarrow \Omega$, and $\Delta \Vdash^{\text{ela}} \rho_1 : \omega$ and $\Delta \Vdash^{\text{ela}} \rho_2 : \omega$, and $[\Delta]\rho_1 = \rho_1$ and $[\Delta]\rho_2 = \rho_2$, if $[\Omega]\rho_1 = [\Omega]\rho_2$, then there exists $\Theta$ and $\Omega'$ such that $\Theta \longrightarrow \Omega'$, and $\Omega \longrightarrow \Omega'$ and $\Delta \Vdash^{\mu} \rho_1 \approx \rho_2 \dashv \Theta$.*

**Lemma F.64** (Completeness of Instantiation). *Given $\Delta \longrightarrow \Omega$, and $\Delta \Vdash^{\text{ela}} \rho : \eta$ and $\Delta \Vdash^{\text{ela}} \omega : \star$, and $[\Delta]\eta = \eta$ and $[\Delta]\omega = \omega$, if $[\Omega]\Delta \vdash^{\text{inst}} [\Omega]\rho_1 : [\Omega]\eta \sqsubseteq [\Omega]\omega \leadsto \rho_2$, then there exists $\rho'_2$, $\Theta$ and $\Omega'$ such that $\Theta \longrightarrow \Omega'$, and $\Omega \longrightarrow \Omega'$ and $\Delta \Vdash^{\text{inst}} \rho_1 : \eta \sqsubseteq \omega \leadsto \rho'_2 \dashv \Theta$, and $[\Omega']\rho'_2 = \rho_2$.*

**Lemma F.65** (Principality of Kinding).

- *Given $\Delta \longrightarrow \Omega$, if $[\Omega]\Delta \vdash^k \sigma : \eta \rightsquigarrow \mu$, and $\Delta \Vdash^k \sigma : \eta' \rightsquigarrow \mu' \dashv \Theta$, then there exists $\Omega'$ such that $\Theta \longrightarrow \Omega'$, and $\Omega \longrightarrow \Omega'$. Moreover, $[\Omega']\eta' = \eta$. Furthermore, if $\mu$ and $\mu'$ are monotypes, then $[\Omega']\mu' = \mu$.*
- *Given $\Delta \longrightarrow \Omega$, if $[\Omega]\Delta \vdash^{kc} \sigma \Leftarrow [\Omega]\eta \rightsquigarrow \mu$, and $\Delta \Vdash^{kc} \sigma \Leftarrow \eta \rightsquigarrow \mu' \dashv \Theta$, then there exists $\Omega'$ such that $\Theta \longrightarrow \Omega'$, and $\Omega \longrightarrow \Omega'$. Furthermore, if $\mu$ and $\mu'$ are monotypes, then $[\Omega']\mu' = \mu$.*
- *Given $\Delta \longrightarrow \Omega$, if $[\Omega]\Delta \vdash^{inst} [\Omega]\rho_1 : [\Omega]\eta \sqsubseteq (\omega_1 \rightarrow \omega_2) \rightsquigarrow \rho_3$, and $[\Omega]\Delta \vdash^{kc} \tau \Leftarrow \omega_1 \rightsquigarrow \rho_4$ and $\Delta \Vdash^{kapp} (\rho_1 : \eta) \bullet \tau : \omega \rightsquigarrow \rho_2 \dashv \Theta$, then there exists $\Omega'$ such that $\Theta \longrightarrow \Omega'$, and $\Omega \longrightarrow \Omega'$. Moreover, $[\Omega']\omega = \omega_2$. Further, $[\Omega']\rho_2 = \rho_3 \rho_4$.*

**Lemma F.66** (Principality of Typing Data Constructor Declaration). *Given $\Delta \longrightarrow \Omega$, if $[\Omega]\Delta \vdash^{dc}_\rho \mathcal{D} \rightsquigarrow \mu_1$, and $\Delta \Vdash^{dc}_\rho \mathcal{D} \rightsquigarrow \mu_2 \dashv \Theta$, then there exists $\Omega'$ such that $\Theta \longrightarrow \Omega'$, and $\Omega \longrightarrow \Omega'$.*

**Lemma F.67** (Principality of Typing Datatype Declaration). *Given $\Delta \longrightarrow \Omega$, if $[\Omega]\Delta \vdash^{dt} \mathcal{T} \rightsquigarrow \Psi$, and $\Delta \Vdash^{dt} \mathcal{T} \rightsquigarrow \Gamma \dashv \Theta$, then there exists $\Omega'$ such that $\Theta \longrightarrow \Omega'$, and $\Omega \longrightarrow \Omega'$.*

**Theorem F.68** (Principality of Typing a Datatype Declaration Group). *If $\Omega \Vdash^{grp} \mathbf{rec}\ \overline{\mathcal{T}_i}^i \rightsquigarrow \overline{\eta_i}^i; \overline{\Gamma_i}^i$, then whenever $[\Omega]\Omega \vdash^{grp} \mathbf{rec}\ \overline{\mathcal{T}_i}^i \rightsquigarrow \overline{\eta'_i}^i; \overline{\Psi_i}^i$ holds, we have $[\Omega]\Omega \vdash [\Omega]\eta_i \leq \eta'_i$.*

### F.2 Proofs

**Lemma F.1** (Well-formedness of Declarative Instantiation). *If $\Sigma \vdash^{ela} \mu_1 : \eta_1$, and $\Sigma \vdash^{inst} \mu_1 : \eta_1 \sqsubseteq \eta_2 \rightsquigarrow \mu_2$, then $\Sigma \vdash^{ela} \mu_2 : \eta_2$.*

Proof. By induction on the derivation.

- Case

$$\frac{}{\Sigma \vdash^{inst} \mu : \omega \sqsubseteq \omega \rightsquigarrow \mu} \text{\scriptsize INST-REFL}$$

  The goal follows trivially.
- Case

$$\frac{\Sigma \vdash^{ela} \rho : \omega_1 \qquad \Sigma \vdash^{inst} \mu_1 @\rho : \eta[a \mapsto \rho] \sqsubseteq \omega_2 \rightsquigarrow \mu_2}{\Sigma \vdash^{inst} \mu_1 : \forall a : \omega_1.\eta \sqsubseteq \omega_2 \rightsquigarrow \mu_2} \text{\scriptsize INST-FORALL}$$

| | |
|---|---|
| $\Sigma \vdash^{ela} \mu_1 : \forall a : \omega_1.\eta$ | Given |
| $\Sigma \vdash^{ela} \rho : \omega_1$ | Given |
| $\Sigma \vdash^{ela} \mu_1 @\rho : \eta[a \mapsto \rho]$ | By rule ELA-KAPP |
| $\Sigma \vdash^{ela} \mu_2 : \eta_2$ | I.H. |

- Case

$$\frac{\Sigma \vdash^{ela} \rho : \omega_1 \qquad \Sigma \vdash^{inst} \mu_1 @\rho : \eta[a \mapsto \rho] \sqsubseteq \omega_2 \rightsquigarrow \mu_2}{\Sigma \vdash^{inst} \mu_1 : \forall\{a : \omega_1\}.\eta \sqsubseteq \omega_2 \rightsquigarrow \mu_2} \text{\scriptsize INST-FORALL-INFER}$$

  Similar as the previous case.

□

**Lemma F.2** (Well-formedness of Declarative Kinding). *We have:*

- *if $\Sigma \vdash^k \sigma : \eta \rightsquigarrow \mu$, then $\Sigma \vdash^{ela} \mu : \eta$;*

- *if* $\Sigma \vdash^{kc} \sigma \Leftarrow \eta \rightsquigarrow \mu$, *then* $\Sigma \vdash^{ela} \mu : \eta$.

Proof. By induction on the derivation.

**Part 1**
- Case for rules KTT-STAR, KTT-NAT, KTT-VAR, KTT-TCON, and KTT-ARROW holds trivially.
- Case

  KTT-APP
  $$\dfrac{\Sigma \vdash^{k} \tau_1 : \eta_1 \rightsquigarrow \rho_1 \quad \Sigma \vdash^{inst} \rho_1 : \eta_1 \sqsubseteq (\omega_1 \rightarrow \omega_2) \rightsquigarrow \rho_2 \quad \Sigma \vdash^{kc} \tau_2 \Leftarrow \omega_1 \rightsquigarrow \rho_3}{\Sigma \vdash^{k} \tau_1\ \tau_2 : \omega_2 \rightsquigarrow \rho_2\ \rho_3}$$

| | |
|---|---|
| $\Sigma \vdash^{ela} \rho_1 : \eta_1$ | I.H. |
| $\Sigma \vdash^{ela} \rho_2 : \omega_1 \rightarrow \omega_2$ | By Lemma F.1 |
| $\Sigma \vdash^{ela} \rho_3 : \omega_1$ | By part 2 |
| $\Sigma \vdash^{ela} \rho_2\ \rho_3 : \omega_2$ | By rule ELA-APP |

- The rest cases are similar, following directly from I.H. and part 2.

**Part 2**

KC-SUB
$$\dfrac{\Sigma \vdash^{k} \sigma : \eta \rightsquigarrow \mu_1 \quad \Sigma \vdash^{inst} \mu_1 : \eta \sqsubseteq \omega \rightsquigarrow \mu_2}{\Sigma \vdash^{kc} \sigma \Leftarrow \omega \rightsquigarrow \mu_2}$$

| | |
|---|---|
| $\Sigma \vdash^{ela} \mu_1 : \eta$ | By part 1 |
| $\Sigma \vdash^{ela} \mu_2 : \omega$ | By Lemma F.1 |

□

**Lemma F.3** (Well-formedness of Declarative Elaborated Kinding). *If $\Sigma$ ok, and $\Sigma \vdash^{ela} \mu : \eta$, then $\Sigma \vdash^{ela} \eta : \star$.*

Proof. By a straightforward induction on the judgment, utilizing the substitution lemma.

□

**Lemma F.4** (Well-formedness of Declarative Typing Signature). *If $\Sigma$ ok, and $\Sigma \vdash^{sig} S \rightsquigarrow T : \eta$, then $\Sigma \vdash^{ela} \eta : \star$.*

Proof. We have

SIG-TT
$$\dfrac{\rceil\sigma\lceil \quad \phi \in Q(\sigma) \quad \phi_1^c \in Q(\forall \phi^c.\ \eta) \quad \Sigma, \phi_1^c \vdash^{k} \forall \phi.\ \sigma : \star \rightsquigarrow \forall \phi^c.\ \eta \quad |\phi| = |\phi^c|}{\Sigma \vdash^{sig} \mathbf{data}\ T : \sigma \rightsquigarrow T : \forall\{\phi_1^c\}.\forall\{\phi^c\}.\eta}$$

| | |
|---|---|
| $\Sigma, \phi_1^c \vdash^{k} \forall \phi.\ \sigma : \star \rightsquigarrow \forall \phi^c.\ \eta$ | Given |
| $\Sigma, \phi_1^c \vdash^{ela} \forall \phi^c.\ \eta : \star$ | By Lemma F.2 |
| $\Sigma, \phi_1^c \vdash^{ela} \forall \{\phi^c\}.\eta : \star$ | By rule ELA-FORALL-INFER |
| $\phi_1^c$ is well-formed | |
| $\Sigma \vdash^{ela} \forall\{\phi_1^c\}.\forall\{\phi^c\}.\eta : \star$ | By rule ELA-FORALL-INFER |

□

**Lemma F.5** (Well-formedness of Declarative Typing Data Constructor Declaration). *If $\Sigma$ ok, and $\Sigma \vdash^{dc}_\rho \mathcal{D} \rightsquigarrow \mu$, then $\Sigma \vdash^{ela} \mu : \star$.*

Proof. We have

$$\frac{\phi^c \in Q(\mu\backslash_{\Sigma,\overline{\tau_i}^i}) \quad \Sigma, \phi^c \vdash^k \forall \phi.\overline{\tau_i}^i \to \rho : \star \leadsto \mu}{\Sigma \vdash^{dc}_\rho \forall \phi.D\,\overline{\tau_i}^i \leadsto \forall\{\phi^c\}.\mu} \text{ DC-TT}$$

| | |
|---|---|
| $\Sigma, \phi^c \vdash^k \forall \phi.\overline{\tau_i}^i \to \rho : \star \leadsto \mu$ | Given |
| $\Sigma, \phi^c \vdash^{ela} \mu : \star$ | By Lemma F.2 |
| $\phi^c$ is well-formed | |
| $\Sigma \vdash^{ela} \forall\{\phi^c\}.\mu : \star$ | By rule ELA-FORALL-INFER |

□

**Lemma F.6** (Well-formedness of Declarative Typing Datatype Declaration). *If $\Sigma$ ok, and $\Sigma \vdash^{dt} \mathcal{T} \leadsto \Psi$, then $\Sigma \vdash \Psi$.*

Proof. We have

$$\frac{(T : \forall\{\phi_1^c\}.\forall\phi_2^c.\overline{\omega_i}^i \to \star) \in \Sigma \quad \overline{\Sigma, \phi_1^c, \phi_2^c, \overline{a_i : \omega_i}^i \vdash^{dc}_{(T\,@\phi_1^c\,@\phi_2^c\,\overline{a_i}^i)} \mathcal{D}_j \leadsto \mu_j}^j}{\Sigma \vdash^{dt} \text{data } T\,\overline{a_i}^i = \overline{\mathcal{D}_j}^j \leadsto \overline{D_j : \forall\{\phi_1^c\}.\forall\phi_2^c.\,\forall\overline{a_i : \omega_i}^i.\mu_j}^j} \text{ DT-TT}$$

| | |
|---|---|
| $\overline{\Sigma, \phi_1^c, \phi_2^c, \overline{a_i : \omega_i}^i \vdash^{ela} \mu_j : \star}^j$ | By Lemma F.5 |
| $\overline{\Sigma \vdash^{ela} \forall\{\phi_1^c\}.\forall\phi_2^c.\,\forall\overline{a_i : \omega_i}^i.\mu_j : \star}^j$ | By rule ELA-FORALL |
| $\Sigma \vdash \overline{D_j : \forall\{\phi_1^c\}.\forall\phi_2^c.\,\forall\overline{a_i : \omega_i}^i.\mu_j}^j$ | By rule ECTX-DCON |

□

**Lemma F.7** (Well-formedness of Declarative Generalization). *If $\Sigma$ ok, and $\Sigma \vdash \Psi_1$ and $\Sigma \vdash^{gen}_{\phi^c} \Psi_1 \leadsto \Psi_2$, then $\Sigma \vdash \Psi_2$.*

Proof. Follows directly from rule ECTX-DCON-TT and rule ELA-FORALL-INFER.

□

### F.2.1 Well-formedness of Algorithmic Type System.

**Lemma F.8** (Well-formedness of Promotion). *If $\Delta_1, \widehat{\alpha} : \omega, \Delta_2$ ok, and $\Delta_1, \widehat{\alpha} : \omega, \Delta_2 \Vdash^{ela} \rho_1 : \omega_2$, and $\Delta_1, \widehat{\alpha} : \omega, \Delta_2 \Vdash^{pr}_{\widehat{\alpha}} \rho_1 \leadsto \rho_2 \dashv \Theta$, then $\Theta = \Theta_1, \widehat{\alpha} : \omega, \Theta_2$, and $\Delta_1, \widehat{\alpha} : \omega, \Delta_2 \longrightarrow \Theta$, and $\Theta_1 \Vdash^{ela} \rho_2 : [\Theta]\omega_2$, and $\Theta$ ok. By weakening, there is also $\Theta \Vdash^{ela} \rho_2 : [\Theta]\omega_2$. Similar lemma holds when in the input context, $\widehat{\alpha} : \omega$ is in a local scope.*

Proof. For most cases, the goal follows directly.
The case for rule A-PR-KAPP is similar as rule A-PR-APP.

- Case

$$\frac{\Delta \Vdash^{pr}_{\widehat{\alpha}} \omega_1 \leadsto \rho_1 \dashv \Delta_1 \quad \Delta_1 \Vdash^{pr}_{\widehat{\alpha}} [\Delta_1]\omega_2 \leadsto \rho_2 \dashv \Theta}{\Delta \Vdash^{pr}_{\widehat{\alpha}} \omega_1\,\omega_2 \leadsto \rho_1\,\rho_2 \dashv \Theta} \text{ A-PR-APP}$$

| | |
|---|---|
| $\Delta \Vdash^{ela} \omega_1\,\omega_2 : \omega_2'$ | Given |
| $\Delta \Vdash^{ela} \omega_1 : \omega_1' \to \omega_2' \wedge \Delta \Vdash^{ela} \omega_2 : \omega_1'$ | By inversion |
| $\Delta \longrightarrow \Delta_1 \wedge \Delta_1 = \Delta_{11}, \widehat{\alpha} : \rho, \Delta_{12} \wedge \Delta_1 \text{ ok} \wedge \Delta_{11} \Vdash^{\text{ela}} \rho_1 : [\Delta_{11}]\omega_1' \to [\Delta_{11}]\omega_2'$ | I.H. |
| $\Delta_1 \Vdash^{\text{ela}} \omega_2 : [\Delta_1]\omega_1'$ | By Lemma F.22 |
| $\Delta_1 \Vdash^{\text{ela}} [\Delta_1]\omega_2 : [\Delta_1]\omega_1'$ | By Lemma F.24 |
| $\Delta_1 \longrightarrow \Theta \wedge \Theta = \Theta_1, \widehat{\alpha} : \rho, \Theta_2 \wedge \Theta \text{ ok} \wedge \Theta_1 \Vdash^{\text{ela}} \rho_2 : [\Theta]([\Delta_1]\omega_1')$ | I.H. |
| $\Delta_{11} \longrightarrow \Theta_1$ | By Lemma F.29 |
| $\Delta \longrightarrow \Theta$ | By Lemma F.32 |
| $\Theta_1 \Vdash^{\text{ela}} \rho_1 : [\Theta_1]([\Delta_{11}]\omega_1') \to [\Theta_1]([\Delta_{11}]\omega_2')$ | By Lemma F.22 |
| $\Theta_1 \Vdash^{\text{ela}} \rho_1 : [\Theta_1]\omega_1' \to [\Theta_1]\omega_2'$ | By Lemma F.30 |
| $\Theta_1 \Vdash^{\text{ela}} \rho_2 : [\Theta]\omega_1'$ | By Lemma F.30 |
| $\Theta_1 \Vdash^{\text{ela}} \rho_1 \rho_2 : [\Theta]\omega_2'$ | By rule a-ela-app |

- Case

$$\dfrac{\Delta \Vdash^{\text{pr}}_{\widehat{\alpha}} [\Delta]\rho \rightsquigarrow \rho_1 \dashv \Theta[\widehat{\alpha}][\widehat{\beta} : \rho]}{\Delta[\widehat{\alpha}][\widehat{\beta} : \rho] \Vdash^{\text{pr}}_{\widehat{\alpha}} \widehat{\beta} \rightsquigarrow \widehat{\beta}_1 \dashv \Theta[\widehat{\beta}_1 : \rho_1, \widehat{\alpha}][\widehat{\beta} : \rho = \widehat{\beta}_1]} \text{a-pr-kuvarR-tt}$$

| | |
|---|---|
| $\Delta \text{ ok}$ | Given |
| $\Delta \Vdash^{\text{ela}} \rho : \star$ | By inversion |
| $\Theta[\widehat{\beta}_1 : \rho_1, \widehat{\alpha}][\widehat{\beta} : \rho = \widehat{\beta}_1] = \Theta_1, \widehat{\beta}_1 : \rho_1, \widehat{\alpha} : \rho_2, \Theta_2, \widehat{\beta} : \rho = \widehat{\beta}_1, \Theta_3$ | Suppose |
| $\Theta[\widehat{\alpha}][\widehat{\beta} : \rho] \text{ ok} \wedge \Delta \longrightarrow \Theta[\widehat{\alpha}][\widehat{\beta} : \rho] \wedge \Theta_1 \Vdash^{\text{ela}} \rho_1 : \star$ | I.H. |
| $\Delta \longrightarrow \Theta[\widehat{\beta}_1 : \rho_1, \widehat{\alpha}][\widehat{\beta} : \rho]$ | By Lemma F.35, Lemma F.32 |
| $\Theta_1, \widehat{\beta}_1 : \rho_1, \widehat{\alpha} : \rho_2, \Theta_2 \Vdash^{\text{ela}} \widehat{\beta}_1 : [\Theta_1]\rho_1$ | By rule a-ela-var and Lemma F.31 |
| $\Theta_1, \widehat{\beta}_1 : \rho_1, \widehat{\alpha} : \rho_2, \Theta_2 \Vdash^{\text{ela}} \widehat{\beta}_1 : [\Theta[\widehat{\alpha}][\widehat{\beta} : \rho]]\rho_1$ | By Lemma F.31 |
| $\Theta_1, \widehat{\beta}_1 : \rho_1, \widehat{\alpha} : \rho_2, \Theta_2 \Vdash^{\text{ela}} \widehat{\beta}_1 : [\Theta[\widehat{\alpha}][\widehat{\beta} : \rho]]([\Delta]\rho)$ | By Lemma F.53 |
| $\Theta_1, \widehat{\beta}_1 : \rho_1, \widehat{\alpha} : \rho_2, \Theta_2 \Vdash^{\text{ela}} \widehat{\beta}_1 : [\Theta[\widehat{\alpha}][\widehat{\beta} : \rho]]\rho$ | By Lemma F.30 |
| $\Theta_1, \widehat{\beta}_1 : \rho_1, \widehat{\alpha} : \rho_2, \Theta_2 \Vdash^{\text{ela}} \widehat{\beta}_1 : [\Theta_1, \widehat{\alpha} : \rho_2, \Theta_2]\rho$ | By Lemma F.53 |
| $\Delta \longrightarrow \Theta[\widehat{\beta}_1 : \rho_1, \widehat{\alpha}][\widehat{\beta} : \rho = \widehat{\beta}_1]$ | By Lemma F.33, Lemma F.32 |

□

**Lemma F.9** (Well-formedness of Moving). *If $\Delta_1 \mathbin{+\!\!+}^{\text{mv}} \Delta_2 \rightsquigarrow \Theta$, then $\text{topo}(\Delta_1, \Delta_2) = \Theta$.*

Proof. By a straightforward induction on the moving judgment.

□

**Lemma F.10** (Well-formedness of Unification). *If $\Delta$ ok, and $\Delta \Vdash^{\text{u}} \kappa_1 \approx \kappa_2 \dashv \Theta$, then $\Delta \longrightarrow \Theta$, and $\Theta$ ok.*

Proof. By induction on the derivation.

- The case for rule a-u-refl-tt follows directly from Lemma F.26.
- Case

$$\dfrac{\Delta \Vdash^{\text{pr}}_{\widehat{\alpha}} \rho_1 \rightsquigarrow \rho_2 \dashv \Theta_1, \widehat{\alpha} : \omega_1, \Theta_2 \quad \Theta_1 \Vdash^{\text{ela}} \rho_2 : \omega_2 \quad \Theta_1 \Vdash^{\text{u}} [\Theta_1]\omega_1 \approx \omega_2 \dashv \Theta_3}{\Delta \Vdash^{\text{u}} \widehat{\alpha} \approx \rho_1 \dashv \Theta_3, \widehat{\alpha} : \omega_1 = \rho_2, \Theta_2} \text{a-u-kvarL-tt}$$

| | |
|---|---|
| $\Delta \longrightarrow \Theta_1, \widehat{\alpha} : \omega_1, \Theta_2 \wedge \Theta_1, \widehat{\alpha} : \omega_1, \Theta_2 \text{ ok}$ | By Lemma F.8 |
| $\Theta_1 \longrightarrow \Theta_3$ | I.H. |
| $\Theta_1 \Vdash^{\text{ela}} \rho_2 : \omega_2$ | Given |
| $\Theta_3 \Vdash^{\text{ela}} \rho_2 : [\Theta_3]\omega_2$ | By Lemma F.22 |
| $[\Theta_3]\omega_2 = [\Theta_3]([\Theta_1]\omega_1)$ | By Lemma F.54 |
| $[\Theta_3]\omega_2 = [\Theta_3]\omega_1$ | By Lemma F.30 |
| $\Theta_3 \Vdash^{\text{ela}} \rho_2 : [\Theta_3]\omega_1$ | By substituting equations |
| $\Theta_1, \widehat{\alpha} : \omega_1, \Theta_2 \longrightarrow \Theta_3, \widehat{\alpha} : \omega_1, \Theta_2$ | By extension rules |
| $\Theta_1, \widehat{\alpha} : \omega_1, \Theta_2 \longrightarrow \Theta_3, \widehat{\alpha} : \omega_1 = \rho_2, \Theta_2$ | Lemma F.33 |
| $\Delta \longrightarrow \Theta_3, \widehat{\alpha} : \omega_1 = \rho_2, \Theta_2$ | Lemma F.32 |

- The case for rule a-u-kvarR-tt is similar as the previous case.
- Case

$$\frac{\Delta_1, \Delta_2 \Vdash^{\text{mv}} \widehat{\alpha} : \omega_1 \rightsquigarrow \Theta \quad \Delta[\{\Theta\}] \Vdash^{\text{pr}}_{\widehat{\alpha}} \rho_1 \rightsquigarrow \rho_2 \dashv \Theta_1, \{\Theta_2, \widehat{\alpha} : \omega_1, \Theta_3\}, \Theta_4 \quad \Theta_1, \{\Theta_2\} \Vdash^{\text{ela}} \rho_2 : \omega_2 \quad \Theta_1, \{\Theta_2\} \Vdash^{\mu} [\Theta_1, \Theta_2]\omega_1 \approx \omega_2 \dashv \Theta_5, \{\Theta_6\}}{\Delta[\{\Delta_1, \widehat{\alpha} : \omega_1, \Delta_2\}] \Vdash^{\mu} \widehat{\alpha} \approx \rho_1 \dashv \Theta_5, \{\Theta_6, \widehat{\alpha} : \omega_1 = \rho_2, \Theta_3\}, \Theta_4} \text{a-u-kvarL-lo-tt}$$

| | |
|---|---|
| $\text{topo}(\Delta_1, \widehat{\alpha} : \omega_1, \Delta_2) = \Theta$ | by Lemma F.9 |
| $\Delta[\{\Delta_1, \widehat{\alpha} : \omega_1, \Delta_2\}] \longrightarrow \Delta[\{\Theta\}]$ | By definition |
| $\Delta[\{\Theta\}] \longrightarrow \Theta_1, \{\Theta_2, \widehat{\alpha} : \omega_1, \Theta_3\}, \Theta_4$ | Lemma F.8 |
| $\Theta_1, \{\Theta_2\} \longrightarrow \Theta_5, \{\Theta_6\}$ | I.H. |
| $\Theta_1, \text{topo}(\Theta_2) \longrightarrow \Theta_5, \Theta_6$ | By inversion |
| $\Theta_1, \text{topo}(\Theta_2), \widehat{\alpha} : \omega_1, \Theta_3 \longrightarrow \Theta_5, \Theta_6, \widehat{\alpha} : \omega_1, \Theta_3$ | By definition |
| $\Theta_1, \{\Theta_2, \widehat{\alpha} : \omega_1, \Theta_3\} \longrightarrow \Theta_5, \{\Theta_6, \widehat{\alpha} : \omega_1, \Theta_3\}$ | By rule a-ctxe-lo |
| $\Theta_1, \{\Theta_2, \widehat{\alpha} : \omega_1, \Theta_3\}, \Theta_4 \longrightarrow \Theta_5, \{\Theta_6, \widehat{\alpha} : \omega_1, \Theta_3\}, \Theta_4$ | By definition |
| $\Theta_1, \{\Theta_2\} \Vdash^{\text{ela}} \rho_2 : \omega_2$ | Given |
| $\Theta_5, \{\Theta_6\} \Vdash^{\text{ela}} \rho_2 : [\Theta_5, \Theta_6]\omega_2$ | By Lemma F.22 |
| $[\Theta_5, \Theta_6]\omega_2 = [\Theta_5, \Theta_6]([\Theta_1, \Theta_2]\omega_1)$ | Lemma F.54 |
| $[\Theta_5, \Theta_6]\omega_2 = [\Theta_5, \Theta_6]\omega_1$ | Lemma F.30 |
| $\Theta_5, \{\Theta_6, \widehat{\alpha} : \omega_1, \Theta_3\}, \Theta_4 \longrightarrow \Theta_5, \{\Theta_6, \widehat{\alpha} : \omega_1 = \rho_2, \Theta_3\}, \Theta_4$ | Lemma F.33 |
| $\Delta[\{\Delta_1, \widehat{\alpha} : \omega_1, \Delta_2\}] \longrightarrow \Theta_5, \{\Theta_6, \widehat{\alpha} : \omega_1 = \rho_2, \Theta_3\}, \Theta_4$ | By Lemma F.32 |

- The case for rule a-u-kvarR-lo-tt is similar as the previous case.
- The case for rule a-u-app follows directly from I.H. and Lemma F.32.
- The case for rule a-u-kapp follows directly from I.H. and Lemma F.32.

□

**Lemma F.11** (Well-formedness of Instantiation). *If $\Delta \Vdash^{\text{inst}} \rho_1 : \eta_1 \sqsubseteq \eta_2 \rightsquigarrow \rho_2 \dashv \Theta$, and $\Delta \Vdash^{\text{ela}} \rho_1 : \eta_1$, then $\Delta \longrightarrow \Theta$, and $\Theta$ ok, and $\Theta \Vdash^{\text{ela}} \rho_2 : [\Theta]\eta_2$.*

Proof. By induction on the derivation.

- Case

$$\frac{\Delta \Vdash^{\mu} \omega_1 \approx \omega_2 \dashv \Theta}{\Delta \Vdash^{\text{inst}} \mu : \omega_1 \sqsubseteq \omega_2 \rightsquigarrow \mu \dashv \Theta} \text{a-inst-refl}$$

| | |
|---|---|
| $\Delta \longrightarrow \Theta$ | Lemma F.10 |
| $[\Theta]\omega_1 = [\Theta]\omega_2$ | Lemma F.54 |
| $\Delta \Vdash^{\text{ela}} \mu : \omega_1$ | Given |
| $\Theta \Vdash^{\text{ela}} \mu : [\Theta]\omega_1$ | Lemma F.22 |
| $\Theta \Vdash^{\text{ela}} \mu : [\Theta]\omega_2$ | By equations |

- Case

$$\frac{\Delta, \widehat{\alpha} : \omega_1 \Vdash^{\text{inst}} \mu_1 \ @ \widehat{\alpha} : \eta[a \mapsto \widehat{\alpha}] \sqsubseteq \omega_2 \leadsto \mu_2 \dashv \Theta}{\Delta \Vdash^{\text{inst}} \mu_1 : \forall a : \omega_1.\eta \sqsubseteq \omega_2 \leadsto \mu_2 \dashv \Theta} \text{A-INST-FORALL}$$

| | |
|---|---|
| $\Delta \longrightarrow \Delta, \widehat{\alpha} : \omega_1$ | rule A-CTXE-ADD-TT |
| $\Delta \Vdash^{\text{ela}} \mu_1 : \forall a : \omega_1.\eta$ | Given |
| $[\Delta](\forall a : \omega_1.\eta) = \forall a : \omega_1.\eta$ | By Lemma F.16 |
| $[\Delta]\omega_1 = \omega_1$ | By inversion |
| $\Delta, \widehat{\alpha} : \omega_1 \Vdash^{\text{ela}} \widehat{\alpha} : [\Delta]\omega_1$ | By rule A-ELA-KUVAR |
| $\Delta, \widehat{\alpha} : \omega_1 \Vdash^{\text{ela}} \widehat{\alpha} : \omega_1$ | By equation |
| $\Delta, \widehat{\alpha} : \omega_1 \Vdash^{\text{ela}} \mu_1 : [\Delta, \widehat{\alpha} : \omega_1](\forall a : \omega_1.\eta)$ | By Lemma F.22 |
| $\Delta, \widehat{\alpha} : \omega_1 \Vdash^{\text{ela}} \mu_1 : [\Delta](\forall a : \omega_1.\eta)$ | $\widehat{\alpha}$ fresh |
| $\Delta, \widehat{\alpha} : \omega_1 \Vdash^{\text{ela}} \mu_1 : (\forall a : \omega_1.\eta)$ | by equation |
| $\Delta, \widehat{\alpha} : \omega_1 \Vdash^{\text{ela}} \mu_1 \ @ \widehat{\alpha} : \eta[a \mapsto [\Delta, \widehat{\alpha} : \omega_1]\widehat{\alpha}]$ | By rule A-ELA-KAPP |
| $\Delta, \widehat{\alpha} : \omega_1 \Vdash^{\text{ela}} \mu_1 \ @ \widehat{\alpha} : \eta[a \mapsto \widehat{\alpha}]$ | By definition |
| $\Delta, \widehat{\alpha} : \omega_1 \longrightarrow \Theta \wedge \Theta \Vdash^{\text{ela}} \mu_2 : [\Theta]\omega_2$ | I.H. |
| $\Delta \longrightarrow \Theta$ | Lemma F.32 |

- The case for rule A-INST-FORALL-INFER is similar to the previous case.

□

**Lemma F.12** (Well-formedness of Quantification Check). *If* $\Delta_1, a : \omega, \Delta_2$ ok, *and* $\Delta_2 \hookleftarrow a$, *then* $\Delta_1, \Delta_2$ ok.

PROOF. All items in $\Delta_2$ are well-formed by strengthening on elaborated kinding.

□

**Lemma F.13** (Well-formedness of Unsolved). *If* $\Delta_1, \Delta_2$ ok, *and* $\Delta_2$ **soft**, *then* $\Delta_1, \text{unsolved}(\Delta_2)$ ok.

PROOF. All unification variables in unsolved($\Delta_2$) are well-formed, which can be derived similarly as Lemma F.24, and strengthening on elaborated kinding.

□

**Lemma F.14** (Well-formedness of topo). *If* $\Delta_1, \Delta_2$ ok, *then* $\Delta_1, \text{topo}(\Delta_2)$ ok.

PROOF. As $\Delta_1, \text{topo}(\Delta_2)$ ok preserves a well-formed ordering, by strengthening and weakening we can prove $\Delta_1, \text{topo}(\Delta_2)$ ok.

□

**Lemma F.15** (Well-formedness of Kinding). *Given* $\Delta$ ok,
- *if* $\Delta \Vdash^k \sigma : \eta \leadsto \mu \dashv \Theta$, *then* $\Delta \longrightarrow \Theta$, *and* $\Theta$ ok *and* $\Theta \Vdash^{\text{ela}} \mu : [\Theta]\eta$;
- *if* $\Delta \Vdash^{\text{kc}} \sigma \Leftarrow \eta \leadsto \mu \dashv \Theta$, *then* $\Delta \longrightarrow \Theta$, *and* $\Theta$ ok *and* $\Theta \Vdash^{\text{ela}} \mu : [\Theta]\eta$.
- *if* $\Delta \Vdash^{\text{kapp}} (\rho_1 : \eta) \bullet \tau : \omega \leadsto \rho_2 \dashv \Theta$, *and* $\Delta \Vdash^{\text{ela}} \rho_1 : \eta$, *then* $\Delta \longrightarrow \Theta$, *and* $\Theta$ ok, *and* $\Theta \Vdash^{\text{ela}} \rho_2 : [\Theta]\omega$.

PROOF. By induction on the derivation.

**Part 1** • The case for rules A-KTT-STAR, A-KTT-NAT, A-KTT-VAR, A-KTT-TCON, and A-KTT-ARROW is trivial.

- Case
A-KTT-FORALL
$$\frac{\Delta \Vdash^{\text{kc}} \kappa \Leftarrow \star \leadsto \omega \dashv \Delta_1 \quad \Delta_1, a : \omega \Vdash^{\text{kc}} \sigma \Leftarrow \star \leadsto \mu \dashv \Delta_2, a : \omega, \Delta_3 \quad \Delta_3 \hookleftarrow a}{\Delta \Vdash^k \forall a : \kappa.\sigma : \star \leadsto \forall a : \omega.[\Delta_3]\mu \dashv \Delta_2, \text{unsolved}(\Delta_3)}$$

| | |
|---|---|
| $\Delta \longrightarrow \Delta_1 \wedge \Delta \Vdash^{\text{ela}} \omega : \star \wedge \Delta_1 \text{ ok}$ | Part 2 |
| $\Delta_1 \Vdash^{\text{ela}} \omega : \star$ | By Lemma F.22 |
| $\Delta_1, a : \omega \text{ ok}$ | By rule A-TCTX-TVAR-TT |
| $\Delta_1, a : \omega \longrightarrow \Delta_2, a : \omega, \Delta_3 \wedge \Delta_2, a : \omega, \Delta_3 \text{ ok}$ | Part 2 |
| $\wedge \Delta_2, a : \omega, \Delta_3 \Vdash^{\text{ela}} \mu : \star$ | Above |
| $\Delta_1 \longrightarrow \Delta_2 \wedge \Delta_3 \text{ soft}$ | By Lemma F.29 |
| $\Delta_3 \hookrightarrow a$ | Given |
| $\Delta_2, \Delta_3 \text{ ok}$ | By Lemma F.12 |
| $\Delta_2, \text{unsolved}(\Delta_3) \text{ ok}$ | By Lemma F.13 |
| $\Delta_2 \longrightarrow \Delta_2, \text{unsolved}(\Delta_3)$ | By rules A-CTXE-ADD-TT and A-CTXE-ADDSOLVED-TT |
| $\Delta \longrightarrow \Delta_2, \text{unsolved}(\Delta_3)$ | Lemma F.32 |
| $\Delta_2, \text{unsolved}(\Delta_3) \Vdash^{\text{ela}} \omega : \star.$ | Lemma F.22 |
| $\Delta_2, a : \omega, \Delta_3 \Vdash^{\text{ela}} \mu : \star$ | Known |
| $\Delta_2, a : \omega, \text{unsolved}(\Delta_3) \Vdash^{\text{ela}} [\Delta_3]\mu : \star.$ | By Lemma F.25 |

Because unsolved($\Delta_3$) does not depend on $a$, we can reorder the context to get that $\Delta_2, \text{unsolved}(\Delta_3), a :$ $\omega \Vdash^{\text{ela}} [\Delta_3]\mu : \star$. So by rule A-ELA-FORALL we get $\Delta_2, \text{unsolved}(\Delta_3) \Vdash^{\text{ela}} \forall a : \omega.[\Delta_3]\mu : \star$.

- The case for rule A-KTT-FORALLI is similar as the previous case.
- Case
$$\frac{\Delta \Vdash^{\text{k}} \tau_1 : \eta_1 \rightsquigarrow \rho_1 \dashv \Delta_1 \quad \Delta_1 \Vdash^{\text{kapp}} (\rho_1 : [\Delta_1]\eta_1) \bullet \tau_2 : \omega \rightsquigarrow \rho \dashv \Theta}{\Delta \Vdash^{\text{k}} \tau_1 \tau_2 : \omega \rightsquigarrow \rho \dashv \Theta} \text{A-KTT-APP}$$

| | |
|---|---|
| $\Delta \longrightarrow \Delta_1 \wedge \Delta_1 \Vdash^{\text{ela}} \rho_1 : [\Delta_1]\eta_1$ | I.H. |
| $\Delta_1 \longrightarrow \Theta \wedge \Theta \Vdash^{\text{ela}} \rho : [\Theta]\omega$ | Part 3 |
| $\Delta \longrightarrow \Theta$ | By Lemma F.32 |

- Case
$$\frac{\Delta \Vdash^{\text{k}} \tau_1 : \eta \rightsquigarrow \rho_1 \dashv \Delta_1 \quad [\Delta_1]\eta = \forall a : \omega.\eta_2 \quad \Delta_1 \Vdash^{\text{kc}} \tau_2 \Leftarrow \omega \rightsquigarrow \rho_2 \dashv \Delta_2}{\Delta \Vdash^{\text{k}} \tau_1 @\tau_2 : \eta_2[a \mapsto \rho_2] \rightsquigarrow \rho_1 @\rho_2 \dashv \Delta_2} \text{A-KTT-KAPP}$$

| | |
|---|---|
| $\Delta \longrightarrow \Delta_1 \wedge \Delta_1 \Vdash^{\text{ela}} \rho_1 : [\Delta_1]\eta$ | I.H. |
| $\Delta_1 \longrightarrow \Delta_2 \wedge \Delta_2 \Vdash^{\text{ela}} \rho_2 : [\Delta_2]\omega$ | Part 2 |
| $\Delta \longrightarrow \Delta_2$ | Lemma F.32 |
| $\Delta_1 \Vdash^{\text{ela}} \rho_1 : \forall a : \omega.\eta_2$ | by equations |
| $\Delta_2 \Vdash^{\text{ela}} \rho_1 : \forall a : [\Delta_2]\omega.[\Delta_2]\eta_2$ | Lemma F.22 |
| $\Delta_2 \Vdash^{\text{ela}} \rho_1 @\rho_2 : ([\Delta_2]\eta_2)[a \mapsto [\Delta_2]\rho_2]$ | By rule A-ELA-KAPP |
| $\Delta_2 \Vdash^{\text{ela}} \rho_1 @\rho_2 : [\Delta_2](\eta_2[a \mapsto \rho_2])$ | By substitution |

- Case rule A-KTT-KAPP-INFER is similar as the previous case.

**Part 2** We have
$$\frac{\Delta \Vdash^{\text{k}} \sigma : \eta \rightsquigarrow \mu_1 \dashv \Delta_1 \quad \Delta_1 \Vdash^{\text{inst}} \mu_1 : [\Delta_1]\eta \sqsubseteq [\Delta_1]\omega \rightsquigarrow \mu_2 \dashv \Delta_2}{\Delta \Vdash^{\text{kc}} \sigma \Leftarrow \omega \rightsquigarrow \mu_2 \dashv \Delta_2} \text{A-KC-SUB}$$

| | |
|---|---|
| $\Delta \longrightarrow \Delta_1 \wedge \Delta_1 \Vdash^{\text{ela}} \mu_1 : [\Delta_1]\eta$ | Part 1 |
| $\Delta_1 \longrightarrow \Delta_2 \wedge \Delta_2 \Vdash^{\text{ela}} \mu_2 : [\Delta_2]([\Delta_1]\omega_2)$ | By Lemma F.11 |
| $\Delta \longrightarrow \Delta_2$ | Lemma F.32 |
| $[\Delta_2]([\Delta_1]\omega) = [\Delta_2]\omega$ | Lemma F.30 |
| $\Delta_2 \Vdash^{\text{ela}} \mu_2 : [\Delta_2]\omega$ | by equations |

**Part 3** By induction on the judgment.

- Case

$$\frac{\Delta \Vdash^{\text{kc}} \tau \Leftarrow \omega_1 \rightsquigarrow \rho_2 \dashv \Theta}{\Delta \Vdash^{\text{kapp}} (\rho_1 : \omega_1 \rightarrow \omega_2) \bullet \tau : \omega_2 \rightsquigarrow \rho_1\,\rho_2 \dashv \Theta} \text{\scriptsize A-KAPP-TT-ARROW}$$

| | |
|---|---|
| $\Delta \longrightarrow \Theta \wedge \Theta \Vdash^{\text{ela}} \rho_2 : [\Theta]\omega_1$ | By Part 2 |
| $\Delta \Vdash^{\text{ela}} \rho_1 : \omega_1 \rightarrow \omega_2$ | Given |
| $\Theta \Vdash^{\text{ela}} \rho_1 : [\Theta]\omega_1 \rightarrow [\Theta]\omega_2$ | By Lemma F.22 |
| $\Theta \Vdash^{\text{ela}} \rho_1\,\rho_2 : [\Theta]\omega_2$ | By rule A-ELA-APP |

- Case

$$\frac{\Delta, \widehat{\alpha} : \omega_1 \Vdash^{\text{kapp}} (\rho_1 \,@\widehat{\alpha} : \eta[a \mapsto \widehat{\alpha}]) \bullet \tau : \omega \rightsquigarrow \rho \dashv \Theta}{\Delta \Vdash^{\text{kapp}} (\rho_1 : \forall a : \omega_1.\eta) \bullet \tau : \omega \rightsquigarrow \rho \dashv \Theta} \text{\scriptsize A-KAPP-TT-FORALL}$$

| | |
|---|---|
| $\Delta \Vdash^{\text{ela}} \rho_1 : \forall a : \omega_1.\eta$ | Given |
| $\Delta \Vdash^{\text{ela}} \forall a : \omega_1.\eta : \star$ | By Lemma F.16 |
| $\Delta \Vdash^{\text{ela}} \omega_1 : \star$ | By inversion |
| $\Delta \longrightarrow \Delta, \widehat{\alpha} : \omega_1$ | By rule A-CTXE-ADD-TT |
| $\Delta, \widehat{\alpha} : \omega_1 \Vdash^{\text{ela}} \rho_1 : \forall a : \omega_1.\eta$ | By Lemma F.22 and $\widehat{\alpha}$ fresh |
| $\Delta, \widehat{\alpha} : \omega_1 \Vdash^{\text{ela}} \widehat{\alpha} : \omega_1$ | By rule A-ELA-KUVAR, Lemma F.16, and $\widehat{\alpha}$ fresh |
| $\Delta, \widehat{\alpha} : \omega_1 \Vdash^{\text{ela}} \rho_1 \,@\widehat{\alpha} : \eta[a \mapsto \widehat{\alpha}]$ | By rule A-ELA-KAPP |
| $\Delta, \widehat{\alpha} : \omega_1 \longrightarrow \Theta \wedge \Theta \Vdash^{\text{ela}} \rho : [\Theta]\omega$ | I.H. |
| $\Delta \longrightarrow \Theta$ | By Lemma F.32 |

- The case for rule A-KAPP-TT-FORALL-INFER is similar to the previous case.
- Case

$$\frac{\Delta_1, \widehat{\alpha_1} : \star, \widehat{\alpha_2} : \star, \widehat{\alpha} : \omega = (\widehat{\alpha_1} \rightarrow \widehat{\alpha_2}), \Delta_2 \Vdash^{\text{kc}} \tau \Leftarrow \widehat{\alpha_1} \rightsquigarrow \rho_2 \dashv \Theta}{\Delta_1, \widehat{\alpha} : \omega, \Delta_2 \Vdash^{\text{kapp}} (\rho_1 : \widehat{\alpha}) \bullet \tau : \widehat{\alpha_2} \rightsquigarrow \rho_1\,\rho_2 \dashv \Theta} \text{\scriptsize A-KAPP-TT-KUVAR}$$

| | |
|---|---|
| $\Delta_1, \widehat{\alpha} : \kappa, \Delta_2 \longrightarrow \Delta_1, \widehat{\alpha_1} : \star, \widehat{\alpha_2} : \star, \widehat{\alpha} : \kappa = (\widehat{\alpha_1} \rightarrow \widehat{\alpha_2}), \Delta_2$ | By Lemmas F.32, F.33 and F.35 |
| $\Delta_1, \widehat{\alpha_1} : \star, \widehat{\alpha_2} : \star, \widehat{\alpha} : \kappa = (\widehat{\alpha_1} \rightarrow \widehat{\alpha_2}), \Delta_2 \longrightarrow \Theta \wedge \Theta \Vdash^{\text{ela}} \rho_2 : [\Theta]\widehat{\alpha_1}$ | By Part 2 |
| $\Delta_1, \widehat{\alpha} : \kappa, \Delta_2 \longrightarrow \Theta$ | By Lemma F.32 |
| $\Delta_1, \widehat{\alpha} : \kappa, \Delta_2 \Vdash^{\text{ela}} \rho_1 : \widehat{\alpha}$ | Given |
| $\Theta \Vdash^{\text{ela}} \rho_1 : [\Theta]\widehat{\alpha}$ | By Lemma F.22 |
| $\Theta \Vdash^{\text{ela}} \rho_1 : [\Theta]\widehat{\alpha_1} \rightarrow [\Theta]\widehat{\alpha_2}$ | By Lemma F.30 |
| $\Theta \Vdash^{\text{ela}} \rho_1\,\rho_2 : [\Theta]\widehat{\alpha_2}$ | By rule A-ELA-APP |

□

**Lemma F.16** (Well-formedness of Elaborated Kinding). *If $\Delta$ ok, and $\Delta \Vdash^{\text{ela}} \mu : \eta$, then $\Delta \Vdash^{\text{ela}} \eta : \star$, and $[\Delta]\eta = \eta$.*

Proof. By induction on the derivation.

- The case for rules A-ELA-STAR, A-ELA-NAT, A-ELA-ARROW, A-ELA-FORALL, and A-ELA-FORALL-INFER is straightforward.
- The case for rules A-ELA-KUVAR, A-ELA-VAR, and A-ELA-TCON is similar. Consider

  $$\text{A-ELA-KUVAR} \quad \frac{(\widehat{\alpha} : \omega) \in \Delta}{\Delta \Vdash^{\text{ela}} \widehat{\alpha} : [\Delta]\omega}$$

  Given $\Delta$ ok, by inversion and weakening, we have $\Delta \Vdash^{\text{ela}} \omega : \star$. By Lemma F.24, we have $\Delta \Vdash^{\text{ela}} [\Delta]\omega : \star$. And $[\Delta]([\Delta]\omega) = [\Delta]\omega$.

- Case

  $$\text{A-ELA-APP} \quad \frac{\Delta \Vdash^{\text{ela}} \rho_1 : \omega_1 \to \omega_2 \quad \Delta \Vdash^{\text{ela}} \rho_2 : \omega_1}{\Delta \Vdash^{\text{ela}} \rho_1\,\rho_2 : \omega_2}$$

  | | |
  |---|---|
  | $\Delta \Vdash^{\text{ela}} \omega_1 \to \omega_2 : \star \land [\Delta](\omega_1 \to \omega_2) = \omega_1 \to \omega_2$ | I.H. |
  | $\Delta \Vdash^{\text{ela}} \omega_2 : \star \land [\Delta]\omega_2 = \omega_2$ | inversion |

- The case for rules A-ELA-KAPP and A-ELA-KAPP-INFER is similar. Consider

  $$\text{A-ELA-KAPP} \quad \frac{\Delta \Vdash^{\text{ela}} \rho_1 : \forall a : \omega.\eta \quad \Delta \Vdash^{\text{ela}} \rho_2 : \omega}{\Delta \Vdash^{\text{ela}} \rho_1 @ \rho_2 : \eta[a \mapsto [\Delta]\rho_2]}$$

  | | |
  |---|---|
  | $\Delta \Vdash^{\text{ela}} \forall a : \omega.\eta : \star$ | I.H. |
  | $\Delta, a : \omega \Vdash^{\text{ela}} \eta : \star$ | By inversion |
  | $\Delta \Vdash^{\text{ela}} \rho_2 : \omega$ | Given |
  | $\Delta \Vdash^{\text{ela}} [\Delta]\rho_2 : \omega$ | Lemma F.24 |
  | $\Sigma \Vdash^{\text{ela}} \eta[a \mapsto [\Delta]\rho_2] : \star$ | By substitution |
  | $[\Delta](\forall a : \omega.\eta) = \forall a : \omega.\eta$ | by I.H. |
  | $[\Delta]\eta = \eta$ | Follows directly |
  | $[\Delta](\eta[a \mapsto [\Delta]\rho_2])$ | |
  | $= ([\Delta]\eta)[a \mapsto ([\Delta]([\Delta]\rho_2))]$ | |
  | $= \eta[a \mapsto [\Delta]\rho_2]$ | |

□

**Lemma F.17** (Well-formedness of Typing Signature). *If $\Omega$ ok, and $\Omega \Vdash^{\text{sig}} \mathcal{S} \rightsquigarrow T : \mu$, then $\Omega \Vdash^{\text{ela}} \mu : \star$.*

Proof. We have

$$\text{A-SIG-TT} \quad \frac{\rceil \sigma \lceil \quad \overline{a_i}^i = \text{fkv}(\sigma) \quad \Omega, \{\overline{\widehat{\alpha_i} : \star, a_i : \widehat{\alpha_i}}^i\} \Vdash^k \sigma : \star \rightsquigarrow \eta \dashv \Delta \quad \phi_1^c = \text{scoped\_sort}(\overline{a_i : [\Delta]\widehat{\alpha_i}}^i) \quad \widehat{\phi_2^c} = \text{unsolved}(\Delta) \quad \Delta \hookrightarrow \overline{a_i}^i}{\Omega \Vdash^{\text{sig}} \mathbf{data}\, T : \sigma \rightsquigarrow T : \forall\{\phi_2^c\}.((\forall\{\phi_1^c\}.[\Delta]\eta)[\widehat{\phi_2^c} \mapsto \phi_2^c])}$$

| | |
|---|---|
| $\Delta \Vdash^{\text{ela}} \eta : \star \land \Omega \longrightarrow \Delta$ | By Lemma F.15 |
| $\Delta \Vdash^{\text{ela}} [\Delta]\eta : \star$ | By Lemma F.24 |
| $\Delta = \Omega, \{\Delta_1\}, \Delta_2 \land \Delta_2$ **soft** | By Lemma F.29 and properties of kinding |
| $\Delta_1$ contains only unification variables except for $\overline{a_i}^i$ | Above |
| $\Delta \hookrightarrow \overline{a_i}^i$ | Given |
| $\Omega, \widehat{\phi_2^c}$ ok | By Lemma F.12 and Lemma F.13 |
| $\Omega, \widehat{\phi_2^c}, \phi_1^c$ ok | $\phi_1^c$ has no solved unification variables |
| $\Omega, \widehat{\phi_2^c}, \phi_1^c \Vdash^{\text{ela}} [\Delta]\eta : \star$ | By strengthening and reorder of context |
| $\Omega, \widehat{\phi_2^c} \Vdash^{\text{ela}} \forall\{\phi_1^c\}.[\Delta]\eta : \star$ | By rule A-ELA-FORALL-INFER |
| $\Omega, \phi_2^c \Vdash^{\text{ela}} (\forall\{\phi_1^c\}.[\Delta]\eta)[\widehat{\phi_2^c} \mapsto \phi_2^c] : \star$ | By substitution |
| $\Omega \Vdash^{\text{ela}} \forall\{\phi_2^c\}.(\forall\{\phi_1^c\}.[\Delta]\eta)[\widehat{\phi_2^c} \mapsto \phi_2^c] : \star$ | By rule A-ELA-FORALL-INFER |

□

**Lemma F.18** (Well-formedness of Typing Data Constructor Declaration). *If $\Delta$ ok, and $\Delta \Vdash^{\text{dc}}_\rho \mathcal{D} \rightsquigarrow \mu \dashv \Theta$, then $\Delta \longrightarrow \Theta$, and $\Theta \Vdash^{\text{ela}} \mu : \star$.*

Proof. We have

$$\frac{\text{A-DC-TT}}{\Delta, \blacktriangleright_D \Vdash^k \forall\phi.(\overline{\tau_i}^i \rightarrow \rho) : \star \rightsquigarrow \mu \dashv \Theta_1, \blacktriangleright_D, \Theta_2 \qquad \widehat{\phi^c} = \text{unsolved}(\Theta_2)}{\Delta \Vdash^{\text{dc}}_\rho \forall\phi.D\,\overline{\tau_i}^i \rightsquigarrow \forall\{\phi^c\}.(([\Theta_2]\mu)[\widehat{\phi^c} \mapsto \phi^c]) \dashv \Theta_1}$$

| | |
|---|---|
| $\Delta, \blacktriangleright_D$ ok | By rule A-TCTX-MARKER |
| $\Delta, \blacktriangleright_D \longrightarrow \Theta_1, \blacktriangleright_D, \Theta_2 \wedge \Theta_1, \blacktriangleright_D, \Theta_2 \Vdash^{\text{ela}} \mu : \star$ | By Lemma F.15 |
| $\Delta \longrightarrow \Theta_1 \wedge \Theta_2$ **soft** | By Lemma F.29 |
| $\Theta_1, \blacktriangleright_D, \widehat{\phi^c}$ ok | By Lemma F.13 |
| $\Theta_1, \blacktriangleright_D, \widehat{\phi^c} \Vdash^{\text{ela}} [\Theta_2]\mu : \star$ | By Lemma F.25 |
| $\Theta_1, \blacktriangleright_D, \phi^c \Vdash^{\text{ela}} ([\Theta_2]\mu)[\widehat{\phi^c} \mapsto \phi^c] : \star$ | By substitution |
| $\Theta_1, \blacktriangleright_D \Vdash^{\text{ela}} \forall\{\phi^c\}.([\Theta_2]\mu)[\widehat{\phi^c} \mapsto \phi^c] : \star$ | By rule A-ELA-FORALL-INFER |
| $\Theta_1 \Vdash^{\text{ela}} \forall\{\phi^c\}.([\Theta_2]\mu)[\widehat{\phi^c} \mapsto \phi^c] : \star$ | By strengthening |

□

**Lemma F.19** (Well-formedness of Typing Datatype Declaration). *If $\Delta$ ok, and $\Delta \Vdash^{\text{dt}} \mathcal{T} \rightsquigarrow \Gamma \dashv \Theta$, then $\Delta \longrightarrow \Theta$, and $\Theta \Vdash^{\text{ectx}} \Gamma$.*

Proof. We have

$$\frac{\text{A-DT-TT}}{(T : \forall\{\phi_1^c\}.\forall\phi_2^c.\,\omega) \in \Delta \quad \Delta, \phi_1^c, \phi_2^c, \overline{\widehat{\alpha_i} : \star}^i \Vdash^\mu [\Delta]\omega \approx (\overline{\widehat{\alpha_i}}^i \rightarrow \star) \dashv \Theta_1, \phi_1^c, \phi_2^c, \overline{\widehat{\alpha_i} : \star = \omega_i}^i}{\Theta_j, \phi_1^c, \phi_2^c, \overline{a_i : \omega_i}^i \Vdash^{\text{dc}}_{(T\,@\phi_1^c\,@\phi_2^c\,\overline{a_i}^i)} \mathcal{D}_j \rightsquigarrow \mu_j \dashv \Theta_{j+1}, \phi_1^c, \phi_2^c, \overline{a_i : \omega_i}^{i\,j}}$$
$$\Delta \Vdash^{\text{dt}} \textbf{data}\,T\,\overline{a_i}^i = \overline{\mathcal{D}_j}^{j \in 1..n} \rightsquigarrow \overline{D_j : \forall\{\phi_1^c\}.\forall\phi_2^c.\,\forall\overline{a_i : \omega_i}^i.\mu_j}^{j} \dashv \Theta_{n+1}$$

| | |
|---|---|
| $\Delta$ ok $\wedge (T : \forall\{\phi_1^c\}.\forall\phi_2^c.\,\omega) \in \Delta$ | Given |
| $\Delta, \phi_1^c, \phi_2^c$ ok | By inversion and weakening |
| $\Delta, \phi_1^c, \phi_2^c, \overline{\widehat{\alpha_i} : \star}^i$ ok | By rule A-TCTX-KUVAR-TT |
| $\Delta, \phi_1^c, \phi_2^c, \overline{\widehat{\alpha_i} : \star}^i \longrightarrow \Theta_1, \phi_1^c, \phi_2^c, \overline{\widehat{\alpha_i} : \star = \omega_i}^i$ | Lemma F.10 |
| $\overline{\Theta_j, \phi_1^c, \phi_2^c, \overline{a_i : \omega_i}^i \longrightarrow \Theta_{j+1}, \phi_1^c, \phi_2^c, \overline{a_i : \omega_i}^i}^{j}$ | By Lemma F.18 |
| $\overline{\Theta_{j+1}, \phi_1^c, \phi_2^c, \overline{a_i : \omega_i}^i \Vdash^{\text{ela}} \mu_j : \star}^{j}$ | By Lemma F.18 |
| $\Delta \longrightarrow \Theta_{n+1}$ | By Lemma F.29 and Lemma F.32 |

$$\frac{\overline{\Theta_{n+1}, \phi_1^c, \phi_2^c, \overline{a_i : \omega_i}^i \Vdash^{\text{ela}} \mu_j : \star}^j}{\Theta_{n+1} \Vdash^{\text{ela}} \overline{\forall\{\phi_1^c\}.\forall\phi_2^c.\,\forall\overline{a_i : \omega_i}^i.\mu_j : \star}^j} \quad \text{By Lemma F.22}$$
$$\text{By rules A-ELA-FORALL and A-ELA-FORALL-INFER}$$
$$\Theta_{n+1} \Vdash^{\text{ectx}} \overline{D_j : \forall\phi^c.\,\mu_j}^j \quad \text{By rule A-ECTX-DCON-TT}$$

□

**Lemma F.20** (Well-formedness of Generalization). *If $\Delta$ ok, and $\Delta \Vdash^{\text{ectx}} \Gamma_1$, and $\Delta \Vdash^{\text{gen}}_{\phi^c} \Gamma_1 \rightsquigarrow \Gamma_2$, then $\Delta \Vdash^{\text{ectx}} \Gamma_2$.*

Proof. Follows directly from rule A-ECTX-DCON-TT, rule A-ELA-FORALL-INFER and substitution.

□

*F.2.2 Properties of Context Extension.* Proofs for many lemmas are essentially the same as its corresponding lemmas in Haskell98. Therefore in this section, we only give proof for those of lemmas with slightly different reasoning or extra cases that are worth attention.

**Lemma F.22** (Extension Weakening). *Given $\Delta \longrightarrow \Theta$, if $\Delta \Vdash^{\text{ela}} \mu : \eta$, then $\Theta \Vdash^{\text{ela}} \mu : [\Theta]\eta$.*

Proof. By a straightforward induction on the elaborated kinding, making use of Lemma F.30.

□

**Lemma F.25** (Soft Substitution Kinding). *If $\Delta_1, \Delta_2$ ok, and $\Delta_2$ **soft**, and $\Delta_1, \Delta_2 \Vdash^{\text{ela}} \mu : \eta$, then $\Delta_1, \text{unsolved}(\Delta_2) \Vdash^{\text{ela}} [\Delta_2]\mu : \eta$.*

Proof. Similar as the proof for Lemma D.12, making use of weakening.

□

**Lemma F.27** (Well-formedness of Context Extension). *If $\Delta$ ok, and $\Delta \longrightarrow \Theta$, then $\Theta$ ok.*

Proof. Similar as the proof for Lemma D.15.
For the case

$$\frac{\text{A-CTXE-LO}}{\Delta \longrightarrow \Theta \quad \Delta, \text{topo}(\Delta_1) \longrightarrow \Theta, \Theta_1}{\Delta, \{\Delta_1\} \longrightarrow \Theta, \{\Theta_1\}}$$

| $\Delta, \{\Delta_1\}$ ok | Given |
| $\Delta, \Delta_1$ ok | By inversion |
| $\Delta, \text{topo}(\Delta_1)$ ok | By Lemma F.14 |
| $\Theta, \Theta_1$ ok | I.H. |

□

**Lemma F.32** (Transitivity of Context Extension). *If $\Delta'$ ok, and $\Delta' \longrightarrow \Delta$, and $\Delta \longrightarrow \Theta$, then $\Delta' \longrightarrow \Theta$.*

Proof. By induction on $\Delta \longrightarrow \Theta$. The proof is similar as the proof for Lemma D.20.
For the case

$$\frac{\text{A-CTXE-LO}}{\Delta \longrightarrow \Theta \quad \Delta, \text{topo}(\Delta_1) \longrightarrow \Theta, \Theta_1}{\Delta, \{\Delta_1\} \longrightarrow \Theta, \{\Theta_1\}}$$

| | |
|---|---|
| $\Delta' \longrightarrow \Delta, \{\Delta_1\}$ | Given |
| $\Delta' = \Delta_2, \{\Delta_3\} \wedge \Delta_2 \longrightarrow \Delta \wedge \Delta_2, \text{topo}(\Delta_3) \longrightarrow \Delta, \Delta_1$ | By inversion |
| $\Delta_2, \text{topo}(\Delta_3) \longrightarrow \Delta, \text{topo}(\Delta_1)$ | By reordering $\Delta_3$ according to $\text{topo}(\Delta_1)$ |
| $\Delta_2, \text{topo}(\Delta_3) \longrightarrow \Theta, \Theta_1$ | I.H. |
| $\Delta_2, \{\Delta_3\} \longrightarrow \Theta, \{\Theta_1\}$ | By rule A-CTXE-LO |

□

**Lemma F.33** (Solution Admissibility for Extension).
- If $\Delta_1, \widehat{\alpha} : \omega, \Delta_2$ ok and $\Delta_1 \Vdash^{\text{ela}} \rho : [\Delta_1]\omega$, then $\Delta_1, \widehat{\alpha} : \omega, \Delta_2 \longrightarrow \Delta_1, \widehat{\alpha} : \omega = \rho, \Delta_2$.
- If $\Delta_1, \{\Delta_3, \widehat{\alpha} : \omega, \Delta_4\}, \Delta_2$ ok and $\Delta_1, \Delta_3 \Vdash^{\text{ela}} \rho : [\Delta_1, \Delta_3]\omega$, then $\Delta_1, \{\Delta_3, \widehat{\alpha} : \omega, \Delta_4\}, \Delta_2 \longrightarrow \Delta_1, \{\Delta_3, \widehat{\alpha} : \omega = \rho, \Delta_4\}, \Delta_2$.

PROOF.    **Part 1** By induction on $\Delta_2$. The proof is similar as the proof for Lemma D.21.
For the case $\Delta_2 = \Delta'_2, \{\Delta_3\}$. By I.H., we have $\Delta_1, \widehat{\alpha} : \omega \longrightarrow \Delta_1, \widehat{\alpha} : \omega = \rho$. Then by rule A-CTXE-LO we have $\Delta_1, \widehat{\alpha} : \omega, \{\Delta'_2\} \longrightarrow \Delta_1, \widehat{\alpha} : \omega = \rho, \{\Delta'_2\}$.

**Part 2** By induction on $\Delta_2$. Most cases are similar as Part 1. When $\Delta_2$ is empty, we only need to prove $\Delta_1, \Delta_3, \widehat{\alpha} : \omega, \Delta_4 \longrightarrow \Delta_1, \Delta_3, \widehat{\alpha} : \omega = \rho, \Delta_4$. By referring Part 1 we are done.

□

**Lemma F.36** (Parallel Admissibility).
- If $\Delta_1 \longrightarrow \Theta_1$, and $\Delta_1, \Delta_2$ ok, and $\Delta_1, \Delta_2 \longrightarrow \Theta_1, \Theta_2$, and $\Delta_2$ is fresh w.r.t. $\Theta_1$, then:
  - if $\Delta_1 \Vdash^{\text{ela}} \omega : \star$, then $\Delta_1, \widehat{\alpha} : \omega, \Delta_2 \longrightarrow \Theta_1, \widehat{\alpha} : \omega, \Theta_2$;
  - if $\Theta_1 \Vdash^{\text{ela}} \rho : [\Theta_1]\omega$, then $\Delta_1, \widehat{\alpha} : \omega, \Delta_2 \longrightarrow \Theta_1, \widehat{\alpha} : \omega = \rho, \Theta_2$;
  - if $[\Theta_1]\rho_1 = [\Theta_1]\rho_2$, then $\Delta_1, \widehat{\alpha} : \omega = \rho_1, \Delta_2 \longrightarrow \Theta_1, \widehat{\alpha} : \omega = \rho_2, \Theta_2$.
- If $\Delta_1, \{\Delta_3\} \longrightarrow \Theta_1, \{\Theta_3\}$, and $\Delta_1, \{\Delta_3, \Delta_4\}, \Delta_2$ ok, and $\Delta_1, \{\Delta_3, \Delta_4\}, \Delta_2 \longrightarrow \Theta_1, \{\Theta_3, \Theta_4\}, \Theta_2$, and $\Delta_2, \Delta_4$ is fresh w.r.t. $\Theta_1, \Theta_3$, then:
  - if $\Delta_1, \{\Delta_3\} \Vdash^{\text{ela}} \omega : \star$, then $\Delta_1, \{\Delta_3, \widehat{\alpha} : \omega, \Delta_4\}, \Delta_2 \longrightarrow \Theta_1, \{\Theta_3, \widehat{\alpha} : \omega, \Theta_4\}, \Theta_2$;
  - if $\Theta_1, \{\Theta_3\} \Vdash^{\text{ela}} \rho : [\Theta_1, \Theta_3]\omega$, then $\Delta_1, \{\Delta_3, \widehat{\alpha} : \omega, \Delta_4\}, \Delta_2 \longrightarrow \Theta_1, \{\Theta_3, \widehat{\alpha} : \omega = \rho, \Theta_4\}, \Theta_2$;
  - if $[\Theta_1, \Theta_3]\rho_1 = [\Theta_1, \Theta_3]\rho_2$, then $\Delta_1, \{\Delta_3, \widehat{\alpha} : \omega = \rho_1, \Delta_4\}, \Delta_2 \longrightarrow \Theta_1, \{\Theta_3, \widehat{\alpha} : \omega = \rho_2, \Theta_4\}, \Theta_2$.

PROOF.    **Part 1** By induction on the size of $\Theta_2$. Most cases are similar as in Lemma D.24.
For the case where $\Theta_2 = \Theta_3, \{\Theta_4\}$, the derivation of $\Delta_1, \Delta_2 \longrightarrow \Theta_1, \Theta_2$ must conclude with rule A-CTXE-LO. It must be $\Delta_2 = \Delta_{21}, \{\Delta_{22}\}$.

| | |
|---|---|
| $\Delta_1, \Delta_{21}, \{\Delta_{22}\} \longrightarrow \Theta_1, \Theta_3, \{\Theta_4\}$ | Given |
| $\Delta_1, \Delta_{21}, \text{topo}(\Delta_{22}) \longrightarrow \Theta_1, \Theta_3, \Theta_4$ | By inversion |
| $\Delta_1, \widehat{\alpha} : \omega, \Delta_{21}, \text{topo}(\Delta_2) \longrightarrow \Theta_1, \widehat{\alpha} : \omega, \Theta_3, \Theta_4$ | I.H. |
| $\Delta_1, \widehat{\alpha} : \omega, \Delta_{21}, \{\Delta_2\} \longrightarrow \Theta_1, \widehat{\alpha} : \omega, \Theta_3, \{\Theta_4\}$ | By rule A-CTXE-LO |

**Part 2** By induction on $\Theta_2$. Most cases are similar to Part 1. For the first case, when $\Theta_2$ is empty, we know $\Delta_2$ is empty. We have $\Delta_1, \text{topo}(\Delta_3), \text{topo}(\Delta_4) \longrightarrow \Theta_1, \Theta_3, \Theta_4$. By Part 1 we know $\Delta_1, \text{topo}(\Delta_3), \widehat{\alpha} : \omega, \text{topo}(\Delta_4) \longrightarrow \Theta_1, \Theta_3, \widehat{\alpha} : \omega, \Theta_4$. By rule A-CTXE-LO we have $\Delta_1, \{\Delta_3, \widehat{\alpha} : \omega, \Delta_4\} \longrightarrow \Theta_1, \{\Theta_3, \widehat{\alpha} : \omega, \Theta_4\}$.

□

**Lemma F.37** (Parallel Extension Solution).
- If $\Delta_1, \widehat{\alpha} : \omega, \Delta_2 \longrightarrow \Theta_1, \widehat{\alpha} : \omega = \rho_2, \Theta_2$, and $[\Theta_1]\rho_1 = [\Theta_1]\rho_2$, then $\Delta_1, \widehat{\alpha} : \omega = \rho_1, \Delta_2 \longrightarrow \Theta_1, \widehat{\alpha} : \omega = \kappa_2, \Theta_2$.

- If $\Delta_1, \{\Delta_3, \widehat{\alpha} : \omega, \Delta_4\}, \Delta_2 \longrightarrow \Theta_1, \{\Theta_3, \widehat{\alpha} : \omega = \rho_2, \Theta_4\}, \Theta_2$, and $[\Theta_1, \Theta_3]\rho_1 = [\Theta_1, \Theta_3]\rho_2$, then $\Delta_1, \{\Delta_3, \widehat{\alpha} : \omega = \rho_1, \Delta_4\}, \Delta_2 \longrightarrow \Theta_1, \{\Theta_3, \widehat{\alpha} : \omega = \rho_2, \Theta_4\}, \Theta_2$.

Proof. **Part 1** By induction on $\Theta_2$. The proof is similar to Lemma D.25. For the case when $\Theta_2 = \Theta_3, \{\Theta_4\}$, we have $\Delta_2 = \Delta_3, \{\Delta_4\}$. And $\Delta_1, \widehat{\alpha} : \omega, \Delta_3, \{\Delta_4\} \longrightarrow \Theta_1, \widehat{\alpha} : \omega, \Theta_3, \{\Theta_4\}$. By inversion, we have $\Delta_1, \widehat{\alpha} : \omega, \Delta_3, \mathsf{topo}(\Delta_4) \longrightarrow \Theta_1, \widehat{\alpha} : \omega, \Theta_3, \Theta_4$. By I.H., we have $\Delta_1, \widehat{\alpha} : \omega = \rho_1, \Delta_3, \mathsf{topo}(\Delta_4) \longrightarrow \Theta_1, \widehat{\alpha} : \omega = \rho_2, \Theta_3, \Theta_4$. By rule A-CTXE-LO we have $\Delta_1, \widehat{\alpha} : \omega = \rho_1, \Delta_3, \{\Delta_4\} \longrightarrow \Theta_1, \widehat{\alpha} : \omega = \rho_2, \Theta_3, \{\Theta_4\}$.

**Part 2** By induction on $\Theta_2$. Most cases are similar to Part 1. We discuss when $\Theta_2$ is empty. Then $\Delta_2$ must to empty. From givens we know that $\Delta_1, \Delta_5, \widehat{\alpha} : \omega, \Delta_6 \longrightarrow \Theta_1, \Theta_3, \widehat{\alpha} : \omega = \rho_2, \Theta_4$ where $\Delta_5, \widehat{\alpha} : \omega, \Delta_6 = \mathsf{topo}(\Delta_3, \widehat{\alpha} : \omega, \Delta_4)$. By Part 1 we have $\Delta_1, \Delta_5, \widehat{\alpha} : \omega = \rho_1, \Delta_6 \longrightarrow \Theta_1, \Theta_3, \widehat{\alpha} : \omega = \rho_2, \Theta_4$. Since $\Delta_5, \widehat{\alpha} : \omega = \rho_1, \Delta_6 = \mathsf{topo}(\Delta_3, \widehat{\alpha} : \omega = \rho_1, \Delta_4)$, by rule A-CTXE-LO we have $\Delta_1, \{\Delta_3, \widehat{\alpha} : \omega = \rho_1, \Delta_4\} \longrightarrow \Theta_1, \{\Theta_3, \widehat{\alpha} : \omega = \rho_2, \Theta_4\}$. □

### F.2.3 Properties of Complete Context.

**Lemma F.43** (Stability of Complete Contexts). *If $\Delta \longrightarrow \Omega$, then $[\Omega]\Delta = [\Omega]\Omega$.*

Proof. By induction on $\Delta \longrightarrow \Omega$. Most cases are the same as Lemma D.31. For the case

$$\frac{\text{A-CTXE-LO} \quad \Delta \longrightarrow \Theta \quad \Delta, \mathsf{topo}(\Delta_1) \longrightarrow \Theta, \Theta_1}{\Delta, \{\Delta_1\} \longrightarrow \Theta, \{\Theta_1\}}$$

| | |
|---|---|
| $\Delta, \{\Delta_1\} \longrightarrow \Omega, \{\Omega_1\}$ | Given |
| $\Delta, \mathsf{topo}(\Delta_1) \longrightarrow \Omega, \Omega_1$ | Given |
| $[\Omega, \{\Omega_1\}](\Omega, \{\Omega_1\})$ | |
| $= [\Omega, \Omega_1](\Omega, \Omega_1)$ | By definition |
| $= [\Omega, \Omega_1](\Delta, \mathsf{topo}(\Delta_1))$ | I.H. |
| $= [\Omega, \{\Omega_1\}](\Delta, \{\Delta_1\})$ | By definition |

□

**Lemma F.46** (Finishing Completions). *If $\Omega$ ok, and $\Omega \longrightarrow \Omega'$, then $[\Omega']\Omega' = \mathit{topo}([\Omega]\Omega)$.*

Proof. By induction on $\Omega \longrightarrow \Omega'$. Most cases are the same as Lemma D.34.
For the case

$$\frac{\text{A-CTXE-LO} \quad \Delta \longrightarrow \Theta \quad \Delta, \mathsf{topo}(\Delta_1) \longrightarrow \Theta, \Theta_1}{\Delta, \{\Delta_1\} \longrightarrow \Theta, \{\Theta_1\}}$$

| | |
|---|---|
| $\Omega, \{\Omega_1\} \longrightarrow \Omega', \{\Omega'_1\} \wedge \Omega, \mathsf{topo}(\Omega_1) \longrightarrow \Omega', \Omega'_1 \wedge \Omega \longrightarrow \Omega'$ | Given |
| $[\Omega', \{\Omega'_1\}](\Omega', \{\Omega'_1\})$ | |
| $= [\Omega', \Omega'_1](\Omega', \Omega'_1)$ | By definition |
| $= \mathsf{topo}([\Omega, \mathsf{topo}(\Omega_1)](\Omega, \mathsf{topo}(\Omega_1)))$ | I.H. |
| $= \mathsf{topo}([\Omega, \Omega_1](\Omega, \Omega_1))$ | Follows |

□

### F.2.4 Decidability.

**Lemma F.47** (Promotion Preserves $\langle \Delta \rangle$). *If $\Delta \Vdash^{\text{pr}}_{\widehat{\alpha}} \omega_1 \leadsto \omega_2 \dashv \Theta$, then $\langle \Delta \rangle = \langle \Theta \rangle$.*

Proof. By a straightforward induction on the derivation.

□

**Lemma F.48** (Unification Makes Progress). *If $\Delta \Vdash^{\text{u}} \omega_1 \approx \omega_2 \dashv \Theta$, then either $\Theta = \Delta$, or $\langle \Theta \rangle < \langle \Delta \rangle$.*

Proof. By induction on the derivation.

- In rule A-U-REFL-TT, the goal holds trivially.
- Case
  $$\frac{\Delta \Vdash^{\text{pr}}_{\widehat{\alpha}} \rho_1 \leadsto \rho_2 \dashv \Theta_1, \widehat{\alpha}:\omega_1, \Theta_2 \quad \Theta_1 \Vdash^{\text{ela}} \rho_2 : \omega_2 \quad \Theta_1 \Vdash^{\text{u}} [\Theta_1]\omega_1 \approx \omega_2 \dashv \Theta_3}{\Delta \Vdash^{\text{u}} \widehat{\alpha} \approx \rho_1 \dashv \Theta_3, \widehat{\alpha}:\omega_1 = \rho_2, \Theta_2} \text{A-U-KVARL-TT}$$

| | |
|---|---|
| $\langle \Theta_1, \widehat{\alpha}:\omega_1, \Theta_2 \rangle = \langle \Delta \rangle$ | Lemma F.47 |
| $\Theta_3 = \Theta_1 \cup \langle \Theta_3 \rangle < \langle \Theta_1 \rangle$ | I.H. |
| $\langle \Theta_3 \rangle \leqslant \langle \Theta_1 \rangle$ | Follows |
| $\langle \Theta_3, \widehat{\alpha}:\omega_1 = \rho_2, \Theta_2 \rangle < \langle \Theta_3, \widehat{\alpha}:\omega_1, \Theta_2 \rangle$ | |
| $\leqslant \langle \Theta_1, \widehat{\alpha}:\omega_1, \Theta_2 \rangle$ | |
| $= \langle \Delta \rangle$ | |

- The case for rule A-U-KVARR-TT is similar as the previous case.
- Case
  $$\frac{\Delta_1, \Delta_2 \Vdash^{\text{mv}} \widehat{\alpha}:\omega_1 \leadsto \Theta \quad \Delta[\{\Theta\}] \Vdash^{\text{pr}}_{\widehat{\alpha}} \rho_1 \leadsto \rho_2 \dashv \Theta_1, \{\Theta_2, \widehat{\alpha}:\omega_1, \Theta_3\}, \Theta_4}{\Theta_1, \{\Theta_2\} \Vdash^{\text{ela}} \rho_2 : \omega_2 \quad \Theta_1, \{\Theta_2\} \Vdash^{\text{u}} [\Theta_1, \Theta_2]\omega_1 \approx \omega_2 \dashv \Theta_5, \{\Theta_6\}}{\Delta[\{\Delta_1, \widehat{\alpha}:\omega_1, \Delta_2\}] \Vdash^{\text{u}} \widehat{\alpha} \approx \rho_1 \dashv \Theta_5, \{\Theta_6, \widehat{\alpha}:\omega_1 = \rho_2, \Theta_3\}, \Theta_4}} \text{A-U-KVARL-LO-TT}$$

| | |
|---|---|
| $\langle \Theta \rangle = \langle \Delta_1, \widehat{\alpha}:\omega, \Delta_2 \rangle$ | By moving |
| $\langle \Delta[\{\Theta\}] \rangle = \langle \Theta_1, \{\Theta_2, \widehat{\alpha}:\omega_1, \Theta_3\}, \Theta_4 \rangle$ | By Lemma F.47 |
| $\Theta_5, \{\Theta_6\} = \Theta_1, \{\Theta_2\} \cup \Theta_5, \{\Theta_6\} < \langle \Theta_1, \{\Theta_2\} \rangle$ | I.H. |
| $\langle \Theta_5, \{\Theta_6\} \rangle \leqslant \langle \Theta_1, \{\Theta_2\} \rangle$ | Follows |
| $\langle \Theta_5, \{\Theta_6, \widehat{\alpha}:\omega_1 = \rho_2, \Theta_3\}, \Theta_4 \rangle$ | |
| $< \langle \Theta_5, \{\Theta_6, \widehat{\alpha}:\omega_1, \Theta_3\}, \Theta_4 \rangle$ | |
| $\leqslant \langle \Theta_1, \{\Theta_2, \widehat{\alpha}:\omega_1, \Theta_3\}, \Theta_4 \rangle$ | |
| $= \langle \Delta[\{\Theta\}] \rangle$ | |
| $= \langle \Delta[\{\Delta_1, \widehat{\alpha}:\omega_1, \Delta_2\}] \rangle.$ | |

- The case for rule A-U-KVARR-LO-TT is similar as the previous case.
- Case
  $$\frac{\Delta \Vdash^{\text{u}} \rho_1 \approx \rho_3 \dashv \Delta_1 \quad \Delta_1 \Vdash^{\text{u}} [\Delta_1]\rho_2 \approx [\Delta_1]\rho_4 \dashv \Theta}{\Delta \Vdash^{\text{u}} \rho_1\,\rho_2 \approx \rho_3\,\rho_4 \dashv \Theta} \text{A-U-APP}$$

| | |
|---|---|
| $\Delta_1 = \Delta \cup \langle \Delta_1 \rangle < \langle \Delta \rangle$ | I.H. |
| $\Delta_1 = \Theta \cup \langle \Theta \rangle < \langle \Delta_1 \rangle$ | I.H. |
| If $\Delta_1 = \Delta$ and $\Delta_1 = \Theta$ | |
| $\Delta = \Theta$ | Follows directly |
| Otherwise | |
| $\langle \Theta \rangle < \langle \Delta \rangle$ | Follows directly |

- The case for rule A-U-KAPP is similar as the previous case.

□

**Lemma F.49** (Promotion Preserves $|\rho|$). *Given a context $\Delta[\widehat{\alpha}]$ ok, if $\Delta \Vdash^{\text{pr}}_{\widehat{\alpha}} \omega_1 \rightsquigarrow \omega_2 \dashv \Theta$, then for all $\rho$, we have $|[\Delta]\rho| = |[\Theta]\rho|$.*

Proof. By a straightforward induction on the promotion judgment.
- Most cases we have $\Delta = \Theta$. So the goal follows trivially.
- Case

$$\frac{\Delta \Vdash^{\text{pr}}_{\widehat{\alpha}} \omega_1 \rightsquigarrow \rho_1 \dashv \Delta_1 \quad \Delta_1 \Vdash^{\text{pr}}_{\widehat{\alpha}} [\Delta_1]\omega_2 \rightsquigarrow \rho_2 \dashv \Theta}{\Delta \Vdash^{\text{pr}}_{\widehat{\alpha}} \omega_1\, \omega_2 \rightsquigarrow \rho_1\, \rho_2 \dashv \Theta} \text{A-PR-APP}$$

The goal follows directly from I.H..

- The case for rule A-PR-KAPP is the same as the previous case.
- Case

$$\frac{\Delta \Vdash^{\text{pr}}_{\widehat{\alpha}} [\Delta]\rho \rightsquigarrow \rho_1 \dashv \Theta[\widehat{\alpha}][\widehat{\beta}:\rho]}{\Delta[\widehat{\alpha}][\widehat{\beta}:\rho] \Vdash^{\text{pr}}_{\widehat{\alpha}} \widehat{\beta} \rightsquigarrow \widehat{\beta}_1 \dashv \Theta[\widehat{\beta}_1:\rho_1, \widehat{\alpha}][\widehat{\beta}:\rho = \widehat{\beta}_1]} \text{A-PR-KUVARR-TT}$$

From I.H., forall $\rho'$, we have $|[\Theta[\widehat{\alpha}][\widehat{\beta}:\rho]]\rho'| = |[\Delta]\rho'|$. As compared to $\Theta[\widehat{\alpha}][\widehat{\beta}:\rho]$, $\Theta[\widehat{\beta}_1:\rho_1, \widehat{\alpha}][\widehat{\beta}:\rho = \widehat{\beta}_1]$ only substituted $\widehat{\beta}$ with $\widehat{\beta}_1$, which preserves the size. Therefore $|[\Theta[\widehat{\alpha}][\widehat{\beta}:\rho]]\rho'| = |[\Theta[\widehat{\beta}_1:\rho_1, \widehat{\alpha}][\widehat{\beta}:\rho = \widehat{\beta}_1]]\rho'|$. So $|[\Theta[\widehat{\beta}_1:\rho_1, \widehat{\alpha}][\widehat{\beta}:\rho = \widehat{\beta}_1]]\rho'| = |[\Delta]\rho'|$.

□

**Theorem F.50** (Promotion Terminates). *Given a context $\Delta[\widehat{\alpha}]$ ok, and a kind $\rho_1$ with $[\Delta]\rho_1 = \rho_1$, it is decidable whether there exists $\Theta$ such that $\Delta \Vdash^{\text{pr}}_{\widehat{\alpha}} \omega_1 \rightsquigarrow \omega_2 \dashv \Theta$.*

Proof. Draw the dependency graph of the input context. We measure the promotion process $\Delta \Vdash^{\text{pr}}_{\widehat{\alpha}} \omega_1 \rightsquigarrow \omega_2 \dashv \Theta$ by the lexicographic order of

(1) the maximal height of the being promoted types in the dependency graph;
(2) $|\omega_1|$.

We prove the measurement always get smaller from the conclusion to the hypothesis.

We first prove (1) gets no larger from the conclusion to the premises. This can be done via a straightforward induction on the promotion judgment.

Now we induction on the promotion judgment.

- Most cases do not have hypothesis.
- Case

$$\frac{\Delta \Vdash^{\text{pr}}_{\widehat{\alpha}} \omega_1 \rightsquigarrow \rho_1 \dashv \Delta_1 \quad \Delta_1 \Vdash^{\text{pr}}_{\widehat{\alpha}} [\Delta_1]\omega_2 \rightsquigarrow \rho_2 \dashv \Theta}{\Delta \Vdash^{\text{pr}}_{\widehat{\alpha}} \omega_1\, \omega_2 \rightsquigarrow \rho_1\, \rho_2 \dashv \Theta} \text{A-PR-APP}$$

| | |
|---|---|
| $|\omega_1| < |\omega_1\, \omega_2|$ | Follows directly |
| $|[\Delta_1]\omega_2| = |[\Delta]\omega_2|$ | By Lemma F.49 |
| $= |\omega_2|$ | Given the equation |
| $< |\omega_1\, \omega_2|$ | Follows |

- The case for rule A-PR-KAPP is the same as the previous case.

- Case
$$\frac{\Delta \Vdash^{\text{pr}}_{\widehat{\alpha}} [\Delta]\rho \leadsto \rho_1 \dashv \Theta[\widehat{\alpha}][\widehat{\beta} : \rho]}{\Delta[\widehat{\alpha}][\widehat{\beta} : \rho] \Vdash^{\text{pr}}_{\widehat{\alpha}} \widehat{\beta} \leadsto \widehat{\beta}_1 \dashv \Theta[\widehat{\beta}_1 : \rho_1, \widehat{\alpha}][\widehat{\beta} : \rho = \widehat{\beta}_1]} \text{ a-pr-kuvarR-tt}$$

In the dependency graph, there are edges from $\widehat{\beta}$ to $[\Delta]\rho$. So the height gets decreased from the conclusion to the hypothesis.

□

**Theorem F.51** (Unification Terminates). *Given a context $\Delta$ ok, and kinds $\rho_1$ and $\rho_2$, where $[\Delta]\rho_1 = \rho_1$, and $[\Delta]\rho_2 = \rho_2$, it is decidable whether there exists $\Theta$ such that $\Delta \Vdash^{\text{u}} \rho_1 \approx \rho_2 \dashv \Theta$.*

PROOF. We measure the unification derivation by the lexicographic order on:
(1) $\langle \Delta \rangle$
(2) $|\rho_1|$

We case analyze the derivation.

- The case for rule A-U-REFL-TT is decidable.
- Case
$$\frac{\Delta \Vdash^{\text{pr}}_{\widehat{\alpha}} \rho_1 \leadsto \rho_2 \dashv \Theta_1, \widehat{\alpha} : \omega_1, \Theta_2 \quad \Theta_1 \Vdash^{\text{ela}} \rho_2 : \omega_2 \quad \Theta_1 \Vdash^{\text{u}} [\Theta_1]\omega_1 \approx \omega_2 \dashv \Theta_3}{\Delta \Vdash^{\text{u}} \widehat{\alpha} \approx \rho_1 \dashv \Theta_3, \widehat{\alpha} : \omega_1 = \rho_2, \Theta_2} \text{ a-u-kvarL-tt}$$

| | |
|---|---|
| $\langle \Theta_1, \widehat{\alpha} : \omega_1, \Theta_2 \rangle = \langle \Delta \rangle$ | Lemma F.47 |
| $\langle \Theta_1 \rangle < \langle \Theta_1, \widehat{\alpha} : \omega_1, \Theta_2 \rangle = \langle \Delta \rangle$ | Follows |

- The case for rule A-U-KVARR-TT is similar as the previous case.
- Case
$$\frac{\Delta_1, \Delta_2 \Vdash^{\text{mv}} \widehat{\alpha} : \omega_1 \leadsto \Theta \quad \Delta[\{\Theta\}] \Vdash^{\text{pr}}_{\widehat{\alpha}} \rho_1 \leadsto \rho_2 \dashv \Theta_1, \{\Theta_2, \widehat{\alpha} : \omega_1, \Theta_3\}, \Theta_4}{\Theta_1, \{\Theta_2\} \Vdash^{\text{ela}} \rho_2 : \omega_2 \quad \Theta_1, \{\Theta_2\} \Vdash^{\text{u}} [\Theta_1, \Theta_2]\omega_1 \approx \omega_2 \dashv \Theta_5, \{\Theta_6\}} \text{ a-u-kvarL-lo-tt}$$
$$\Delta[\{\Delta_1, \widehat{\alpha} : \omega_1, \Delta_2\}] \Vdash^{\text{u}} \widehat{\alpha} \approx \rho_1 \dashv \Theta_5, \{\Theta_6, \widehat{\alpha} : \omega_1 = \rho_2, \Theta_3\}, \Theta_4$$

| | |
|---|---|
| $\langle \Theta \rangle = \langle \Delta_1, \widehat{\alpha} : \omega_1, \Delta_2 \rangle$ | By moving |
| $\langle \Delta[\{\Theta\}] \rangle = \langle \Delta[\{\Delta_1, \widehat{\alpha} : \omega_1, \Delta_2\}] \rangle$ | Follows |
| $\langle \Delta[\{\Theta\}] \rangle = \langle \Theta_1, \{\Theta_2, \widehat{\alpha} : \omega_1, \Theta_3\}, \Theta_4 \rangle$ | Lemma F.47 |
| $\langle \Theta_1, \{\Theta_2\} \rangle$ | |
| $< \langle \Theta_1, \{\Theta_2, \widehat{\alpha} : \omega_1, \Theta_3\}, \Theta_4 \rangle$ | |
| $= \langle \Delta[\{\Delta_1, \widehat{\alpha} : \omega_1, \Delta_2\}] \rangle$ | |

- The case for rule A-U-KVARR-LO-TT is similar as the previous case.
- Case
$$\frac{\Delta \Vdash^{\text{u}} \rho_1 \approx \rho_3 \dashv \Delta_1 \quad \Delta_1 \Vdash^{\text{u}} [\Delta_1]\rho_2 \approx [\Delta_1]\rho_4 \dashv \Theta}{\Delta \Vdash^{\text{u}} \rho_1 \rho_2 \approx \rho_3 \rho_4 \dashv \Theta} \text{ a-u-app}$$

For the first condition, we know that $\langle \Delta \rangle = \langle \Delta \rangle$ and the size of the expression decreases. For the second condition, from Lemma F.48, we know that either $\Delta_1 = \Delta$, or $\langle \Delta_1 \rangle < \langle \Delta \rangle$. In the former case, we know that $[\Delta_1]\rho_2 = \rho_2$. So the size of the expression decreases. In the latter case, we have $\langle \Delta_1 \rangle < \langle \Delta \rangle$ so we are done.

- The case for rule A-U-KAPP is similar as the previous one.

□

### F.2.5 Source of Unification Variables.

**Lemma F.52** (Source of Unification Variables). *If $\Delta \Vdash^k \sigma : \eta \rightsquigarrow \mu \dashv \Theta$, then for any $\widehat{\alpha} \in$ unsolved($\Theta$), either $\widehat{\alpha} \in$ fkv($[\Theta]\mu$), or there exists $\widehat{\beta} \in$ unsolved($\Delta$) such that $\widehat{\alpha} \in$ fkv($[\Theta]\widehat{\beta}$).*

PROOF. This lemma depends on the similar lemma on many judgments, including kind checking, instantiation, and unification. We prove them one by one.

When the input context is the same as the output context, the lemma holds trivially, as all unsolved unification variables in $\Theta$ are in $\Delta$. So we will skip the discussion of those cases.

**Part 1: Kinding** By induction on the judgment.

- Case
  A-KTT-FORALL
  $$\frac{\Delta \Vdash^{kc} \kappa \Leftarrow \star \rightsquigarrow \omega \dashv \Delta_1 \quad \Delta_1, a : \omega \Vdash^{kc} \sigma \Leftarrow \star \rightsquigarrow \mu \dashv \Delta_2, a : \omega, \Delta_3 \quad \Delta_3 \hookrightarrow a}{\Delta \Vdash^k \forall a : \kappa.\sigma : \star \rightsquigarrow \forall a : \omega.[\Delta_3]\mu \dashv \Delta_2, \text{unsolved}(\Delta_3)}$$

  Given $\widehat{\alpha} \in$ unsolved($\Delta_2$, unsolved($\Delta_3$)), we know that $\widehat{\alpha} \in$ unsolved($\Delta_2, a : \omega, \Delta_3$).
  Then by the lemma on kind checking. we have two cases.
  (1) $\widehat{\alpha} \in$ fkv($[\Delta_2, a : \omega, \Delta_3]\mu$). Then
    (a) $\widehat{\alpha} \in$ fkv($\mu$), and $\widehat{\alpha}$ is unsolved in $\Delta_2, a : \omega, \Delta_3$.
      Therefore $\widehat{\alpha} \in$ fkv($[\Delta_3]\mu$).
      Since $\widehat{\alpha} \in$ unsolved($\Delta_2$, unsolved($\Delta_3$)), we have $\widehat{\alpha} \in$ fkv($[\Delta_2,$ unsolved($\Delta_3$)]$([\Delta_3]\mu$)) so we are done.
    (b) there exists a $\widehat{\beta_2} \in$ fkv($\mu$), such that $\widehat{\alpha} \in$ fkv($[\Delta_2, a : \omega, \Delta_3]\widehat{\beta_2}$).
      Now the goal is to prove $\widehat{\alpha} \in$ fkv($[\Delta_2,$ unsolved($\Delta_3$)]$([\Delta_3]\widehat{\beta_2})$).
      Notice that $[\Delta_2,$ unsolved($\Delta_3$)]$([\Delta_3]\widehat{\beta_2}) = [\Delta_2]([\Delta_3]\widehat{\beta_2}) = [\Delta_2, a : \omega, \Delta_3]\widehat{\beta_2}$.
      So we are done.
  (2) there exists $\widehat{\beta_1} \in$ unsolved($\Delta_1, a : \omega$) such that $\widehat{\alpha} \in$ fkv($[\Delta_2, a : \omega, \Delta_3]\widehat{\beta_1}$).
    Because $\widehat{\beta_1}$ is in $\Delta_1, a : \omega$, then it must be $\widehat{\beta_1}$ in $\Delta_2, a : \omega$ by Lemma F.29 and Lemma F.21.
    Therefore $[\Delta_2, a : \omega, \Delta_3]\widehat{\beta_1} = [\Delta_2]\widehat{\beta_1}$.
    So we have $\widehat{\alpha} \in$ fkv($[\Delta_2]\widehat{\beta_1}$).
    Also, it must be $\widehat{\beta_1} \in$ unsolved($\Delta_1$). Then by the lemma on kind checking. we have two subcases.
    (a) $\widehat{\beta_1} \in$ fkv($[\Delta_1]\omega$).
      We know that $\Delta_1 \longrightarrow \Delta_2$ by Lemma F.15 and Lemma F.29.
      So $[\Delta_2,$ unsolved($\Delta_3$)]$\omega = [\Delta_2]\omega = [\Delta_2]([\Delta_1]\omega)$.
      We already know that $\widehat{\beta_1} \in$ fkv($[\Delta_1]\omega$) and $\widehat{\alpha} \in$ fkv($[\Delta_2]\widehat{\beta_1}$), so we know $\widehat{\alpha} \in$ fkv($[\Delta_2]([\Delta_1]\omega)$) and we are done.
    (b) there exists $\widehat{\beta_3} \in$ unsolved($\Delta$) such that $\widehat{\beta_1} \in$ fkv($[\Delta_1]\widehat{\beta_3}$).
      Similar as the previous subcase, we have $[\Delta_2,$ unsolved($\Delta_3$)]$\widehat{\beta_3} = [\Delta_2]\widehat{\beta_3} = [\Delta_2]([\Delta_1]\widehat{\beta_3})$.
      We already know that $\widehat{\beta_1} \in$ fkv($[\Delta_1]\widehat{\beta_3}$) and $\widehat{\alpha} \in$ fkv($[\Delta_2]\widehat{\beta_1}$), so we know $\widehat{\alpha} \in$ fkv($[\Delta_2]([\Delta_1]\widehat{\beta_3})$) and we are done.

- The case for rule A-KTT-FORALLI is similar as the previous case.
- Case
  A-KTT-APP
  $$\frac{\Delta \Vdash^k \tau_1 : \eta_1 \rightsquigarrow \rho_1 \dashv \Delta_1 \quad \Delta_1 \Vdash^{kapp} (\rho_1 : [\Delta_1]\eta_1) \bullet \tau_2 : \omega \rightsquigarrow \rho \dashv \Theta}{\Delta \Vdash^k \tau_1 \tau_2 : \omega \rightsquigarrow \rho \dashv \Theta}$$

  Given $\widehat{\alpha} \in$ unsolved($\Theta$), by the lemma on application kinding part we have two cases.

(1) $\widehat{\alpha} \in \mathsf{fkv}([\Theta]\rho)$. Then the goal follows directly.
(2) there exists $\widehat{\beta_1} \in \mathsf{unsolved}(\Delta_1)$ such that $\widehat{\alpha} \in \mathsf{fkv}([\Theta]\widehat{\beta_1})$.
Because $\widehat{\beta_1} \in \mathsf{unsolved}(\Delta_1)$, by I.H., we have two subcases.
  (a) $\widehat{\beta_1} \in \mathsf{fkv}([\Delta_1]\rho_1)$.
  Then by Lemma F.30 we have $[\Theta]\rho_1 = [\Theta]([\Delta_1]\rho_1)$.
  We already know that $\widehat{\beta_1} \in \mathsf{fkv}([\Delta_1]\rho_1)$ and $\widehat{\alpha} \in \mathsf{fkv}([\Theta]\widehat{\beta_1})$ so we must have $\widehat{\alpha} \in \mathsf{fkv}([\Theta]\rho_1)$.
  By the lemma on application kinding, we have $\widehat{\alpha} \in \mathsf{fkv}([\Theta]\rho)$, so we are done.
  (b) there exists $\widehat{\beta_2} \in \mathsf{unsolved}(\Delta_1)$, such that $\widehat{\beta_1} \in \mathsf{fkv}([\Delta_1]\widehat{\beta_2})$
  By Lemma F.30 we have $[\Theta]\widehat{\beta_2} = [\Theta]([\Delta_1]\widehat{\beta_2})$.
  And we must have $\widehat{\alpha} \in \mathsf{fkv}([\Theta]([\Delta_1]\widehat{\beta_2}))$.
- The case for rules A-KTT-KAPP and A-KTT-KAPP-INFER is similar as the previous case.

**Instantiation** The statement for instantiation is: if $\Delta \Vdash^{\mathsf{inst}} \mu_1 : \eta_1 \sqsubseteq \eta_2 \rightsquigarrow \mu_2 \dashv \Theta$, then for any $\widehat{\alpha} \in \mathsf{unsolved}(\Theta)$, either $\widehat{\alpha} \in \mathsf{fkv}([\Theta]\mu_2)$, or there exists $\widehat{\beta} \in \mathsf{unsolved}(\Delta)$, such that $\widehat{\alpha} \in \mathsf{fkv}([\Theta]\widehat{\beta})$. Moreover, $\mu_2$ contains all the unification variables in $\mu_1$.
We prove it by induction on the derivation.
- Case
$$\frac{\Delta \Vdash^{\mathsf{u}} \omega_1 \approx \omega_2 \dashv \Theta}{\Delta \Vdash^{\mathsf{inst}} \mu : \omega_1 \sqsubseteq \omega_2 \rightsquigarrow \mu \dashv \Theta} \text{A-INST-REFL}$$

The first half of the goal follows directly from the lemma on unification part, and the second goal holds trivially.
- Case
$$\frac{\Delta, \widehat{\alpha} : \omega_1 \Vdash^{\mathsf{inst}} \mu_1 @\widehat{\alpha} : \eta[a \mapsto \widehat{\alpha}] \sqsubseteq \omega_2 \rightsquigarrow \mu_2 \dashv \Theta}{\Delta \Vdash^{\mathsf{inst}} \mu_1 : \forall a : \omega_1.\eta \sqsubseteq \omega_2 \rightsquigarrow \mu_2 \dashv \Theta} \text{A-INST-FORALL}$$

The second half of the goal follows directly from I.H.. Given $\widehat{\alpha_1} \in \mathsf{unsolved}(\Theta)$, by I.H., we have two cases.
(1) $\widehat{\alpha_1} \in \mathsf{fkv}([\Theta]\mu_2)$. So the first half of the goal holds directly.
(2) there exists $\widehat{\beta} \in \mathsf{unsolved}(\Delta, \widehat{\alpha} : \omega_1)$, such that $\widehat{\alpha_1} \in \mathsf{fkv}([\Theta]\widehat{\beta})$.
  Then we have either $\widehat{\beta} = \widehat{\alpha}$, or $\widehat{\beta} \in \mathsf{unsolved}(\Delta)$. In the former case, as $\mu_1 @\widehat{\alpha}$ contains $\widehat{\alpha}$, we have $\mu_2$ contains $\widehat{\alpha}$. Therefore $\widehat{\alpha_1} \in \mathsf{fkv}([\Theta]\mu_2)$ and we are done. In the latter case, the goal follows directly.
- The case for rule A-INST-FORALL-INFER is similar as the previous case.

**Application Kinding** The statement for application kinding is: if $\Delta \Vdash^{\mathsf{kapp}} (\rho_1 : \eta) \bullet \tau : \omega \rightsquigarrow \rho_2 \dashv \Theta$, then for any $\widehat{\alpha} \in \mathsf{unsolved}(\Theta)$, either $\widehat{\alpha} \in \mathsf{fkv}([\Theta]\rho_2)$, or there exists $\widehat{\beta} \in \mathsf{unsolved}(\Delta)$, such that $\widehat{\alpha} \in \mathsf{fkv}([\Theta]\widehat{\beta})$. Moreover, $\rho_2$ contains all the unification variables in $\rho_1$.
We prove it by induction on the derivation.
- Case
$$\frac{\Delta \Vdash^{\mathsf{kc}} \tau \Leftarrow \omega_1 \rightsquigarrow \rho_2 \dashv \Theta}{\Delta \Vdash^{\mathsf{kapp}} (\rho_1 : \omega_1 \rightarrow \omega_2) \bullet \tau : \omega_2 \rightsquigarrow \rho_1 \rho_2 \dashv \Theta} \text{A-KAPP-TT-ARROW}$$

The first half of the goal follows directly from the lemma on kind checking part.
The second half of the goal holds trivially.

- Case
  
  a-kapp-tt-forall
  $$\frac{\Delta, \widehat{\alpha} : \omega_1 \Vdash^{\mathsf{kapp}} (\rho_1 \ @\widehat{\alpha} : \eta[a \mapsto \widehat{\alpha}]) \bullet \tau : \omega \leadsto \rho \dashv \Theta}{\Delta \Vdash^{\mathsf{kapp}} (\rho_1 : \forall a : \omega_1.\eta) \bullet \tau : \omega \leadsto \rho \dashv \Theta}$$
  
  The second half of the goal follows directly from I.H..
  Given $\widehat{\alpha}_1 \in \mathsf{unsolved}(\Theta)$, by I.H., we have two cases.
  (1) $\widehat{\alpha}_1 \in \mathsf{fkv}([\Theta]\rho)$. So the first half of the goal holds directly.
  (2) there exists $\widehat{\beta} \in \mathsf{unsolved}(\Delta, \widehat{\alpha} : \omega_1)$, such that $\widehat{\alpha}_1 \in \mathsf{fkv}([\Theta]\widehat{\beta})$.
  Then we have either $\widehat{\beta} = \widehat{\alpha}$, or $\widehat{\beta} \in \mathsf{unsolved}(\Delta)$. In the former case, as $\rho_1 \ @\widehat{\alpha}$ contains $\widehat{\alpha}$, we have $\rho$ contains $\widehat{\alpha}$. Therefore $\widehat{\alpha}_1 \in \mathsf{fkv}([\Theta]\rho)$ and we are done. In the latter case, the goal follows directly.

- The case for rule a-kapp-tt-forall-infer is the same as previous case.
- Case
  
  a-kapp-tt-kuvar
  $$\frac{\Delta_1, \widehat{\alpha}_1 : \star, \widehat{\alpha}_2 : \star, \widehat{\alpha} : \omega = (\widehat{\alpha}_1 \to \widehat{\alpha}_2), \Delta_2 \Vdash^{\mathsf{kc}} \tau \Leftarrow \widehat{\alpha}_1 \leadsto \rho_2 \dashv \Theta}{\Delta_1, \widehat{\alpha} : \omega, \Delta_2 \Vdash^{\mathsf{kapp}} (\rho_1 : \widehat{\alpha}) \bullet \tau : \widehat{\alpha}_2 \leadsto \rho_1 \rho_2 \dashv \Theta}$$
  
  The second half of the goal follows trivially.
  Given $\widehat{\alpha}_3 \in \mathsf{unsolved}(\Theta)$, by I.H., we have two cases.
  (1) $\widehat{\alpha}_3 \in \mathsf{fkv}([\Theta]\rho_2)$. So the first half of the goal holds directly.
  (2) there exists $\widehat{\beta} \in \mathsf{unsolved}(\Delta_1, \widehat{\alpha}_1 : \star, \widehat{\alpha}_2 : \star, \widehat{\alpha} : \omega = (\widehat{\alpha}_1 \to \widehat{\alpha}_2), \Delta_2)$, such that $\widehat{\alpha}_3 \in \mathsf{fkv}([\Theta]\widehat{\beta})$.
  Then we have either $\widehat{\beta} = \widehat{\alpha}_1$, or $\widehat{\beta} = \widehat{\alpha}_2$, or $\widehat{\beta} \in \mathsf{unsolved}(\Delta_1, \widehat{\alpha} : \omega = \widehat{\alpha}_1 \to \widehat{\alpha}_2, \Delta_2)$. In the former two cases, we pick $\widehat{\alpha}$ from the input context. And $[\Theta]\widehat{\alpha} = [\Theta]([\Delta_1, \widehat{\alpha}_1 : \star, \widehat{\alpha}_2 : \star, \widehat{\alpha} : \kappa = (\widehat{\alpha}_1 \to \widehat{\alpha}_2), \Delta_2]\widehat{\alpha}) = [\Theta](\widehat{\alpha}_1 \to \widehat{\alpha}_2)$ by Lemma F.30. Therefore $\widehat{\alpha}_3 \in \mathsf{fkv}([\Theta]\widehat{\alpha})$.
  In the later case, then it must be $\widehat{\beta} \in \mathsf{unsolved}(\Delta_1, \widehat{\alpha} : \omega, \Delta_2)$ So we are done.

**Kind Checking** The statement for kind checking is: if $\Delta \Vdash^{\mathsf{kc}} \sigma \Leftarrow \eta \leadsto \mu \dashv \Theta$, then for any $\widehat{\alpha} \in \mathsf{unsolved}(\Theta)$, either $\widehat{\alpha} \in \mathsf{fkv}([\Theta]\mu)$, or there exists $\widehat{\beta} \in \mathsf{unsolved}(\Delta)$, such that $\widehat{\alpha} \in \mathsf{fkv}([\Theta]\widehat{\beta})$.

To prove the lemma, we have

a-kc-sub
$$\frac{\Delta \Vdash^{\mathsf{k}} \sigma : \eta \leadsto \mu_1 \dashv \Delta_1 \qquad \Delta_1 \Vdash^{\mathsf{inst}} \mu_1 : [\Delta_1]\eta \sqsubseteq [\Delta_1]\omega \leadsto \mu_2 \dashv \Delta_2}{\Delta \Vdash^{\mathsf{kc}} \sigma \Leftarrow \omega \leadsto \mu_2 \dashv \Delta_2}$$

Given $\widehat{\alpha} \in \mathsf{unsolved}(\Delta_2)$, by the lemma on the instantiation part, we have two cases.
(1) $\widehat{\alpha} \in \mathsf{fkv}([\Delta_2]\mu_2)$. Then the goal follows directly.
(2) there exists $\widehat{\beta} \in \mathsf{unsolved}(\Delta_1)$, such that $\widehat{\alpha} \in \mathsf{fkv}([\Delta_2]\widehat{\beta})$. Then because $\widehat{\beta} \in \mathsf{unsolved}(\Delta_1)$, by the lemma on the kinding part, we have two subcases.
  (a) $\widehat{\beta} \in \mathsf{fkv}([\Delta_1]\mu_1)$. Then by the lemma on the instantiation part, we know that $\widehat{\beta} \in \mathsf{fkv}([\Delta_1]\mu_2)$. By Lemma F.30, we have $[\Delta_2]\mu_2 = [\Delta_2]([\Delta_1]\mu_2)$. So we have $\widehat{\alpha} \in \mathsf{fkv}([\Delta_2]([\Delta_1]\mu_2))$.
  (2) there exists $\widehat{\beta}_2 \in \mathsf{unsolved}(\Delta)$, such that $\widehat{\beta} \in \mathsf{fkv}([\Delta_1]\widehat{\beta}_2)$. By Lemma F.30, we have $[\Delta_2]\widehat{\beta}_2 = [\Delta_2]([\Delta_1](\widehat{\beta}_2))$. So we have $\widehat{\alpha} \in \mathsf{fkv}([\Delta_2]([\Delta_1]\widehat{\beta}_2))$.

**Promotion** The statement for promotion is: if $\Delta \Vdash^{\mathsf{pr}}_{\widehat{\alpha}} \omega_1 \leadsto \omega_2 \dashv \Theta$, then for any $\widehat{\alpha}' \in \mathsf{unsolved}(\Theta)$, there exists $\widehat{\beta} \in \mathsf{unsolved}(\Delta)$, such that $\widehat{\alpha}' \in \mathsf{fkv}([\Theta]\widehat{\beta})$.

The only interesting case here is

$$\frac{\Delta \Vdash^{\text{pr}}_{\widehat{\alpha}} [\Delta]\rho \rightsquigarrow \rho_1 \dashv \Theta[\widehat{\alpha}][\widehat{\beta}:\rho]}{\Delta[\widehat{\alpha}][\widehat{\beta}:\rho] \Vdash^{\text{pr}}_{\widehat{\alpha}} \widehat{\beta} \rightsquigarrow \widehat{\beta}_1 \dashv \Theta[\widehat{\beta}_1:\rho_1,\widehat{\alpha}][\widehat{\beta}:\rho=\widehat{\beta}_1]} \text{A-PR-KUVARR-TT}$$

Given $\widehat{\alpha}' \in \text{unsolved}(\Theta[\widehat{\beta}_1:\rho_1,\widehat{\alpha}][\widehat{\beta}:\rho=\widehat{\beta}_1])$, we have two cases:
- $\widehat{\alpha}'$ is not $\widehat{\beta}_1$.
  Then we have $\widehat{\alpha}' \in \text{unsolved}(\Theta[\widehat{\alpha}][\widehat{\beta}:\rho])$, and by I.H. we are done.
- $\widehat{\alpha}'$ is $\widehat{\beta}_1$.
  Then we pick $\widehat{\beta}$ from the input context, and we have that $[\Theta[\widehat{\beta}_1:\rho_1,\widehat{\alpha}][\widehat{\beta}:\rho=\widehat{\beta}_1]]\widehat{\beta} = \widehat{\beta}_1$ so we are done.

**Unification** The statement for unification is: if $\Delta \Vdash^{\text{u}} \omega_1 \approx \omega_2 \dashv \Theta$, then for any $\widehat{\alpha} \in \text{unsolved}(\Theta)$, there exists $\widehat{\beta} \in \text{unsolved}(\Delta)$, such that $\widehat{\alpha} \in \text{fkv}([\Theta]\widehat{\beta})$.

Here, all cases are essentially the same. We discuss two of them and the rest can be proved in a similar way.
- Case
$$\frac{\Delta \Vdash^{\text{u}} \rho_1 \approx \rho_3 \dashv \Delta_1 \qquad \Delta_1 \Vdash^{\text{u}} [\Delta_1]\rho_2 \approx [\Delta_1]\rho_4 \dashv \Theta}{\Delta \Vdash^{\text{u}} \rho_1\,\rho_2 \approx \rho_3\,\rho_4 \dashv \Theta} \text{A-U-APP}$$

  Given $\widehat{\alpha} \in \text{unsolved}(\Theta)$, by I.H., we know that there exists $\widehat{\beta} \in \text{unsolved}(\Delta_1)$, such that $\widehat{\alpha} \in \text{fkv}([\Theta]\widehat{\beta})$.
  And because $\widehat{\beta} \in \text{unsolved}(\Delta_1)$, by I.H., we know that there exists $\widehat{\beta}_2 \in \text{unsolved}(\Delta)$, such that $\widehat{\beta} \in \text{fkv}([\Delta_1]\widehat{\beta}_2)$.
  By Lemma F.30 we know that $[\Theta]\widehat{\beta}_2 = [\Theta]([\Delta_1]\widehat{\beta}_2)$. So we must have $\widehat{\alpha} \in \text{fkv}([\Theta]([\Delta_1]\widehat{\beta}_2))$.

- Case
$$\begin{array}{c} \text{A-U-KVARL-LO-TT} \\ \Delta_1, \Delta_2 \Vdash^{\text{mv}} \widehat{\alpha}:\omega_1 \rightsquigarrow \Theta \qquad \Delta[\{\Theta\}] \Vdash^{\text{pr}}_{\widehat{\alpha}} \rho_1 \rightsquigarrow \rho_2 \dashv \Theta_1, \{\Theta_2, \widehat{\alpha}:\omega_1, \Theta_3\}, \Theta_4 \\ \Theta_1, \{\Theta_2\} \Vdash^{\text{ela}} \rho_2 : \omega_2 \qquad \Theta_1, \{\Theta_2\} \Vdash^{\text{u}} [\Theta_1, \Theta_2]\omega_1 \approx \omega_2 \dashv \Theta_5, \{\Theta_6\} \\ \hline \Delta[\{\Delta_1, \widehat{\alpha}:\omega_1, \Delta_2\}] \Vdash^{\text{u}} \widehat{\alpha} \approx \rho_1 \dashv \Theta_5, \{\Theta_6, \widehat{\alpha}:\omega_1=\rho_2, \Theta_3\}, \Theta_4 \end{array}$$

  Given $\widehat{\alpha}_1 \in \text{unsolved}(\Theta_5, \{\Theta_6, \widehat{\alpha}:\omega_1=\rho_2, \Theta_3\}, \Theta_4)$, we have two cases to discuss.
  (1) $\widehat{\alpha}_1 \in \text{unsolved}(\Theta_5, \{\Theta_6\})$.
  Then by I.H., we know that there is $\widehat{\beta}_1 \in \text{unsolved}(\Theta_1, \{\Theta_2\})$ such that $\widehat{\alpha}_1 \in \text{fkv}([\Theta_5, \{\Theta_6\}]\widehat{\beta}_1)$.
  By the definition, we know that $\widehat{\beta}_1 \in \text{unsolved}(\Theta_1, \{\Theta_2, \widehat{\alpha}:\omega_1, \Theta_3\}, \Theta_4)$.
  Then by the lemma on the promotion part, we know that there exists a $\widehat{\beta}_2 \in \text{unsolved}(\Delta[\{\Theta\}])$ such that $\widehat{\beta}_1 \in \text{fkv}([\Theta_1, \{\Theta_2, \widehat{\alpha}:\omega_1, \Theta_3\}, \Theta_4]\widehat{\beta}_2)$.
  By the definition of moving, we know that all unsolved unification in $\text{unsolved}(\Delta_1, \widehat{\alpha}:\omega_1, \Delta_2)$ are in $\text{unsolved}(\Theta)$. Therefore we have $\widehat{\beta}_2 \in \text{unsolved}(\Delta[\{\Theta\}])$.
  We have that $[\Theta_5, \{\Theta_6, \widehat{\alpha}:\omega_1=\rho_2, \Theta_3\}, \Theta_4]\widehat{\beta}_2 = [\Theta_5, \{\Theta_6, \widehat{\alpha}:\omega_1=\rho_2, \Theta_3\}, \Theta_4]([\Theta_1, \{\Theta_2, \widehat{\alpha}:\omega_1, \Theta_3\}, \Theta_4]\widehat{\beta}_2)$ by Lemma F.30, as $\Theta_1, \{\Theta_2, \widehat{\alpha}:\omega_1, \Theta_3\}, \Theta_4 \longrightarrow \Theta_5, \{\Theta_6, \widehat{\alpha}:\omega_1=\rho_2, \Theta_3\}, \Theta_4$, whose derivation can be found in the proof of Lemma F.10.
  Then we must have $\widehat{\alpha}_1 \in \text{fkv}([\Theta_5, \{\Theta_6, \widehat{\alpha}:\omega_1=\rho_2, \Theta_3\}, \Theta_4]([\Theta_1, \{\Theta_2, \widehat{\alpha}:\omega_1, \Theta_3\}, \Theta_4]\widehat{\beta}_2))$.
  (2) $\widehat{\alpha}_1$ is in the domain of $\Theta_3$ and $\Theta_4$.
  Then it must be in $\text{unsolved}(\Theta_1, \{\Theta_2, \widehat{\alpha}:\omega_1, \Theta_3\}, \Theta_4)$.
  Then by the lemma on the promotion part, we know that there exists a $\widehat{\beta} \in \text{unsolved}(\Delta[\{\Theta\}])$ such that $\widehat{\alpha}_1 \in \text{fkv}([\Theta_1, \{\Theta_2, \widehat{\alpha}:\omega_1, \Theta_3\}, \Theta_4]\widehat{\beta})$.

By moving, we know that all unsolved unification in $\text{unsolved}(\Delta[\{\Theta\}])$ are in $\text{unsolved}(\Delta[\{\Delta_1, \widehat{\alpha} : \omega_1, \Delta_2\}])$.

Therefore we have $\widehat{\beta} \in \text{unsolved}(\Delta[\{\Delta_1, \widehat{\alpha} : \omega_1, \Delta_2\}])$.

We have that $[\Theta_5, \{\Theta_6, \widehat{\alpha} : \omega_1 = \rho_2, \Theta_3\}, \Theta_4]\widehat{\beta} = [\Theta_5, \{\Theta_6, \widehat{\alpha} : \omega_1 = \rho_2, \Theta_3\}, \Theta_4]([\Theta_1, \{\Theta_2, \widehat{\alpha} : \omega_1, \Theta_3\}, \Theta_4]\widehat{\beta})$ by Lemma F.30, as $\Theta_1, \{\Theta_2, \widehat{\alpha} : \omega_1, \Theta_3\}, \Theta_4 \longrightarrow \Theta_5, \{\Theta_6, \widehat{\alpha} : \omega_1 = \rho_2, \Theta_3\}, \Theta_4$, whose derivation can be found in the proof of Lemma F.10.

Then we must have $\widehat{\alpha}_1 \in \text{fkv}([\Theta_5, \{\Theta_6, \widehat{\alpha} : \omega_1 = \rho_2, \Theta_3\}, \Theta_4]([\Theta_1, \{\Theta_2, \widehat{\alpha} : \omega_1, \Theta_3\}, \Theta_4]\widehat{\beta}))$.

□

### F.2.6 Soundness of Algorithm.

**Lemma F.53** (Soundness of Promotion). *If $\Delta$ ok, and $[\Delta]\omega_1 = \omega_1$, and $\Delta \Vdash^{\text{pr}}_{\widehat{\alpha}} \omega_1 \rightsquigarrow \omega_2 \dashv \Theta$, then $[\Theta]\omega_1 = [\Theta]\omega_2 = \omega_2$. If $\Theta \longrightarrow \Omega$, then $[\Omega]\omega_1 = [\Omega]\omega_2$.*

Proof. The first half follows directly from a straightforward induction on promotion.
The second half of the goal follows directly from and Lemma F.30.

□

**Lemma F.54** (Soundness of Unification). *If $\Delta$ ok, and $\Delta \Vdash^{\text{u}} \omega_1 \approx \omega_2 \dashv \Theta$, then $[\Theta]\omega_1 = [\Theta]\omega_2$. If $\Theta \longrightarrow \Omega$, then $[\Omega]\omega_1 = [\Omega]\omega_2$.*

Proof. By Lemma F.30, we only need to prove the first half of the lemma.

The case for rule A-U-REFL-TT holds trivially. And the case for rule A-U-APP and rule A-U-KAPP follows from I.H. and Lemma F.30. As rule A-U-KVARL-TT and rule A-U-KVARR-TT, rule A-U-KVARL-LO-TT and rule A-U-KVARR-LO-TT are symmetric, we only prove one of them.

- Case

  A-U-KVARL-TT
  $$\frac{\Delta \Vdash^{\text{pr}}_{\widehat{\alpha}} \rho_1 \rightsquigarrow \rho_2 \dashv \Theta_1, \widehat{\alpha} : \omega_1, \Theta_2 \qquad \Theta_1 \Vdash^{\text{ela}} \rho_2 : \omega_2 \qquad \Theta_1 \Vdash^{\text{u}} [\Theta_1]\omega_1 \approx \omega_2 \dashv \Theta_3}{\Delta \Vdash^{\text{u}} \widehat{\alpha} \approx \rho_1 \dashv \Theta_3, \widehat{\alpha} : \omega_1 = \rho_2, \Theta_2}$$

  | | |
  |---|---|
  | $\Theta_1, \widehat{\alpha} : \omega_1, \Theta_2 \longrightarrow \Theta_3, \widehat{\alpha} : \omega_1 = \rho_2, \Theta_2$ | We have proved in Lemma F.10 |
  | $[\Theta_1, \widehat{\alpha} : \omega_1, \Theta_2]\rho_1 = [\Theta_1, \widehat{\alpha} : \omega_1, \Theta_2]\rho_2$ | By Lemma F.53 |
  | $[\Theta_3, \widehat{\alpha} : \omega_1 = \rho_2, \Theta_2]\rho_1 = [\Theta_3, \widehat{\alpha} : \omega_1 = \rho_2, \Theta_2]\rho_2$ | By Lemma F.30 |
  | $[\Theta_3, \widehat{\alpha} : \omega_1 = \rho_2, \Theta_2]\widehat{\alpha} = [\Theta_3, \widehat{\alpha} : \omega_1 = \rho_2, \Theta_2]\rho_2$ | By definition |
  | $[\Theta_3, \widehat{\alpha} : \omega_1 = \rho_2, \Theta_2]\widehat{\alpha} = [\Theta_3, \widehat{\alpha} : \omega_1 = \rho_2, \Theta_2]\rho_1$ | By equations |

- Case

  A-U-KVARL-LO-TT
  $$\frac{\Delta_1, \Delta_2 \Vdash^{\text{mv}} \widehat{\alpha} : \omega_1 \rightsquigarrow \Theta \qquad \Delta[\{\Theta\}] \Vdash^{\text{pr}}_{\widehat{\alpha}} \rho_1 \rightsquigarrow \rho_2 \dashv \Theta_1, \{\Theta_2, \widehat{\alpha} : \omega_1, \Theta_3\}, \Theta_4 \qquad \Theta_1, \{\Theta_2\} \Vdash^{\text{ela}} \rho_2 : \omega_2 \qquad \Theta_1, \{\Theta_2\} \Vdash^{\text{u}} [\Theta_1, \Theta_2]\omega_1 \approx \omega_2 \dashv \Theta_5, \{\Theta_6\}}{\Delta[\{\Delta_1, \widehat{\alpha} : \omega_1, \Delta_2\}] \Vdash^{\text{u}} \widehat{\alpha} \approx \rho_1 \dashv \Theta_5, \{\Theta_6, \widehat{\alpha} : \omega_1 = \rho_2, \Theta_3\}, \Theta_4}$$

  | | |
  |---|---|
  | $\Theta_1, \{\Theta_2, \widehat{\alpha} : \omega_1, \Theta_3\}, \Theta_4 \longrightarrow \Theta_5, \{\Theta_6, \widehat{\alpha} : \omega_1 = \rho_2, \Theta_3\}, \Theta_4$ | We have proved in Lemma F.10 |
  | $[\Theta_1, \{\Theta_2, \widehat{\alpha} : \omega_1, \Theta_3\}, \Theta_4]\rho_1 = [\Theta_1, \{\Theta_2, \widehat{\alpha} : \omega_1, \Theta_3\}, \Theta_4]\rho_2$ | By Lemma F.53 |
  | $[\Theta_5, \{\Theta_6, \widehat{\alpha} : \omega_1 = \rho_2, \Theta_3\}, \Theta_4]\rho_1 = [\Theta_5, \{\Theta_6, \widehat{\alpha} : \omega_1 = \rho_2, \Theta_3\}, \Theta_4]\rho_2$ | By Lemma F.30 |
  | $[\Theta_5, \{\Theta_6, \widehat{\alpha} : \omega_1 = \rho_2, \Theta_3\}, \Theta_4]\widehat{\alpha} = [\Theta_5, \{\Theta_6, \widehat{\alpha} : \omega_1 = \rho_2, \Theta_3\}, \Theta_4]\rho_2$ | By definition |
  | $[\Theta_5, \{\Theta_6, \widehat{\alpha} : \omega_1 = \rho_2, \Theta_3\}, \Theta_4]\rho_1 = [\Theta_5, \{\Theta_6, \widehat{\alpha} : \omega_1 = \rho_2, \Theta_3\}, \Theta_4]\widehat{\alpha}$ | By equations |

□

**Lemma F.55** (Soundness of Instantiation). *If $\Delta$ ok, and $\Delta \Vdash^{\mathsf{ela}} \mu_1 : \eta$, and $\Delta \Vdash^{\mathsf{ela}} \omega : \star$, and $\Delta \Vdash^{\mathsf{inst}} \mu_1 : \eta \sqsubseteq \omega \leadsto \mu_2 \dashv \Theta$, and $\Theta \longrightarrow \Omega$, then $[\Omega]\Delta \vdash^{\mathsf{inst}} [\Omega]\mu_1 : [\Omega]\eta \sqsubseteq [\Omega]\omega \leadsto [\Omega]\mu_2$.*

- Case
$$\frac{\Delta \Vdash^{\mathsf{u}} \omega_1 \approx \omega_2 \dashv \Theta}{\Delta \Vdash^{\mathsf{inst}} \mu : \omega_1 \sqsubseteq \omega_2 \leadsto \mu \dashv \Theta} \text{{\sc a-inst-refl}}$$

| | |
|---|---|
| $[\Omega]\omega_1 = [\Omega]\omega_2$ | By Lemma F.54 |
| $[\Omega]\Delta \vdash^{\mathsf{inst}} [\Omega]\mu : [\Omega]\omega_1 \sqsubseteq [\Omega]\omega_2 \leadsto [\Omega]\mu$ | By rule {\sc inst-refl} |

- Case
$$\frac{\Delta, \widehat{\alpha} : \omega_1 \Vdash^{\mathsf{inst}} \mu_1 \, @\widehat{\alpha} : \eta[a \mapsto \widehat{\alpha}] \sqsubseteq \omega_2 \leadsto \mu_2 \dashv \Theta}{\Delta \Vdash^{\mathsf{inst}} \mu_1 : \forall a : \omega_1.\eta \sqsubseteq \omega_2 \leadsto \mu_2 \dashv \Theta} \text{{\sc a-inst-forall}}$$

| | |
|---|---|
| $\Theta \longrightarrow \Omega$ | Given |
| $[\Omega](\Delta, \widehat{\alpha} : \omega_1) \vdash^{\mathsf{inst}} [\Omega]\mu_1 \, @([\Omega]\widehat{\alpha}) : [\Omega](\eta[a \mapsto \widehat{\alpha}]) \sqsubseteq [\Omega]\omega_2 \leadsto [\Omega]\mu_2$ | I.H. |
| $[\Omega](\Delta, \widehat{\alpha} : \omega_1) \vdash^{\mathsf{inst}} [\Omega]\mu_1 \, @([\Omega]\widehat{\alpha}) : ([\Omega]\eta)[a \mapsto [\Omega]\widehat{\alpha}] \sqsubseteq [\Omega]\omega_2 \leadsto [\Omega]\mu_2$ | By substitution |
| $\Delta \Vdash^{\mathsf{ela}} \mu_1 : \forall a : \omega_1.\eta$ | Given |
| $\Delta \Vdash^{\mathsf{ela}} \omega_1 : \star$ | By inversion |
| $\Delta \longrightarrow \Delta, \widehat{\alpha} : \omega_1$ | By rule {\sc -a-ctxe-add-tt} |
| $\Delta, \widehat{\alpha} : \omega_1 \longrightarrow \Theta$ | By Lemma F.11 |
| $\Theta \longrightarrow \Omega$ | Given |
| $\Delta, \widehat{\alpha} : \omega_1 \longrightarrow \Omega \wedge \Delta \longrightarrow \Omega$ | By Lemma F.32 |
| $[\Omega](\Delta, \widehat{\alpha} : \omega_1) = [\Omega]\Delta$ | By Lemma F.45 |
| $[\Omega]\Delta \vdash^{\mathsf{inst}} [\Omega]\mu_1 \, @([\Omega]\widehat{\alpha}) : ([\Omega]\eta)[a \mapsto [\Omega]\widehat{\alpha}] \sqsubseteq [\Omega]\omega_2 \leadsto [\Omega]\mu_2$ | By equation |
| $\Delta, \widehat{\alpha} : \omega_1 \Vdash^{\mathsf{ela}} \widehat{\alpha} : [\Delta]\omega_1$ | By rule {\sc a-ela-kuvar} |
| $[\Omega](\Delta, \widehat{\alpha} : \omega_1) \vdash^{\mathsf{ela}} [\Omega]\widehat{\alpha} : [\Omega]([\Delta]\omega_1)$ | By Lemma F.57 |
| $[\Omega]\Delta \vdash^{\mathsf{ela}} [\Omega]\widehat{\alpha} : [\Omega]([\Delta]\omega_1)$ | By equation |
| $[\Omega]\Delta \vdash^{\mathsf{ela}} [\Omega]\widehat{\alpha} : [\Omega]\omega_1$ | By Lemma F.30 |
| $[\Omega]\Delta \vdash^{\mathsf{inst}} [\Omega]\mu_1 : \forall a : [\Omega]\omega_1.[\Omega]\eta \sqsubseteq [\Omega]\omega_2 \leadsto [\Omega]\mu_2$ | By rule {\sc inst-forall} |

- The case for rule {\sc a-inst-foralli} is similar as the previous one.

□

**Lemma F.56** (Soundness of Kinding). *If $\Delta$ ok, we have*
- *if $\Delta \Vdash^{\mathsf{k}} \sigma : \eta \leadsto \mu \dashv \Theta$, and $\Theta \longrightarrow \Omega$, then $[\Omega]\Delta \vdash^{\mathsf{k}} \sigma : [\Omega]\eta \leadsto [\Omega]\mu$;*
- *if $\Delta \Vdash^{\mathsf{kc}} \sigma \Leftarrow \eta \leadsto \mu \dashv \Theta$, and $\Theta \longrightarrow \Omega$, then $[\Omega]\Delta \vdash^{\mathsf{kc}} \sigma \Leftarrow [\Omega]\eta \leadsto [\Omega]\mu$.*
- *if $\Delta \Vdash^{\mathsf{kapp}} (\rho_1 : \eta) \bullet \tau : \omega \leadsto \rho_2 \dashv \Theta$, and $\Delta \Vdash^{\mathsf{ela}} \rho_1 : \eta$, and $\Theta \longrightarrow \Omega$, then $[\Omega]\Delta \vdash^{\mathsf{inst}} [\Omega]\rho_1 : [\Omega]\eta \sqsubseteq (\omega_1 \rightarrow [\Omega]\omega) \leadsto \rho_3$, and $[\Omega]\Delta \vdash^{\mathsf{kc}} \tau \Leftarrow \omega_1 \leadsto \rho_4$. and $[\Omega]\rho_2 = \rho_3\,\rho_4$.*

Proof. By induction on the derivation.

**Part 1**
- The case for rules {\sc a-ktt-star}, {\sc a-ktt-nat}, {\sc a-ktt-var}, {\sc a-ktt-tcon}, and {\sc a-ktt-arrow} are straightforward.
- Case
$$\frac{\Delta \Vdash^{\mathsf{k}} \tau_1 : \eta_1 \leadsto \rho_1 \dashv \Delta_1 \quad \Delta_1 \Vdash^{\mathsf{kapp}} (\rho_1 : [\Delta_1]\eta_1) \bullet \tau_2 : \omega \leadsto \rho \dashv \Theta}{\Delta \Vdash^{\mathsf{k}} \tau_1\,\tau_2 : \omega \leadsto \rho \dashv \Theta} \text{{\sc a-ktt-app}}$$

By well-formedness of the judgments, we know every output context is an extension of the input context and by transitivity we have that output context is an extension of all the previous input contexts.

| | |
|---|---|
| $[\Omega]\Delta \Vdash^k \tau_1 : [\Omega]\eta_1 \rightsquigarrow [\Omega]\rho_1$ | I.H. |
| $\Delta_1 \Vdash^{ela} \rho_1 : [\Delta_1]\eta_1$ | By Lemma F.15 |
| $[\Omega]\Delta_1 \Vdash^{inst} [\Omega]\rho_1 : [\Omega]([\Delta_1]\eta_1) \sqsubseteq \omega_1 \rightarrow [\Omega]\omega \rightsquigarrow \rho_2$ | By Part 3 |
| $\wedge [\Omega]\Delta_1 \Vdash^{kc} \tau_2 \Leftarrow \omega_1 \rightsquigarrow \rho_3$ | |
| $\wedge [\Omega]\rho = \rho_2\,\rho_3$ | |
| $[\Omega]\Delta = [\Omega]\Delta_1$ | By Lemma F.45 |
| $[\Omega]\Delta \Vdash^{inst} [\Omega]\rho_1 : [\Omega]\eta_1 \sqsubseteq \omega_1 \rightarrow [\Omega]\omega \rightsquigarrow \rho_2$ | By equations and Lemma F.30 |
| $[\Omega]\Delta \Vdash^{kc} \tau_2 \Leftarrow \omega_1 \rightsquigarrow \rho_3$ | By equations |
| $[\Omega]\Delta \Vdash^k \tau_1\,\tau_2 : [\Omega]\omega \rightsquigarrow [\Omega]\rho$ | By rule KTT-APP |

- Case
  A-KTT-KAPP
  $$\dfrac{\Delta \Vdash^k \tau_1 : \eta \rightsquigarrow \rho_1 \dashv \Delta_1 \quad [\Delta_1]\eta = \forall a : \omega.\eta_2 \quad \Delta_1 \Vdash^{kc} \tau_2 \Leftarrow \omega \rightsquigarrow \rho_2 \dashv \Delta_2}{\Delta \Vdash^k \tau_1\,@\tau_2 : \eta_2[a \mapsto \rho_2] \rightsquigarrow \rho_1\,@\rho_2 \dashv \Delta_2}$$

By well-formedness of the judgments, we know every output context is an extension of the input context and by transitivity we have that output context is an extension of all the previous input contexts.

| | |
|---|---|
| $[\Omega]\Delta \Vdash^k \tau_1 : [\Omega]\eta \rightsquigarrow [\Omega]\rho_1$ | I.H. |
| $[\Omega]\eta = [\Omega]([\Delta_1]\eta) = \forall a : [\Omega]\omega.[\Omega]\eta_2$. | By Lemma F.30 |
| $[\Omega]\Delta \Vdash^k \tau_1 : \forall a : [\Omega]\omega.[\Omega]\eta_2 \rightsquigarrow [\Omega]\rho_1$ | By equations |
| $[\Omega]\Delta_1 \Vdash^{kc} \tau_2 \Leftarrow [\Omega]\omega \rightsquigarrow [\Omega]\rho_2$ | By Part 2 |
| $[\Omega]\Delta \Vdash^{kc} \tau_2 \Leftarrow [\Omega]\omega \rightsquigarrow [\Omega]\rho_2$ | By Lemma F.45 |
| $[\Omega]\Delta \Vdash^k \tau_1\,@\tau_2 : ([\Omega]\eta_2)[a \mapsto [\Omega]\rho_2] \rightsquigarrow [\Omega]\rho_1\,@[\Omega]\rho_2$ | By rule KTT-KAPP |
| $[\Omega]\Delta \Vdash^k \tau_1\,@\tau_2 : [\Omega](\eta_2[a \mapsto \rho_2]) \rightsquigarrow [\Omega](\rho_1\,@\rho_2)$ | By substitutions |

- The case for rule A-KTT-KAPP-INFER is similar to the previous case.
- Case
  A-KTT-FORALL
  $$\dfrac{\Delta \Vdash^{kc} \kappa \Leftarrow \star \rightsquigarrow \omega \dashv \Delta_1 \quad \Delta_1, a : \omega \Vdash^{kc} \sigma \Leftarrow \star \rightsquigarrow \mu \dashv \Delta_2, a : \omega, \Delta_3 \quad \Delta_3 \hookrightarrow a}{\Delta \Vdash^k \forall a : \kappa.\sigma : \star \rightsquigarrow \forall a : \omega.[\Delta_3]\mu \dashv \Delta_2, \mathsf{unsolved}(\Delta_3)}$$

| | |
|---|---|
| $\Delta_1, a : \omega \longrightarrow \Delta_2, a : \omega, \Delta_3$ | Lemma F.15 |
| $\Delta_1 \longrightarrow \Delta_2 \wedge \Delta_3$ soft | By Lemma F.29 |
| $\Delta_2 \longrightarrow \Delta_2, \mathsf{unsolved}(\Delta_3)$ | As proved in Lemma F.15 |
| $\Delta_2, \mathsf{unsolved}(\Delta_3) \longrightarrow \Omega$ | Given |
| $\Delta_1 \longrightarrow \Omega$ | by Lemma F.32 |
| $[\Omega]\Delta \Vdash^{kc} \kappa \Leftarrow \star \rightsquigarrow [\Omega]\omega$ | By Part 2 |
| $\Delta_2 \longrightarrow \Omega_1 \wedge \Omega = \Omega_1, \Omega_2 \wedge \Omega_2$ soft | By Lemma F.29 |
| $\Delta_2, a : \omega \longrightarrow \Omega_1, a : \omega$ | By rule A-CTXE-TVAR-TT |
| Construct a $\Omega_3$ such that | |
| $\Delta_2, a : \omega, \Delta_3 \longrightarrow \Omega_1, a : \omega, \Omega_3$ | |
| $\wedge \forall \widehat{\alpha} \in \mathsf{unsolved}(\Delta_3)$, the solution for $\widehat{\alpha}$ in $\Omega_3$ is $[\Omega_2]\widehat{\alpha}$ | |
| $[\Omega_1, a : \omega, \Omega_3](\Delta_2, a : \omega, \Delta_3) \Vdash^{kc} \sigma \Leftarrow \star \rightsquigarrow [\Omega_1, a : \omega, \Omega_3]\mu$ | Part 2 |
| $[\Omega_1, a : \omega, \Omega_3](\Delta_2, a : \omega, \Delta_3)$ | |
| $= [\Omega_1, a : \omega](\Delta_2, a : \omega)$ | By Lemma F.44 |
| $= [\Omega_1]\Delta_2, a : [\Omega_1]\omega$ | By definition |
| $= [\Omega_1]\Delta_2, a : [\Omega]\omega$ | By Lemma F.31 |
| $= [\Omega_1]\Delta, a : [\Omega]\omega$ | By Lemma F.45 |
| $= [\Omega]\Delta, a : [\Omega]\omega$ | By Lemma F.44 |
| $[\Omega_1, a : \omega, \Omega_3]\mu$ | |
| $= [\Omega_1, \Omega_2]([\Delta_3]\mu)$ | By the way $\Omega_3$ is constructed |
| $= [\Omega]([\Delta_3]\mu)$ | |
| $[\Omega]\Delta, a : [\Omega]\omega \Vdash^{kc} \sigma \Leftarrow \star \leadsto [\Omega]([\Delta_3]\mu)$ | By equations |
| $[\Omega]\Delta \Vdash^{k} \forall a : \kappa.\sigma : \star \leadsto \forall a : [\Omega]\omega.[\Omega]([\Delta_3]\mu)$ | By rule KTT-FORALL |
| $[\Omega]\Delta \Vdash^{k} (\forall a : \kappa.\sigma) : \star \leadsto [\Omega](\forall a : \omega.([\Delta_3]\mu))$ | By substitution |

- The case for rule A-KTT-FORALLI is similar to the previous case.
  The notable thing is that we use the solution of $\widehat{\alpha}$ (as in the rule A-KTT-FORALLI) in $\Omega$ as the $\omega$ in rule KTT-FORALL.

**Part 2** We have

$$\dfrac{\text{A-KC-SUB}}{\Delta \Vdash^{k} \sigma : \eta \leadsto \mu_1 \dashv \Delta_1 \qquad \Delta_1 \Vdash^{inst} \mu_1 : [\Delta_1]\eta \sqsubseteq [\Delta_1]\omega \leadsto \mu_2 \dashv \Delta_2}{\Delta \Vdash^{kc} \sigma \Leftarrow \omega \leadsto \mu_2 \dashv \Delta_2}$$

Follows directly from Part 1 and soundness of instantiation (Lemma F.55).

**Part 3** By induction on the judgment.

- Case

$$\dfrac{\text{A-KAPP-TT-ARROW}}{\Delta \Vdash^{kc} \tau \Leftarrow \omega_1 \leadsto \rho_2 \dashv \Theta}{\Delta \Vdash^{kapp} (\rho_1 : \omega_1 \to \omega_2) \bullet \tau : \omega_2 \leadsto \rho_1 \rho_2 \dashv \Theta}$$

| | |
|---|---|
| $[\Omega]\Delta \Vdash^{inst} [\Omega]\rho_1 : [\Omega]\omega_1 \to [\Omega]\omega_2 \sqsubseteq [\Omega]\omega_1 \to [\Omega]\omega_2 \leadsto [\Omega]\rho_1$ | By rule INST-REFL |
| $[\Omega]\Delta \Vdash^{kc} \tau \Leftarrow [\Omega]\omega_1 \leadsto [\Omega]\rho_2$ | Part 2 |

- Case

$$\dfrac{\text{A-KAPP-TT-FORALL}}{\Delta, \widehat{\alpha} : \omega_1 \Vdash^{kapp} (\rho_1 \,@\widehat{\alpha} : \eta[a \mapsto \widehat{\alpha}]) \bullet \tau : \omega \leadsto \rho \dashv \Theta}{\Delta \Vdash^{kapp} (\rho_1 : \forall a : \omega_1.\eta) \bullet \tau : \omega \leadsto \rho \dashv \Theta}$$

| | |
|---|---|
| $[\Omega](\Delta, \widehat{\alpha} : \omega_1) \Vdash^{inst} [\Omega](\rho_1 \,@\widehat{\alpha}) : [\Omega](\eta[a \mapsto \widehat{\alpha}]) \sqsubseteq \omega' \to [\Omega]\omega \leadsto \rho_3$ | I.H. |
| $[\Omega](\Delta, \widehat{\alpha} : \omega_1) \Vdash^{kc} \tau \Leftarrow \omega' \leadsto \rho_4 \wedge [\Omega]\rho = \rho_3 \rho_4$ | Above |
| $\Delta \longrightarrow \Theta$ | By Lemma F.15 |
| $\Delta \longrightarrow \Omega$ | By Lemma F.32 |
| $\Delta, \widehat{\alpha} : \omega_1 \longrightarrow \Theta$ | By Lemma F.15 |
| $\Delta, \widehat{\alpha} : \omega_1 \longrightarrow \Omega$ | By Lemma F.32 |
| $[\Omega]\Delta = [\Omega](\Delta, \widehat{\alpha} : \omega_1)$ | By Lemma F.45 |
| $[\Omega]\Delta \Vdash^{inst} [\Omega]\rho_1 \,@([\Omega]\widehat{\alpha}) : ([\Omega]\eta[a \mapsto [\Omega]\widehat{\alpha}]) \sqsubseteq \omega' \to [\Omega]\omega \leadsto \rho_3$ | By equations and substitutions |
| $[\Omega]\Delta \Vdash^{kc} \tau \Leftarrow \omega' \leadsto \rho_4$ | By equations |
| $\Delta, \widehat{\alpha} : \omega_1 \Vdash^{ela} \widehat{\alpha} : [\Delta, \widehat{\alpha} : \omega_1]\omega_1$ | By rule A-ELA-KUVAR |
| $[\Omega](\Delta, \widehat{\alpha} : \omega_1) \Vdash^{ela} [\Omega]\widehat{\alpha} : [\Omega]([\Delta, \widehat{\alpha} : \omega_1]\omega_1)$ | By Lemma F.57 |
| $[\Omega]\Delta \Vdash^{ela} [\Omega]\widehat{\alpha} : [\Omega]([\Delta, \widehat{\alpha} : \omega_1]\omega_1)$ | By equations |
| $[\Omega]\Delta \Vdash^{ela} [\Omega]\widehat{\alpha} : [\Omega]\omega_1$ | By Lemma F.30 |
| $[\Omega]\Delta \Vdash^{inst} [\Omega]\rho_1 : \forall a : [\Omega]\omega_1.[\Omega]\eta \sqsubseteq \omega' \to [\Omega]\omega \leadsto \rho_3$ | By rule INST-FORALL |

- The case for rule A-KAPP-TT-FORALL-INFER is similar as the previous case.
- Case

$$\frac{\text{A-KAPP-TT-KUVAR}}{\Delta_1, \widehat{\alpha}_1 : \star, \widehat{\alpha}_2 : \star, \widehat{\alpha} : \omega = (\widehat{\alpha}_1 \to \widehat{\alpha}_2), \Delta_2 \Vdash^{\text{kc}} \tau \Leftarrow \widehat{\alpha}_1 \leadsto \rho_2 \dashv \Theta}{\Delta_1, \widehat{\alpha} : \omega, \Delta_2 \Vdash^{\text{kapp}} (\rho_1 : \widehat{\alpha}) \bullet \tau : \widehat{\alpha}_2 \leadsto \rho_1 \rho_2 \dashv \Theta}$$

| | |
|---|---|
| $\Delta_1, \widehat{\alpha}_1 : \star, \widehat{\alpha}_2 : \star, \widehat{\alpha} : \omega = (\widehat{\alpha}_1 \to \widehat{\alpha}_2), \Delta_2 \longrightarrow \Theta$ | By Lemma F.15 |
| $\Delta_1, \widehat{\alpha}_1 : \star, \widehat{\alpha}_2 : \star, \widehat{\alpha} : \omega = (\widehat{\alpha}_1 \to \widehat{\alpha}_2), \Delta_2 \longrightarrow \Omega$ | By Lemma F.32 |
| $[\Delta_1, \widehat{\alpha}_1 : \star, \widehat{\alpha}_2 : \star, \widehat{\alpha} : \omega = (\widehat{\alpha}_1 \to \widehat{\alpha}_2), \Delta_2]\widehat{\alpha}$ | |
| $[\Delta_1, \widehat{\alpha}_1 : \star, \widehat{\alpha}_2 : \star, \widehat{\alpha} : \omega = (\widehat{\alpha}_1 \to \widehat{\alpha}_2), \Delta_2](\widehat{\alpha}_1 \to \widehat{\alpha}_2)$ | By definition |
| $[\Omega]\widehat{\alpha} = [\Omega](\widehat{\alpha}_1 \to \widehat{\alpha}_2)$ | By Lemma F.22 |
| $[\Omega](\Delta_1, \widehat{\alpha} : \omega, \Delta_2) \Vdash^{\text{inst}} [\Omega]\rho_1 : [\Omega]\widehat{\alpha} \sqsubseteq [\Omega]\widehat{\alpha}_1 \to [\Omega]\widehat{\alpha}_2 \leadsto [\Omega]\rho_1$ | By rule INST-REFL |
| $[\Omega](\Delta_1, \widehat{\alpha}_1 : \star, \widehat{\alpha}_2 : \star, \widehat{\alpha} : \omega = (\widehat{\alpha}_1 \to \widehat{\alpha}_2), \Delta_2) \Vdash^{\text{kc}} \tau \Leftarrow [\Omega]\widehat{\alpha}_1 \leadsto [\Omega]\rho_2$ | Part 2 |
| $\Delta_1, \widehat{\alpha} : \omega, \Delta_2 \longrightarrow \Theta$ | By Lemma F.15 |
| $\Delta_1, \widehat{\alpha} : \omega, \Delta_2 \longrightarrow \Omega$ | By Lemma F.32 |
| $[\Omega](\Delta_1, \widehat{\alpha} : \omega, \Delta_2) \Vdash^{\text{kc}} \tau \Leftarrow [\Omega]\widehat{\alpha}_1 \leadsto [\Omega]\rho_2$ | By Lemma F.45 |

□

**Lemma F.57** (Soundness of Elaborated Kinding). *If $\Delta$ ok, and $\Delta \Vdash^{\text{ela}} \mu : \eta$, and $\Delta \longrightarrow \Omega$, then $[\Omega]\Delta \Vdash^{\text{ela}} [\Omega]\mu : [\Omega]\eta$.*

PROOF. By a straightforward induction on the derivation.

□

**Lemma F.58** (Soundness of Typing Signature). *If $\Delta$ ok, and $\Omega \Vdash^{\text{sig}} \mathcal{S} \leadsto T : \eta$, then $[\Omega]\Omega \Vdash^{\text{sig}} \mathcal{S} \leadsto T : \eta$.*

PROOF. We have

$$\frac{\text{A-SIG-TT}}{\begin{array}{c} \rceil\sigma\lceil \quad \overline{a_i}^i = \text{fkv}(\sigma) \quad \Omega, \overline{\{\widehat{\alpha}_i : \star, a_i : \widehat{\alpha}_i\}}^i \Vdash^{\text{k}} \sigma : \star \leadsto \eta \dashv \Delta \\ \phi_1^c = \text{scoped\_sort}(\overline{a_i : [\Delta]\widehat{\alpha}_i}^i) \quad \widehat{\phi}_2^c = \text{unsolved}(\Delta) \quad \Delta \hookrightarrow \overline{a_i}^i \end{array}}{\Omega \Vdash^{\text{sig}} \textbf{data}\, T : \sigma \leadsto T : \forall\{\phi_2^c\}.((\forall\{\phi_1^c\}.[\Delta]\eta)[\widehat{\phi}_2^c \mapsto \phi_2^c])}$$

From $\Delta \hookrightarrow \overline{a_i}^i$ we know that all unsolved unification variables in $\Delta$ do not depend on $\overline{a_i}^i$.

Given $\widehat{\phi}_2^c = \text{unsolved}(\Delta)$, we further know that $\widehat{\phi}_2^c$ only contains unsolved unification variable that do not depend on $\overline{a_i}^i$.

So $\phi_2^c$ only contains type variables that do not depend on any unification variable or $\overline{a_i}^i$.

By weakening, we can add $\phi_2^c$ into the kinding judgment, so we get $\Omega, \phi_2^c, \overline{\{\widehat{\alpha}_i : \star, a_i : \widehat{\alpha}_i\}}^i \Vdash^{\text{k}} \sigma : \star \leadsto \eta \dashv \Delta_1$, where $\Delta_1$ is identical to $\Delta$ except for the presence of $\phi_2^c$.

Now, we solve all unsolved unification variable in $\Delta_1$ (i.e., the domain of $\widehat{\phi}_2^c$) to its corresponding type variable in $\phi_2^c$. We get a complete context $\Omega_1$ and $\Delta_1 \longrightarrow \Omega_1$.

By by Lemma F.15, we have $\Omega, \phi_2^c, \overline{\{\widehat{\alpha}_i : \star, a_i : \widehat{\alpha}_i\}}^i \longrightarrow \Delta_1$.

So by Lemma F.32 we have $\Omega, \phi_2^c, \overline{\{\widehat{\alpha}_i : \star, a_i : \widehat{\alpha}_i\}}^i \longrightarrow \Omega_1$.

By soundness of kinding (Lemma F.56), we know that $[\Omega_1](\Omega, \phi_2^c, \overline{\{\widehat{\alpha}_i : \star, a_i : \widehat{\alpha}_i\}}^i) \Vdash^{\text{k}} \sigma : \star \leadsto [\Omega_1]\eta$.

$[\Omega_1](\Omega, \phi_2^c, \overline{\{\widehat{\alpha}_i : \star, a_i : \widehat{\alpha}_i\}}^i) = [\Omega]\Omega, \phi_2^c, \phi_3^c[\widehat{\phi}^c \mapsto \phi_2^c]$, where $\phi_3^c$ is a well-formed order of $\phi_1^c$.

And $[\Omega_1]\eta = ([\Delta]\eta)[\widehat{\phi^c} \mapsto \phi_2^c]$, because $\Delta$ contains all the solved unification variable in $\Omega_1$ except for $\widehat{\phi^c}$.

Namely, $[\Omega]\Omega, \phi_2^c, \phi_3^c[\widehat{\phi^c} \mapsto \phi_2^c] \Vdash^k \sigma : \star \rightsquigarrow ([\Delta]\eta)[\widehat{\phi^c} \mapsto \phi_2^c]$. By reordering the context while preserving well-formedness, we have $[\Omega]\Omega, \phi_2^c, \phi_1^c[\widehat{\phi^c} \mapsto \phi_2^c] \Vdash^k \sigma : \star \rightsquigarrow ([\Delta]\eta)[\widehat{\phi^c} \mapsto \phi_2^c]$.

By the kinding rule we can get $[\Omega]\Omega, \phi_2^c \Vdash^k \forall \{\phi_1^c[\widehat{\phi^c} \mapsto \phi_2^c]\}.\sigma : \star \rightsquigarrow \forall \{\phi_1^c[\widehat{\phi^c} \mapsto \phi_2^c]\}.([\Delta]\eta)[\widehat{\phi^c} \mapsto \phi_2^c]$. By substitution we get $[\Omega]\Omega, \phi_2^c \Vdash^k \forall \{\phi_1^c[\widehat{\phi^c} \mapsto \phi_2^c]\}.\sigma : \star \rightsquigarrow (\forall \{\phi_1^c\}.[\Delta]\eta)[\widehat{\phi^c} \mapsto \phi_2^c]$.

To prove the rule, our goal is to prove all preconditions in

$$\frac{\rceil \sigma \lceil \quad \phi \in Q(\sigma) \quad \phi_1^c \in Q(\forall \phi^c.\, \eta) \quad \Sigma, \phi_1^c \Vdash^k \forall \phi.\, \sigma : \star \rightsquigarrow \forall \phi^c.\, \eta \quad |\phi| = |\phi^c|}{\Sigma \vdash^{\text{sig}} \textbf{data}\, T : \sigma \rightsquigarrow T : \forall \{\phi_1^c\}.\forall \{\phi^c\}.\eta} \text{SIG-TT}$$

We have $\rceil \sigma \lceil$ as given. We claim that $\phi_1^c$ fits $\phi$ ($\phi^c$), and $\phi_2^c$ fits $\phi_1^c$.

We first prove $\phi_1^c$ fits $\phi$. Because $\phi_1^c = \text{scoped\_sort}(\overline{a_i : [\Delta]\widehat{\alpha_i}}^i)$, obviously $\phi_1^c$ is one of the well-formed permutation of $\overline{a_i}^i$, namely the free kind binder of $\sigma$.

We then prove $\phi_2^c$ fits $\phi_1^c$. That requires us to prove that $\phi_2^c$ is the free kind binder of $(\forall \{\phi_1^c\}.[\Delta]\eta)[\widehat{\phi^c} \mapsto \phi_2^c]$. Because $\widehat{\phi_2^c} = \text{unsolved}(\Delta)$, by Lemma F.52, we know every unsolved unification variable in $\widehat{\phi_2^c}$ either appears in $[\Delta]\eta$, or appears in $\phi_1^c$. For sure $[\Delta]\eta$ and $\phi_1^c$ cannot contain more unsolved unification variable than $\widehat{\phi_2^c}$ or otherwise it would be ill-formed. Namely, $\widehat{\phi_2^c}$ *are* the free unification variables of $[\Delta]\eta$ and $\phi_1^c$. By substituting $\widehat{\phi_2^c}$ with $\phi_2^c$, we know that $\phi_2^c$ are the free kind binder in $(\forall \{\phi_1^c\}.[\Delta]\eta)[\widehat{\phi_2^c} \mapsto \phi_2^c]$.

By now we have proved all the preconditions and we conclude that $[\Omega]\Omega \vdash^{\text{sig}} \mathcal{S} \rightsquigarrow T : \eta$.
□

**Lemma F.59** (Soundness of Typing Data Constructor Decl.). *If $\Delta$ ok, and $\Delta \Vdash^{\text{dc}}_\rho \mathcal{D} \rightsquigarrow \mu \dashv \Theta$, and $\Theta \longrightarrow \Omega$, then $[\Omega]\Delta \vdash^{\text{dc}}_{([\Omega]\rho)} \mathcal{D} \rightsquigarrow [\Omega]\mu$.*

Proof. We have

$$\frac{\Delta, \blacktriangleright_D \Vdash^k \forall \phi.(\overline{\tau_i}^i \to \rho) : \star \rightsquigarrow \mu \dashv \Theta_1, \blacktriangleright_D, \Theta_2 \quad \widehat{\phi^c} = \text{unsolved}(\Theta_2)}{\Delta \Vdash^{\text{dc}}_\rho \forall \phi.D\, \overline{\tau_i}^i \rightsquigarrow \forall \{\phi^c\}.(([\Theta_2]\mu)[\widehat{\phi^c} \mapsto \phi^c]) \dashv \Theta_1} \text{A-DC-TT}$$

To prove our goal, we claim that $\phi^c$ fits the $\phi^c$ in

$$\frac{\phi^c \in Q(\mu \backslash_{\Sigma, \overline{\tau_i}^i}) \quad \Sigma, \phi^c \Vdash^k \forall \phi.\overline{\tau_i}^i \to \rho : \star \rightsquigarrow \mu}{\Sigma \vdash^{\text{dc}}_\rho \forall \phi.D\, \overline{\tau_i}^i \rightsquigarrow \forall \{\phi^c\}.\mu} \text{DC-TT}$$

We prove this by Lemma F.52 and the similar reasoning as in Lemma F.58.

The important thing to note is $\widehat{\phi^c}$ only contains unsolved unification variables in $\mu$.

Note that $\mu$ might contain unsolved unification variables in $\Theta_1$. Then they must be the dependency of unsolved unification variables in $\Delta$. And those are not unification variables that we should generalize over.
□

**Lemma F.60** (Soundness of Typing Datatype Decl.). *If $\Delta$ ok, and $\Delta \Vdash^{\text{dt}} \mathcal{T} \rightsquigarrow \Gamma \dashv \Theta$, and $\Theta \longrightarrow \Omega$, then $[\Omega]\Delta \vdash^{\text{dt}} \mathcal{T} \rightsquigarrow [\Omega]\Gamma$.*

Proof. We have

A-DT-TT

$$\dfrac{(T : \forall\{\phi_1^c\}.\forall \phi_2^c.\, \omega) \in \Delta \quad \Delta, \phi_1^c, \phi_2^c, \overline{\widehat{\alpha_i} : \star}^i \Vdash^u [\Delta]\omega \approx (\overline{\widehat{\alpha_i}}^i \to \star) \dashv \Theta_1, \phi_1^c, \phi_2^c, \overline{\widehat{\alpha_i} : \star = \omega_i}^i \quad \overline{\Theta_j, \phi_1^c, \phi_2^c, \overline{a_i : \omega_i}^i \Vdash^{dc}_{(T\, @\phi_1^c\, @\phi_2^c\, \overline{a_i}^i)} \mathcal{D}_j \leadsto \mu_j \dashv \Theta_{j+1}, \phi_1^c, \phi_2^c, \overline{a_i : \omega_i}^i}^j}{\Delta \Vdash^{dt} \mathbf{data}\, T\, \overline{a_i}^i = \overline{\mathcal{D}_j}^{j\in 1..n} \leadsto \overline{D_j : \forall\{\phi_1^c\}.\forall \phi_2^c.\, \forall \overline{a_i : \omega_i}^i.\mu_j}^j \dashv \Theta_{n+1}}$$

| | |
|---|---|
| $\Delta, \phi_1^c, \phi_2^c, \overline{\widehat{\alpha_i} : \star}^i \longrightarrow \Theta_1, \phi_1^c, \phi_2^c, \overline{\widehat{\alpha_i} : \star = \omega_i}^i$ | By Lemma F.10 |
| $\Delta \longrightarrow \Theta_1$ | By Lemma F.29 |
| $\overline{\Theta_j, \phi_1^c, \phi_2^c, \overline{a_i : \omega_i}^i \longrightarrow \Theta_{j+1}, \phi_1^c, \phi_2^c, \overline{a_i : \omega_i}^i}^j$ | By Lemma F.18 |
| $\overline{\Theta_j \longrightarrow \Theta_{j+1}}^j$ | By Lemma F.29 |
| $\Theta_{n+1} \longrightarrow \Omega$ | Given |
| $\Theta_1 \longrightarrow \Omega$ | By Lemma F.32 |
| $\Delta \longrightarrow \Omega$ | By Lemma F.32 |
| $\Theta_1, \phi_1^c, \phi_2^c, \overline{\widehat{\alpha_i} : \star = \omega_i}^i \longrightarrow \Omega, \phi_1^c, \phi_2^c, \overline{\widehat{\alpha_i} : \star = \omega_i}^i$ | By definition |
| $[\Omega, \phi_1^c, \phi_2^c, \overline{\widehat{\alpha_i} : \star = \omega_i}^i]([\Delta]\omega) = [\Omega, \phi_1^c, \phi_2^c, \overline{\widehat{\alpha_i} : \star = \omega_i}^i](\overline{\widehat{\alpha_i}}^i \to \star)$ | By Lemma F.54 |
| $[\Omega]([\Delta]\omega) = \overline{[\Omega]\omega_i}^i \to \star$ | By definition and freshness |
| $[\Omega]\omega = \overline{[\Omega]\omega_i}^i \to \star$ | By Lemma F.30 |
| $(T : \forall\{\phi_1^c\}.\forall \phi_2^c.\, \overline{[\Omega]\omega_i}^i \to \star) \in [\Omega]\Delta$ | By Lemma F.39 |
| $\overline{\Theta_{j+1} \longrightarrow \Omega}^j$ | By Lemma F.32 |
| $\overline{\Theta_{j+1}, \phi_1^c, \phi_2^c, \overline{a_i : \omega_i}^i \longrightarrow \Omega, \phi_1^c, \phi_2^c, \overline{a_i : \omega_i}^i}^j$ | By definition |
| $\overline{[\Omega]\Delta, \phi_1^c, \phi_2^c, \overline{a_i : [\Omega]\omega_i}^i \Vdash^{dc}_{(T\, @\phi_1^c\, @\phi_2^c\, \overline{a_i}^i)} \mathcal{D}_j \leadsto [\Omega]\mu_j}^j$ | By Lemma F.59, and Lemma F.45 |
| $[\Omega]\Delta \vdash^{dt} \mathbf{data}\, T\, \overline{a_i}^i = \overline{\mathcal{D}_j}^j \leadsto \overline{D_j : [\Omega](\forall\{\phi_1^c\}.\forall \phi_2^c.\, \forall \overline{a_i : \omega_i}^i.\mu_j)}^j$ | By rule DT-TT |

□

**Lemma F.61** (Soundness of Typing Program). *If* $\Omega; \Gamma \Vdash^{pgm} pgm : \mu$, *then* $[\Omega]\Omega; [\Omega]\Gamma \vdash^{pgm} pgm : [\Omega]\mu$.

Proof. By induction on the derivation.

- Case

  A-PGM-EXPR
  $$\dfrac{[\Omega]\Omega; [\Omega]\Gamma \vdash e : \sigma}{\Omega; \Gamma \Vdash^{pgm} e : \sigma}$$

  The goal holds directly.

- Case

  A-PGM-SIG
  $$\dfrac{\Omega \Vdash^{sig} \mathcal{S} \leadsto T : \eta \quad \Omega, T : \eta; \Gamma \Vdash^{pgm} pgm : \mu}{\Omega; \Gamma \Vdash^{pgm} \mathbf{sig}\, \mathcal{S}; pgm : \mu}$$

  The goal holds directly from soundness of typing signature (Lemma F.58) and I.H..

- Case
  A-PGM-DT-TT

$$\cfrac{\Theta_1 = \Omega, \overline{\widehat{\alpha_i} : \star}^i, \overline{T_i : \widehat{\alpha_i}}^i \quad \overline{\Theta_i \Vdash^{\mathsf{dt}} \mathcal{T}_i \rightsquigarrow \Gamma_i \dashv \Theta_{i+1}}^i \qquad \overline{\widehat{\phi_i^c} = \mathsf{unsolved}([\Theta_{n+1}]\widehat{\alpha_i})}^i \quad \overline{\Theta_{n+1} \Vdash^{\mathsf{gen}}_{\phi_i^c} ([\Theta_{n+1}](\Gamma_i[\overline{\widehat{\phi_i^c} \mapsto \phi_i^c}^i])) \rightsquigarrow \Gamma_i'}^i \qquad \overline{\Omega, T_i : \forall\{\overline{\phi_i^c}\}.(([\Theta_{n+1}]\widehat{\alpha_i})[\overline{\widehat{\phi_i^c} \mapsto \phi_i^c}^i])}^i; \Gamma, \overline{\Gamma_i'[\overline{T_i \mapsto T_i \,@\phi_i^c}^i]}^i \Vdash^{\mathsf{pgm}} pgm : \mu}{\Omega; \Gamma \Vdash^{\mathsf{pgm}} \mathbf{rec}\, \overline{\mathcal{T}_i}^{i\in 1..n}; pgm : \mu}$$

The key is to prove that $\phi_i^c$ corresponds to the $\phi_i^c$ in rule PGM-DT-TT. The reasoning is similar to the one in Lemma F.58.

The key observation here is that, in typing datatype decl (rule A-DT-TT), the result context does not have new unification variables at the end. Therefore, all unsolved unification variable in $\Theta_{n+1}$ is in one of the free kind variable in $[\Theta_{n+1}]\widehat{\alpha_i}$. Once we have all the $\phi_i^c$, the rest of preconditions follow straightforwardly. □

### F.2.7 Principality.

**Lemma F.62** (Completeness of Promotion). *Given $\Delta$ ok, and $\Delta \longrightarrow \Omega$, and $\widehat{\alpha} \in \Delta$, and $\Delta \Vdash^{\mathsf{ela}} \rho : \omega$, and $[\Delta]\widehat{\alpha} = \widehat{\alpha}$, and $[\Delta]\rho = \rho$, if $\rho$ does not depend on $\widehat{\alpha}$ in the dependency graph, then there exists $\rho_2$, $\Theta$ and $\Omega'$ such that $\Theta \longrightarrow \Omega'$, and $\Omega \longrightarrow \Omega'$, and $\Delta \Vdash^{\mathsf{pr}}_{\widehat{\alpha}} \rho \rightsquigarrow \rho_2 \dashv \Theta$.*

Proof. By induction on the lexicographic order indicated in the proof of Theorem F.50.
The proof is essentially the same as Lemma D.45.
For case $\rho = \widehat{\beta}$, and the context $\Delta[\widehat{\alpha} : \omega][\widehat{\beta} : \rho_1]$, we have

| | |
|---|---|
| $\Delta[\widehat{\alpha} : \omega][\widehat{\beta} : \rho_1] \Vdash^{\mathsf{pr}}_{\widehat{\alpha}} [\Delta]\rho_1 \rightsquigarrow \rho_2 \dashv \Delta_1[\widehat{\alpha} : \omega][\widehat{\beta} : \rho_1]$ | I.H. |
| $\wedge \Delta_1[\widehat{\alpha} : \omega][\widehat{\beta} : \rho_1] \longrightarrow \Omega_1 \wedge \Omega \longrightarrow \Omega_1$ | Above |
| $\Delta[\widehat{\alpha} : \omega][\widehat{\beta} : \rho_1] \Vdash^{\mathsf{pr}}_{\widehat{\alpha}} \widehat{\beta} \rightsquigarrow \widehat{\beta_2} \dashv \Delta_1[\widehat{\beta_2} : \rho_2, \widehat{\alpha} : \omega][\widehat{\beta} : \rho_1 = \widehat{\beta_2}]$ | By rule A-PR-KUVARR-TT |
| $\Delta_1[\widehat{\alpha} : \omega][\widehat{\beta} : \rho_1] = \Delta_{11}, \widehat{\alpha} : \omega, \Delta_{12}, \widehat{\beta} : \rho_1, \Delta_{13}$ | Let |
| $\Delta_{11} \Vdash^{\mathsf{ela}} \rho_2 : \star$ | By Lemma F.8 |
| $\Omega_1 = \Omega_{11}, \widehat{\alpha} : \omega = \rho_3, \Omega_{12}, \widehat{\beta} : \rho_1 = \rho_4, \Omega_{13} \wedge \Delta_{11} \longrightarrow \Omega_{11}$ | By Lemma F.29 |
| $[\Delta_1]\rho_2 = [\Delta_1]([\Delta]\rho_1)$ | By Lemma F.53 |
| $\quad = [\Delta_1]\rho_1$ | By Lemma F.30 |
| $[\Omega_1]\rho_2 = [\Omega_1]\rho_1$ | By Lemma F.30 |
| $\Omega_1 \Vdash^{\mathsf{ela}} \rho_4 : [\Omega_1]\rho_1$ | By inversion and weakening |
| $\Omega_1 \Vdash^{\mathsf{ela}} [\Omega_1]\rho_4 : [\Omega_1]\rho_1$ | By Lemma F.24 |
| $\Omega_1 \Vdash^{\mathsf{ela}} [\Omega_1]\rho_4 : [\Omega_1]\rho_2$ | By equation |
| $\Omega_{11} \Vdash^{\mathsf{ela}} [\Omega_1]\rho_4 : [\Omega_1]\rho_2$ | By strengthening |
| $\Omega_{11} \Vdash^{\mathsf{ela}} \rho_2 : \star$ | By Lemma F.22 |
| $\Omega_{11} \Vdash^{\mathsf{ela}} [\Omega_1]\rho_4 : [\Omega_{11}]\rho_2$ | By Lemma F.31 |
| $\Delta_1[\widehat{\alpha} : \omega][\widehat{\beta} : \rho_1] \longrightarrow \Omega_1[\widehat{\alpha} : \omega][\widehat{\beta} : \rho_1 = \rho_4]$ | Given |
| $\Delta_1[\widehat{\beta_2} : \rho_2, \widehat{\alpha} : \omega][\widehat{\beta} : \rho_1] \longrightarrow \Omega_1[\widehat{\beta_2} : \rho_2 = [\Omega_1]\rho_4, \widehat{\alpha} : \omega][\widehat{\beta} : \rho_1 = \rho_4]$ | By Lemma F.36 |
| $\Delta_1[\widehat{\beta_2} : \rho_2, \widehat{\alpha} : \omega][\widehat{\beta} : \rho_1 = \widehat{\beta_2}] \longrightarrow \Omega_1[\widehat{\beta_2} : \rho_2 = [\Omega_1]\rho_4, \widehat{\alpha} : \omega][\widehat{\beta} : \rho_1 = \rho_4]$ | By Lemma F.37 |
| $\Omega \longrightarrow \Omega_1[\widehat{\beta_2} : \rho_2 = [\Omega_1]\rho_4, \widehat{\alpha} : \omega][\widehat{\beta} : \rho_1 = \rho_4]$ | By Lemmas F.32 and F.34 |

□

**Lemma F.63** (Completeness of Unification). *Given $\Delta$ ok, and $\Delta \longrightarrow \Omega$, and $\Delta \Vdash^{\text{ela}} \rho_1 : \omega$ and $\Delta \Vdash^{\text{ela}} \rho_2 : \omega$, and $[\Delta]\rho_1 = \rho_1$ and $[\Delta]\rho_2 = \rho_2$, if $[\Omega]\rho_1 = [\Omega]\rho_2$, then there exists $\Theta$ and $\Omega'$ such that $\Theta \longrightarrow \Omega'$, and $\Omega \longrightarrow \Omega'$ and $\Delta \Vdash^{\mu} \rho_1 \approx \rho_2 \dashv \Theta$.*

Proof. By induction on the lexicographic order indicated in the proof of Theorem F.51. Then case analysis on $\rho_1$ and $\rho_2$.

The proof is essentially the same as Lemma D.46.

For case $\rho_1 = \widehat{\alpha}$, and $\rho_2$ does not depend on $\widehat{\alpha}$ in the dependency graph, we have

| | |
|---|---|
| $\Delta \Vdash^{\text{pr}}_{\widehat{\alpha}} \rho_2 \rightsquigarrow \rho_3 \dashv \Theta_1 \land \Theta_1 \longrightarrow \Omega_1 \land \Omega \longrightarrow \Omega_1$ | By Lemma F.62 |
| $\Theta_1 = \Theta_{11}, \widehat{\alpha} : \omega_3, \Theta_{12} \land \Theta_{11} \Vdash^{\text{ela}} \rho_3 : [\Theta_{11}]\omega$ | By Lemma F.8 |
| $[\Omega_1]\rho_2 = [\Omega_1]\rho_3$ | By Lemma F.53 |
| $\Omega_1 = \Omega_{11}, \widehat{\alpha} : \omega_3 = \rho_4, \Omega_{12} \land \Theta_{11} \longrightarrow \Omega_{11}$ | Lemma F.29 |
| $[\Omega]\widehat{\alpha} = [\Omega]\rho_2$ | Given |
| $[\Omega_1]\widehat{\alpha} = [\Omega_1]\rho_2$ | By Lemma F.30 |
| $[\Omega_1]\rho_4 = [\Omega_1]\rho_2$ | By definition |
| $[\Omega_1]\rho_4 = [\Omega_1]\rho_3$ | By equations |
| $\Delta \longrightarrow \Omega_1$ | By Lemma F.8 and Lemma F.32 |
| $\Delta \Vdash^{\text{ela}} \widehat{\alpha} : \omega$ | Given |
| $\Omega_1 \Vdash^{\text{ela}} \widehat{\alpha} : [\Omega_1]\omega$ | By Lemma F.22 |
| $[\Omega_1]\omega_3 = [\Omega_1]\omega$ | By inversion |
| $[\Omega_{11}]\omega_3 = [\Omega_{11}]\omega$ | By Lemma F.31 |
| $[\Omega_{11}]([\Theta_{11}]\omega_3) = [\Omega_{11}]([\Theta_{11}]\omega)$ | By Lemma F.30 |
| $\Theta_{11} \Vdash^{\mu} [\Theta_{11}]\omega_3 \approx [\Theta_{11}]\omega \dashv \Theta_2 \land \Theta_2 \longrightarrow \Omega_2 \land \Omega_{11} \longrightarrow \Omega_2$ | I.H. |
| $\Delta \Vdash^{\mu} \widehat{\alpha} \approx \rho_2 \dashv \Theta_2, \widehat{\alpha} : \omega_3 = \rho_3, \Theta_{12}$ | By rule A-U-KVARL-TT |
| $\Theta_2, \widehat{\alpha} : \omega_3 = \rho_3, \Theta_{12} \longrightarrow \Omega_2, \widehat{\alpha} : \omega_3 = \rho_4, \Omega_{12}$ | By Lemma F.36, Lemma F.38, Lemma F.37 |
| $\Omega \longrightarrow \Omega_2, \widehat{\alpha} : \omega_3 = \rho_4, \Omega_{12}$ | Similarly |

What if $\rho_2$ depends on $\widehat{\alpha}$? For that to be possible, according to the dependency graph, we have either $\widehat{\alpha} = \star$ or $\widehat{\alpha} = \rightarrow$. Then for the unification constrain to be solvable, we have either $\rho_2 = \star$, $\rho_2 = \rightarrow$, or $\rho_2 = \widehat{\beta}$. When $\rho_2 = \star$ or $\rho_2 = \rightarrow$, we know $\rho_2$ does not depend on $\widehat{\alpha}$ at all. When $\rho_2 = \widehat{\beta}$, because we know that the context is well-formed, if $\widehat{\beta}$ depends on $\widehat{\alpha}$, we must have $\widehat{\alpha}$ not depending on $\widehat{\beta}$. So we can solve the case using rule A-U-KVARR-TT.

The case when the variable in a local scope is similar. □

**Lemma F.64** (Completeness of Instantiation). *Given $\Delta \longrightarrow \Omega$, and $\Delta \Vdash^{\text{ela}} \rho : \eta$ and $\Delta \Vdash^{\text{ela}} \omega : \star$, and $[\Delta]\eta = \eta$ and $[\Delta]\omega = \omega$, if $[\Omega]\Delta \vdash^{\text{inst}} [\Omega]\rho_1 : [\Omega]\eta \sqsubseteq [\Omega]\omega \rightsquigarrow \rho_2$, then there exists $\rho'_2$, $\Theta$ and $\Omega'$ such that $\Theta \longrightarrow \Omega'$, and $\Omega \longrightarrow \Omega'$ and $\Delta \Vdash^{\text{inst}} \rho_1 : \eta \sqsubseteq \omega \rightsquigarrow \rho'_2 \dashv \Theta$, and $[\Omega']\rho'_2 = \rho_2$.*

Proof. By induction on the declarative instantiation.

- Case

$$\frac{}{\Sigma \vdash^{\text{inst}} \mu : \omega \sqsubseteq \omega \rightsquigarrow \mu} \text{ INST-REFL}$$

Follows directly from rule A-INST-REFL and Lemma F.63.

- Case

$$\frac{\Sigma \vdash^{\text{ela}} \rho : \omega_1 \quad \Sigma \vdash^{\text{inst}} \mu_1 @\rho : \eta[a \mapsto \rho] \sqsubseteq \omega_2 \rightsquigarrow \mu_2}{\Sigma \vdash^{\text{inst}} \mu_1 : \forall a : \omega_1.\eta \sqsubseteq \omega_2 \rightsquigarrow \mu_2} \text{ INST-FORALL}$$

We case analyze $\eta$, and it can only be of the shape $\forall a : \omega_2.\eta_2$, and $[\Omega]\omega_2 = \omega_1$ and $[\Omega]\eta_2 = \eta$.
From hypothesis we get $[\Omega]\Delta \vdash^{\text{inst}} ([\Omega]\mu_1 @[\Omega]\rho) : [\Omega]\mu[a \mapsto [\Omega]\rho] \sqsubseteq [\Omega]\omega_2 \leadsto [\Omega]\mu_2$.
By substitution, $[\Omega]\Delta \vdash^{\text{inst}} [\Omega](\mu_1 @\rho) : [\Omega](\mu[a \mapsto \rho]) \sqsubseteq [\Omega]\omega_2 \leadsto [\Omega]\mu_2$.
By definition, $[\Omega, \widehat{\alpha} : \omega_1 = \rho]\Delta \vdash^{\text{inst}} [\Omega, \widehat{\alpha} : \omega_1 = \rho](\mu_1 @\widehat{\alpha}) : [\Omega, \widehat{\alpha} : \omega_1 = \rho](\mu[a \mapsto \widehat{\alpha}]) \sqsubseteq [\Omega, \widehat{\alpha} : \omega_1 = \rho]\omega_2 \leadsto [\Omega, \widehat{\alpha} : \omega_1 = \rho]\mu_2$.
The goal follows directly from I.H., and rule A-INST-FORALL.

- The case for rule INST-FORALL-INFER is similar to the previous case.

□

**Lemma F.65** (Principality of Kinding).

- *Given $\Delta \longrightarrow \Omega$, if $[\Omega]\Delta \vdash^{\text{k}} \sigma : \eta \leadsto \mu$, and $\Delta \Vdash^{\text{k}} \sigma : \eta' \leadsto \mu' \dashv \Theta$, then there exists $\Omega'$ such that $\Theta \longrightarrow \Omega'$, and $\Omega \longrightarrow \Omega'$. Moreover, $[\Omega']\eta' = \eta$. Furthermore, if $\mu$ and $\mu'$ are monotypes, then $[\Omega']\mu' = \mu$.*
- *Given $\Delta \longrightarrow \Omega$, if $[\Omega]\Delta \vdash^{\text{kc}} \sigma \Leftarrow [\Omega]\eta \leadsto \mu$, and $\Delta \Vdash^{\text{kc}} \sigma \Leftarrow \eta \leadsto \mu' \dashv \Theta$, then there exists $\Omega'$ such that $\Theta \longrightarrow \Omega'$, and $\Omega \longrightarrow \Omega'$. Furthermore, if $\mu$ and $\mu'$ are monotypes, then $[\Omega']\mu' = \mu$.*
- *Given $\Delta \longrightarrow \Omega$, if $[\Omega]\Delta \vdash^{\text{inst}} [\Omega]\rho_1 : [\Omega]\eta \sqsubseteq (\omega_1 \to \omega_2) \leadsto \rho_3$, and $[\Omega]\Delta \vdash^{\text{kc}} \tau \Leftarrow \omega_1 \leadsto \rho_4$ and $\Delta \Vdash^{\text{kapp}} (\rho_1 : \eta) \bullet \tau : \omega \leadsto \rho_2 \dashv \Theta$, then there exists $\Omega'$ such that $\Theta \longrightarrow \Omega'$, and $\Omega \longrightarrow \Omega'$. Moreover, $[\Omega']\omega = \omega_2$. Further, $[\Omega']\rho_2 = \rho_3\ \rho_4$.*

Proof. From this lemma, we make use of *Any* to ensure every algorithmic context can be extended to a complete context. The existence of *Any* does not affect at all how this lemma is used.

By induction on the algorithmic kinding.

**Part 1**
- The case for rules A-KTT-STAR, A-KTT-NAT, A-KTT-VAR, A-KTT-TCON, and A-KTT-ARROW follows trivially by picking $\Theta = \Delta$, and $\Omega' = \Omega$.
- Case
A-KTT-FORALL
$$\frac{\Delta \Vdash^{\text{kc}} \kappa \Leftarrow \star \leadsto \omega \dashv \Delta_1 \quad \Delta_1, a : \omega \Vdash^{\text{kc}} \sigma \Leftarrow \star \leadsto \mu \dashv \Delta_2, a : \omega, \Delta_3 \quad \Delta_3 \hookrightarrow a}{\Delta \Vdash^{\text{k}} \forall a : \kappa.\sigma : \star \leadsto \forall a : \omega.[\Delta_3]\mu \dashv \Delta_2, \text{unsolved}(\Delta_3)}$$

| | |
|---|---|
| $[\Omega]\Delta \vdash^{\text{k}} \forall a : \kappa.\sigma : \star \leadsto \forall a : \omega_1.\mu_1$ | Given |
| $[\Omega]\Delta \vdash^{\text{kc}} \kappa \Leftarrow \star \leadsto \omega_1 \wedge [\Omega]\Delta, a : \omega_1 \vdash^{\text{kc}} \sigma \Leftarrow \star \leadsto \mu_1$ | By inversion |
| $\Delta_1 \longrightarrow \Omega_1 \wedge \Omega \longrightarrow \Omega_1 \wedge [\Omega_1]\omega = \omega_1$ | I.H. |
| $[\Omega_1, a : \omega](\Delta_1, a : \omega)$ | |
| $= [\Omega_1]\Delta_1, a : \omega_1$ | By definition |
| is a well-formed permutation of $[\Omega]\Delta$ | By Lemma F.45 and Lemma F.46 |
| $[\Omega_1, a : \omega](\Delta_1, a : \omega) \vdash^{\text{kc}} \sigma \Leftarrow \star \leadsto \mu_1$ | Follows |
| $\Delta_2, a : \omega, \Delta_3 \longrightarrow \Omega_2 \wedge \Omega_1, a : \omega \longrightarrow \Omega_2$ | I.H. |
| $\Delta_1, a : \omega \longrightarrow \Delta_2, a : \omega, \Delta_3$ | By Lemma F.15 |
| $\Delta_1 \longrightarrow \Delta_2 \wedge \Delta_3$ **soft** | By Lemma F.29 |
| $\Omega_2 = \Omega_{21}, a : \omega, \Omega_{22} \wedge \Delta_2 \longrightarrow \Omega_{21} \wedge \Omega_1 \longrightarrow \Omega_{21} \wedge \Omega_{22}$ **soft** | By Lemma F.29 |
| construct $\Omega_{23}$ which contain same domain of unsolved($\Delta_3$) | |
| $\Omega' = \Omega_{21}, \Omega_{23}$ | Let |
| $\Delta_2, \text{unsolved}(\Delta_3) \longrightarrow \Omega'$ | By rule A-CTXE-SOLVE-TT |
| $\Omega_1 \longrightarrow \Omega'$ | By rule A-CTXE-ADDSOLVED-TT |

- The case for rule A-KTT-FORALLI is similar as the previous case.

- Case
  
  A-KTT-APP
  $$\frac{\Delta \Vdash^k \tau_1 : \eta_1 \leadsto \rho_1 \dashv \Delta_1 \qquad \Delta_1 \Vdash^{kapp} (\rho_1 : [\Delta_1]\eta_1) \bullet \tau_2 : \omega \leadsto \rho \dashv \Theta}{\Delta \Vdash^k \tau_1 \tau_2 : \omega \leadsto \rho \dashv \Theta}$$

| | |
|---|---|
| $[\Omega]\Delta \vdash^k \tau_1 \tau_2 : \omega_2 \leadsto \rho_2 \rho_3$ | Given |
| $[\Omega]\Delta \vdash^k \tau_1 : \eta_2 \leadsto \rho_4 \land [\Omega]\Delta \vdash^{inst} \rho_4 : \eta_2 \sqsubseteq \omega_1 \to \omega_2 \leadsto \rho_2$ | |
| $\land [\Omega]\Delta \vdash^{kc} \tau_2 \Leftarrow \omega_1 \leadsto \rho_3$ | By inversion |
| $\Delta_1 \longrightarrow \Omega_1 \land \Omega \longrightarrow \Omega_1 \land [\Omega_1]\rho_1 = \rho_4 \land [\Omega_1]\eta_1 = \eta_2$ | I.H. |
| $[\Omega_1]\Delta$ is a well-formed permutation of $[\Omega]\Delta$ | By Lemma F.43 and Lemma F.46 |
| $[\Omega_1]\Delta \vdash^{inst} [\Omega_1]\rho_1 : [\Omega]\eta_1 \sqsubseteq \omega_1 \to \omega_2 \leadsto \rho_2$ | By equations |
| $[\Omega_1]\Delta \vdash^{kc} \tau_2 \Leftarrow \omega_1 \leadsto \rho_3$ | By equations |
| $\Theta \longrightarrow \Omega' \land \Omega_1 \longrightarrow \Omega' \land [\Omega']\rho = \rho_2 \rho_3 \land [\Omega']\omega = \omega_2$ | By Part 3 |
| $\Omega \longrightarrow \Omega'$ | By Lemma F.32 |

- Case
  
  A-KTT-KAPP
  $$\frac{\Delta \Vdash^k \tau_1 : \eta \leadsto \rho_1 \dashv \Delta_1 \qquad [\Delta_1]\eta = \forall a : \omega.\eta_2 \qquad \Delta_1 \Vdash^{kc} \tau_2 \Leftarrow \omega \leadsto \rho_2 \dashv \Delta_2}{\Delta \Vdash^k \tau_1 @\tau_2 : \eta_2[a \mapsto \rho_2] \leadsto \rho_1 @\rho_2 \dashv \Delta_2}$$

| | |
|---|---|
| $[\Omega]\Delta \vdash^k \tau_1 @\tau_2 : \mu_3[a \mapsto \rho_3] \leadsto \rho_4 @\rho_3$ | Given |
| $[\Omega]\Delta \vdash^k \tau_1 : \forall a : \omega_2.\mu_3 \leadsto \rho_4 \land [\Omega]\Delta \vdash^{kc} \tau_2 \Leftarrow \omega_2 \leadsto \rho_3$ | By inversion |
| $\Delta_1 \longrightarrow \Omega_1 \land \Omega \longrightarrow \Omega_1 \land [\Omega_1]\eta = \forall a : \omega_2.\mu_3 \land [\Omega_1]\rho_1 = \rho_4$ | I.H. |
| $[\Omega_1]\Delta$ is a well-formed permutation of $[\Omega]\Delta$ | By Lemma F.43 and Lemma F.46 |
| $[\Omega_1]\eta = [\Omega_1]([\Delta_1]\eta)$ | By Lemma F.30 |
| $= \forall a : [\Omega_1]\omega.[\Omega_1]\eta_2$ | By definition |
| $= [\Omega_1]\eta = \forall a : \omega_2.\mu_3$ | Given |
| $[\Omega_1]\omega = \omega_2$ | Follows |
| $[\Omega_1]\eta_2 = \mu_3$ | Follows |
| $[\Omega_1]\Delta \vdash^{kc} \tau_2 \Leftarrow [\Omega_1]\omega \leadsto \rho_3$ | By equations |
| $\Delta_2 \longrightarrow \Omega' \land \Omega_1 \longrightarrow \Omega' \land [\Omega']\rho_2 = \rho_3$ | By Part 2 |
| $\Omega \longrightarrow \Omega'$ | By Lemma F.32 |
| $[\Omega'](\eta_2[a \mapsto \rho_2])$ | |
| $= ([\Omega']\eta_2)[a \mapsto [\Omega']\rho_2]$ | By substitution |
| $= (\mu_3)[a \mapsto \rho_3]$ | By Lemma F.41 |

- The case for rule A-KTT-KAPP-INFER is similar as the previous case.

**Part 2** We have

A-KC-SUB
$$\frac{\Delta \Vdash^k \sigma : \eta \leadsto \mu_1 \dashv \Delta_1 \qquad \Delta_1 \Vdash^{inst} \mu_1 : [\Delta_1]\eta \sqsubseteq [\Delta_1]\omega \leadsto \mu_2 \dashv \Delta_2}{\Delta \Vdash^{kc} \sigma \Leftarrow \omega \leadsto \mu_2 \dashv \Delta_2}$$

| | |
|---|---|
| $[\Omega]\Delta \vdash^{kc} \sigma \Leftarrow [\Omega]\omega \leadsto \mu_4$ | Given |
| $[\Omega]\Delta \vdash^k \sigma : \eta_3 \leadsto \mu_3$ | By inversion |
| $[\Omega]\Delta \vdash^{inst} \mu_3 : \eta_3 \sqsubseteq [\Omega]\omega \leadsto \mu_4$ | By inversion |
| $\Delta_1 \longrightarrow \Omega_1 \land \Omega \longrightarrow \Omega_1 \land [\Omega_1]\eta = \eta_3$ | I.H. |
| If $\mu_1$ and $\mu_3$ are monotypes, then $[\Omega_1]\mu_1 = \mu_3$ | I.H. |
| $[\Omega_1]\Delta$ is a well-formed permutation of $[\Omega]\Delta$ | By Lemma F.43 and Lemma F.46 |
| $[\Omega_1]\omega = [\Omega]\omega$ | By Lemma F.41 |
| $[\Omega_1]\Delta_1 \vdash^{\text{inst}} \mu_3 : [\Omega_1]\eta \sqsubseteq [\Omega_1]\omega \rightsquigarrow \mu_4$ | Follows |
| If $\mu_1$ and $\mu_3$ are monotypes | |
| $[\Omega_1]\Delta_1 \vdash^{\text{inst}} [\Omega_1]\mu_1 : [\Omega_1]\eta \sqsubseteq [\Omega_1]\omega \rightsquigarrow \mu_4$ | Follows |
| $[\Omega_1]\Delta_1 \vdash^{\text{inst}} [\Omega_1]\mu_1 : [\Omega_1]([\Delta_1]\eta) \sqsubseteq [\Omega_1]([\Delta_1]\omega) \rightsquigarrow \mu_4$ | By Lemma F.30 |
| $\Omega_1 \longrightarrow \Omega' \land \Delta_2 \longrightarrow \Omega' \land [\Omega']\mu_2 = \mu_4$ | By Lemma F.64 |
| If $\mu_1$ and $\mu_3$ are polytypes | |
| then only rule INST-REFL and rule A-INST-REFL can apply | |
| $\Delta_2 = \Delta_1$ | Follows |
| $\Omega' = \Omega_1$ | Let |

**Part 3** • Case

$$\frac{\Delta \Vdash^{\text{kc}} \tau \Leftarrow \omega_1 \rightsquigarrow \rho_2 \dashv \Theta}{\Delta \Vdash^{\text{kapp}} (\rho_1 : \omega_1 \rightarrow \omega_2) \bullet \tau : \omega_2 \rightsquigarrow \rho_1\,\rho_2 \dashv \Theta} \text{ A-KAPP-TT-ARROW}$$

| | |
|---|---|
| $[\Omega]\Delta \vdash^{\text{inst}} [\Omega]\rho_1 : [\Omega]\omega_1 \rightarrow [\Omega]\omega_2 \sqsubseteq [\Omega]\omega_1 \rightarrow [\Omega]\omega_2 \rightsquigarrow [\Omega]\rho_1$ | Given |
| $[\Omega]\Delta \vdash^{\text{kc}} \tau \Leftarrow [\Omega]\omega_1 \rightsquigarrow \rho_4$ | Given |
| The goal follows directly from Part 2 | |

• Case

$$\frac{\Delta, \widehat{\alpha} : \omega_1 \Vdash^{\text{kapp}} (\rho_1\,@\widehat{\alpha} : \eta[a \mapsto \widehat{\alpha}]) \bullet \tau : \omega \rightsquigarrow \rho \dashv \Theta}{\Delta \Vdash^{\text{kapp}} (\rho_1 : \forall a : \omega_1.\eta) \bullet \tau : \omega \rightsquigarrow \rho \dashv \Theta} \text{ A-KAPP-TT-FORALL}$$

| | |
|---|---|
| $[\Omega]\Delta \vdash^{\text{inst}} [\Omega]\rho_1 : \forall a : [\Omega]\omega_1.[\Omega]\eta \sqsubseteq \omega_3 \rightarrow \omega_4 \rightsquigarrow \rho_3$ | Given |
| $[\Omega]\Delta \vdash^{\text{kc}} \tau \Leftarrow \omega_3 \rightsquigarrow \rho_4$ | Given |
| $[\Omega]\Delta \vdash^{\text{ela}} \rho_5 : [\Omega]\omega_1$ | By inversion |
| $[\Omega]\Delta \vdash^{\text{inst}} [\Omega]\rho_1 \,@\rho_5 : ([\Omega]\eta)[a \mapsto \rho_5] \sqsubseteq \omega_3 \rightarrow \omega_4 \rightsquigarrow \rho_3$ | By inversion |
| $[\Omega, \widehat{\alpha} : \omega_1 = \rho_5](\Delta, \widehat{\alpha} : \omega_1) \vdash^{\text{inst}}$ | |
| $\quad [\Omega, \widehat{\alpha} : \omega_1 = \rho_5](\rho_1\,@\widehat{\alpha}) : [\Omega, \widehat{\alpha} : \omega_1 = \rho_5](\eta[a \mapsto \widehat{\alpha}]) \sqsubseteq \omega_3 \rightarrow \omega_4 \rightsquigarrow \rho_3$ | By definition |
| The goal follows from I.H. | |

- The case for rule A-KAPP-TT-FORALL-INFER is similar as the previous case.
- Case

$$\frac{\Delta_1, \widehat{\alpha}_1 : \star, \widehat{\alpha}_2 : \star, \widehat{\alpha} : \omega = (\widehat{\alpha}_1 \rightarrow \widehat{\alpha}_2), \Delta_2 \Vdash^{\text{kc}} \tau \Leftarrow \widehat{\alpha}_1 \rightsquigarrow \rho_2 \dashv \Theta}{\Delta_1, \widehat{\alpha} : \omega, \Delta_2 \Vdash^{\text{kapp}} (\rho_1 : \widehat{\alpha}) \bullet \tau : \widehat{\alpha}_2 \rightsquigarrow \rho_1\,\rho_2 \dashv \Theta} \text{ A-KAPP-TT-KUVAR}$$

As $\widehat{\alpha}$ can only be instantiated with monotypes, obviously the declarative instantiation judgment must be rule INST-REFL. Then the goal follows directly from Part 2.

□

**Lemma F.66** (Principality of Typing Data Constructor Declaration). *Given $\Delta \longrightarrow \Omega$, if $[\Omega]\Delta \vdash^{\text{dc}}_\rho \mathcal{D} \rightsquigarrow \mu_1$, and $\Delta \Vdash^{\text{dc}}_\rho \mathcal{D} \rightsquigarrow \mu_2 \dashv \Theta$, then there exists $\Omega'$ such that $\Theta \longrightarrow \Omega'$, and $\Omega \longrightarrow \Omega'$.*

PROOF. We have

$$\frac{\Delta, \blacktriangleright_D \Vdash^{\text{k}} \forall \phi.(\overline{\tau_i}^i \rightarrow \rho) : \star \rightsquigarrow \mu \dashv \Theta_1, \blacktriangleright_D, \Theta_2 \qquad \widehat{\phi^c} = \text{unsolved}(\Theta_2)}{\Delta \Vdash^{\text{dc}}_\rho \forall \phi.D\,\overline{\tau_i}^i \rightsquigarrow \forall\{\phi^c\}.(([\Theta_2]\mu)[\widehat{\phi^c} \mapsto \phi^c]) \dashv \Theta_1} \text{ A-DC-TT}$$

| | |
|---|---|
| $[\Omega]\Delta \vdash^{dc}_\rho \mathcal{D} \leadsto \mu_1$ | Given |
| $[\Omega]\Delta, \phi^c \vdash^k \forall\phi.\overline{\tau_i}^i \to \rho : \star \leadsto \mu_1$ | By inversion |
| $\Delta, \blacktriangleright_D \Vdash^k \forall\phi.(\overline{\tau_i}^i \to \rho) : \star \leadsto \mu \dashv \Theta_1, \blacktriangleright_D, \Theta_2$ | Given |
| $\Delta, \blacktriangleright_D, \phi^c \Vdash^k \forall\phi.(\overline{\tau_i}^i \to \rho) : \star \leadsto \mu \dashv \Theta_1, \blacktriangleright_D, \phi^c, \Theta_2$ | By weakening |
| $\Delta \longrightarrow \Omega$ | Given |
| $\Delta, \blacktriangleright_D, \phi^c \longrightarrow \Omega, \blacktriangleright_D, \phi^c$ | By definition |
| $[\Omega, \blacktriangleright_D, \phi^c](\Delta, \blacktriangleright_D, \phi^c) \vdash^k \forall\phi.\overline{\tau_i}^i \to \rho : \star \leadsto \mu_1$ | By definition |
| $\Theta_1, \blacktriangleright_D, \phi^c, \Theta_2 \longrightarrow \Omega_1 \wedge \Omega, \blacktriangleright_D, \phi^c \longrightarrow \Omega_1$ | By Lemma F.65 |
| $\Omega_1 = \Omega', \blacktriangleright_D, \Omega_{12} \wedge \Theta_1 \longrightarrow \Omega_1 \wedge \Theta_1 \longrightarrow \Omega' \wedge \Omega \longrightarrow \Omega'$ | By Lemma F.29 |

□

**Lemma F.67** (Principality of Typing Datatype Declaration). *Given $\Delta \longrightarrow \Omega$, if $[\Omega]\Delta \vdash^{dt} \mathcal{T} \leadsto \Psi$, and $\Delta \Vdash^{dt} \mathcal{T} \leadsto \Gamma \dashv \Theta$, then there exists $\Omega'$ such that $\Theta \longrightarrow \Omega'$, and $\Omega \longrightarrow \Omega'$.*

Proof. We have

A-DT-TT

$$\frac{(T : \forall\{\phi_1^c\}.\forall\phi_2^c.\,\omega) \in \Delta \quad \Delta, \phi_1^c, \phi_2^c, \overline{\widehat{\alpha_i} : \star}^i \Vdash^\mu [\Delta]\omega \approx (\overline{\widehat{\alpha_i}}^i \to \star) \dashv \Theta_1, \phi_1^c, \phi_2^c, \overline{\widehat{\alpha_i} : \star = \omega_i}^i \quad \overline{\Theta_j, \phi_1^c, \phi_2^c, \overline{a_i : \omega_i}^i \Vdash^{dc}_{(T\,@\phi_1^c\,@\phi_2^c\,\overline{a_i}^i)} \mathcal{D}_j \leadsto \mu_j \dashv \Theta_{j+1}, \phi_1^c, \phi_2^c, \overline{a_i : \omega_i}^i}^j}{\Delta \Vdash^{dt} \mathbf{data}\ T\,\overline{a_i}^i = \overline{\mathcal{D}_j}^{j\in 1..n} \leadsto \overline{D_j : \forall\{\phi_1^c\}.\forall\phi_2^c.\,\forall\overline{a_i : \omega_i}^i.\mu_j}^j \dashv \Theta_{n+1}}$$

| | |
|---|---|
| $[\Omega]\Delta \vdash^{dc}_\rho \mathcal{D} \leadsto \mu_1$ | Given |
| $(T : (\forall\{\phi_1^c\}.\forall\phi_2^c.\,[\Omega]\omega)) \in [\Omega]\Delta$ | Inversion |
| $[\Omega]\omega = \overline{\omega'_i}^i \to \star$ | Inversion |
| $[\Omega]\Delta, \phi_1^c, \phi_2^c, \overline{a_i : \omega'_i}^i \vdash^{dc}_{(T\,@\phi_1^c\,@\phi_2^c\,\overline{a_i}^i)} \mathcal{D}_j \leadsto \mu_j$ | Inversion |
| $\Delta \longrightarrow \Omega$ | Given |
| $\Delta, \phi_1^c, \phi_2^c, \overline{\widehat{\alpha_i} : \star}^i \longrightarrow \Omega, \phi_1^c, \phi_2^c, \overline{\widehat{\alpha_i} : \star = \omega'_i}^i$ | By definition |
| $[\Omega, \phi_1^c, \phi_2^c, \overline{\widehat{\alpha_i} : \star = \omega'_i}^i]\omega = [\Omega, \phi_1^c, \phi_2^c, \overline{\widehat{\alpha_i} : \star = \omega'_i}^i](\overline{\widehat{\alpha_i}}^i \to \star)$ | By substitution |
| $\Omega, \phi_1^c, \phi_2^c, \overline{\widehat{\alpha_i} : \star = \omega'_i}^i \longrightarrow \Omega_1$ | By Lemma F.63 |
| $\Theta_1, \phi_1^c, \phi_2^c, \overline{\widehat{\alpha_i} : \star = \omega_i}^i \longrightarrow \Omega_1$ | By Lemma F.63 |
| $\Omega_1 = \Omega_2, \Omega_3 \wedge \Omega \longrightarrow \Omega_2 \wedge \Theta_1 \longrightarrow \Omega_2$ | By Lemma F.29 |
| $\Theta_1, \phi_1^c, \phi_2^c, \overline{a_i : \omega_i}^i \longrightarrow \Omega_2, \phi_1^c, \phi_2^c, \overline{a_i : \omega_i}^i$ | By definition |
| $[\Omega_2, \phi_1^c, \phi_2^c, \overline{a_i : \omega_i}^i](\Theta_1, \phi_1^c, \phi_2^c, \overline{a_i : \omega_i}^i)$ | |
| $= [\Omega_2]\Theta_1, \phi_1^c, \phi_2^c, \overline{a_i : [\Omega_2]\omega_i}^i$ | By definition |
| $= [\Omega_2]\Theta_1, \phi_1^c, \phi_2^c, \overline{a_i : \omega'_i}^i$ | By Lemma F.30 |
| is a well-formed permutation of $[\Omega]\Delta, \phi_1^c, \phi_2^c, \overline{a_i : \omega'_i}^i$ | By Lemma F.46 |
| $[\Omega]\Delta, \phi_1^c, \phi_2^c, \overline{a_i : \omega'_i}^i \vdash^{dc}_{(T\,@\phi_1^c\,@\phi_2^c\,\overline{a_i}^i)} \mathcal{D}_j \leadsto \mu_j$ | Given |
| $[\Omega_2, \phi_1^c, \phi_2^c, \overline{a_i : \omega_i}^i](\Theta_1, \phi_1^c, \phi_2^c, \overline{a_i : \omega_i}^i) \vdash^{dc}_{(T\,@\phi_1^c\,@\phi_2^c\,\overline{a_i}^i)} \mathcal{D}_j \leadsto \mu_j$ | Follows |
| $(\Theta_2, \phi_1^c, \phi_2^c, \overline{a_i : \omega_i}^i) \longrightarrow \Omega_3$ | By Lemma F.66 |
| $(\Omega_2, \phi_1^c, \phi_2^c, \overline{a_i : \omega_i}^i) \longrightarrow \Omega_3$ | By Lemma F.66 |

Repeating the process for each $j$, we can finally get $(\Theta_{n+1}, \phi_1^c, \phi_2^c, \overline{a_i : \omega_i}^i) \longrightarrow \Omega''$, and $(\Omega_{n+1}, \phi_1^c, \phi_2^c, \overline{a_i : \omega_i}^i) \longrightarrow \Omega''$.

$$\Omega'' = \Omega', \Omega_0 \wedge \Theta_{n+1} \longrightarrow \Omega' \wedge \Omega_{n+1} \longrightarrow \Omega' \quad \text{By Lemma F.29}$$
$$\Omega \longrightarrow \Omega' \quad \text{By Lemma F.32}$$

□

**Theorem F.68** (Principality of Typing a Datatype Declaration Group). *If* $\Omega \Vdash^{\text{grp}} \textbf{rec}\ \overline{\mathcal{T}_i}^i \rightsquigarrow \overline{\eta_i}^i; \overline{\Gamma_i}^i$, *then whenever* $[\Omega]\Omega \Vdash^{\text{grp}} \textbf{rec}\ \overline{\mathcal{T}_i}^i \rightsquigarrow \overline{\eta_i'}^i; \overline{\Psi_i}^i$ *holds, we have* $[\Omega]\Omega \vdash [\Omega]\eta_i \leq \eta_i'$.

Proof. Given

PGM-DT-TT
$$\frac{\overline{\Sigma, \phi_i^c \Vdash^{\text{ela}} \omega_i : \star}^i \quad \overline{\phi_i^c \in Q(\omega_i)}^i \quad \overline{\Sigma, \cup\overline{\phi_i^c}^i, \overline{T_i : \omega_i}^i \Vdash^{\text{dt}} \mathcal{T}_i \rightsquigarrow \Psi_i}^i}{\overline{\Sigma, \cup\overline{\phi_i^c}^i, \overline{T_i : \omega_i}^i \Vdash^{\text{gen}}_{\phi_i^c} \Psi_i \rightsquigarrow \Psi_i'}^i \quad \Sigma, \overline{T_i : \forall\{\phi_i^c\}.\omega_i}^i; \Psi, \overline{\Psi_i'[\overline{T_i \mapsto T_i @\phi_i^c}^i]}^i \Vdash^{\text{pgm}} pgm : \sigma}{\Sigma; \Psi \Vdash^{\text{pgm}} \textbf{rec}\ \overline{\mathcal{T}_i}^i; pgm : \sigma}$$

A-PGM-DT-TT
$$\frac{\Theta_1 = \Omega, \overline{\widehat{\alpha_i} : \star}^i, \overline{T_i : \widehat{\alpha_i}}^i \quad \overline{\Theta_i \Vdash^{\text{dt}} \mathcal{T}_i \rightsquigarrow \Gamma_i \dashv \Theta_{i+1}}^i}{\overline{\widehat{\phi_i^c} = \text{unsolved}([\Theta_{n+1}]\widehat{\alpha_i})}^i \quad \overline{\Theta_{n+1} \Vdash^{\text{gen}}_{\phi_i^c} ([\Theta_{n+1}](\Gamma_i[\overline{\widehat{\phi_i^c} \mapsto \phi_i^c}^i])) \rightsquigarrow \Gamma_i'}^i}{\Omega, \overline{T_i : \forall\{\phi_i^c\}.(([\Theta_{n+1}]\widehat{\alpha_i})[\overline{\widehat{\phi_i^c} \mapsto \phi_i^c}^i])}^i; \Gamma, \overline{\Gamma_i'[\overline{T_i \mapsto T_i @\phi_i^c}^i]}^i \Vdash^{\text{pgm}} pgm : \mu}{\Omega; \Gamma \Vdash^{\text{pgm}} \textbf{rec}\ \overline{\mathcal{T}_i}^{i \in 1..n}; pgm : \mu}$$

Our goal is to prove that $[\Omega]\Delta \vdash \forall\{\phi_i^c\}.(([\Theta_{n+1}]\widehat{\alpha_i})[\overline{\widehat{\phi_i^c} \mapsto \phi_i^c}^i]) \leq \forall\{\phi_i^c\}.\omega_i$.

Similar as the proof in Lemma F.67, we can weaken the context $\Theta_i$ by adding $\cup\overline{\phi_i^c}^i$. By weakening we can get $\Theta_{n+1}'$, which is exactly the same as $\Theta_{n+1}$, except for the addition of $\cup\overline{\phi_i^c}^i$.

Let $\Omega_1$ be $\Omega, \overline{\cup\phi_i^c}^i, \overline{\widehat{\alpha_i} : \star = \omega_i}^i, \overline{T_i : \widehat{\alpha_i}}^i$.

According to the definition, our goal is equivalent to prove that for some $\Omega'$, we have $\Theta_{n+1}' \longrightarrow \Omega'$, and $[\Omega']\widehat{\alpha_i} = \omega_i$. According to Lemma F.67, we can prove there is indeed a $\Omega'$, such that $\Omega_1 \longrightarrow \Omega'$ and $\Theta_{n+1}' \longrightarrow \Omega'$. Moreover $[\Omega']\widehat{\alpha_i} = [\Omega_1]\widehat{\alpha_i} = \omega_i$ by Lemma F.41, so we are done.

□

Kind Inference for Datatypes: Technical Supplement

53:88  Ningning Xie, Richard A. Eisenberg, and Bruno C. d. S. Oliveira